\documentclass[12pt]{article}

\usepackage{smile}
\usepackage{graphicx,amsmath, subfiles}
\usepackage{epstopdf,epsfig,subfigure, bm}
\usepackage{geometry,authblk,upgreek}
\usepackage{enumitem,hyperref,nameref}
\hypersetup{hidelinks}
\usepackage{parskip}
\setenumerate[1]{itemsep=0pt,partopsep=0pt,parsep=0pt,topsep=0pt}
\setitemize[1]{itemsep=0pt,partopsep=0pt,parsep=0pt,topsep=0pt}
\usepackage[numbers,sort&compress,super]{natbib}
\usepackage{lineno}
\newcommand{\beginsupplement}{%
        \captionsetup[figure]{labelfont={bf},name={Supplementary Fig.},labelsep=colon}
    \captionsetup[table]{labelfont={bf},name={Supplementary Table},labelsep=colon}
        \setcounter{figure}{0}
        \setcounter{equation}{0}
        \def\theequation{S\arabic{equation}}
     }
\usepackage[section]{placeins}
\usepackage{caption}
\newcommand{\eq}[1]{Eq.\eqref{#1}}
\newcommand{\fig}[1]{Fig.~\ref{#1}}

\usepackage[dvipsnames]{xcolor}
\usepackage{xparse,xcoffins}
\ExplSyntaxOn
\NewCoffin\imagecoffin
\NewCoffin\labelcoffin

\keys_define:nn { miguel/label }
{
	label   .tl_set:N = \l_miguel_label_tl,
	labelbox .bool_set:N = \l_miguel_label_box_bool,
	labelbox .default:n = true,
	fontsize .tl_set:N = \l_miguel_label_size_tl,
	fontsize .initial:n = \normalsize,
	pos .choice:,
	pos/nw .code:n = \tl_set:Nn \l_miguel_label_pos_tl { left,up },
	pos/ne .code:n = \tl_set:Nn \l_miguel_label_pos_tl { right,up },
	pos/sw .code:n = \tl_set:Nn \l_miguel_label_pos_tl { left,down },
	pos/se .code:n = \tl_set:Nn \l_miguel_label_pos_tl { right,down },
	pos/n .code:n = \tl_set:Nn \l_miguel_label_pos_tl { hc,up },
	pos/w .code:n = \tl_set:Nn \l_miguel_label_pos_tl { left,vc },
	pos/s .code:n = \tl_set:Nn \l_miguel_label_pos_tl { hc,down },
	pos/e .code:n = \tl_set:Nn \l_miguel_label_pos_tl { right,vc },
	pos .initial:n = nw,
	unknown .code:n   = \clist_put_right:Nx \l_miguel_label_clist
	{ \l_keys_key_tl = \exp_not:n { #1 } }
}
\clist_new:N \l_miguel_label_clist
\box_new:N \l_miguel_label_box
\box_new:N \l_miguel_label_image_box

\NewDocumentCommand{\xincludegraphics}{O{}m}
{
	\group_begin:
	\tl_clear:N \l_miguel_label_tl
	\clist_clear:N \l_miguel_label_clist
	\keys_set:nn { miguel/label } { #1 }
	\tl_if_empty:NTF \l_miguel_label_tl
	{
		\miguel_includegraphics:Vn \l_miguel_label_clist { #2 }
	}
	{
		\SetHorizontalCoffin\imagecoffin
		{
			\miguel_includegraphics:Vn \l_miguel_label_clist { #2 }
		}
		\SetHorizontalCoffin\labelcoffin
		{
			\raisebox{\depth}
			{
				\bool_if:NTF \l_miguel_label_box_bool
				{ \fcolorbox{white}{white}{\l_miguel_label_size_tl\l_miguel_label_tl} }
				{ \l_miguel_label_size_tl\l_miguel_label_tl }
			}
		}
		\SetVerticalPole\imagecoffin{left}{3pt+\CoffinWidth\labelcoffin/2}
		\SetVerticalPole\imagecoffin{right}{\Width-3pt-\CoffinWidth\labelcoffin/2}
		\SetHorizontalPole\imagecoffin{up}{\Height-3pt-\CoffinHeight\labelcoffin/2}
		\SetHorizontalPole\imagecoffin{down}{3pt+\CoffinHeight\labelcoffin/2}
		\use:x{\JoinCoffins\imagecoffin[\l_miguel_label_pos_tl]\labelcoffin[vc,hc]} 
		\TypesetCoffin\imagecoffin
	}
	\group_end:
}
\NewDocumentCommand{\setlabel}{m}
{
	\keys_set:nn { miguel/label } { #1 }
}

\cs_new_protected:Nn \miguel_includegraphics:nn
{
	\includegraphics[#1]{#2}
}
\cs_generate_variant:Nn \miguel_includegraphics:nn { V }

\ExplSyntaxOff

\title{\Large Self-Directed Online Machine Learning for Topology Optimization}
\author[1]{Changyu Deng}
\author[2]{Yizhou Wang}
\author[2]{Can Qin}
\author[2]{Yun Fu}
\author[1,3,*]{Wei Lu}
\date{}

\affil[1]{\small\raggedright Department of Mechanical Engineering, University of Michigan, Ann Arbor, MI 48109, United States}
\affil[2]{Department of Electrical and Computer Engineering, Northeastern University, Boston, MA 02115, United States}
\affil[3]{Department of Materials Science and Engineering, University of Michigan, Ann Arbor, MI 48109, United States}
\affil[*]{Corresponding author: weilu@umich.edu}

\begin{document}
	\maketitle
\vspace*{-2em}This paper is published on Nature Communications (\url{https://rdcu.be/cFHgJ})

\begin{abstract}
Topology optimization by optimally distributing materials in a given domain requires non-gradient optimizers to solve highly complicated problems. However, with hundreds of design variables or more involved, solving such problems would require millions of Finite Element Method (FEM) calculations whose computational cost is huge and impractical. Here we report Self-directed Online Learning Optimization (SOLO) which integrates Deep Neural Network (DNN) with FEM calculations. A DNN learns and substitutes the objective as a function of design variables. A small number of training data is generated dynamically based on the DNN's prediction of the optimum. The DNN adapts to the new training data and gives better prediction in the region of interest until convergence. The optimum predicted by the DNN is proved to converge to the true global optimum through iterations. Our algorithm was tested by four types of problems including compliance minimization, fluid-structure optimization, heat transfer enhancement and truss optimization. It reduced the computational time by $2\sim5$ orders of magnitude compared with directly using heuristic methods,  and outperformed all state-of-the-art algorithms tested in our experiments. This approach enables solving large multi-dimensional optimization problems.
	\end{abstract}
\section*{Main}

Distributing materials in a domain to optimize performance is a significant topic in many fields, such as solid mechanics, heat transfer, acoustics, fluid mechanics, materials design and various multiphysics disciplines\cite{deaton2014}. Many numerical approaches\cite{bendse1988} have been developed since 1988, where the problems are formulated by density, level set, phase field, topological derivative or other methods\cite{rozvany2009}. Typically, these approaches use gradient-based optimizers, such as the Method of Moving Asymptotes (MMA), and thus have various restrictions on the properties of governing equations and optimization constraints to allow for fast computation of gradients. Because of the intrinsic limitation of gradient-based algorithms, the majority of existing approaches have only been applied to simple problems, since they would fail as soon as the problem becomes complicated such as involving varying signs on gradients or non-linear constraints\cite{sigmund2013}. To address these difficulties, non-gradient methods have been developed which play a significant role in overcoming the tendency to be trapped in a local minimum\cite{sigmund2011}.

Non-gradient optimizers, also known as gradient-free or derivative-free methods, do not use the gradient or derivative of the objective function and has been attempted by several researchers, most of which are stochastic and heuristic methods. For instance, Hajela et al. applied Genetic Algorithm (GA) to a truss structure optimization problem to reduce weight\cite{hajela1995}.  Shim and Manoochehri minimized the material use subject to maximum stress constraints by a Simulated Annealing (SA) approach\cite{shim1997}. Besides these two popular methods, other algorithms have been investigated as well, such as ant colonies\cite{kaveh2008,luh2009}, particle swarms\cite{luh2011}, harmony search\cite{lee2004}, and bacterial foraging\cite{georgiou2014}. Non-gradient methods have four advantages over gradient-based methods\cite{sigmund2011}: better optima, applicable to discrete designs, free of gradients and efficient to parallelize. However, the major disadvantage of the methods is their high computational cost from calling the objective functions, which becomes prohibitively expensive for large systems\cite{rozvany2009}.   As a trade-off, sometimes searching space can be reduced in order for less computation. For instance, pattern search has been applied \cite{guirguis2018high,guirguis2016derivative} which is a non-heuristic method with a smaller searching space but is more likely to be trapped in local minima. 

Machine learning has been used in sequential model-based optimization (SMBO) to target at expensive objective function evaluation \cite{bartz2016survey,hutter2011sequential}. For instance,  Bayesian optimization (BO) \cite{frazier2018tutorial} uses a Gaussian prior to approximate the conditional probability distribution of an objective $p(y|x)$ where $y=F(x)$ is the objective and $x$ is the design variable (vector); then the unknown regions can be estimated by the probability model. In Covariance Matrix Adaptation Evolution Strategy (CMA-ES)\cite{hansen2016cma}, a multivariable Gaussian distribution is used to sample new queries. However, as demonstrated later in the paper, these methods are not designed for large-scale and high-dimensional problems, and thus do not perform well in topology optimization for slow convergence \cite{bujny2016hybrid} or requirement of shrinking design space \cite{luo2020topology}. Despite some improvement to scale up these algorithms \cite{jin2020improved,wang2016bayesian}, none of them has shown superior performance in topology optimization to the best of our knowledge.

There are some reports on leveraging machine learning to reduce the computational cost of topology optimization\cite{lei2018,banga20183d,oh2019,sosnovik2019,rawat2019,jang2020generative,shen2019new,yu2019,sasaki2019topology}. Most of them are generative models which predict solutions of the same problem under different conditions, after being trained by optimized solutions from gradient-based methods. For example, Yu et al.\cite{yu2019} used 100,000 optimal solutions to a simple compliance problem with various boundary forces and the optimal mass fractions to train a neural network consisting of Convolutional Neural Network (CNN) and conditional Generative Adversarial Network (cGAN), which can predict near-optimal designs for any given boundary forces. However, generative models are not topology optimization algorithms: they rely on existing optimal designs as the training data. The predictions are restricted by the coverage of the training datasets. To consider different domain geometries or constraints, new datasets and networks would be required. Besides, the designs predicted by the networks are close to, but still different from the optimal designs. An offline learning method \cite{sasaki2019topology} replaces some FEM calculations during the optimization process with DNN's prediction, yet gives limited improvement especially considering that it requires the solutions to similar problems for training.

Here we propose an algorithm called Self-directed Online Learning Optimization (SOLO) to accelerate non-gradient topology optimization. A DNN is used to map designs to objectives as a surrogate model to approximate and replace the original function which is expensive to calculate. A heuristic optimization algorithm finds the possible optimal design according to DNN’s prediction. Based on the optimum, new query points are dynamically generated and evaluated by the Finite Element Method (FEM) to serve as additional training data. The loop of such self-directed online learning is repeated until convergence. This iterative learning scheme, which can be categorized as an SMBO algorithm, takes advantage of the searching abilities of heuristic methods and the high computational speed of DNN. Theoretical convergence rate is derived under some assumptions. In contrast to gradient-based methods, this algorithm does not rely on gradient information of objective functions of the topology optimization problems. This property allows it to be applied to binary and discrete design variables in addition to continuous ones. To show its performance, we test the algorithm by two compliance minimization problems (designing solid so that the structure achieves maximum stiffness for a given loading), two fluid-structure optimization problems (designing fluid tunnel to minimize the fluid pressure loss for a given inlet speed), a heat transfer enhancement problem (designing a copper structure to reduce the charging time of a heat storage system), and three truss optimization problems (choosing the cross-sectional areas of bars in a truss). Our algorithm reduces the computational cost by at least two orders of magnitude compared with directly applying heuristic methods including Generalized Simulated Annealing (GSA), Binary Bat Algorithm (BBA) and Bat Algorithm (BA). It also outperforms an offline version (where all training data are randomly generated), BO, CMA-ES, and a recent algorithm based on reinforcement learning \cite{gaymann2019deep}.

\subsection*{Results}

\textbf{Formulation and overview.} Consider the following topology optimization problem: in a design domain $\Omega$, find the material distribution $\rho(\mathbf{x})$ that could take either 0 (void) or 1 (solid) at point $\mathbf{x}$ to minimize the objective function $F$, subject to a volume constraint $G_0\le 0$ and possibly $M$ other constraints $G_j\leq0(j=1,...,M) $ \cite{sigmund2013}. Mathematically, this problem can be written as looking for a function $\rho$ defined on the domain $\Omega$,
\begin{equation}
\begin{array}{c}
	\min\limits_{\mathrm{dom}(\rho)=\Omega} F(\rho) \\
\begin{cases}
G_0(\rho) =\int_\Omega{\rho ( \mathbf{x} )\, \mathrm{d} \mathbf{x}}-V_0\le 0\\
G_j(\rho) \le 0,\quad j=1,...,M\\
\rho (\mathbf{x})=0 \text{ or } 1,\quad \forall \mathbf{x}\in \Omega\\
\end{cases}
\end{array},
\end{equation}
where $V_0$ denotes the given volume. To solve such a problem numerically, the domain $\Omega $ is discretized into finite elements to describe the density distribution by $N$ nodal or elemental values,
\begin{equation}
\begin{array}{c}
	\min\limits_{\bm{\rho}=[ \rho_1,\rho_2,...,\rho_N]^T } F(\bm{\rho}) 
	\\
	\begin{cases}
	G_0(\bm{\rho}) =\sum\limits_{i=1}^N{w_i\rho _i}-V_0\le 0\\
	G_j(\bm{\rho})\le 0, \quad j=1,...,M\\
	    \rho_i \in S, \quad i=1,...,N 
	\end{cases} 
	\end{array},
	\label{eq2}
\end{equation}
where $w_i$ denotes the weight of integration. The domain of $\rho_i$ is usually binary ($S=\{0,1\}$), but more generally may take other values such as discrete ($S=\{a_0, a_1,...,a_K\}$) or continuous  ($S=[0,1]$). 

Our algorithm can be applied to \eq{eq2} with binary, discrete or continuous design variables. In this section, we  discuss the case of continuous design variables since it is most general.

In many applications, the objective function is quite complicated and time-consuming to calculate, since it requires solving partial differential equations by, for instance, FEM. To reduce the number of FEM calculations and accelerate non-gradient optimization, we build a DNN to evaluate the objective function. In a naive way, the entire domain of the objective function should be explored to generate the training data. This would incur a huge number of FEM calculations. However, we only care about the function values close to the global optimum and do not require precise predictions in irrelevant regions. In other words, most information about the objective function in the domain is unnecessary except the details around the optimum. So we do not need to generate data to train in those irrelevant regions. 

\begin{figure}[!t]
	\centering
		\begin{minipage}{0.46\textwidth}
		\hspace{1em}
	\xincludegraphics[height=7.5cm,label=	\hspace{-1em}\textbf{a}]{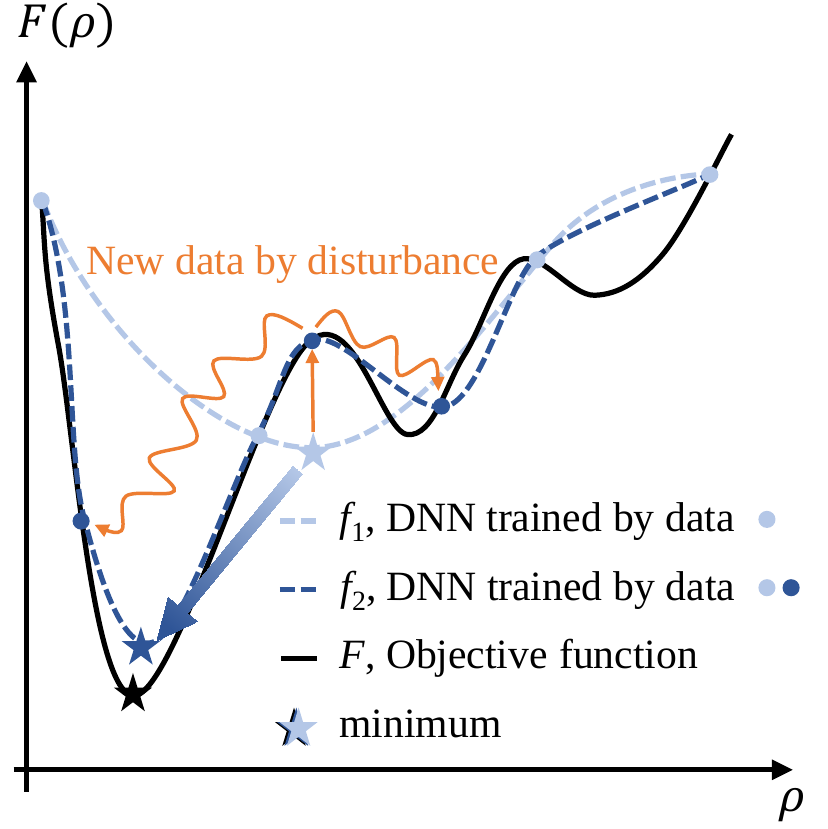}
	\end{minipage}%
		\begin{minipage}{0.54\textwidth}
	\xincludegraphics[height=7.5cm,label=\textbf{b}]{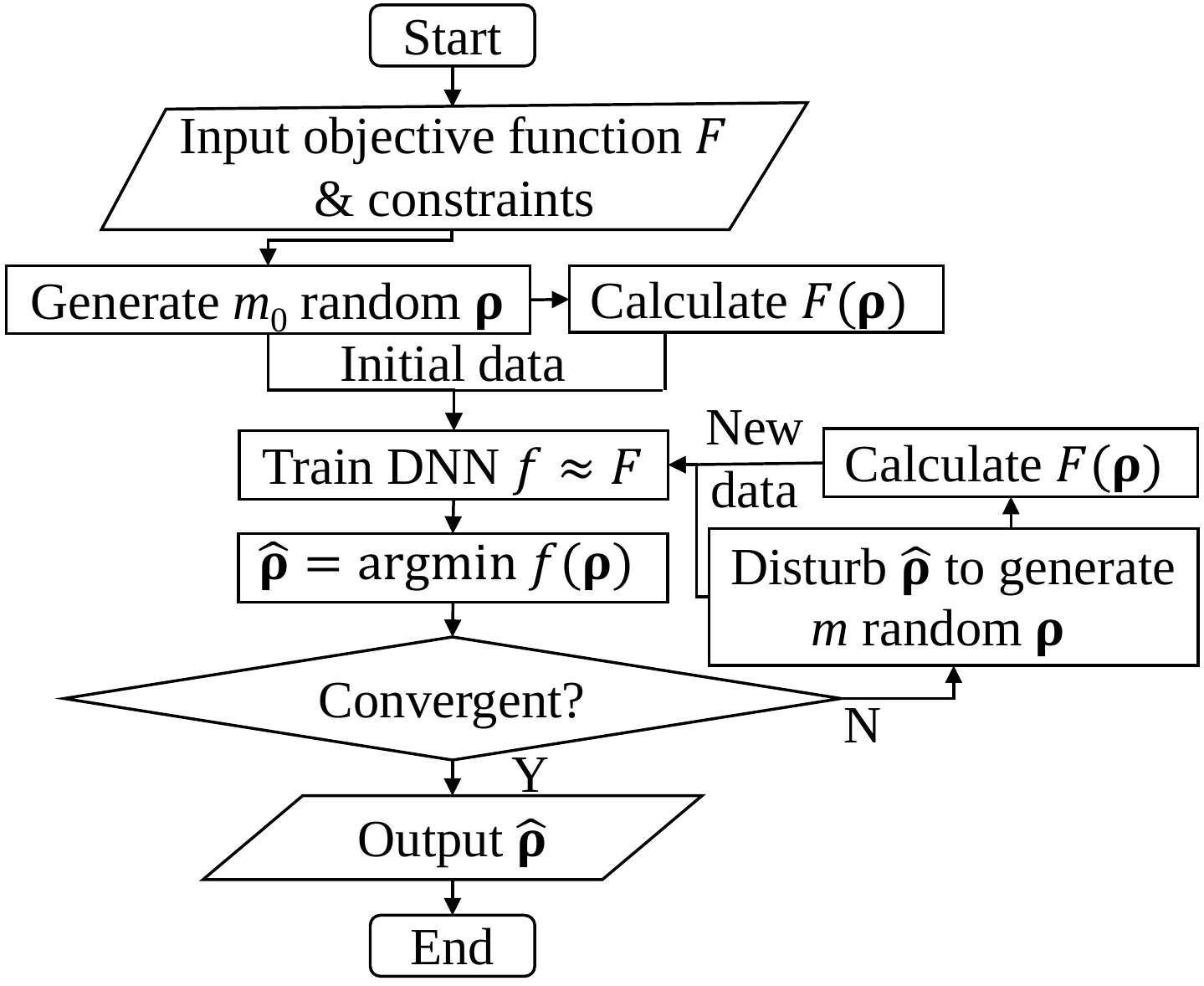}
	\end{minipage}
\caption{\textbf{Schematics of the proposed self-directed online learning optimization.} \textbf{a}, Schematic illustration of self-directed online training. The initial batch of training data (light blue dots) is randomly located. The  DNN $f_1$ (dashed light-blue line) trained on first batch of data only gives a rough representation of the true objective function $F$ (solid black line). The second batch training data (dark blue dots) are generated by adding disturbance (orange curve) to the minimum of $f_1$. After trained with two batches, the DNN $f_2$ (dashed dark-blue line) is more refined around the minimum (the region of interest), while remains almost the same at other locations such as the right convex part. $f_2$ is very close to finding the exact global minimum point. \textbf{b}, Flow diagram of the algorithm.}
\label{fig1}
\end{figure}

An intuitive explanation is shown in \fig{fig1}a. In a 1D minimization example, we can generate a small dataset to train the DNN and refine the mesh around the minimum obtained from the current prediction to achieve higher resolution at the place of interest in the next iteration. After several batches, the minimum of the predicted function would converge to that of the objective function.

\fig{fig1}b shows the flow diagram of the proposed algorithm. A small batch of random vectors (or arrays) $\bm{\rho}$ satisfying the constraints in \eq{eq2} is generated. The corresponding objective values $F(\bm{\rho})$ are calculated by FEM. Then, $\bm{\rho}$ and $F(\bm{\rho})$ are inputted into the DNN as the training data so that the DNN has a certain level of ability to predict the function values based on the design variables. Namely, the output of the DNN $f(\bm{\rho})$ approximates the objective function $F(\bm{\rho})$. Next, the global minimum of the objective function $f(\bm{\rho})$ is calculated by a heuristic algorithm. After obtaining the optimized array $\hat{\bm{\rho}}$, more training data are generated accordingly. Inspired by the concept of GA~\cite{whitley1994}, the disturbance we add to the array is more than a small perturbation, and is categorized as mutation, crossover and convolution. Mutation means replacing one or several design variables with random numbers; crossover means exchanging several values in the array; convolution means applying a convolution filter to the variables (see Methods for details). Then constraints are checked and enforced. The self-directed learning and optimization process stops when the value of the objective function $F(\hat{\bm{\rho}})$ does not change anymore or the computation budget is exhausted.

This algorithm can converge provably under some mild assumptions.
Given the total number of training data $n_{train}$, for any trained DNN  with small training error, we have
    \begin{equation}\label{eq: theory}
        [F(\hat{\bm{\rho}}) - F^*]^2 \le \tilde{O}\left(\frac{C}{\sqrt{n_{train}}}\right),
    \end{equation}
where $C$ is a constant related to some inherent properties of $F$ and DNN, $F^*$ is the global minimum of $F$, and $\tilde{O}$ omits $\log$ terms. 
This result states that when our trained DNN can fit the training data well, then our algorithm can converge to the  global optimal value. We provide convergence guarantee with concrete convergence rate for our proposed algorithm, and to the best of our knowledge, this is the first non-asymptotic convergence result for heuristic optimization methods using DNN as a surrogate model. 
The detailed theory and its derivation are elaborated in Supplementary Section 2.

In the following, we will apply the algorithm to  eight classic examples of four types (covering binary, discrete and continuous variables): two compliance minimization problems, two fluid-structure optimization problems, a heat transfer enhancement problem and three truss optimization problems.

\textbf{Compliance minimization.} We first test the algorithm on two simple continuous compliance minimization problems. We show that our algorithm can converge to global optimum and is faster than other non-gradient methods.

\begin{figure}[!thb]
	\centering
	\begin{minipage}{0.24\textwidth}\centering
		\xincludegraphics[height=5.1cm,label=\textbf{a}]{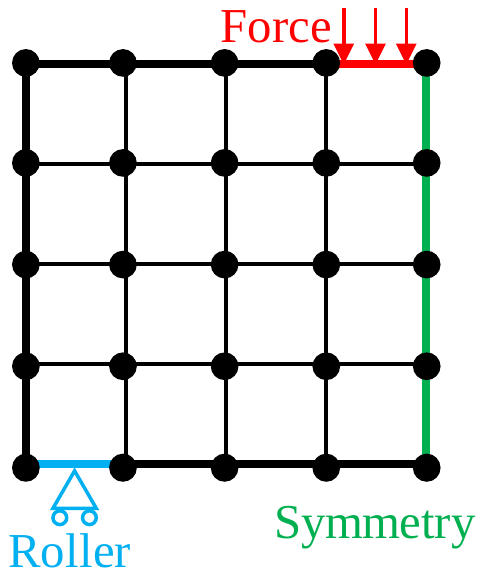}
	\end{minipage}%
	\begin{minipage}{0.38\textwidth}\centering
		\xincludegraphics[height=5.1cm,label=\negthinspace \textbf{b}]{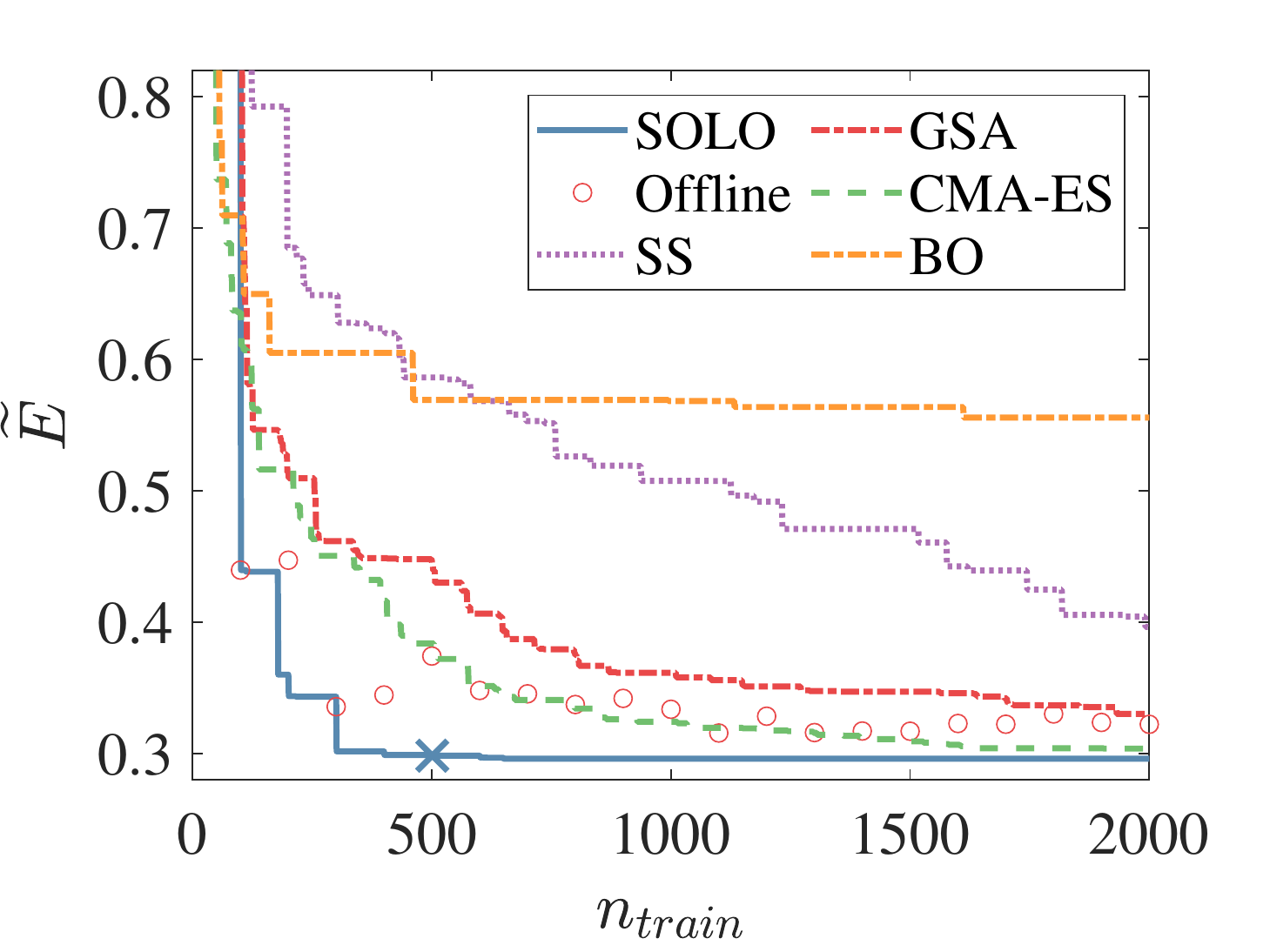}
	\end{minipage}%
	\begin{minipage}{0.38\textwidth}\centering
	\xincludegraphics[height=5.1cm,label=\negthinspace \textbf{c}]{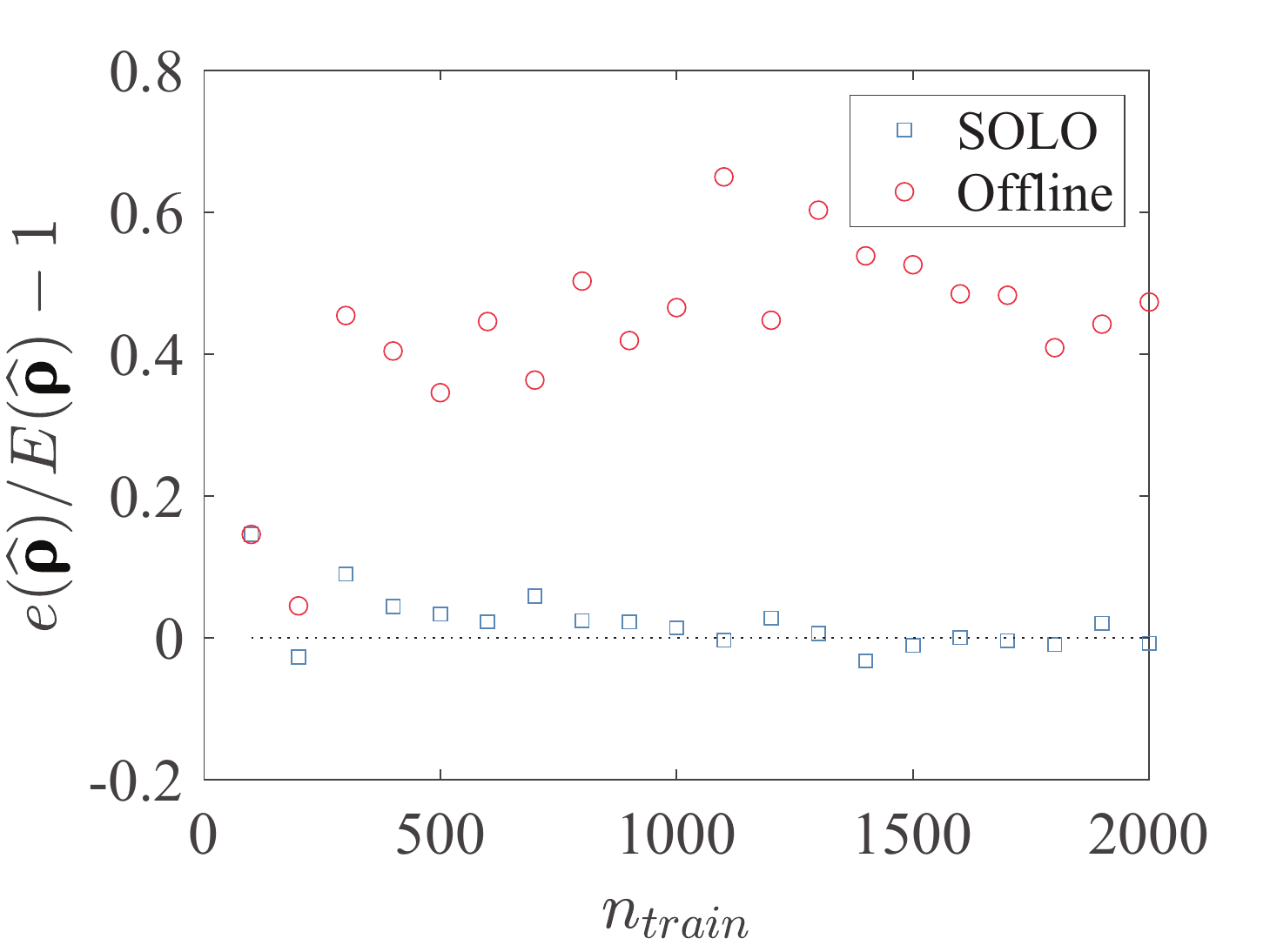}
	\end{minipage}\\
	\vspace{0.2cm}
	\begin{minipage}{0.31\textwidth}
	\parbox[c][1.5em]{\textwidth}{\textbf{d} Gradient-based $\widetilde{E}  =  0.293$}
	\includegraphics[height=4.9cm]{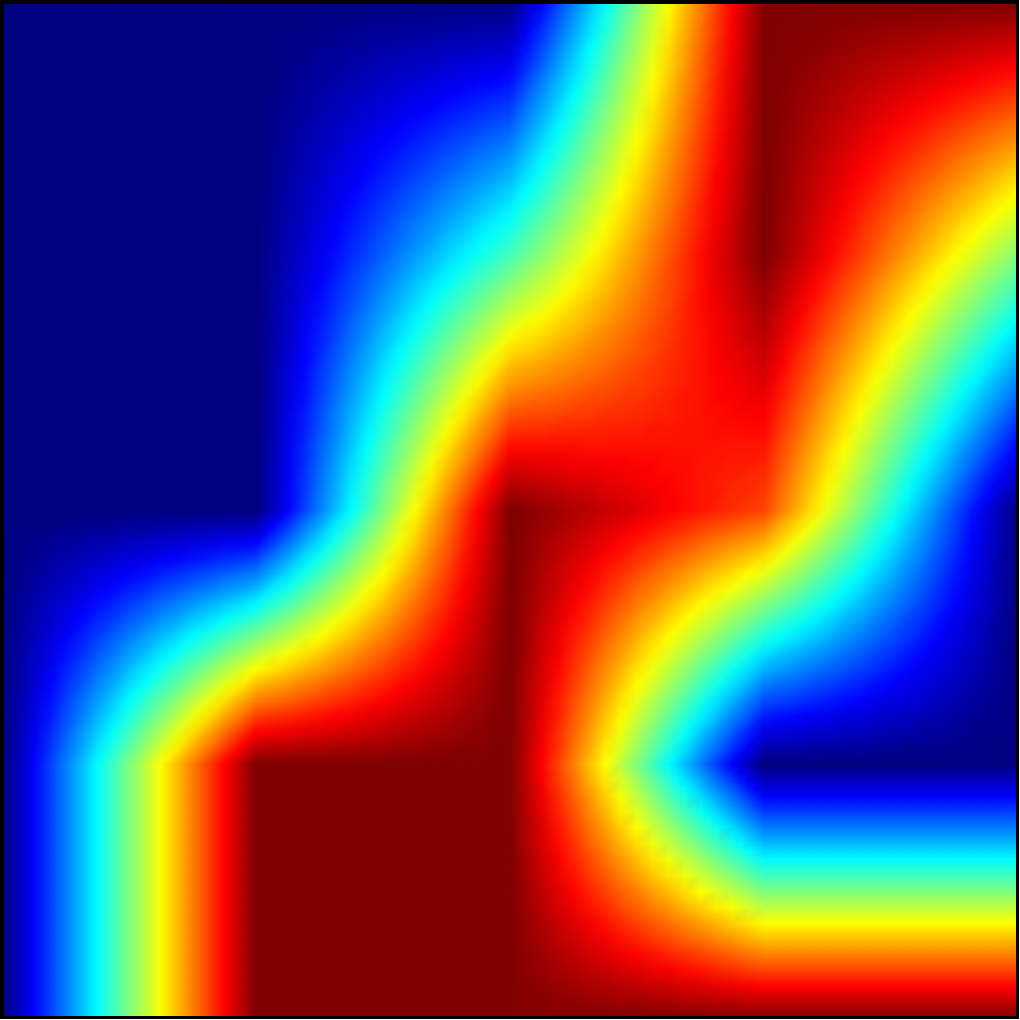}
	\end{minipage}
	\begin{minipage}{0.31\textwidth}
		\parbox[c][1.5em]{\textwidth}{\textbf{e} SOLO@501 $\widetilde{E}  =  0.298$}
	\xincludegraphics[height=4.9cm]{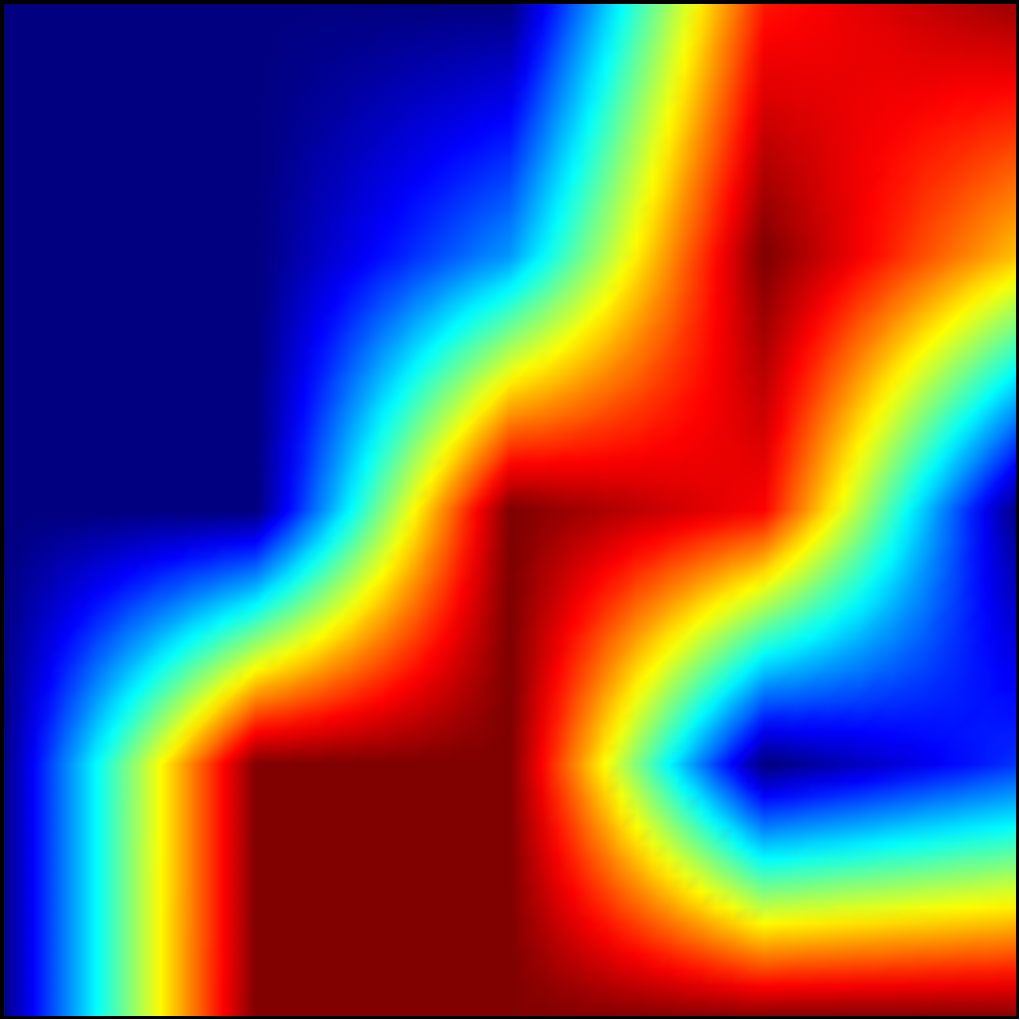}
\end{minipage}
	\begin{minipage}{0.36\textwidth}
		\parbox[c][1.5em]{\textwidth}{\textbf{f} SOLO@5,782 $\widetilde{E}  =  0.293$}
	\xincludegraphics[height=4.9cm]{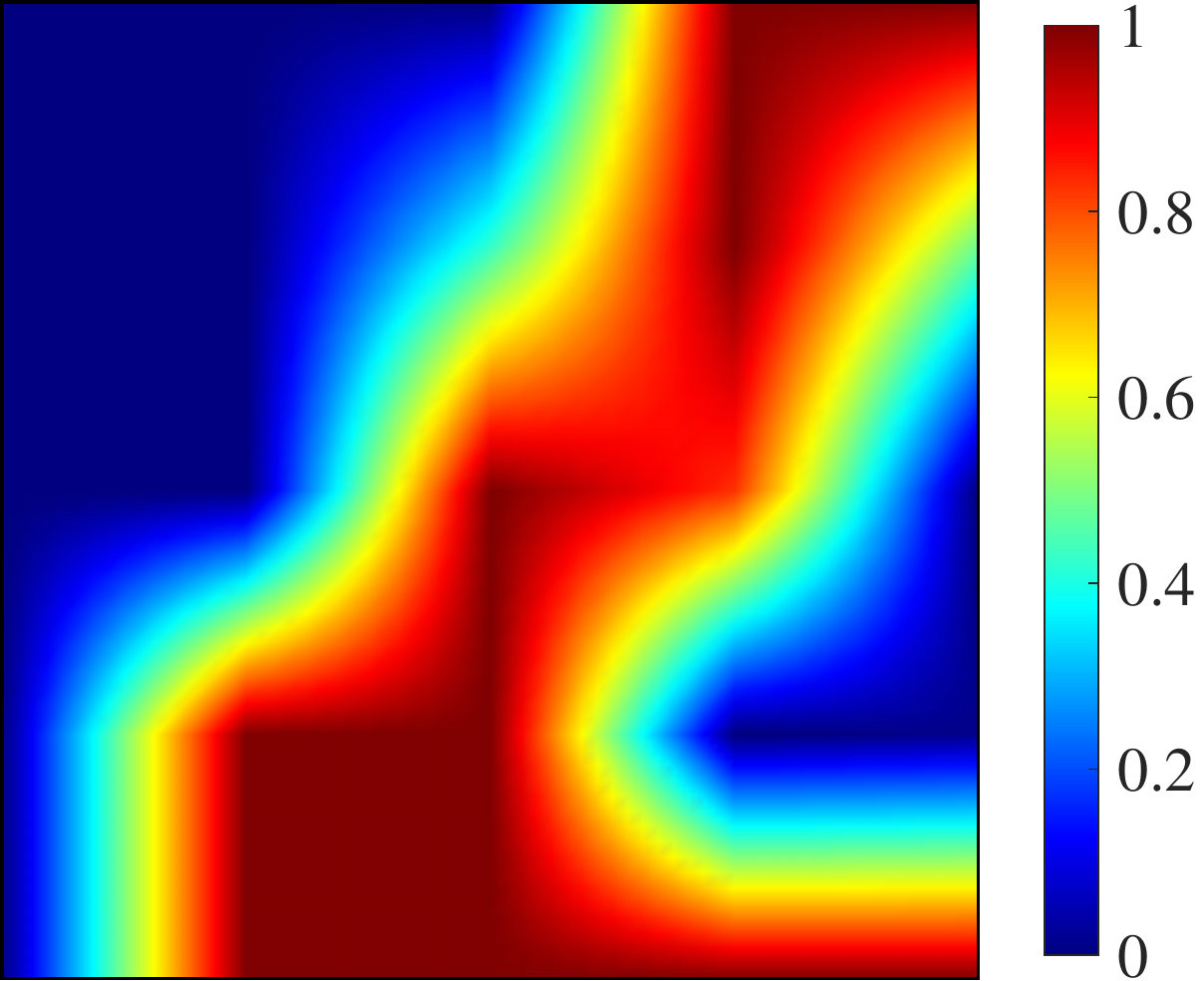}
\end{minipage}
	\caption{\textbf{Setup and results of a compliance minimization problem with 5$\times$5 design variables.} \textbf{a}, Problem setup:  minimizing compliance subject to maximum volume constraint. \textbf{b}, Best dimensionless energy with a total of $n_{train}$ accumulated training samples. SOLO denotes our proposed method where the cross ``X" denotes the convergence point (presented in \textbf{e}), ``Offline'' denotes training a DNN offline and then uses GSA to search for the optimum without updating the DNN, whose results are independent so they are plotted as circles instead of a curve, SS denotes Stochastic Search, which is the same as SOLO except that $\hat{\bm{\rho}}$ in each loop is obtained by the minimum of existing samples, CMA-ES denotes Covariance Matrix Adaptation Evolution Strategy, BO denotes Bayesian Optimization. SOLO converges the fastest among these methods.  \textbf{c}, Energy  prediction  error  of $\hat{\bm{\rho}}$ relative to FEM calculation of the same material distribution. $e\hat{\bm{\rho}}$ denotes DNN's prediction, $E\hat{\bm{\rho}}$ denotes FEM's result. \textbf{d}, Optimized design produced by the gradient-based method. $\widetilde{E}  =  0.293$. \textbf{e}, Optimized design produced by SOLO. $n_{train}=501$ and $\widetilde{E}=0.298$. \textbf{f}, Optimized design produced by SOLO. $n_{train}=5,782$ and $\widetilde{E}=0.293$. In \textbf{d}-\textbf{f}, dark red denotes $\rho=1$ and dark blue denotes $\rho=0$, as indicated by the right color scale bar.}
	\label{fig2}
\end{figure}

As shown in \fig{fig2}a, a square domain is divided evenly by a $4\times 4$ mesh. A force downward is applied at the top right edge; the bottom left edge is set as a roller (no vertical displacement); the right boundary is set to be symmetric. There are 25 nodal design variables to control the material distribution, i.e. density $\bm{\rho}$.  Our goal is to find the density $\rho_i(i=1,2,...,25)$, subject to a volume constraint of 0.5, such that the elastic energy $E$ of the structure is minimized, equivalent to minimizing compliance or the vertical displacement where the external force is applied. Formally, 
\begin{equation}\label{eq5}
	\min\limits_{\bm{\rho}\in[0,1]^N } \widetilde{E}( \bm{\rho} ) =\frac{E( \bm{\rho} )}{E( \bm{\rho} _O )}, 
\end{equation} 
where $\bm{\rho} _O=[0.5,0.5,...,0.5]^T$. The constraint is
\begin{equation}
	\mathbf{w}\cdot\bm{\rho}\le 0.5,
\end{equation} 
where $\mathbf{w}$ denotes the vector  of  linear  Gaussian  quadrature. In \eq{eq5} we use the dimensionless elastic energy $\widetilde{E}( \bm{\rho} )$, defined as the ratio of elastic energy of the structure with given material distribution to that of the reference uniform distribution (the material density is 0.5 everywhere in the domain). The elastic energy is calculated by FEM from the Young’s modulus in the domain, which is related to density by the popular Simplified Isotropic Material with Penalization (SIMP) method,\cite{bendse2004}
\begin{equation}
	Y( \rho (\mathbf{x})) =Y_0\rho(\mathbf{x})^3  +\varepsilon\left[ 1-\rho (\mathbf{x})^3\right],
\end{equation}
where $Y$ and $Y_0$ denote the Young's moduli as a variable and a constant respectively, $\varepsilon$ is a small number to avoid numerical singularity and $\rho(\mathbf{x})$  is the material density at a given location $\mathbf{x}$ interpolated linearly by the nodal values of the element.

For benchmark, we use a traditional gradient-based algorithm, the Method of Moving Asymptotes (MMA), to find the optimized solution (\fig{fig2}d). 

For our proposed method, we use 100 random arrays to initialize the DNN. Then  Generalized Simulated Annealing (GSA) is used to obtain the minimum $\hat{\bm{\rho}}$ based on the DNN's prediction. Afterwards, 100 additional samples will be generated by adding disturbance to $\hat{\bm{\rho}}$ including mutation and crossover. Such a loop continues until convergence. 

We compare our proposed method, Self-directed Online Learning Optimization (SOLO), with five other algorithms.  In \fig{fig2}b, SOLO converges at $n_{train}=501$. ``Offline'' denotes a naive implementation to couple DNN with GSA, which trains a DNN offline by $n_{train}$ random samples and then uses GSA to search for the optimum, without updating the DNN. As expected, the elastic energy decreases with the number of accumulated training samples $n_{train}$. This is because more training data will make the DNN estimate the elastic energy  more accurately. Yet it converges much slower than SOLO and does not work well even with $n_{train}=2,000$. 
More results are shown in Supplementary Fig.~1. 
SS denotes Stochastic Search, which uses current minimum (the minimum of existing samples) to generate new searching samples; the setup is the same as SOLO except that the base design $\hat{\bm{\rho}}$ is obtained from the current minimum instead of a DNN. Comparing SS with SOLO, we can conclude that the DNN in SOLO gives a better searching direction than using existing optima.   CMA-ES denotes Covariance Matrix Adaptation Evolution Strategy with a multi-variable Gaussian prior. BO denotes Bayesian Optimization with Gaussian distribution as the prior and expected improvement (EI) as the acquisition function. Our method outperforms all these methods in terms of convergence speed. CMA-ES ranks the second with an objective value 3\% higher than SOLO at $n_{train}=2,000$.


To assess inference accuracy in online and offline learning, we compare the DNN-predicted energy with  that calculated by FEM on the same material distribution. The relative error is defined by
$[{e( \hat{\bm{\rho}} ) -E(  \hat{\bm{\rho}} )}]/{E(  \hat{\bm{\rho}} )}$
where $e( \hat{\bm{\rho}} )$  and $E( \hat{\bm{\rho}} )$  denote energy calculated by DNN and FEM respectively. The energy prediction error is shown in \fig{fig2}c. When $n_{train}$ is small, both networks overestimate the energy since their training datasets, composed of randomly distributed density values, correspond to higher energy. As $n_{train}$ increases, the error of SOLO fluctuates around zero since solutions with low energy are fed back to the network.

\begin{figure}[!thb]
	\centering
	\begin{minipage}{0.24\textwidth}\centering
		\xincludegraphics[height=5.1cm,label=\textbf{a}]{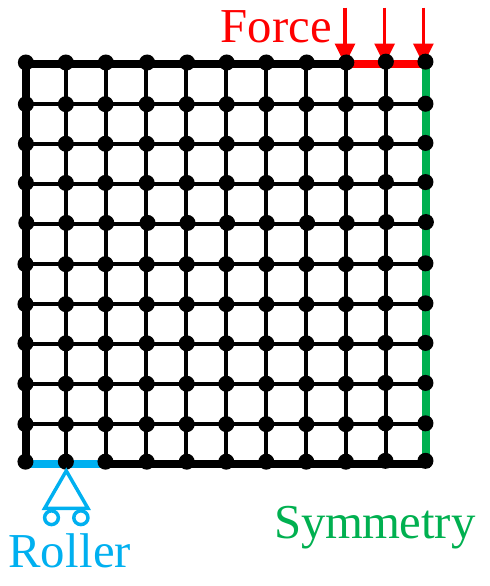}
	\end{minipage}%
	\begin{minipage}{0.38\textwidth}\centering
		\xincludegraphics[height=5.1cm,label=\negthinspace \textbf{b}]{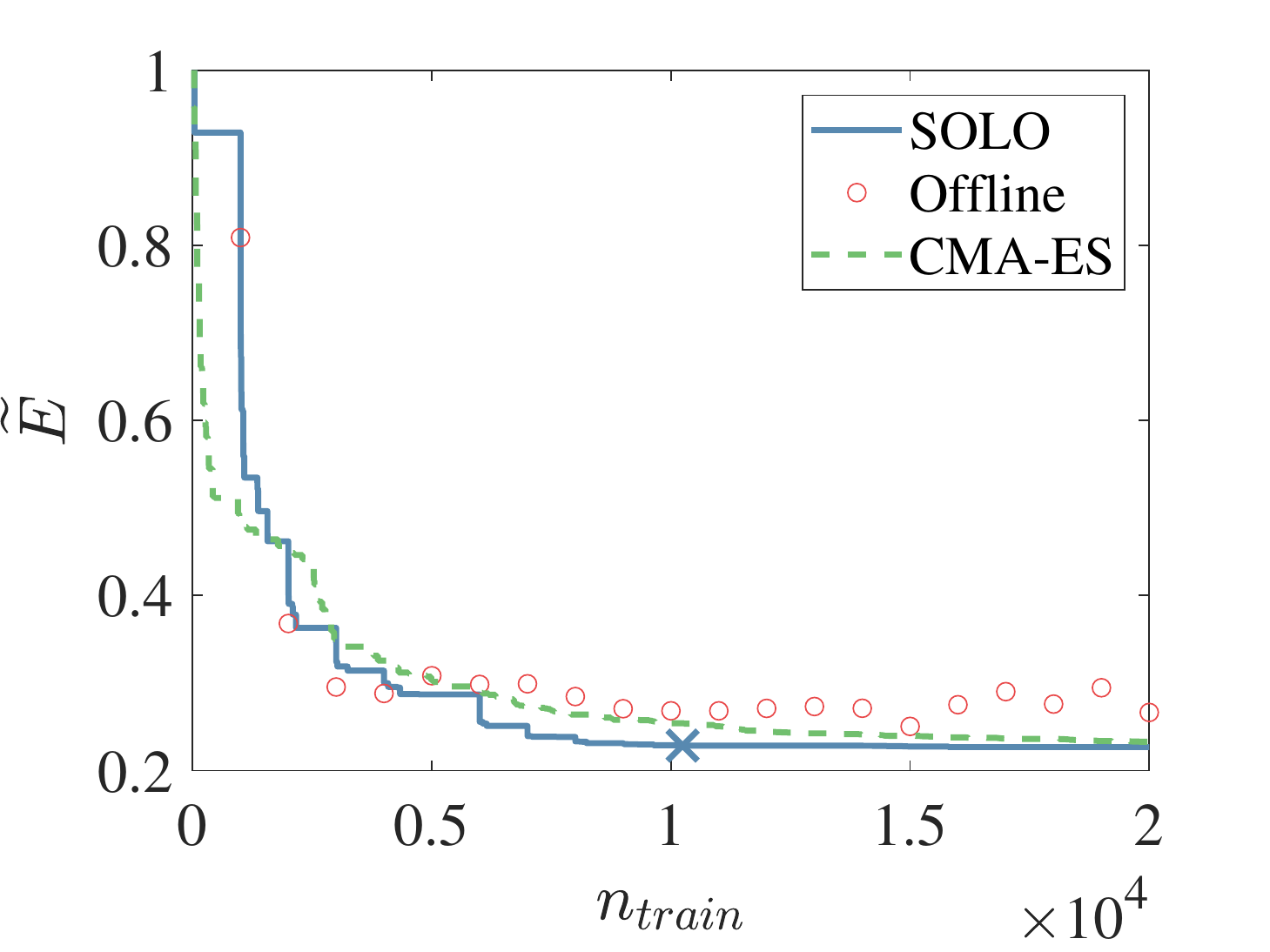}
	\end{minipage}%
	\begin{minipage}{0.38\textwidth}\centering
		\xincludegraphics[height=5.1cm,label=\negthinspace \textbf{c}]{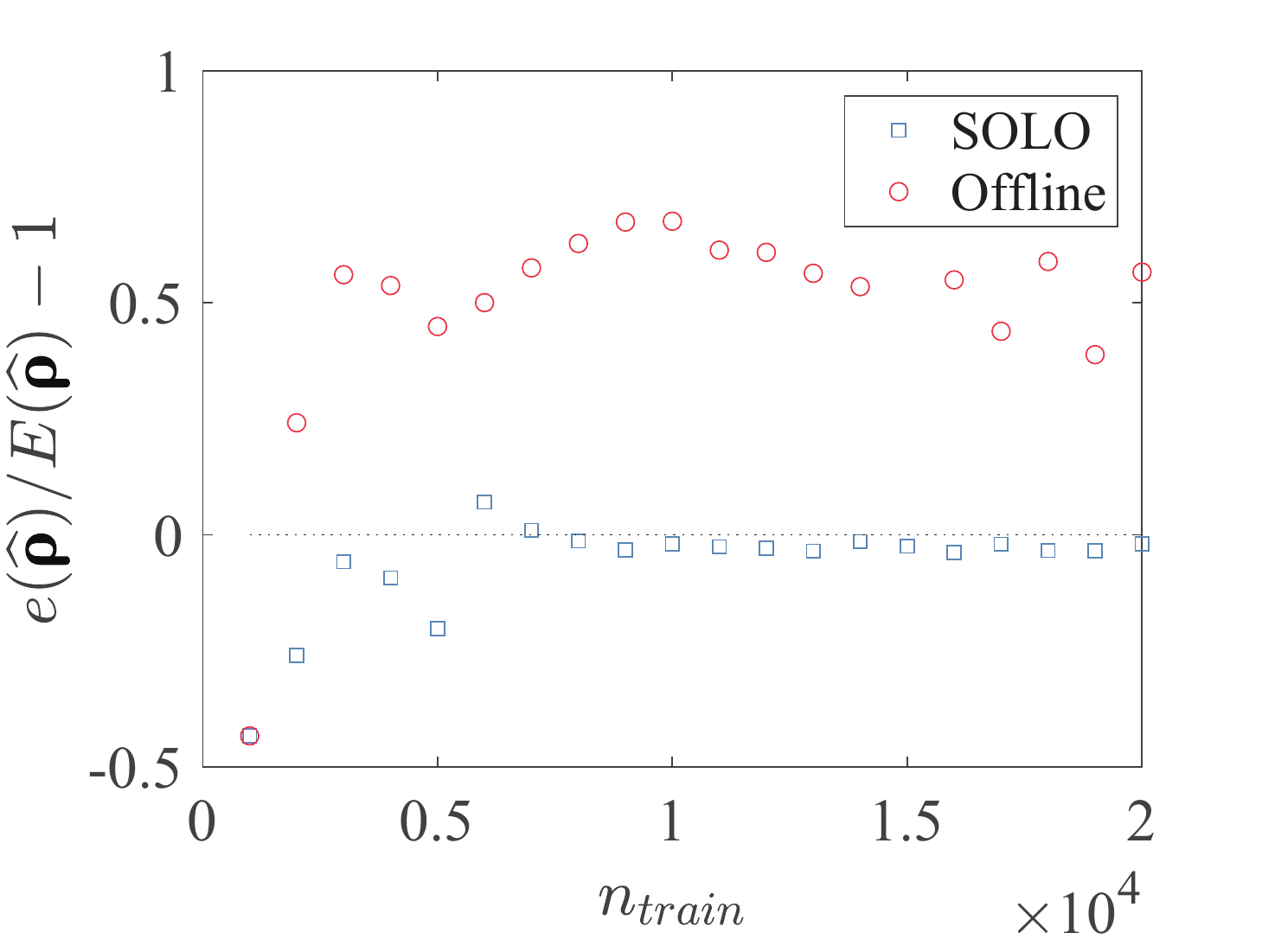}
	\end{minipage}\\
	\vspace{0.2cm}
	\begin{minipage}{0.31\textwidth}\raggedright
	\parbox[c][1.5em]{\textwidth}{\textbf{d} \small{Gradient-based $\widetilde{E}=0.222$}}
	\includegraphics[height=4.9cm]{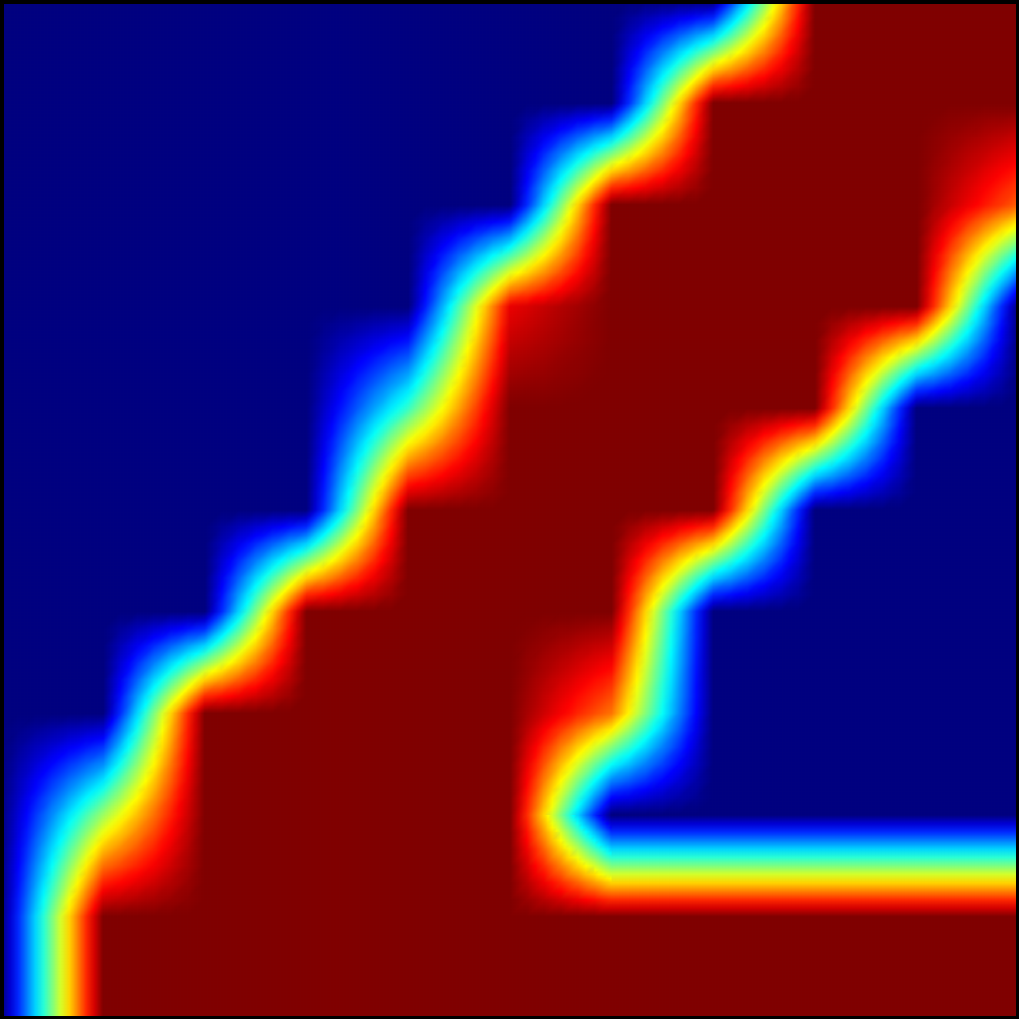}
	\end{minipage}
	\begin{minipage}{0.31\textwidth}\raggedright
	\parbox[c][1.5em]{\textwidth}{\textbf{e} \small{SOLO@10,243 $\widetilde{E}=0.228$}}
	\includegraphics[height=4.9cm]{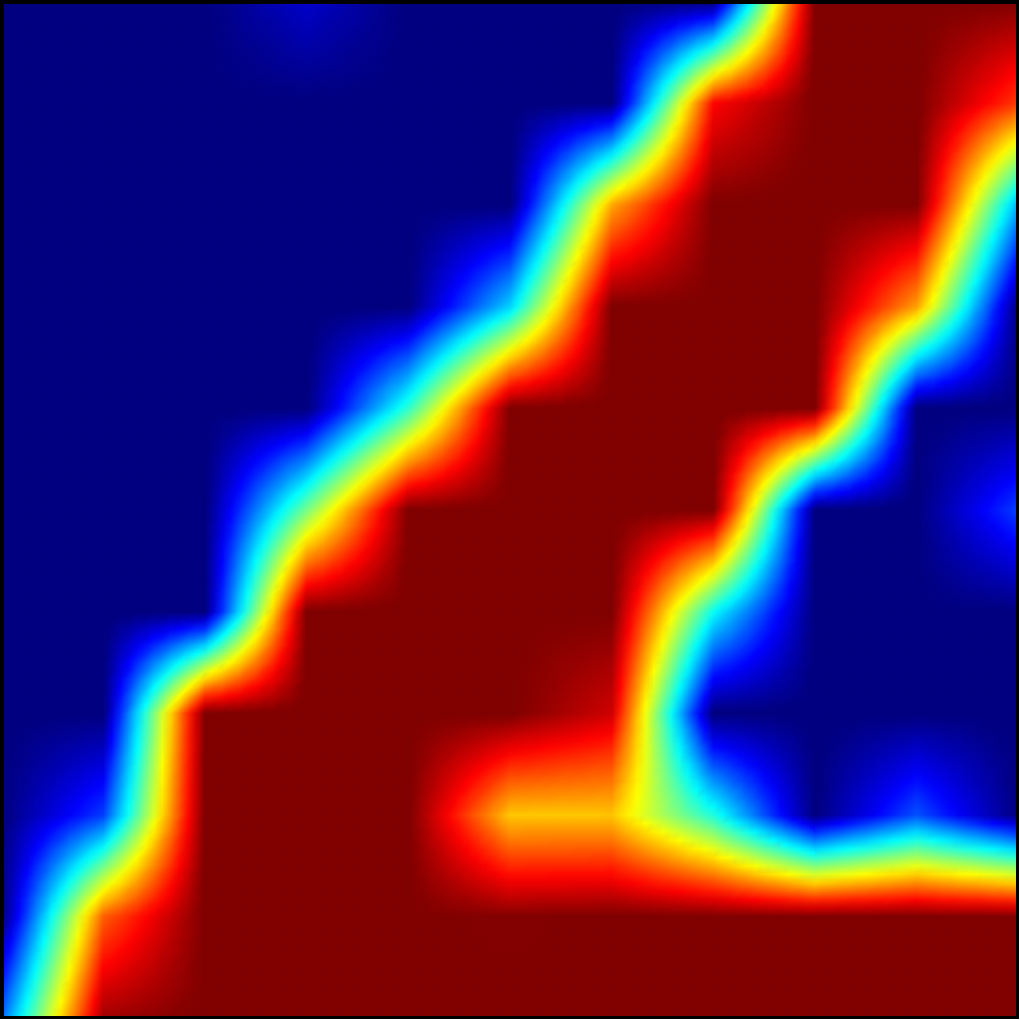}
	\end{minipage}
	\begin{minipage}{0.36\textwidth}\raggedright
	\parbox[c][1.5em]{\textwidth}{\textbf{f} \small{SOLO@77,691 $\widetilde{E}=0.222$}}
	\includegraphics[height=4.9cm]{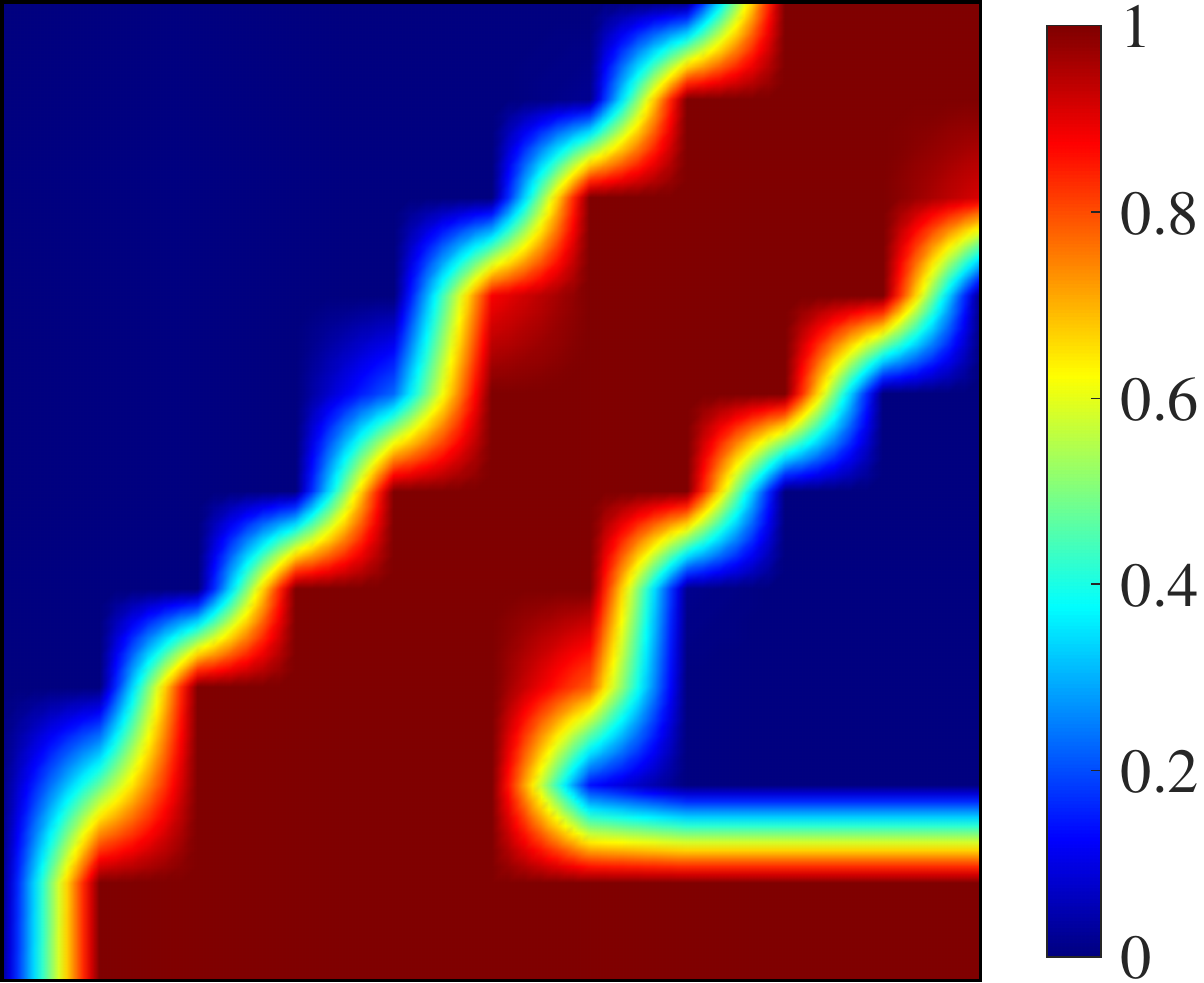}
	\end{minipage}
	\caption{\textbf{Setup and results of a compliance minimization problem with 11$\times$11 design variables.} \textbf{a}, Problem setup:  minimizing compliance subject to maximum volume constraint. \textbf{b},  Best dimensionless energy with a total of $n_{train}$ accumulated training samples. SOLO denotes our proposed method where the cross ``X" denotes the convergence point (presented in \textbf{e}), ``Offline'' denotes training a DNN offline and then uses GSA to search for the optimum without updating the DNN, whose results are independent so they are plotted as circles instead of a curve, CMA-ES denotes Covariance Matrix Adaptation Evolution Strategy. SOLO converges the fastest among these methods. \textbf{c}, Energy prediction error of $\hat{\bm{\rho}}$ relative to FEM calculation of the same material distribution. $e\hat{\bm{\rho}}$ denotes DNN's prediction, $E\hat{\bm{\rho}}$ denotes FEM's result. \textbf{d}, Optimized design produced by the gradient-based method. $\widetilde{E}=0.222$. \textbf{e}, Optimized design produced by SOLO. $n_{train}=10,243$ and $\widetilde{E}=0.228$. \textbf{f}, Optimized design produced by SOLO. $n_{train}=77,691$ and $\widetilde{E}=0.222$. In \textbf{d}-\textbf{f}, dark red denotes $\rho=1$ and dark blue denotes $\rho=0$, as indicated by the right bar.}
	\label{fig3}
\end{figure}

The solution of SOLO using 501 samples is presented in \fig{fig2}e, whose energy is 0.298, almost the same as that of the benchmark in \fig{fig2}d. With higher $n_{train}$, the solution from SOLO becomes closer to that of the benchmark (the evolution of optimized structures is shown in Supplementary Fig.~2). In \fig{fig2}f, the energy is the same as the benchmark. The material distribution in \fig{fig2}f does not differ much from that in \fig{fig2}e. In fact, using only 501 samples is sufficient for the online training to find the optimized material distribution. We find that in our problem and optimization setting, the GSA needs about 2$\times{10}^5$ function evaluations to obtain the minimum of DNN. Since the DNN approximates the objective function, we estimate GSA needs the same number of evaluations when applying to the objective, then it means 2$\times{10}^5$ FEM calculations are required if directly using GSA. From this perspective, SOLO reduces the number of FEM calculations to 1/400.

A similar problem with a finer mesh having 121 (11$\times$11) design variables is shown in \fig{fig3}a. The benchmark solution from MMA is shown in \fig{fig3}d, whose energy is 0.222. The trends in \fig{fig3}b and c are similar to those in \fig{fig2} with a coarse mesh. \fig{fig3}b shows that SOLO converges at  $n_{train}=10,243$, giving $\widetilde{E}=0.228$. Our method again outperforms CMA-ES, the second best algorithm according to \fig{fig2}b.  The material distribution solutions are shown in \fig{fig3}e and f. The configuration of SOLO is the same as that for the coarse mesh except that each loop has 1,000 incremental samples and GSA performs 4$\times{10}^6$ function evaluations. Compared with directly using GSA, SOLO reduces the number of FEM calculations to 1/400 as well. 
The evolution of optimized structures is shown in Supplementary Fig.~3.

\textbf{Fluid-structure optimization.} In the following two problems, we leverage our algorithm to address binary fluid-structure optimization. We want to show that our method outperforms the gradient-based method and a recent algorithm based on reinforcement learning\cite{gaymann2019deep}. 

\begin{figure}[!htb]
	\centering
	\begin{minipage}{0.5\textwidth}\centering
		\xincludegraphics[height=4 cm,label=\textbf{a}]{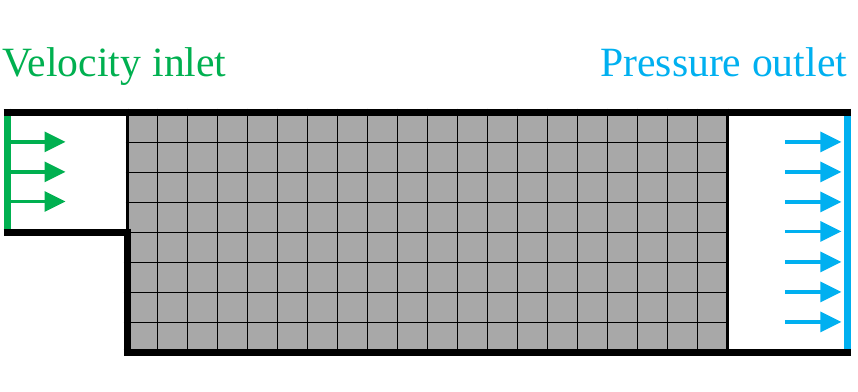}
	\end{minipage}%
	\begin{minipage}{0.5\textwidth}\centering
		\xincludegraphics[height=4 cm,label=\hspace{-0.3cm}\textbf{b}]{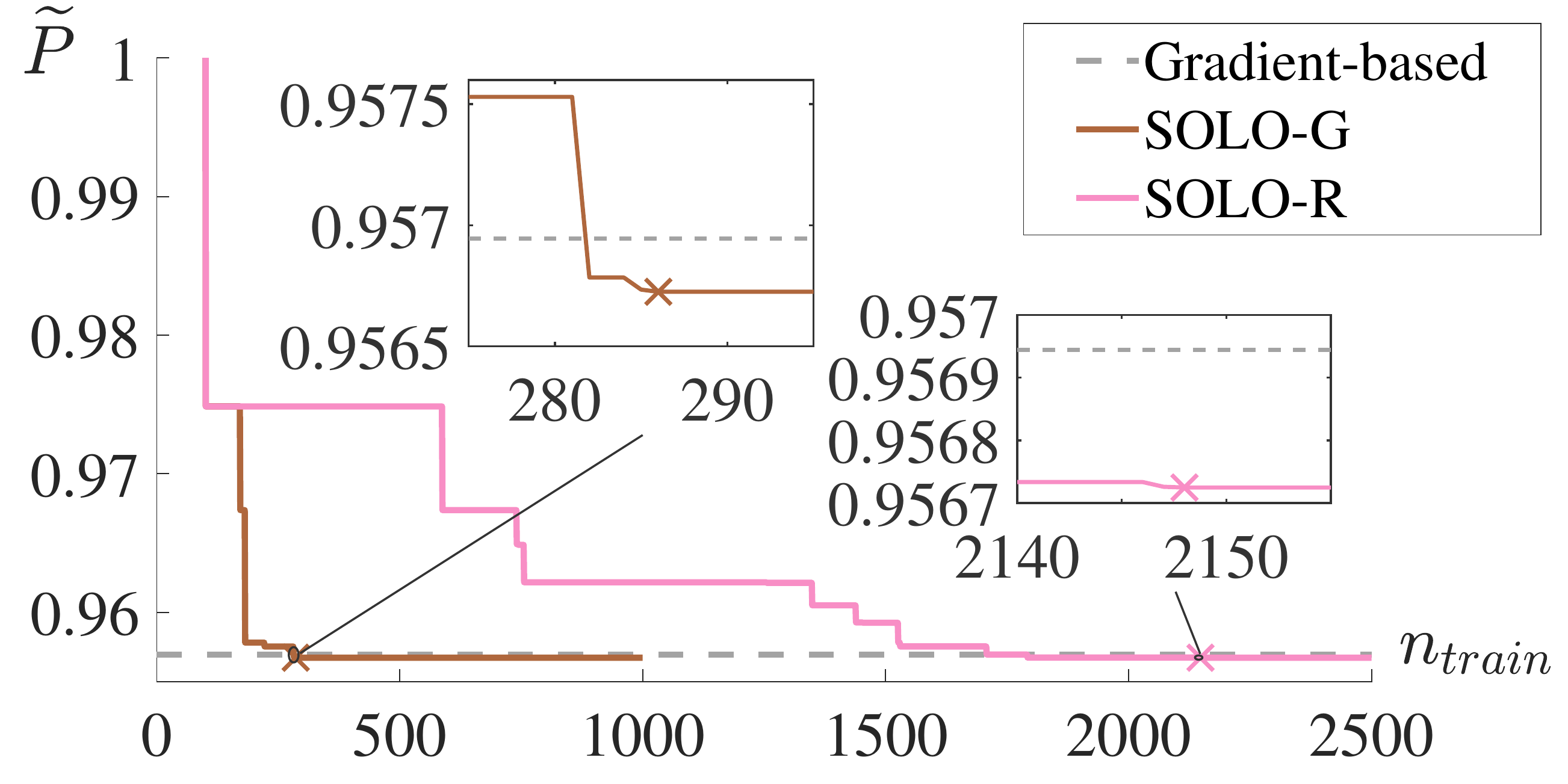}
	\end{minipage}\\
	\vspace{0.2cm}
	\begin{minipage}{0.33\textwidth}\centering
				\parbox[c][1em]{\textwidth}{\textbf{c} \small{Gradient-based $\widetilde{P}=0.9569$}}
		\includegraphics[width=\textwidth]{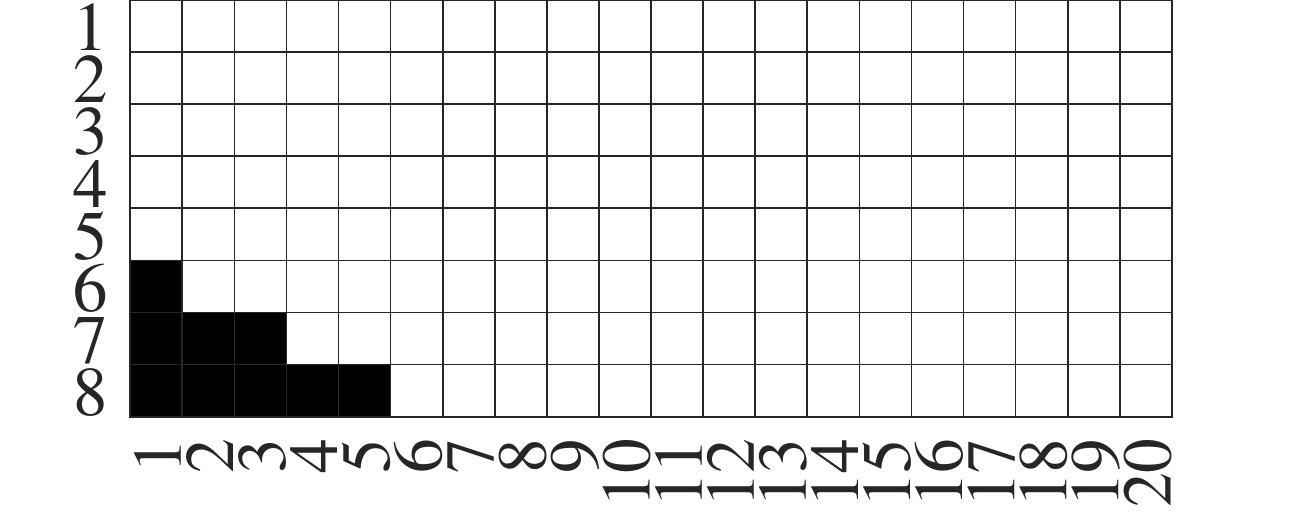}
	\end{minipage}%
	\begin{minipage}{0.33\textwidth}\centering
		\parbox[c][1em]{\textwidth}{\textbf{d} \small{SOLO-G@286 $\widetilde{P}=0.9567$}}
		\includegraphics[width=\textwidth]{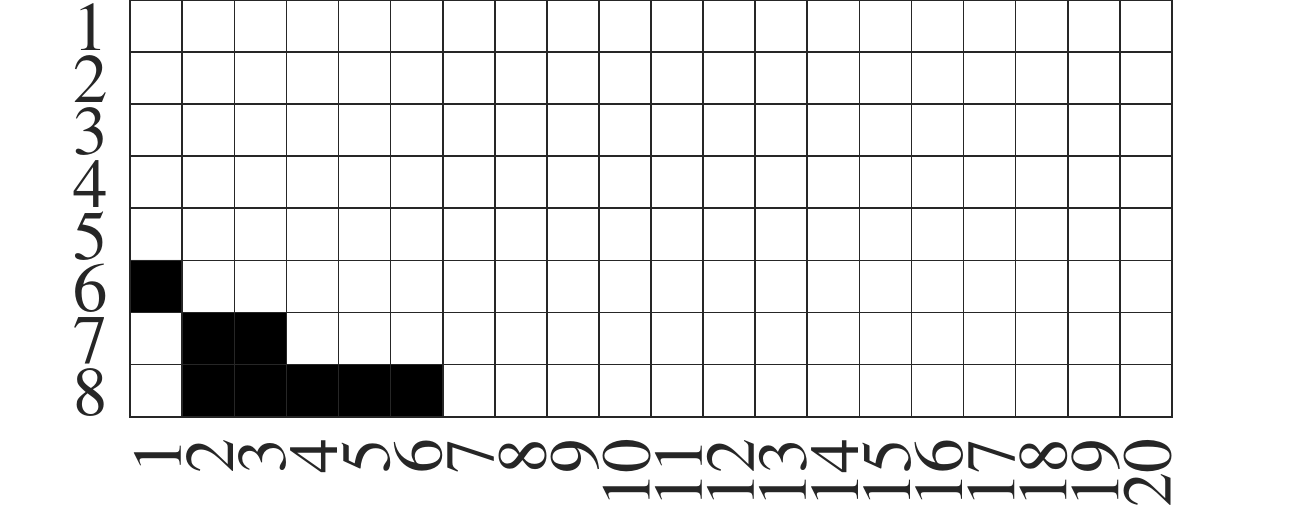}
	\end{minipage}%
	\begin{minipage}{0.33\textwidth}\centering
			\parbox[c][1em]{\textwidth}{\textbf{e} \small{SOLO-R@2,148 $\widetilde{P}=0.9567$}}
		\includegraphics[width=\textwidth]{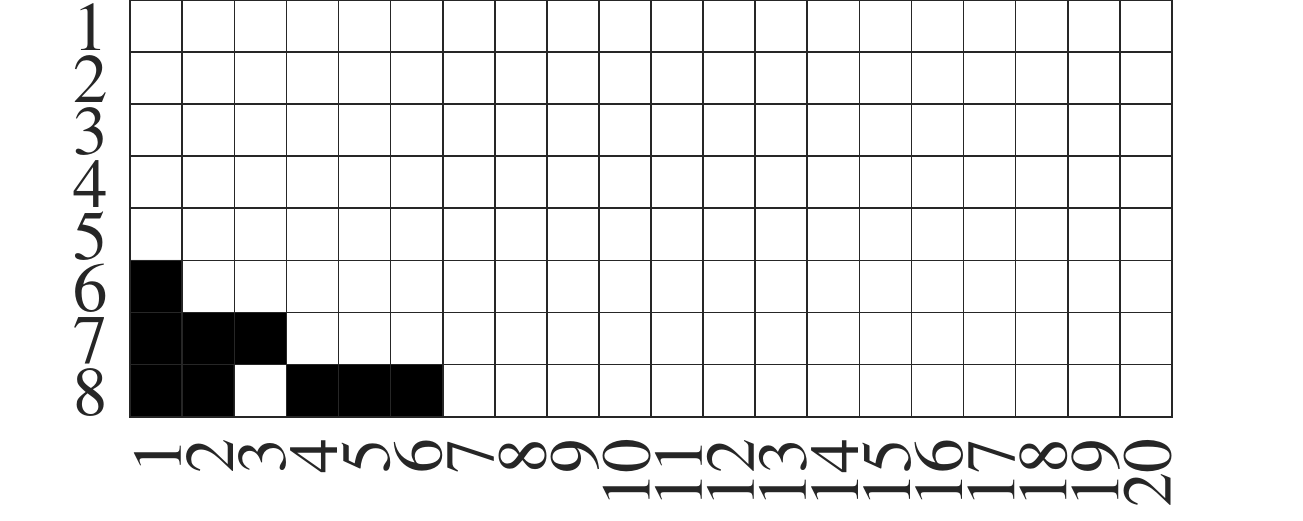}
	\end{minipage}
	\caption{\textbf{Setup and results of a fluid-structure optimization problem with 20$\times$8 design variables.} \textbf{a}, Problem setup: minimizing pressure drop through the tunnel. The vertical green line on the left denotes the inlet while the vertical blue line on the right denotes the outlet.  \textbf{b},  Dimensionless inlet pressure  versus $n_{train}$, the number of accumulated training samples. SOLO-G denotes a greedy version of our proposed method, SOLO-R denotes the regular version of our proposed method. The horizontal dashed line denotes the solution from the gradient-based method. The cross ``X" denotes the convergence point (presented in \textbf{d} and \textbf{e}, respectively). \textbf{c}, Optimized design obtained by the gradient-based method. $\widetilde{P}=0.9569$. \textbf{d}, Optimized design obtained by SOLO-G. $n_{train}=286$ and $\widetilde{P}=0.9567$. \textbf{e}, Optimized design obtained by SOLO-R. $n_{train}=2,148$ and $\widetilde{P}=0.9567$. In \textbf{c}-\textbf{e}, black denotes $\rho=1$ (solid) and white denotes $\rho=0$ (void). The solutions in \textbf{d} and \textbf{e} are equivalent since the flow is blocked by the black squares forming the ramp surface and the white squares within the ramp at the left bottom corner are irrelevant.}
	\label{fig4}
\end{figure}

 As shown in \fig{fig4}a, the fluid enters the left inlet at a given velocity perpendicular to the inlet, and flows through the channel bounded by walls to the outlet where the pressure is set as zero. In the $20\times 8$ mesh, we add solid blocks to change the flow field to minimize the friction loss when the fluid flows through the channel. Namely, we want to minimize the normalized inlet pressure
\begin{equation}
  	\min\limits_{\bm{\rho}\in\{0,1\}^N }  \widetilde{P}(\bm{\rho})=\frac{P(\bm{\rho})}{P(\bm{\rho}_O)},
\end{equation}
where $P$ denotes the average inlet pressure and $\bm{\rho}_O=[0,0,...,0]^T$ indicates no solid in the domain. As for the fluid properties, we select a configuration with a low Reynolds number for stable steady solution \cite{deng2019numerical}, specifically, \begin{equation}
    \mathrm{Re}=\frac{DvL}{\mu}=40,
\end{equation}
where $D$ denotes fluid density, $\mu$ denotes viscosity, $v$ denotes inlet velocity and $L$ denotes inlet width (green line).

For the benchmark, we use a typical gradient-based algorithm which adds an impermeable medium to change binary variables to continuous ones \cite{olesen2006high}. It uses the adjoint method to derive gradients and MMA as the solver. The solution is presented in \fig{fig4}c. The solid blocks form a ramp at the left bottom corner for a smooth flow expansion.

We use two variants of our algorithm. One is denoted as SOLO-G, a greedy version of SOLO where additional 10 samples produced in each loop are all from the DNN's prediction. The initial batch is composed of a solution filled with zeros and 160 solutions each of which has a single element equal to one and others equal to zero. The pressure values corresponding to these designs are calculated by FEM. These 161 samples are used to train a DNN. Next, Binary Bat Algorithm (BBA) is used to find the minimum of the DNN. The top 10 solutions (after removing repeated ones) encountered during BBA searching will be used as the next batch of training data. The other variant, denoted as SOLO-R, is a regular version of SOLO where each loop has 100 incremental samples. 10 of them are produced in the same way as SOLO-G whereas the rest 90 are generated by adding disturbance to the best solution predicted by the DNN. Similar to the compliance minimization problems, the disturbance includes mutation and crossover.

\begin{figure}[!tb]
	\centering
	\begin{minipage}{0.5\textwidth}\centering
		\xincludegraphics[height=4 cm,label=\textbf{a}]{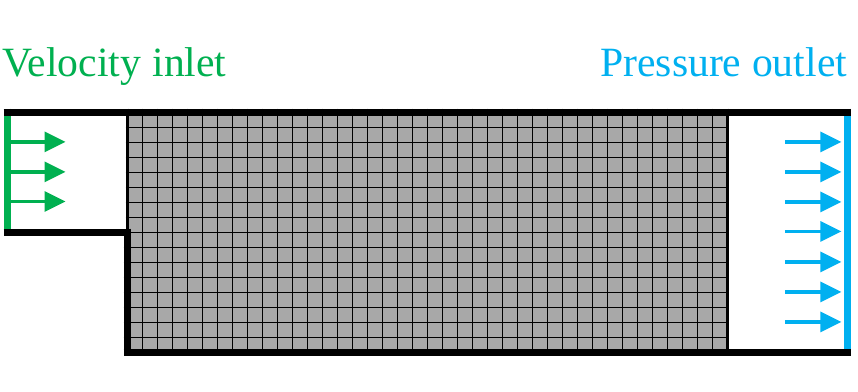}
	\end{minipage}%
	\begin{minipage}{0.5\textwidth}\centering
		\xincludegraphics[height=4 cm,label=\hspace{-0.3cm}\textbf{b}]{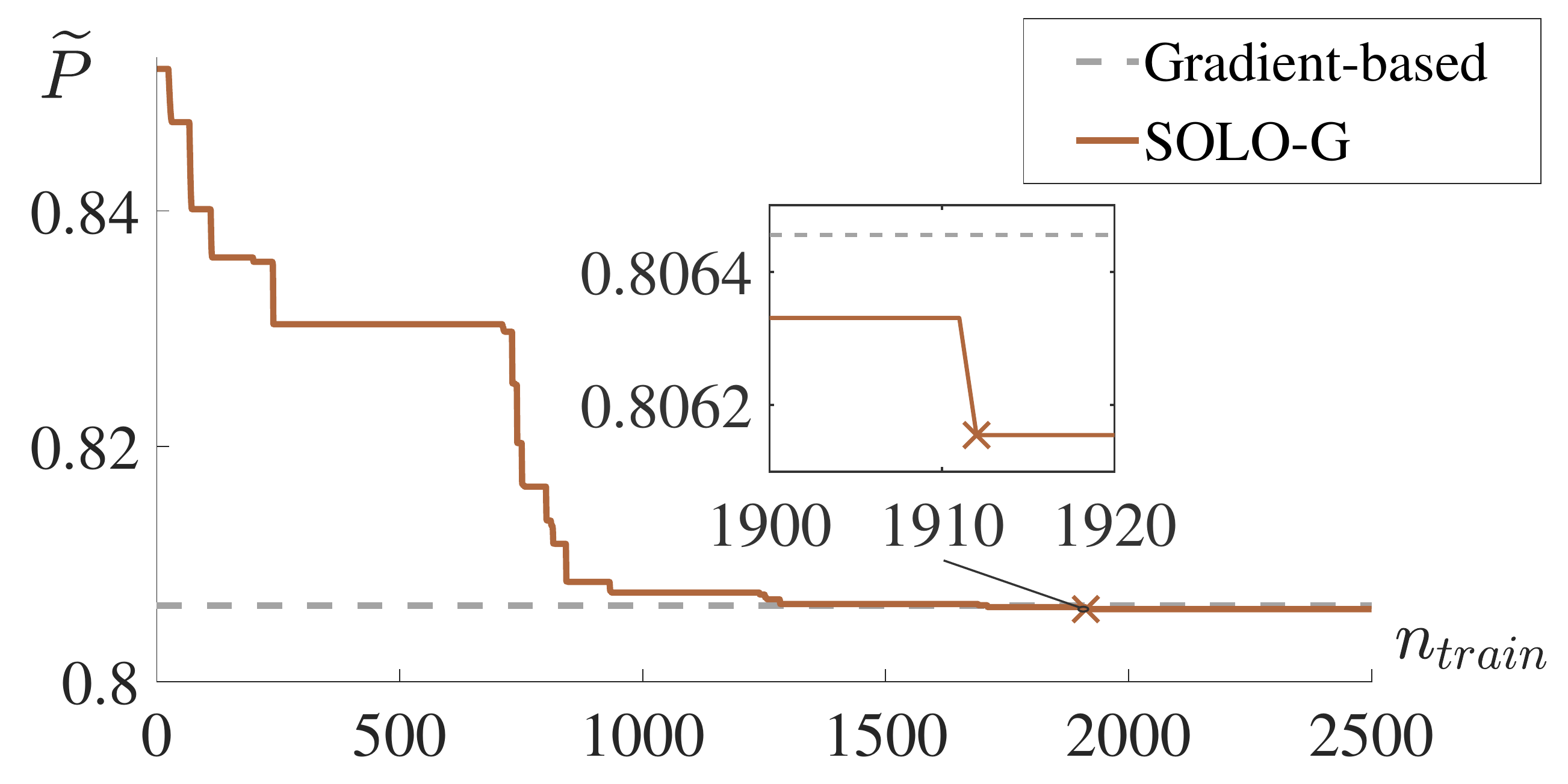}
	\end{minipage}\\
		\vspace{0.2cm}
	\begin{minipage}{0.5\textwidth}\centering
		\parbox[c][1em]{\textwidth}{\textbf{c} \small{Gradient-based $\widetilde{P}=0.8065$}}
		\includegraphics[width=\textwidth]{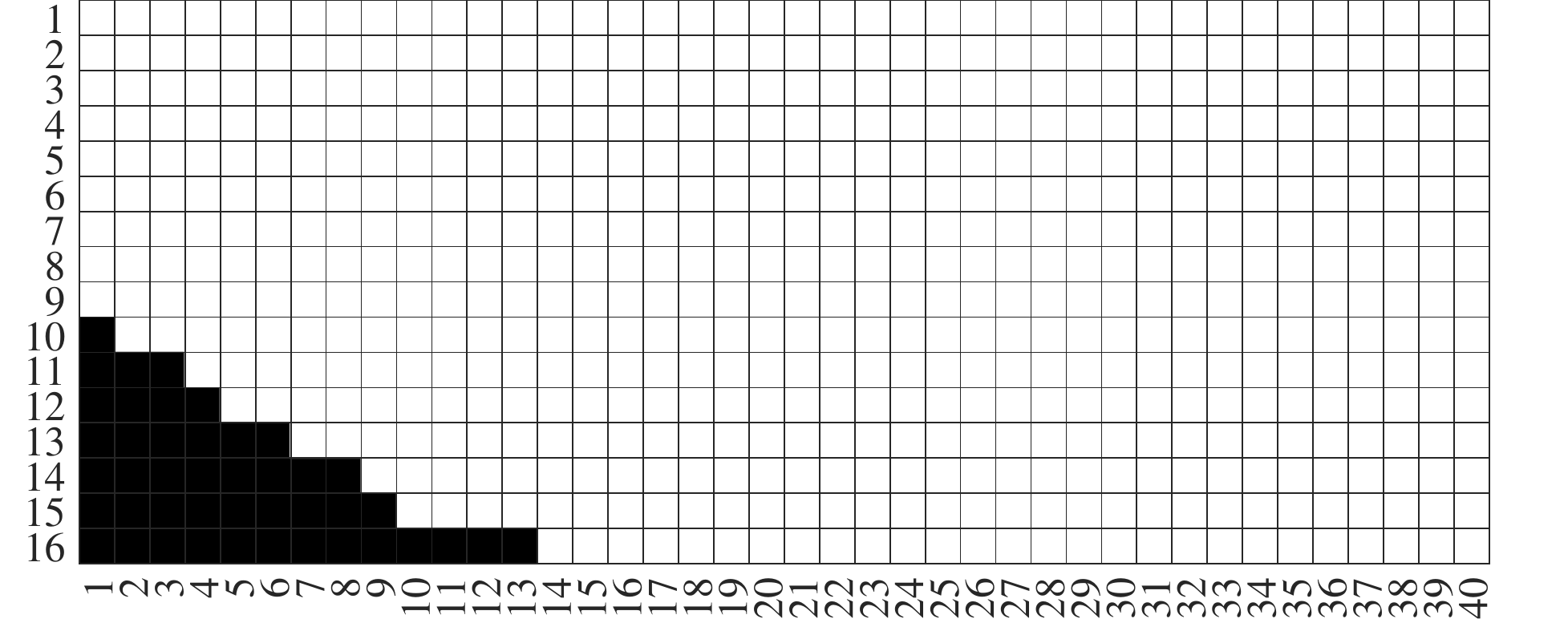}
	\end{minipage}%
	\begin{minipage}{0.5\textwidth}\centering
			\parbox[c][1em]{\textwidth}{\textbf{d} \small{SOLO-G@1,912 $\widetilde{P}=0.8062$}}
		\includegraphics[width=\textwidth]{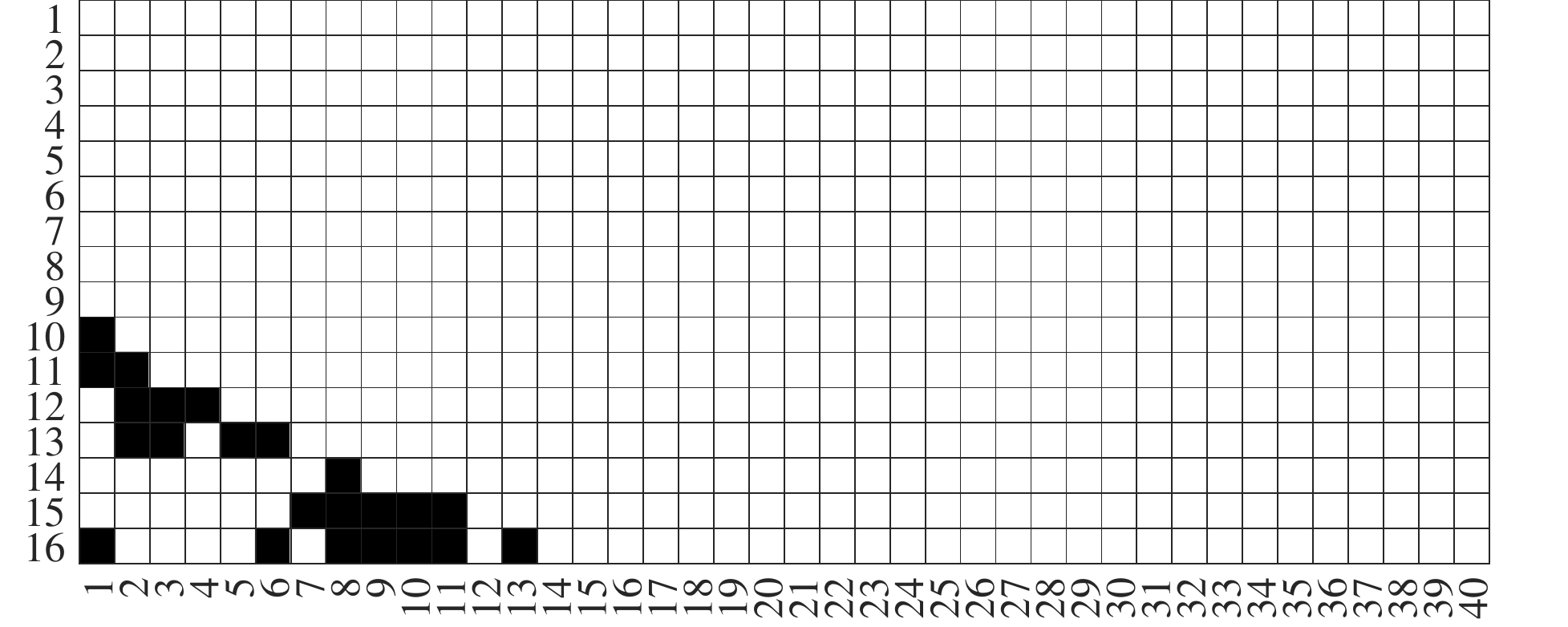}
	\end{minipage}%
	\caption{\textbf{Setup and results of a fluid-structure optimization problem with 40$\times$16 design variables.} \textbf{a}, Problem setup: minimizing pressure drop through the tunnel. \textbf{b},  Dimensionless inlet pressure  versus $n_{train}$, the number of accumulated training samples. SOLO-G denotes a greedy version of our proposed method, where the cross ``X" denotes the convergence point (presented in \textbf{d}). The horizontal dashed line denotes the solution from the gradient-based method.  \textbf{c}, Optimized design obtained by the gradient-based method. $\widetilde{P}=0.8065$. \textbf{d}, Optimized design obtained by SOLO-G. $n_{train}=1,912$ and $\widetilde{P}=0.8062$. In \textbf{c},\textbf{d}, black denotes $\rho=1$ (solid) and white denotes $\rho=0$ (void). The SOLO-G result in \textbf{d} has two gaps at the 7th and 12th columns, while the gradient-based result in \textbf{c} gives a smooth ramp. We try filling the gaps and find that their existence indeed reduces pressure, which demonstrates the powerfulness of our method.}
	\label{fig5}
\end{figure}

As shown in \fig{fig4}b, SOLO-G and SOLO-R converge to the same objective function value $\widetilde{P}=0.9567$ at $n_{train}=286$ and $n_{train}=2,148$ respectively. Their solutions are equivalent, shown in \fig{fig4}d and e. Intermediate solutions from SOLO-G are shown in Supplementary Fig.~4. We obtain the optimum better than the gradient-based method ($\widetilde{P}=0.9569$) after only 286 FEM calculations. For comparison, a recent topology optimization work based on reinforcement learning used the same geometry setup and obtained the same solution as the gradient-based method after thousands of iterations \cite{gaymann2019deep}; our approach demonstrates better performance. Compared with directly using BBA which requires $10^8$ evaluations, SOLO-G reduces FEM calculations by orders of magnitude to about $1/(3\times10^5)$. To account for randomness, we repeat the experiments another four times and the results are similar to \fig{fig4}b (see Supplementary Figs.~5 and 6).

We also apply our algorithm to a finer mesh, with $40\times 16$ design variables (\fig{fig5}a). SOLO-G converges at $n_{train}=1,912$, shown in \fig{fig5}b. Our design (\fig{fig5}d, $\widetilde{P}=0.8062$) is found to be better than the solution from the gradient-based algorithm (\fig{fig5}c, $\widetilde{P}=0.8065$). Intermediate solutions from SOLO-G are shown in Supplementary Fig.~7. Compared with directly using BBA which needs $2\times10^8$ evaluations, SOLO-G reduces the number of FEM calculations to $1/10^5$. Similar trends can be observed when repeating the experiments (see Supplementary Fig.~7). It is interesting to note that the optimum in \fig{fig5}d has two gaps at the 7th and 12th columns. It is a little counterintuitive, since the gradient-based method gives a smooth ramp (\fig{fig5}c). We try filling the gaps and find that their existence indeed reduces pressure (see Supplementary Fig.~8), which demonstrates how powerful our method is.

\textbf{Heat transfer enhancement.} In this example, we would like to solve a complicated problem that gradient-based methods are difficult to address. Phase change materials are used for energy storage by absorbing and releasing latent heat when the materials change phases, typically between solid and liquid. Due to their simple structure and high heat storage capacity, they are widely used in desalination, buildings, refrigeration, solar system, electronic cooling, spacecraft and so forth\cite{kamkari2014experimental}. However, commonly used non-metallic materials  suffer from very low thermal conductivity. A popular solution is to add high conductivity material (such as copper) as fins to enhance heat transfer \cite{desai2020numerical}. Topology optimization is implemented to optimize the geometry of fins. To deal with such transient problems, current gradient-based methods have to simplify the problem by using predetermined time period and fixed boundary conditions \cite{chen2020topology,pizzolato2017topology,iradukunda2020transient,zhao2020topology}. By contrast, in real applications, these conditions depend on user demand and environment, or even couple with the temperature field of the energy storage system \cite{li2018optimization,weng2019optimization,yan2016experimental,arici2020pcm,xu2020numerical}. Therefore, problems with more complex settings need to be addressed.

\begin{figure}[!tb]
	\centering
	\begin{minipage}{0.3275\textwidth}\centering
		\xincludegraphics[height=5 cm,label=\hspace{-0.5em}\textbf{a}]{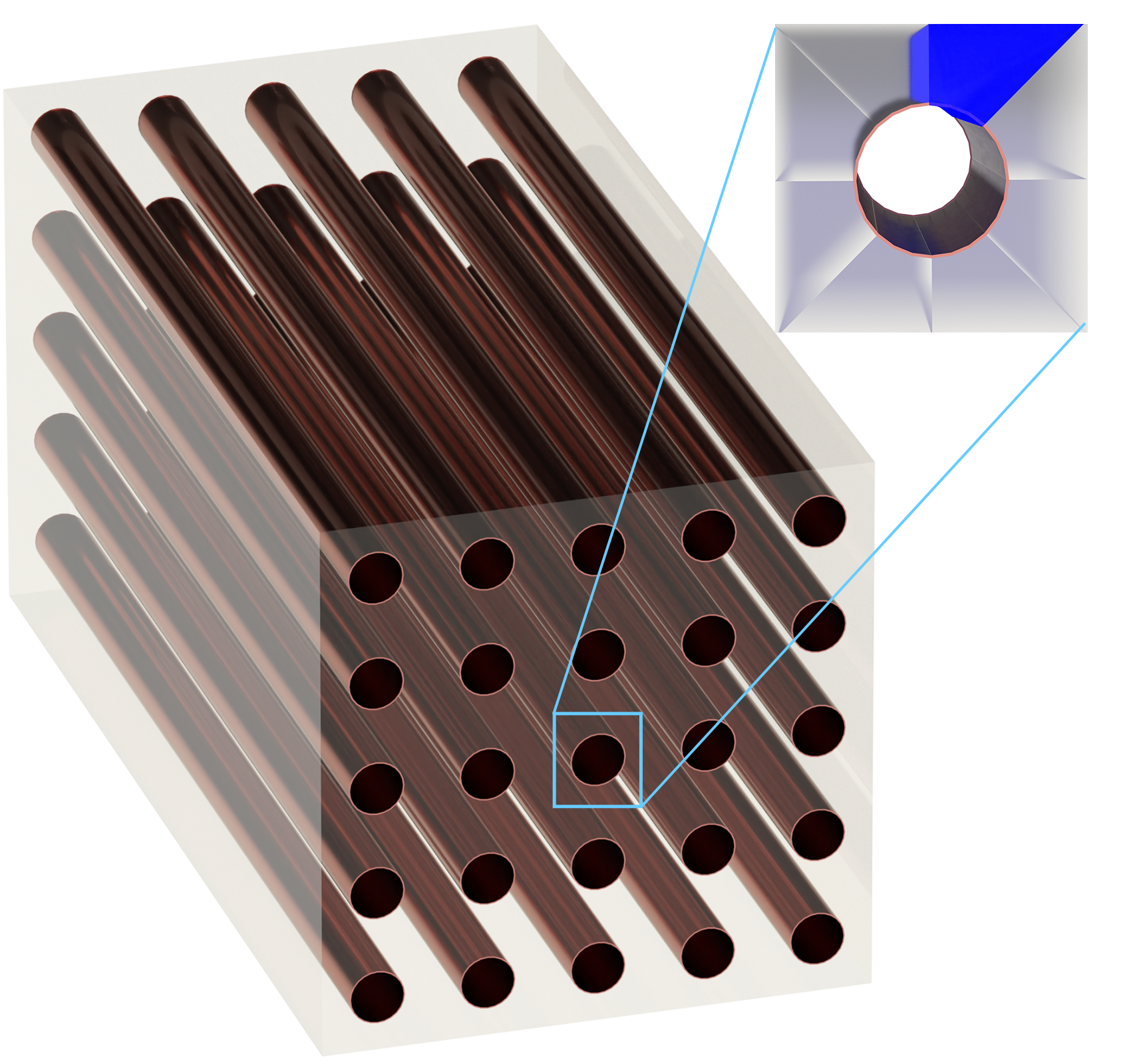}
	\end{minipage}%
		\begin{minipage}{0.3275\textwidth}\centering
		\xincludegraphics[height=5 cm,label=\hspace{-0.1em}\textbf{b}]{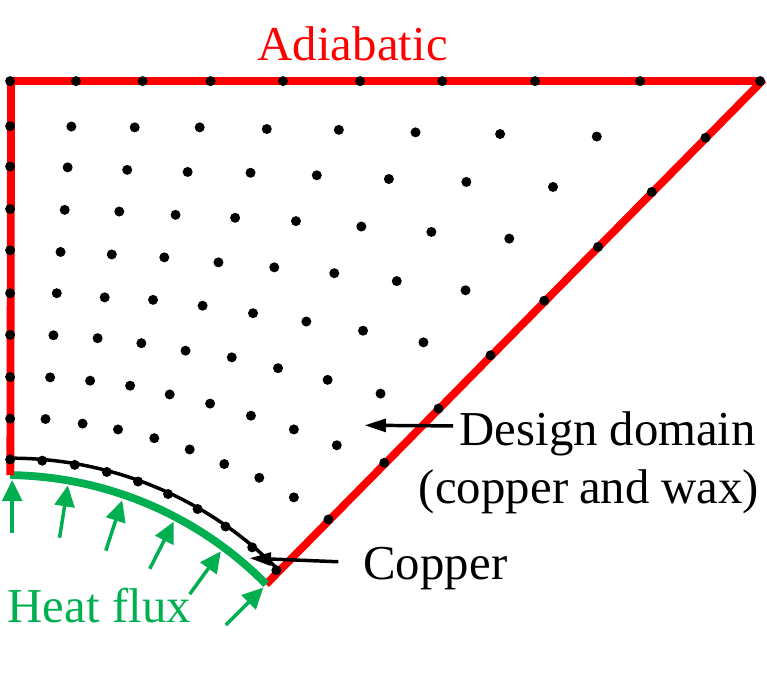}
	\end{minipage}%
	\begin{minipage}{0.345\textwidth}\centering
		\xincludegraphics[height=5 cm,label=\textbf{c}]{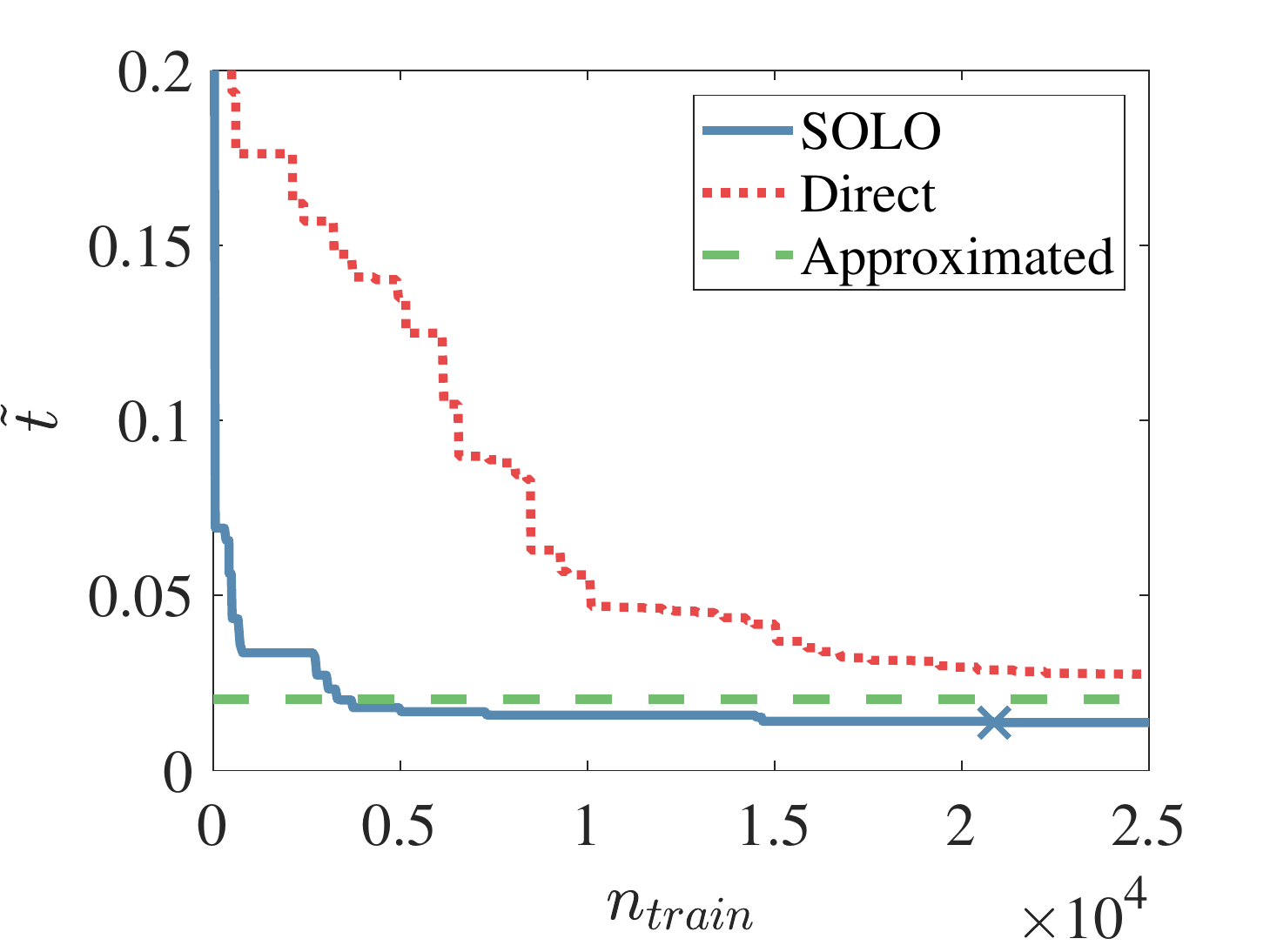}
	\end{minipage}\\
		\vspace{0.2cm}
	\begin{minipage}{0.33\textwidth}\centering
		\parbox[c][1em]{\textwidth}{\textbf{d} \small{SOLO@20,860 $\widetilde{t}=0.0137$}}
		\includegraphics[width=\textwidth]{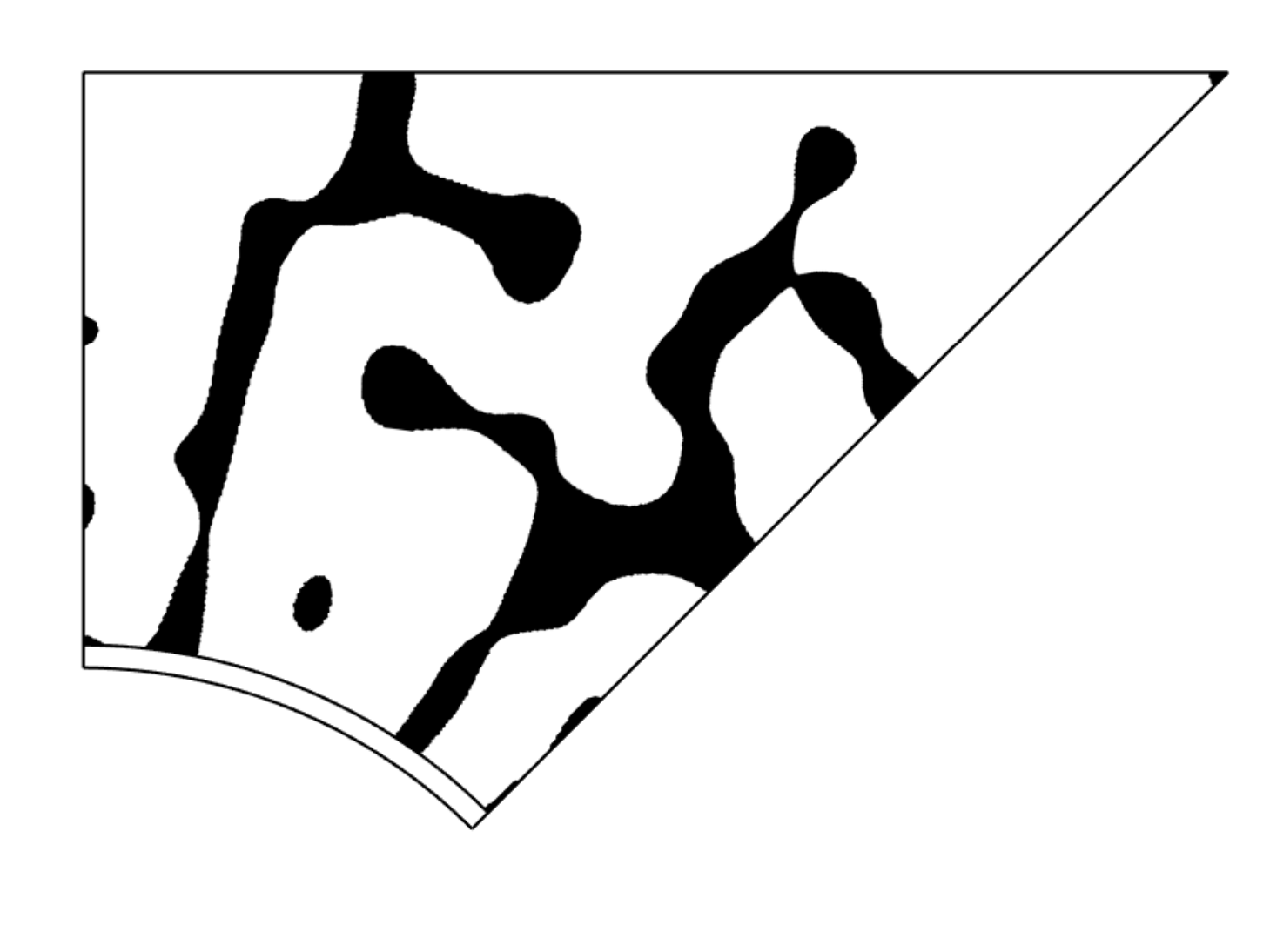}
	\end{minipage}%
	\begin{minipage}{0.33\textwidth}\centering
			\parbox[c][1em]{\textwidth}{\textbf{e} \small{Direct@24,644 $\widetilde{t}=0.0275$}}
		\includegraphics[width=\textwidth]{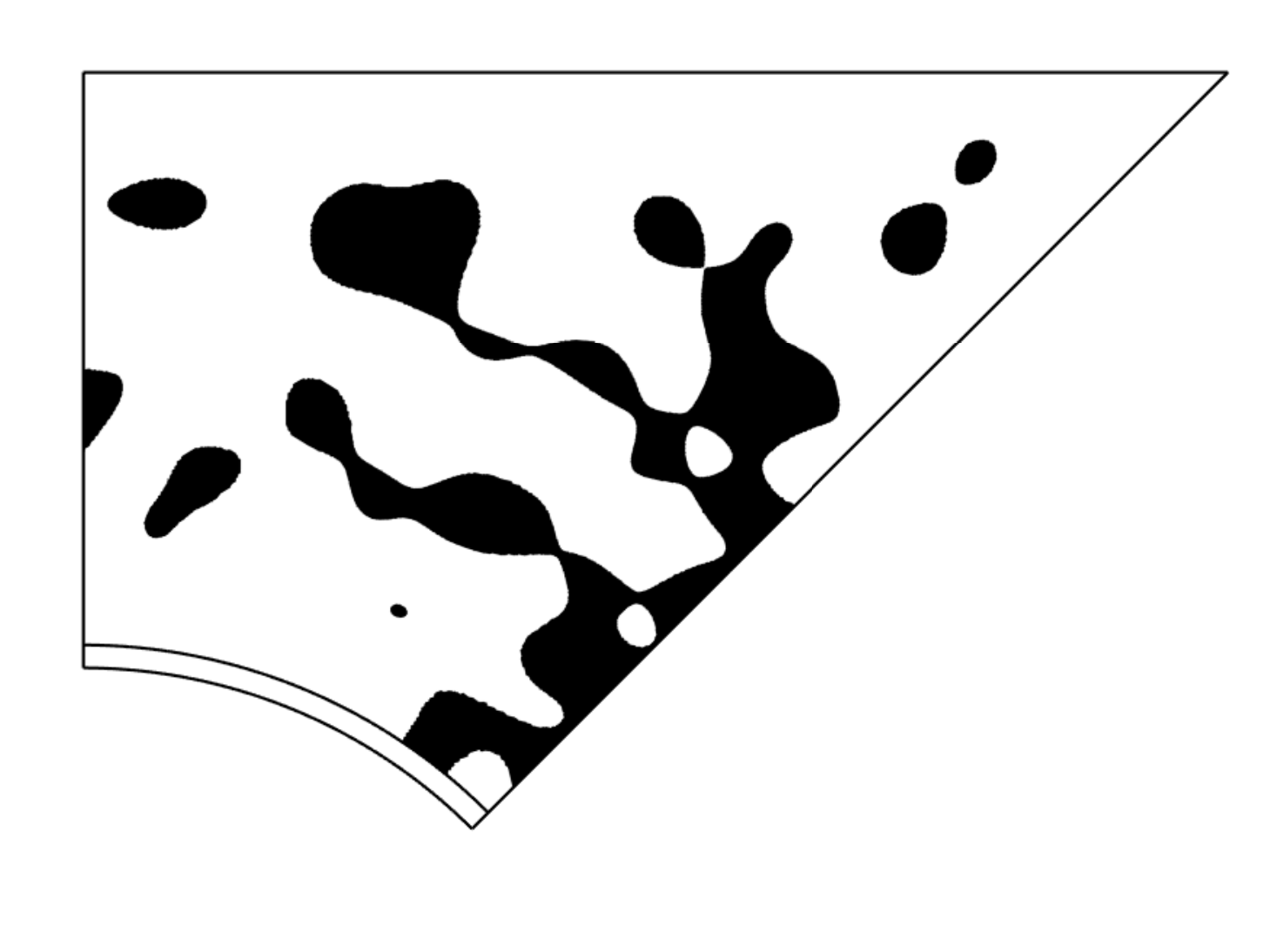}
	\end{minipage}%
		\begin{minipage}{0.33\textwidth}\centering
			\parbox[c][1em]{\textwidth}{\textbf{f} \small{Approximated $\widetilde{t}=0.0203$}}
		\includegraphics[width=\textwidth]{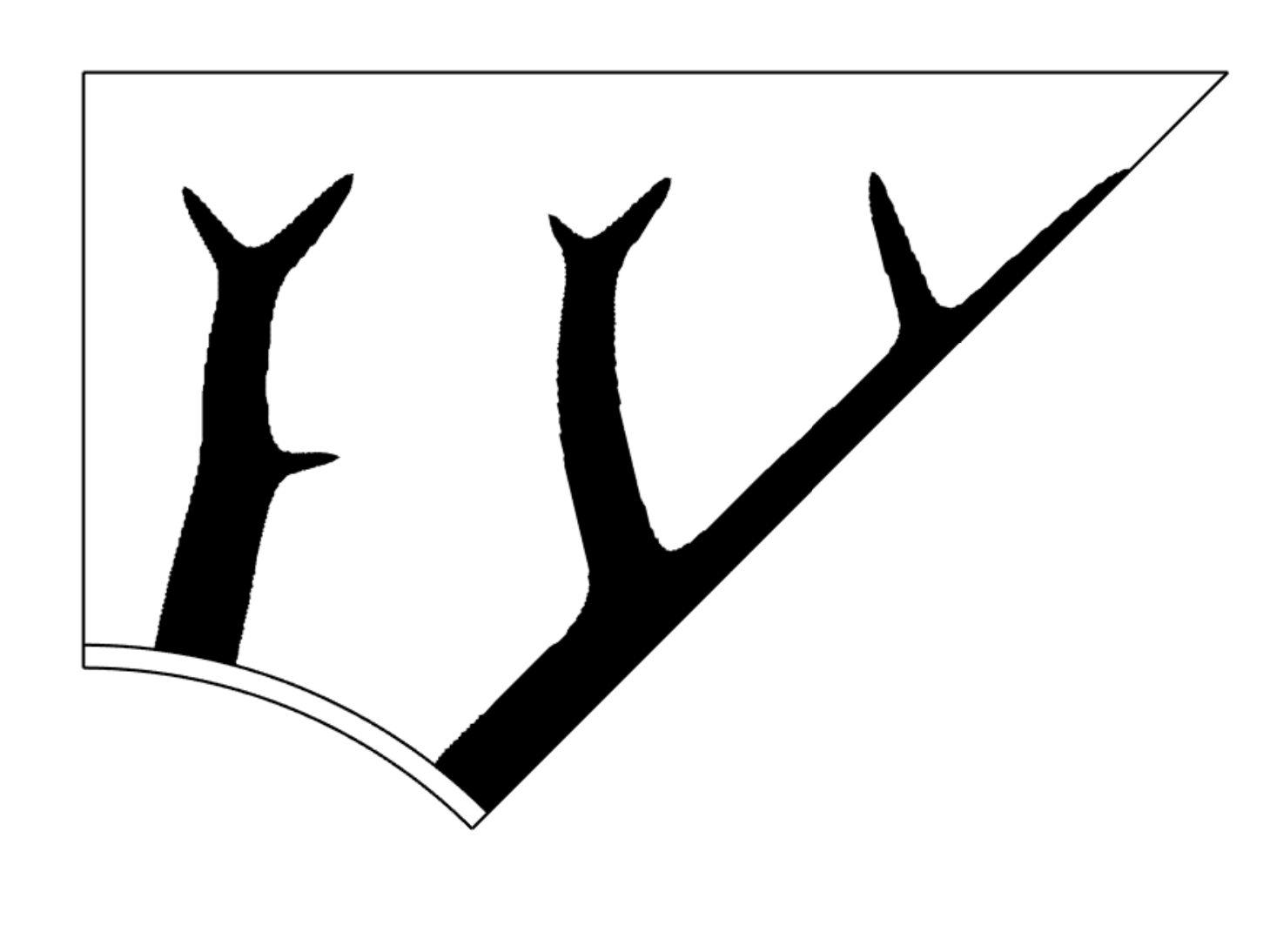}
	\end{minipage}%
	\caption{\textbf{Setup and results of a heat transfer enhancement problem with 10$\times$10 design variables.} \textbf{a}, Engineering background: a group of copper pipes are inserted in a phase change material. Because of symmetry, we only need to consider 1/8 of the unit cell (dark blue area in the top right corner). \textbf{b}, Problem setup: minimizing the time to charge the system with a given amount of heat, subject to heat flux, temperature and volume constraints. The black dots denote locations of design variables.  \textbf{c},  Dimensionless charging time  versus $n_{train}$, the number of accumulated training samples. SOLO denotes our proposed method, where the cross ``X" denotes the convergence point (presented in \textbf{d}). ``Direct" denotes solving the problem directly by gradient descent. ``Approximated" denotes simplifying this problem to a steady state problem.  \textbf{d}, Optimized design obtained by SOLO. $\widetilde{t}=0.0137$. \textbf{e}, Optimized design obtained by ``Direct". $n_{train}=24,644$ and $\widetilde{t}=0.0275$. \textbf{f}, Optimized design obtained by ``Approximated". $\widetilde{t}=0.0203$. In \textbf{d}-\textbf{f}, black denotes $\rho=1$ (copper) and white denotes $\rho=0$ (wax). The SOLO result in \textbf{d} has islands isolated from major branches, while the ``Approximated" result in \textbf{f} gives a connected structure. We try combining the islands to be part of major branches and find that the existence of isolated islands indeed reduces time, which demonstrates the powerfulness of our method. }
	\label{fig6}
\end{figure}

We consider a heat absorption scenario where time is variant and the boundary condition is coupled with the temperature field. As shown in \fig{fig6}a, copper pipes containing heat source are inserted in a phase change material, paraffin wax RT54HC \cite{yu2018effect}; the heat source can be fast-charging batteries for electric vehicles or hot water for residential buildings. Considering symmetry, the problem is converted to a 2D problem in \fig{fig6}b. We fill the domain with wax to store heat and with copper to enhance heat transfer. The material distribution $\rho(\mathbf{x})\in\{0,1\}$ (1 being copper and 0 being wax) is represented by a 10$\times$10 mesh. Specifically, a continuous function is interpolated by Gaussian basis functions from the 10$\times$10 design variables and then converted to binary values by a threshold (see Methods for details). Our goal is to find the optimal $\bm{\rho}$ to minimize the time to charge the system with a given amount of heat
\begin{equation}\label{eq:heat-problem}
  	\min\limits_{\bm{\rho}\in[0,1]^N}  \widetilde{t}(\bm{\rho})=\frac{t(\bm{\rho})}{t(\bm{\rho}_O)},
\end{equation}
where $N=100$, $\bm{\rho}_O=[0,0,..,0]^T$ means no copper inside the design domain, and $t(\bm{\rho})$ is the time to charge the system with $Q_0$ amount of heat, expressed by
\begin{equation}\label{eq:10}
  	\int_0^{t(\bm{\rho})} q(\bm{\rho},t) \mathrm{d}t = Q_0,
\end{equation}
subject to the maximum heat flux constraint at the boundary (green curve in \fig{fig6}b)
\begin{equation}\label{eq:11}
  	q(\bm{\rho},t) \leq q_0,
\end{equation}
the constraint of maximum temperature of the domain,
\begin{equation}
  	T(\bm{\rho},q,\mathbf{x},t)\leq T_0,
\end{equation}
and given copper usage, i.e., the volume constraint of copper,
\begin{equation}\label{eq:13}
  	\frac{\int_{\Omega}\rho(\mathbf{x})\mathrm{d}\mathbf{x}}{\int_{\Omega}\mathrm{d}\mathbf{x}}= 0.2.
\end{equation}

Here $Q_0$, $q_0$ and $T_0$ are preset constants. Obviously, the bottom left boundary (inner side of copper pipes) has the highest temperature during charging, thus we only need to consider the temperature constraint at this boundary. Physically, there are one or two charging steps: the system is charged at heat flux $q_0$ until the boundary temperature reaches $T_0$ or the total heat flow reaches $Q_0$ (whichever first), and if it is the former case, the heat flux is reduced to maintain the boundary temperature at $T_0$ until the total heat flow requirement is satisfied. In practice, we choose parameters such that the system will go through two steps for sure. 

To solve the problem with objective \eq{eq:heat-problem} and constraints in Eqs.\eqref{eq:11}-\eqref{eq:13}, our method SOLO is initialized by 500 random samples to train a DNN. Bat Algorithm (BA) is then used to find the minimum of the DNN, based on which additional 200 samples are generated in each loop by mutation and convolution. Two gradient-based methods are used as baselines to compare with our algorithm: one is to solve Problem \eqref{eq:heat-problem}-\eqref{eq:13} directly by gradient descent, denoted as ``Direct"; the other is to simplify this problem to a steady state problem \cite{zhao2020topology}, denoted as ``Approximated". In \fig{fig6}c, SOLO converges at $n_{train}=20,860$ (marked by a cross ``X") with lower $\widetilde{t}$ than other methods. It appears counter-intuitive that the solution of SOLO, shown in \fig{fig6}d, has some copper islands isolated from major branches. We tried removing these islands and adding more copper materials to the major branches to maintain copper volume, yet the variants showed worse performance, as shown in Supplementary Fig.~10. ``Direct" gives the worst solution in \fig{fig6}e. ``Approximated" yields a good solution with a tree structure, as shown in \fig{fig6}f; since it does not solve the same problem as the other two methods, we do not consider its relation with $n_{train}$ and represent it by a horizontal line in \fig{fig6}c. 

Our method gives a good solution after 20,860 FEM calculations, while BA is estimated to need $4\times10^8$ calculations. In summary, our method outperforms the other two methods and reduces the number of FEM calculations by over four orders of magnitude compared with BA. 


\textbf{Truss optimization.} In this example, we test the scalability of SOLO with over a thousand design variables. Also, we will compare it with a heuristic method, BA, to provide direct evidence that SOLO can reduce the number of FEM computations by over two orders of magnitude. 

\begin{figure}[!tb]
	\centering
		\begin{minipage}{0.16\textwidth}
	    		\xincludegraphics[width=\textwidth,label=\textbf{a}]{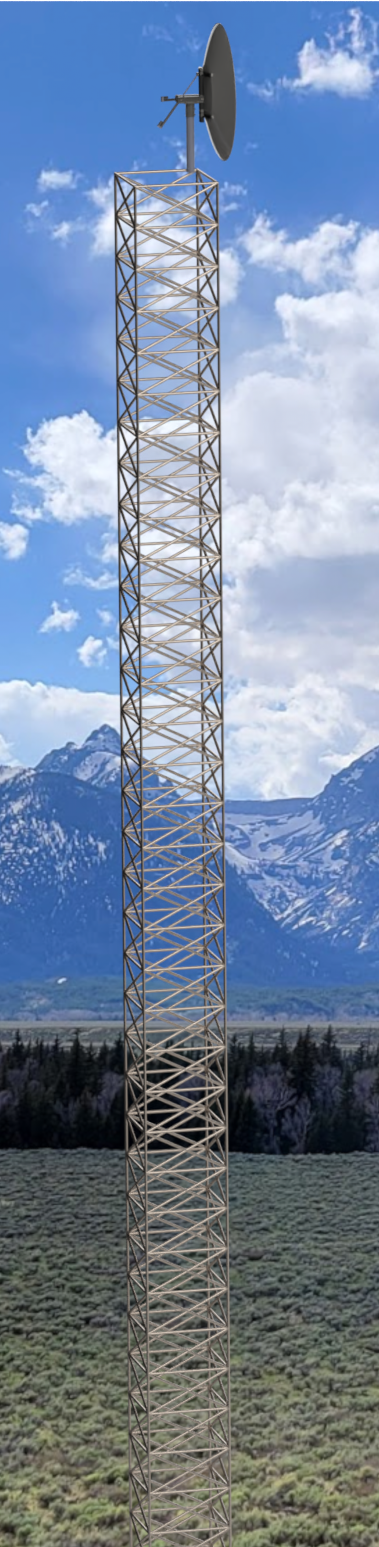}
	\end{minipage}
	\begin{minipage}{0.246\textwidth}
	    		\xincludegraphics[width=\textwidth,label=\textbf{b}]{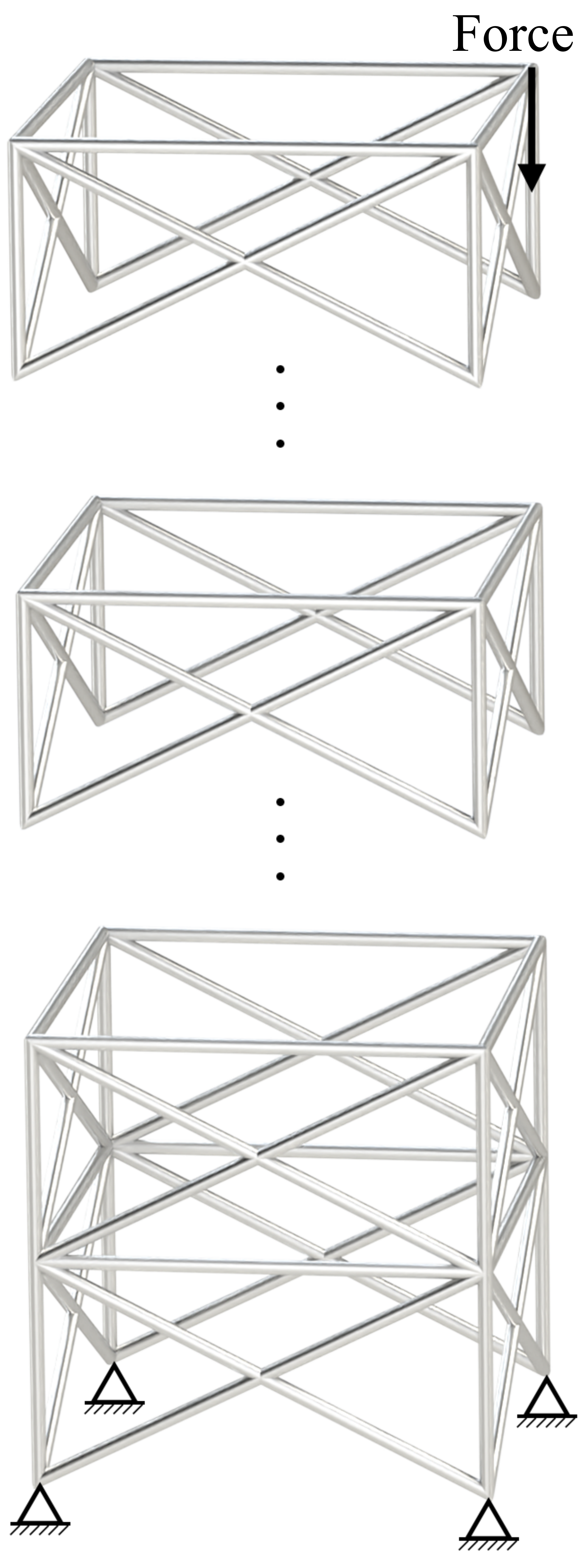}
	\end{minipage}
	\begin{minipage}{0.5\textwidth}\centering
		\parbox[c][1em]{\textwidth}{\textbf{c} \small{72 bars}}
		\includegraphics[width=\textwidth]{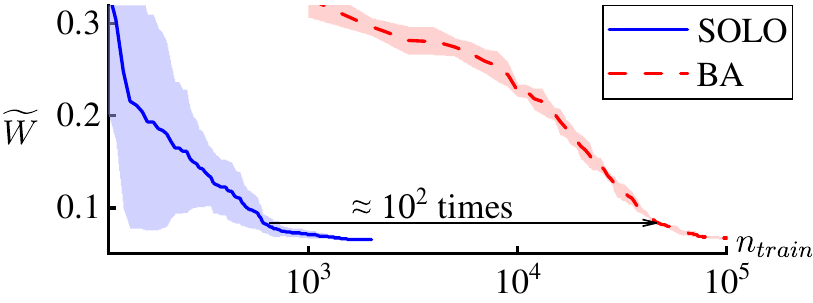}
			\parbox[c][1em]{\textwidth}{\textbf{d} \small{432 bars}}
		\includegraphics[width=\textwidth]{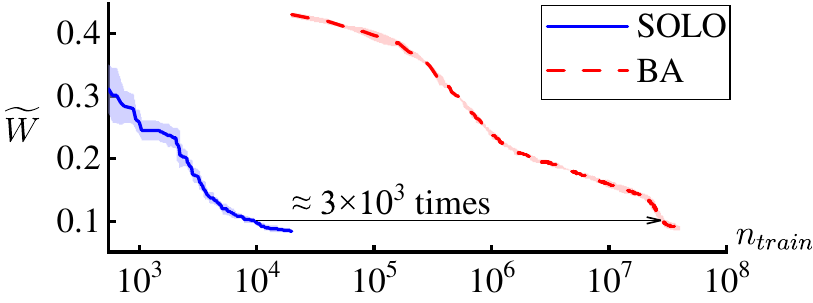}
			\parbox[c][1em]{\textwidth}{\textbf{e} \small{1,008 bars}}
		\includegraphics[width=\textwidth]{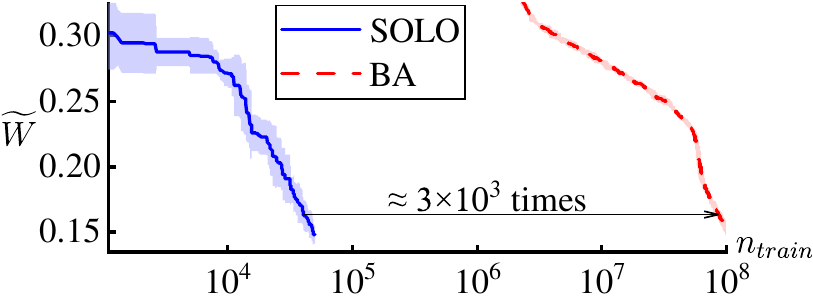}
	\end{minipage}%
	\caption{\textbf{Setup and results of three truss optimization problems with different numbers of bars (equal to the numbers of design variables).} \textbf{a}, Illustration of an antenna tower, an exemplary application of truss structures. \textbf{b}, Illustration of the problem setup: minimizing total weight through changing the size of each bar, subject to stress and displacement constraints. The block is repeated until the given number of bars is reached. \textbf{c}-\textbf{e}, Dimensionless weight $\widetilde{W}$  versus the number of accumulated training samples $n_{train}$. SOLO denotes our proposed method. BA denotes Bat Algorithm. The numbers of bars for these three sub-figures are 72, 432 and 1,008, respectively. Each experiment is repeated five times; the curves denote the mean and the shadows denote the standard deviation. }
	\label{fig7}
\end{figure}

Truss structures are widely used in bridges, towers, buildings and so forth. An exemplary application, an antenna tower, is shown in \fig{fig7}a. Researchers have been working on optimizing truss structures from different perspectives. 
A classic truss optimization benchmark problem is to optimize a structure with 72 bars \cite{gomes2011truss,farshchin2016multi,perez2007particle,camp2014design}, as shown in \fig{fig7}b with four repeated blocks, so as to minimize the weight of the bars subject to displacement and tension constraint. Following this benchmark problem, we set the goal to optimize the size of each bar (the bars can all have different sizes) to minimize total dimensionless weight
\begin{equation}\label{eq:14}
  	\min\limits_{\bm{\rho}\in\{a_1,a_2,...,a_{16}\}^N}  \widetilde{W}(\bm{\rho})=\frac{W(\bm{\rho})}{W(\bm{\rho}_{max})}=\frac{\sum_{i=1}^N\rho_iL_i\gamma_i}{W(\bm{\rho}_{max})},
\end{equation}
where $\rho_i$, $L_i$ and $\gamma_i$ are the cross-sectional area, length, and unit weight of the $i$-th bar, respectively; $\bm{\rho}_{max}$ uses the largest cross-sectional area for all bars; $N=72$ is the number of bars. Each bar is only allowed to choose from 16 discrete cross-sectional area values $a_1,a_2,...,a_{16}$, to represent standardized components in engineering applications. The tension constraint requires all bars to not exceed the maximum stress
\begin{equation}\label{eq:15}
  	|\sigma_i|\leq \sigma_0,\quad i=1,2,...,N.
\end{equation}
The displacement constraint is applied to the connections of the bars: the displacement in any direction is required to be lower than a threshold
\begin{equation}\label{eq:16}
  	||\Delta \mathbf{x}_i||_{\infty}\leq \delta_0,\quad i=1,2,...,N_c,
\end{equation}
where $N_c$ is the number of connections. 

Now we have an optimization problem with objective \eq{eq:14} subject to stress constraint \eq{eq:15} and displacement constraint \eq{eq:16}. In addition to the popular 72-bar problem, we add more repeated blocks to the structure to generate two more problems, with 432 and 1,008 bars. Geometric symmetry is not considered while solving the problems. Therefore, the design space goes up to $16^{1,008}\approx 10^{1,214}$, which is extremely huge. For the three problems, SOLO is initialized by 100, 500 and 1,000 samples, respectively. The number of incremental samples per loop is 10\% of the initialization samples. 10\% of incremental samples are the optima obtained by BA based on the DNN's prediction, and the rest 90\% are generated by mutation of the best solution predicted by the DNN.  

The results are shown in \fig{fig7}c-e. To reach the same objective weight, BA needs over $10^2$ times of calculations of SOLO. The difference becomes even larger when the number of variables increases. These examples demonstrate the scalability of SOLO by showing higher efficiency in computation, especially with a large number of design variables.

\subsection*{Discussion}

Topology optimization is an important problem with broad applications in many scientific and engineering disciplines. Solving non-linear high-dimensional optimization problems requires non-gradient methods, but the high computational cost is a major challenge. We proposed an approach of self-directed online learning optimization (SOLO) to dramatically accelerate the optimization process and make solving complex optimization problems possible.

We demonstrated the effectiveness of the approach in solving eight problems of four types, i.e., two compliance minimization problems, two fluid-structure optimization problems, a heat transfer enhancement problem and three truss optimization problems. For the compliance problems with 25  and 121 continuous design variables, our approach converged and produced optimized solutions same as the known optima with only 501 and 10,243 FEM calculations, respectively, which are about 1/400 of directly using GSA and FEM without DNN based on our estimation. For the fluid problems with 160 and 640 binary variables, our method (SOLO-G) converged after 286 and 1,912 FEM calculations, respectively, with solutions better than the benchmark. It used less than $1/10^5$ of FEM calculations compared with directly applying BBA to FEM, and converged much faster than another work based on reinforcement learning. In the heat transfer enhancement example, we investigated a complicated, transient and non-linear problem. Our method gave a solution that outperformed other baselines after 20,860 FEM calculations, which was estimated to be four orders of magnitude less than BA. Similar to other SMBO methods, overhead computation was introduced (by training DNNs and finding their optima), but it was almost negligible (see the time profile in Supplementary Table 1) which is attractive for real-world applications where new designs want to be developed and tested. In these examples, we estimated the amount of computation of directly using heuristic algorithms, which showed that our approach led to $2\sim5$ orders of magnitude of computation reduction. In addition to this estimation, we applied BA to the original objectives in the three truss optimization problems and observed $2\sim4$ orders of magnitude of calculation reduction using our approach. 


Our algorithm is neat and efficient, and has great potential for large-scale applications. We bring a new perspective for high-dimensional optimization by embedding deep learning in optimization methods. More techniques, such as parallel FEM computation, uncertainty modeling and disturbance based on sensitivity analysis, can be incorporated to enhance the performance.

\section*{Methods}

\textbf{Enforcement of volume constraint.} Compliance and heat transfer problems have volume constraints. The latter will be detailed in Section \textit{Interpolation of design variables}, thus we only discuss the former here. In the two compliance problems, all matrices representing the density distribution $\bm{\rho}$  have the same weighted average $\sum_{i=1}^N{w_i}\rho _i=V_0$  due to the volume constraint where $w_i$  denotes the weight of linear Gaussian quadrature. A matrix from the initial batch is generated by three steps: 
\begin{enumerate}
	\item Generate a random matrix with elements uniformly distributed from 0 to 1.
	\item Rescale the array to enforce the predefined weighted average.
	\item Set the elements greater than one, if any, to one and then adjust those elements less than one to maintain the average. 
\end{enumerate}

Matrices for the second batch and afterwards add random disturbance to optimized solutions $\hat{\bm{\rho}}$  and then go through \textit{Step 2} and \textit{3} above to make sure the volume satisfies the constraint. 
	
\textbf{Finite Element Method (FEM) and gradient-based baselines.} The objective function values of material designs are calculated by FEM as the ground truth to train the DNN. In the compliance and fluid problems, the meshes of FEM are the same as the design variables. In the heat problem, the meshes are finer. Numerical results are obtained by COMSOL Multiphysics 5.4 (except the truss problems). Solutions from gradient-based methods (including ``Approximated" ) are all solved by MMA via COMSOL with optimality tolerance as 0.001. In the fluid problems, the gradient-based baseline method produces a continuous array, and we use multiple thresholds to convert it to binary arrays and recompute their objective (pressure) to select the best binary array. In the heat problem, the ``Approximated" method uses the same resolution as the other two methods (SOLO and ``Direct") for a fair comparison. Specifically, we apply a Helmholtz filter\cite{lazarov2011filters}, whose radius is half of the minimum distance of two design variable locations, to yield a mesh-independent solution. The solution is a continuous array; we use a threshold to convert it to a binary array which satisfies the volume constraint in \eq{eq:14}. 

\textbf{Interpolation of design variables.} In the two compliance problems and the heat problem, we use a  vector (or matrix) $\bm{\rho}$ to represent a spacial function $\rho(\mathbf{x})$. Interpolation is needed to obtain the function $\rho(\mathbf{x})$ for FEM and plotting. Given design variables $\bm{\rho}=[\rho_1,\rho_2,...,\rho_N]^T$, we get the values $\rho(\mathbf{x})$ by two interpolation methods. For the compliance problems, we use bilinear interpolation\cite{han2013comparison}. Suppose $\mathbf{x}=(x,y)$ is within a rectangular element whose nodal coordinates are $(x_1,y_1),(x_1,y_2),(x_2,y_1),(x_2,y_2)$, the interpolated function value can be calculated by
\begin{equation}\label{bilinear}
    \rho(x,y)= \frac{\begin{bmatrix}x_2-x & x-x_1\end{bmatrix}
\begin{bmatrix}
F(x_1,y_1) & F(x_1,y_2)\\ F(x_2,y_1) & F(x_2,y_2) \end{bmatrix}\begin{bmatrix}
y_2-y\\y-y_1
\end{bmatrix}}{(x_2-x_1)(y_2-y_1)}.
\end{equation}

For the heat problem, a continuous function  $\bar{\rho}(\mathbf{x})\in[0,1]$ (which will later be converted to a binary function which takes 0 or 1) is interpolated by Gaussian basis functions \cite{guirguis2018high,luo2020topology}:
\begin{equation}\label{interpolate}
    \bar{\rho}(x,y)= \sum_{i=1}^N \lambda_i \phi(\mathbf{x},\mathbf{x}_i)+a_0+a_1x+a_2y,
\end{equation}
where $\phi(\mathbf{x},\mathbf{x}_i)=e^{-(\mathbf{x}-\mathbf{x}_i)^2/d^2}$ ($d$ is a preset distance), and $\lambda_i, a_0, a_1, a_2$ are parameters to be determined. 
The following constraints are needed to guarantee a unique solution 
\begin{equation}
    \sum_{i=1}^N \lambda_i =0,  \sum_{i=1}^N \lambda_ix_i=0,  \sum_{i=1}^N \lambda_iy_i=0.
\end{equation}

Expressing the above equations by a matrix form, we have

\begin{equation}
	\begin{bmatrix}
	\phi \left( \mathbf{x}_1,\mathbf{x}_1 \right)&		\dots&		\phi \left( \mathbf{x}_1,\mathbf{x}_N \right)&		1&		x_1&		y_1\\
	\vdots&		\ddots&		\vdots&		\vdots&		\vdots&		\vdots\\
	\phi \left( \mathbf{x}_N,\mathbf{x}_1 \right)&		\dots&		\phi \left( \mathbf{x}_N,\mathbf{x}_N \right)&		1&		x_N&		y_N\\
	1&		\dots&		1&		0&		0&		0\\
	x_1&		\dots&		x_N&		0&		0&		0\\
	y_1&		\dots&		y_N&		0&		0&		0\\
\end{bmatrix}
\begin{bmatrix}
	\lambda_1\\
	\vdots\\
	\lambda_N\\
	a_0\\
	a_1\\
	a_2\\
\end{bmatrix}  =\begin{bmatrix}
	\rho _1\\
	\vdots\\
	\rho _N\\
	0\\
	0\\
	0\\
\end{bmatrix},
\end{equation}
abbreviated as $\bm\Phi \bm\lambda = \begin{bmatrix}
\bm{\rho}\\\mathbf{0}
\end{bmatrix}$. We get $\bm\lambda=\bm\Phi^{-1}\begin{bmatrix}
\bm{\rho}\\\mathbf{0}
\end{bmatrix}$ and interpolate $\bar{\rho}(\mathbf{x})$ by \eq{interpolate}. Then we set a threshold $\rho_{thres}$ to convert the continuous function $\bar{\rho}(\mathbf{x})$ to a binary one  $\rho(\mathbf{x})\in\{0,1\}$, i.e., $\rho(\mathbf{x})=1$ if $\bar{\rho}(\mathbf{x})\ge \rho_{thres}$ and $\rho(\mathbf{x})=0$ otherwise. The threshold $\rho_{thres}$ is controlled to satisfy the copper volume constraint \eq{eq:13}.

\begin{figure}[!htb]
	\centering
	\includegraphics[width=0.7\linewidth]{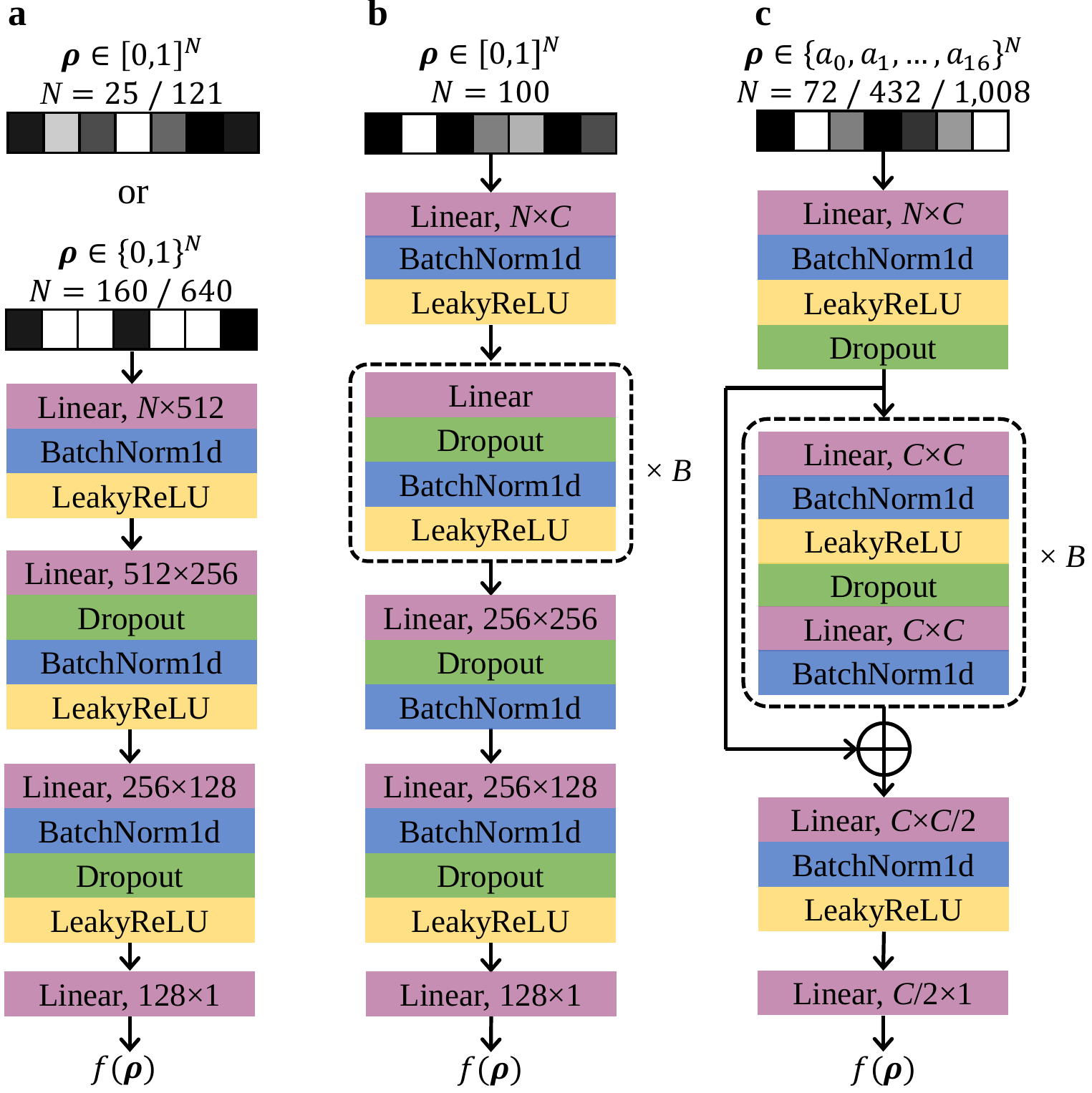}
	\caption{\textbf{Architectures of DNN.} The input is a design vector $\bm{\rho}$ and the output is the predicted objective function value $f(\bm{\rho})$.``Linear'' presents a linear transformation and ``BatchNorm1d'' denotes one-dimensional batch normalization used to avoid internal covariate shift and gradient explosion for stable training~\cite{ioffe2015batch}. 
	``LeakyReLU'' is an activation function extended from ReLU with activated negative values. ``Dropout'' is a regularization method to prevent overfitting by randomly masking nodes~\cite{JMLR:v15:srivastava14a}. \textbf{a}, the DNN in the compliance and fluid problems. \textbf{b}, the DNN in the heat problem. Two architectures are used in this problem. At the 100th loop and before, $B=1$, $C=512$, and the Linear layer in the dashed box is $512\times 256$. At the 101st loop and afterwards, $B=4$, $C=512$ and the 4 Linear layers are $256\times512$, $512\times512$, $512\times512$ and $512\times256$, respectively. \textbf{c}, the DNN in the truss optimization problems. $B=1$. $C=512$ when $N=72/432$; $C=1,024$ when $N=1,008$.}
	\label{fig8}
\end{figure}

\textbf{Deep Neural Network (DNN).} The architectures of the DNN used in this paper is presented in \fig{fig8}. The design variable $\bm{\rho}$ is flattened to a 1D vector as the input to DNN. All inputs are normalized before training and we introduce batch normalization (BN)\cite{ioffe2015batch} within the network as regularization. The output of DNN is reciprocal of the objective function (energy, pressure, charging time or weight) to give better resolution at lower objective values. For the rest of this paper, we regard the DNN to approximate the objective function for simplicity. To optimize the DNN training process, we apply ADAM\cite{kingma2014adam} as the optimizer implemented on the platform of PyTorch 1.8.0\cite{paszke2017automatic}. The learning rate is 0.01. The loss function is set as Mean Square Error (MSE)\cite{lehmann1998}. All models are trained for 1,000 epochs with a batch size of 1,024 (if the number of training data is less than 1,024, all the data will be used as one batch).

\begin{figure}[!t]
	\centering
	\includegraphics[width=0.8\textwidth]{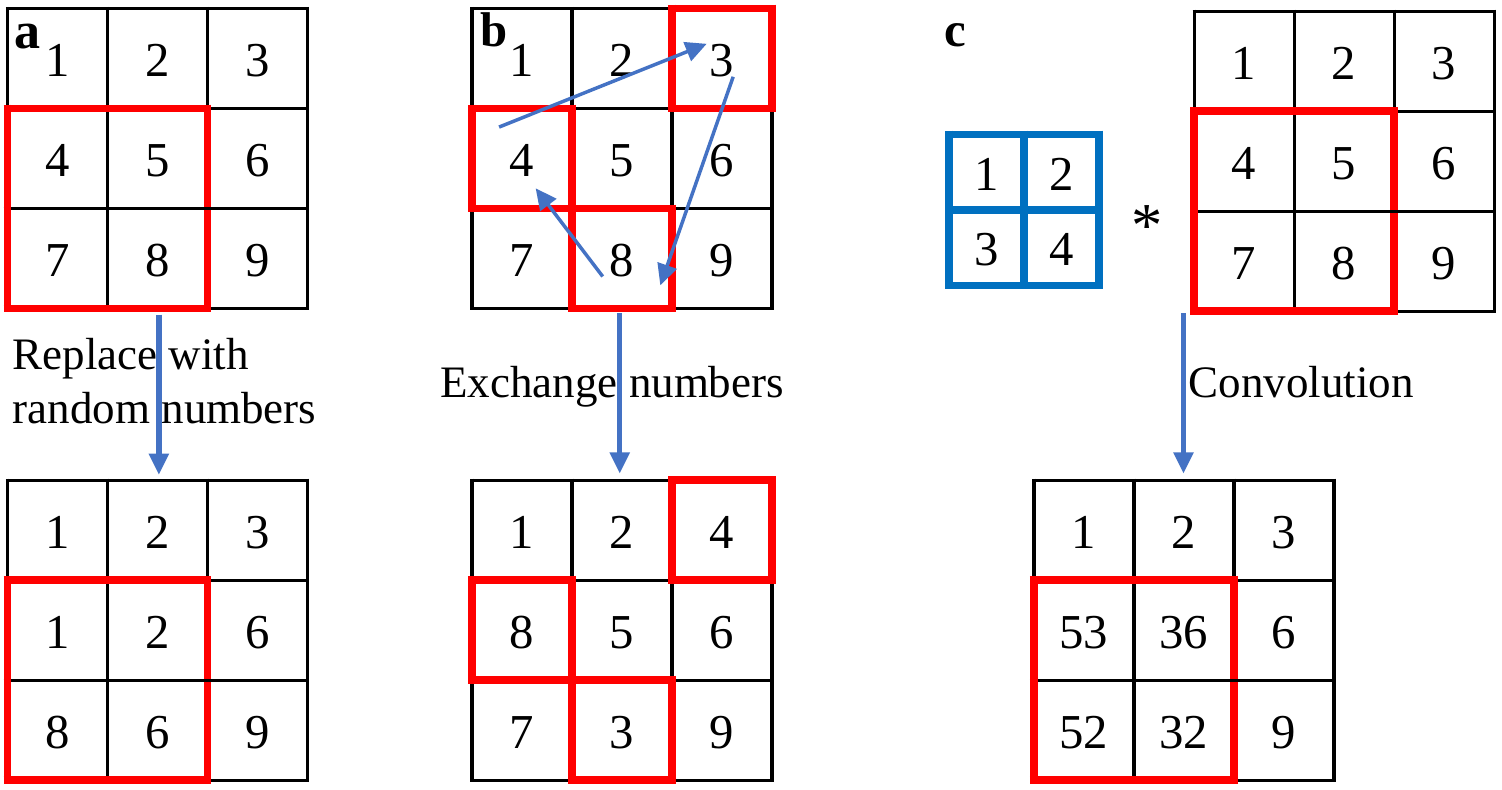}
	\caption{\textbf{Illustration of mutation and crossover}. \textbf{a}, An example of mutation: some adjacent cells (in the red box) are replaced with random numbers. \textbf{b}, An example of crossover: several cells (in the red boxes) are exchanged.  \textbf{c}, An example of convolution: several cells (in the red box) are convoluted with a kernel (in the blue cell). The volume constraint may be enforced at next step, not shown here.}
	\label{fig9}
\end{figure}

\textbf{Random generation of new samples from a base design.} After calculating the optimized array  $\hat{\bm{\rho}}$, more training data are generated by adding disturbance to it. As shown in \fig{fig9}, there are three kinds of disturbance: mutation, crossover and convolution. They are all likely to change the weighted average of an array, thus the enforcement of volume constraint will be applied when necessary. Mutation means mutating several adjacent cells in the optimized array, i.e., generating random numbers from 0 to 1 to replace the original elements. In the 2D example shown in \fig{fig9}a, the numbers in a 2-by-2 box are set as random. Crossover denotes the crossover of cells in the array  $\hat{\bm{\rho}}$ and is achieved by the following steps:
\begin{enumerate}
	\item Assign a linear index to each element in the array.
	\item Randomly pick several indices.
	\item Generate a random sequence of the indices.
	\item Replace the original numbers according to the sequence above. 
	As shown in \fig{fig9}b, indices are assigned sequentially from left to right and from top to bottom. The indices we pick in \textit{Step 2} are 3, 4 and 8; the sequence generated in \textit{Step 3} is 4, 8 and 3. 
\end{enumerate}

In the two compliance problems, the ways to generate a new input matrix based on $\hat{\bm{\rho}}$ and their possibilities are:
\begin{enumerate}[label=(\alph*)]
    \item mutation: mutating one element in $\hat{\bm{\rho}}$ (10\%);
    \item mutation: mutating a 2$\times$2 matrix in $\hat{\bm{\rho}}$ (10\%);
    \item mutation: mutating a 3$\times$3 matrix in $\hat{\bm{\rho}}$ (20\%);
    \item mutation: mutating a 4$\times$4 matrix in $\hat{\bm{\rho}}$ (20\%);
    \item crossover: choosing an integer $n$ from one to the number of total elements, selecting $n$ cells in $\hat{\bm{\rho}}$ and permuting them (20\%);
    \item generating a completely random matrix like the initial batch (20\%).
\end{enumerate}
In the fluid problem with $20\times 8$ mesh, i.e. SOLO-R, the ways are the same as previous ones except a threshold is needed to convert the continuous array into a binary one. The threshold has 50\% probability to be $\beta_1^4$ where $\beta_1$ is uniformly sampled from $[0,1]$, and has 50\% probability to be the element-wise mean of $\hat{\bm{\rho}}$. In the heat problem, crossover is replaced by convolution. It is the same as the compliance problems except that (e) above is replaced by 
\begin{enumerate}[label=(g)]
    \item Convolution: substituting a submatrix of the array, whose size and the corresponding probability is the same as (a-d), with a \textit{same convolution} (the output has the same size as the input submatrix) between the submatrix and $2\times 2$ kernel whose element is uniformly sampled from $[0,1]$.   
\end{enumerate}
In the truss optimization problems, the design variable $\bm{\rho}$ is purely one-dimensional and can no longer be represented as a matrix. Therefore, we only use mutation. First, $\beta_2$ is uniformly sampled from $[0,1]$ to indicate the ratio of elements to be mutated in $\hat{\bm{\rho}}$, and then those elements are randomly selected to add $\gamma$ to themselves; $\gamma$ is uniformly sampled from $[-1,1]$.

\textbf{Generalized Simulated Annealing (GSA).} Simulated Annealing (SA) is a stochastic method to determine the global minimum of a objective function by simulating the annealing process of a molten metal~\cite{xiang2017}. GSA is a type of SA  with specific forms of visiting function and acceptance probability\cite{xiang2013}. Assuming objective
\begin{equation}
\hat{\bm{\rho}}=\underset{\bm{\rho}\in[0,1]^N}{\text{arg}\min}\ h(\bm{\rho}),
\end{equation}
we do the following:
\begin{enumerate}
	\item Generate an initial state $\bm{\rho} ^{(0)}=[\rho _{1}^{(0)},\rho _{2}^{(0)},...,\rho _{N}^{(0)} ]^T$   randomly and obtain its function value  $E^{(0)}=h( \bm{\rho} ^{(0)} ) $. Set parameters $T(0)$, $t_{max}$, $q_v$, $q_a$. 
	\item For artificial time step $t =1$ to $t_{max}$,
	\begin{enumerate}
	\item Generate a new state  $\bm{\rho} ^{(t)}=\bm{\rho} ^{(t-1)}+\Delta \bm{\rho}^{(t)} $, where the probability distribution of $\Delta \bm{\rho}^{(t)}$  follows the visiting function
	\begin{equation}
		g( \Delta \bm{\rho}^{(t)}) \propto \frac{[ T(t) ] ^{-\frac{N}{3-q_v}}}{\left\{ 1+( q_v-1) \frac{ [\Delta \bm{\rho} ^{(t)}] ^2}{[ T( t ) ] ^{\frac{2}{3-q_v}}} \right\} ^{\frac{1}{q_v-1}+\frac{N-1}{2}}}.
	\end{equation}
	where $T$  denotes the artificial temperature calculated by	
	\begin{equation}
	T( t ) =T(0)\frac{2^{q_v-1}-1}{( 1+t ) ^{q_v-1}-1}.
	\end{equation}
	\item Calculate the energy difference
	\begin{equation}
		\Delta E=E^{(t)}-E^{(t-1)}=h( \bm{\rho} ^{(t)} ) -h( \bm{\rho} ^{(t-1)} ) .
	\end{equation}
	\item Calculate the probability to accept the new state
	\begin{equation}
		p=\min \left\{ \text{1,}\left[ 1-\left( 1-q_a \right) \frac{t}{T( t )}\Delta E \right] ^{\frac{1}{1-q_a}} \right\}.
	\end{equation}
	Determine whether to accept the new state based on the probability, if not, $\bm{\rho} ^{(t)}=\bm{\rho} ^{(t-1)}$.
	\end{enumerate}
\item Conduct local search to refine the state.
\end{enumerate}

Since compliance minimization has a volume constraint, the objective function used in the optimization process is written as
\begin{equation}
	h(\bm{\rho})= f( \bm{\rho}) +c(\mathbf{w}\cdot \bm{\rho}-V_0 ) ^2,
\end{equation}
where $c$ is a constant to transform the constrained problem to an unconstrained problem by adding a penalty
term.
GSA is implemented via SciPy package with default parameter setting. For more details please refer to its documentation\cite{community2019}.

\textbf{Bat Algorithm (BA).} Bat Algorithm (BA) is a heuristic optimization algorithm, inspired by the echolocative behavior of bats \cite{yang2010new}. This algorithm carries out the search process using artificial bats mimicking the natural pulse loudness, emission frequency and velocity of real bats. It solves the problem 
\begin{equation}
\hat{\bm{\rho}}=\underset{\bm{\rho}\in[0,1]^N}{\text{arg}\min}\ h(\bm{\rho}).
\end{equation}
We adopt a modification \cite{yilmaz2015new} and implement as follows:
\begin{enumerate}
    \item Generate $M$ vectors $\bm{\rho}^{(0,1)},\bm{\rho}^{(0,2)}, ..., \bm{\rho}^{(0,M)}$. We use $\bm{\rho}^{(t,m)}$ to denote a vector, flattened from the array representing design variables. It is treated as the position of the $m$-th artificial bat, where $m=1,2,...,M$. We use $\rho^{(t,m)}_i\in [0,1]$ to denote the $i$-th dimension of vector $\bm{\rho}^{(t,m)}$, where $i=1,2,...,N$. Thus, $\bm{\rho}^{(0,m)}=[\rho^{(0,m)}_1,\rho^{(0,m)}_2,..\rho^{(0,m)}_N]^T$.
    \item Calculate their function values and find the minimum $\bm{\rho}^*=\textrm{arg}\min h(\bm{\rho}^{(0,m)})$.
    \item Initialize their velocity $\mathbf{v}^{(0,1)},\mathbf{v}^{(0,2)},...,\mathbf{v}^{(0,m)}, ..., \mathbf{v}^{(0,M)}$. 
    \item Determine parameters $q_{min}$, $q_{max}$, $t_{max}$, $\alpha$, $\gamma$, $r^{(0)}$, $A^{(0)}$, $w_{init}$, $w_{final}$.
	\item For artificial time step $t =1$ to $t_{max}$,
	\begin{enumerate}
		    \item Update parameters $A^{(t)}=\alpha A^{(t-1)}$, $r^{(t)}=r^{(0)}(1-e^{-\gamma t})$, $w^{(t)}=(1 - t / t_{max})^2(w_{init} - w_{final}) + w_{final}$.
	    \item 	For $m=1,2,...,M$,
	   \begin{enumerate}
	    \item Calculate sound frequency

	    \begin{equation}
	        q^{(t,m)}=q_{min}+(q_{max}-q_{min})\beta,
	    \end{equation}
	    where $\beta$ is a random number that has a uniform distribution in $[0,1]$.
	    \item Update velocity based on frequency
	    	    \begin{equation}
	        \mathbf{v}^{(t,m)}=w^{(t)}\mathbf{v}^{(t-1,m)}+(\bm{\rho}^{(t-1,m)}-\bm{\rho}^*) q^{(t,m)}.
	    \end{equation}
	    \item Get a (temporary) new solution. Calculate the new position
	    \begin{equation}
	        \bm{\rho}^{(t,m)}=\bm{\rho}^{(t,m-1)}+\mathbf{v}^{(t,m)}.
	    \end{equation}
	    \item Local search. Generate $\beta'_i (i=1,2,...,N)$, a series of random numbers uniformly sampled in $[0,1]$. For those $i$ satisfying $\beta'_i>r^{(t)}$, add noise to current best solution
	    \begin{equation}
	        \rho^{(t,m)}_i = \rho^*_i+\epsilon A^{(t)},
	    \end{equation}
	    where $\epsilon$ is a random variable sampled in Gaussian distribution with zero mean, $\rho^*_i$ is the $i$-th component of $\bm{\rho}^*$. If $\rho_i^{(t,m)}$ goes over the range $[0,1]$, it is thresholded to 0 or 1. 
	    For others, keep them as they are. 
	    \item Determine whether to accept the new solution. Reverse to the previous step $\bm{\rho}^{(t,m)}=\bm{\rho}^{(t-1,m)}$, if $h(\bm{\rho}^{(t,m)})>h(\bm{\rho}^{(t-1,m)})$ or  $\beta''>A^{(t)}$ (where $\beta''$ is random number uniformly sampled in $[0,1]$).
	   \end{enumerate}
	   \item Update $\bm{\rho}^*=\textrm{arg}\min_{m=1,2,...,M} h(\bm{\rho}^{(t,m)})$.
	\end{enumerate}
	        \item Output $\hat{\bm{\rho}}=\bm{\rho}^*$.
	\end{enumerate}
BA is used in the heat and truss problems. In the heat problem, we optimize $f$ without adding penalty terms since the volume constraint is controlled by a threshold, i.e., $h=f$. In the truss optimization problems, we need to choose $\bm{\rho}^{(t,m)}$ in a discrete space since only 16 values are allowed. Before we evaluate $h(\bm{\rho}^{(t,m)})$, we will replace $\rho_i^{(t,m)}$ by the nearest discrete values. To deal with constraints in Eqs. \eqref{eq:15} and \eqref{eq:16}, the objective function is converted to
\begin{equation}
    h(\bm{\rho})=W(\bm{\rho})\left(1+\sum_{|\sigma_i|>\sigma_0}\frac{|\sigma_i|-\sigma_0}{\sigma_0}+\sum_{||\Delta\mathbf{x}_i||_\infty>\delta_0}\frac{||\Delta\mathbf{x}_i||_\infty-\delta_0}{\delta_0}\right)^2.
\end{equation}

\textbf{Binary Bat Algorithm (BBA).}
Binary Bat Algorithm \cite{mirjalili2014binary,ramasamy2018modified} is a binary version of BA. To solve
\begin{equation}
\hat{\bm{\rho}}=\underset{\bm{\rho}\in\{0,1\}^N}{\text{arg}\min}\ h(\bm{\rho}),
\end{equation}
we slightly adjust the original algorithm and implement it as follows:
 
\begin{enumerate}
    \item Generate $M$ vectors $\bm{\rho}^{(0,1)},\bm{\rho}^{(0,2)}, ..., \bm{\rho}^{(0,M)}$. We use $\bm{\rho}^{(t,m)}$ to denote a vector, flattened from the array representing design variables. It is treated as the position of the $m$-th artificial bat, where $m=1,2,...,M$. We use $\rho^{(t,m)}_i\in \{0,1\}$ to denote the $i$-th dimension of vector $\bm{\rho}^{(t,m)}$, where $i=1,2,...,N$. Thus, $\bm{\rho}^{(0,m)}=[\rho^{(0,m)}_1,\rho^{(0,m)}_2,..\rho^{(0,m)}_N]^T$. 
    \item Calculate their function values and find the minimum $\bm{\rho}^*=\textrm{arg}\min h(\bm{\rho}^{(0,m)})$.
    \item Initialize their velocity $\mathbf{v}^{(0,1)},\mathbf{v}^{(0,2)},...,\mathbf{v}^{(0,m)}, ..., \mathbf{v}^{(0,M)}$. 
    \item Determine parameters $q_{min}$, $q_{max}$, $t_{max}$, $\alpha$, $\gamma$, $r^{(0)}$, $A^{(0)}$.
	\item For artificial time step $t =1$ to $t_{max}$,
	\begin{enumerate}
		    \item Update parameters $A^{(t)}=\alpha A^{(t-1)}$, $r^{(t)}=r^{(0)}(1-e^{-\gamma t})$.
	    \item 	For $m=1,2,...,M$,
	   \begin{enumerate}
	    \item Calculate sound frequency

	    \begin{equation}
	        q^{(t,m)}=q_{min}+(q_{max}-q_{min})\beta,
	    \end{equation}
	    where $\beta$ is a random number that has a uniform distribution in $[0,1]$.
	    \item Update velocity based on frequency
	    	    \begin{equation}
	        \mathbf{v}^{(t,m)}=\mathbf{v}^{(t-1,m)}+(\bm{\rho}^{(t-1,m)}-\bm{\rho}^*) q^{(t,m)}.
	    \end{equation}
	    \item Get a (temporary) new solution. Calculate the possibility to change position based on velocity
	    \begin{equation}
	        V_i^{(t,m)}=\left| \frac{2}{\pi}\textrm{arctan}\left(\frac{\pi}{2} v_i^{(t,m)}\right) \right|+\frac{1}{N}.
	    \end{equation}
	    \item Random flip. Generate $\beta'_i (i=1,2,...,N)$, a series of random numbers uniformly in [0,1]. For those $i$ satisfying $\beta'_i<V_i^{(t,m)}$, change the position by flipping the 0/1 values
	    \begin{equation}
	        \rho^{(t,m)}_i = 1-\rho^{(t-1,m)}_i.
	    \end{equation}
	    For others, keep them as they are.
	    \item Accept the local optimum. Generate $\beta''_i (i=1,2,...,N)$, a series of random numbers uniformly sampled in $[0,1]$. For those $i$ satisfying $\beta''_i>r^{(t)}$, set  $\rho^{(t,m)}_i=\rho^*_{i}$.
	    \item Determine whether to accept the new solution. Reverse to the previous step $\bm{\rho}^{(t,m)}=\bm{\rho}^{(t-1,m)}$, if $h(\bm{\rho}^{(t,m)})>h(\bm{\rho}^{(t-1,m)})$ or  $\beta'''>A^{(t)}$ (where $\beta'''$ is random number uniformly sampled in [0,1]).
	   \end{enumerate}
	   \item Update $\bm{\rho}^*=\textrm{arg}\min_{m=1,2,...,M} h(\bm{\rho}^{(t,m)})$.
	\end{enumerate}
	        \item Output $\hat{\bm{\rho}}=\bm{\rho}^*$.
	\end{enumerate}
BBA is used in the fluid problems. Since we do not have constraints in these problems, we can optimize $f$ without adding penalty terms, i.e., $h=f$. 

\section*{Data availability}
The optimization data generated in this study have been deposited in
the Zenodo database~\cite{my_data}.

\section*{Code availability}
All code (MATLAB and Python) used in this paper is deposited in the Zenodo repository~\cite{my_github} or available at  \url{https://github.com/deng-cy/deep_learning_topology_opt}.


\bibliographystyle{naturemag}

\bibliography{readcube}


\section*{Acknowledgement}
The authors gratefully acknowledge the support by the National Science Foundation under Grant No.\ CNS-1446117 (W.L.). 

\section*{Competing interests}
The authors declare no conflict of interests.

\section*{Author contributions}
C.D.\ designed the algorithm and drafted the manuscript. Y.W. derived the convergence theory. C.D.\ and C.Q.\ wrote the code. Y.W., C.Q.\ and Y.F.\ edited the manuscript. W.L.\ supervised this study and revised the manuscript.

\newpage
\section*{Supplementary Information}
\beginsupplement
\begin{center}
	\textbf{Table of Contents}
\end{center}

Section \ref{s1}: \nameref{s1} \dotfill \pageref{s1}

\begin{figure}[H]
	\centering
	\begin{minipage}{\textwidth}\centering
		\xincludegraphics[width=\textwidth,label=\textbf{a}]{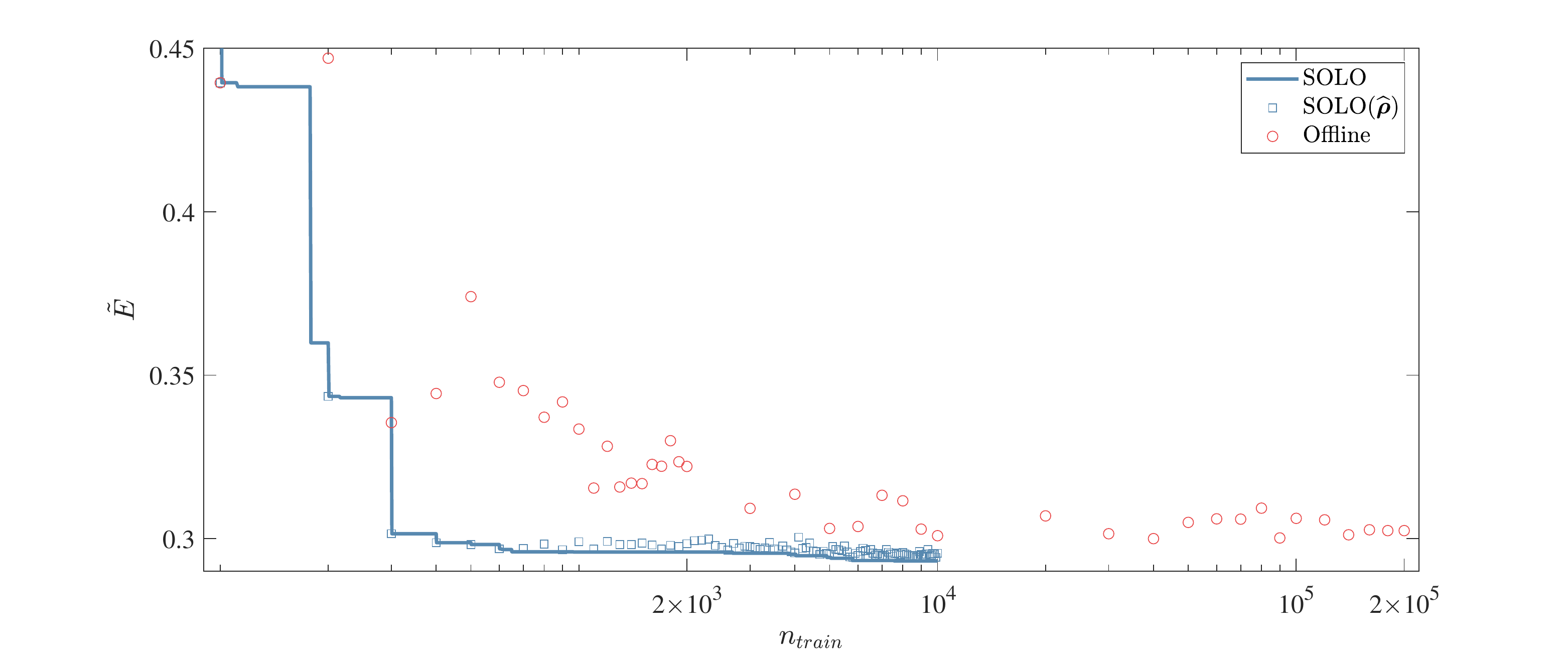}
	\end{minipage}
	\begin{minipage}{\textwidth}\centering
		\xincludegraphics[width=\textwidth,label=\textbf{b}]{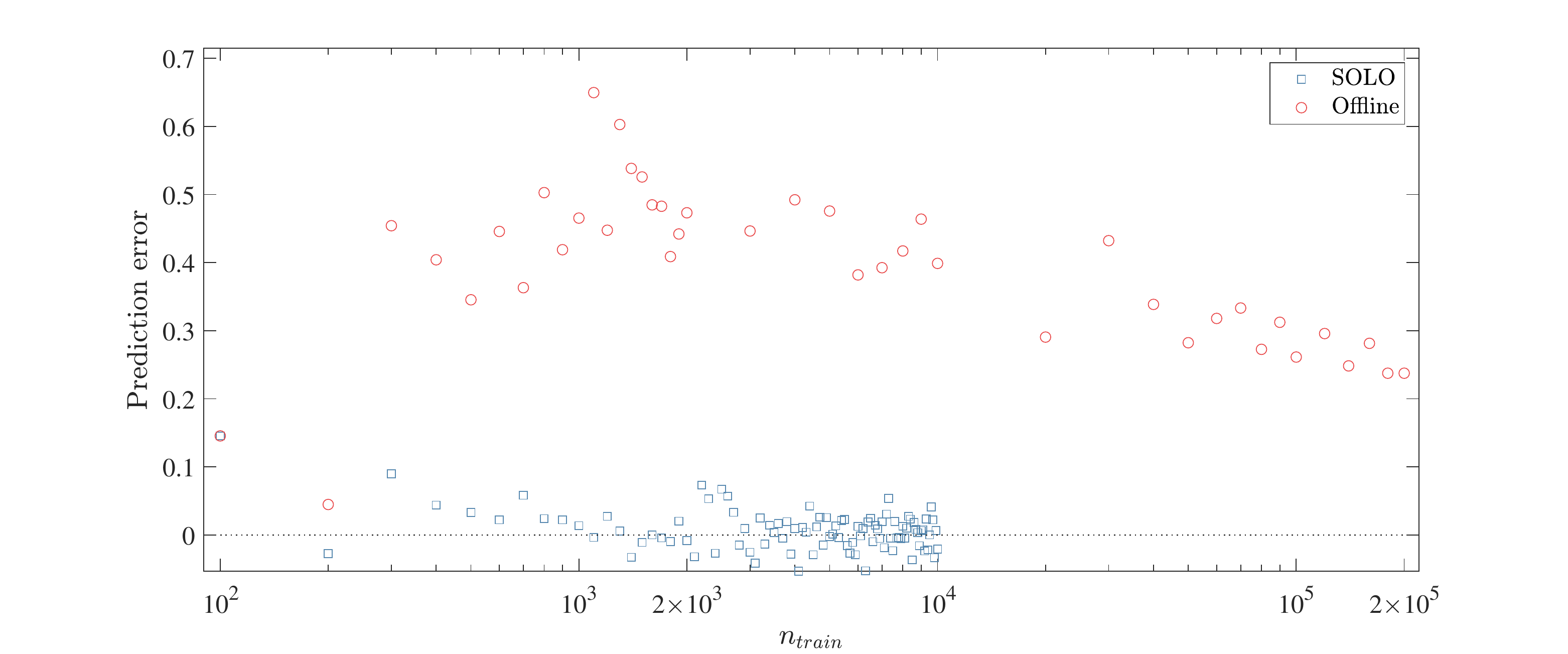}
	\end{minipage}
	\caption{\textbf{Objective (energy) and prediction error of the compliance minimization problem with 5$\times$5 variables.} \textbf{a}, Dimensionless energy as a function of $n_{train}$. For SOLO, the solid line denotes the best objective values and the squares denote $\widetilde{E}(\hat{\bm{\rho}})$. \textbf{b}, Energy prediction error of $\hat{\bm{\rho}}$.}
	\label{figs1}
\end{figure}

\begin{figure}[H]
	\centering
	\begin{minipage}{0.32\textwidth}\centering
		\xincludegraphics[height=5cm,label={\textbf{a}}]{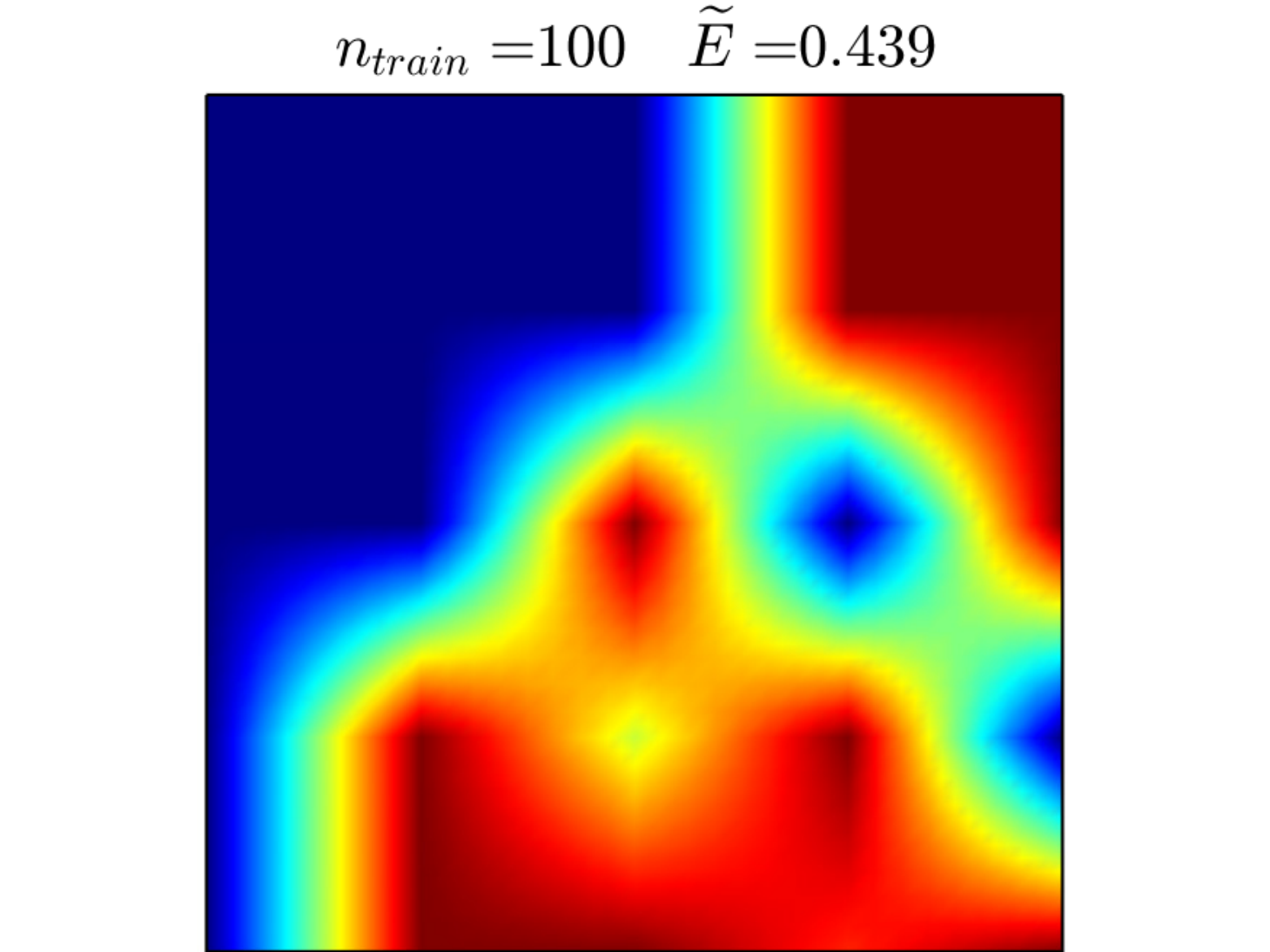}
	\end{minipage}
	\begin{minipage}{0.32\textwidth}\centering
		\xincludegraphics[height=5cm,label={\textbf{b}}]{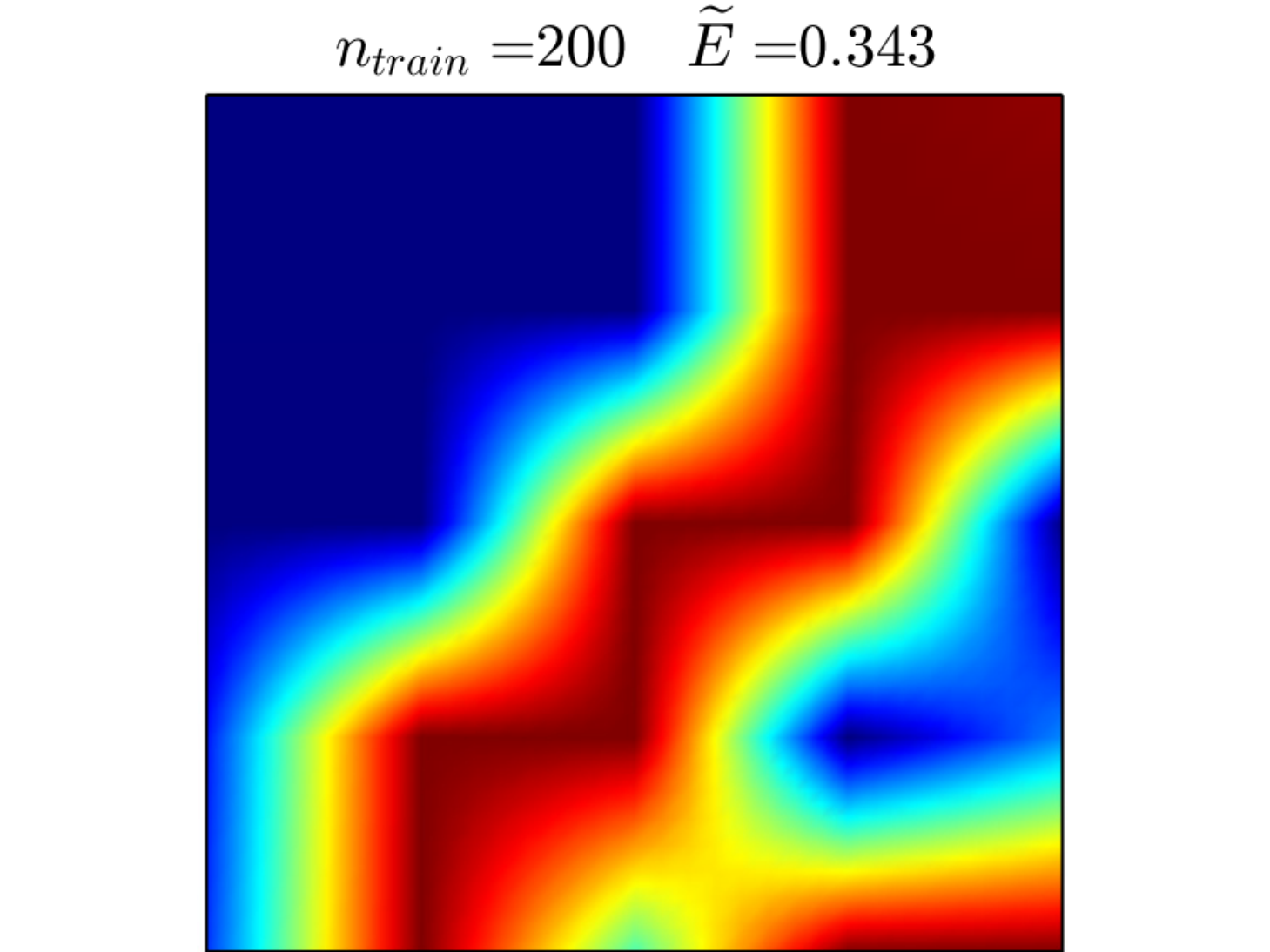}
	\end{minipage}
	\begin{minipage}{0.32\textwidth}\centering
		\xincludegraphics[height=5cm,label={\textbf{c}}]{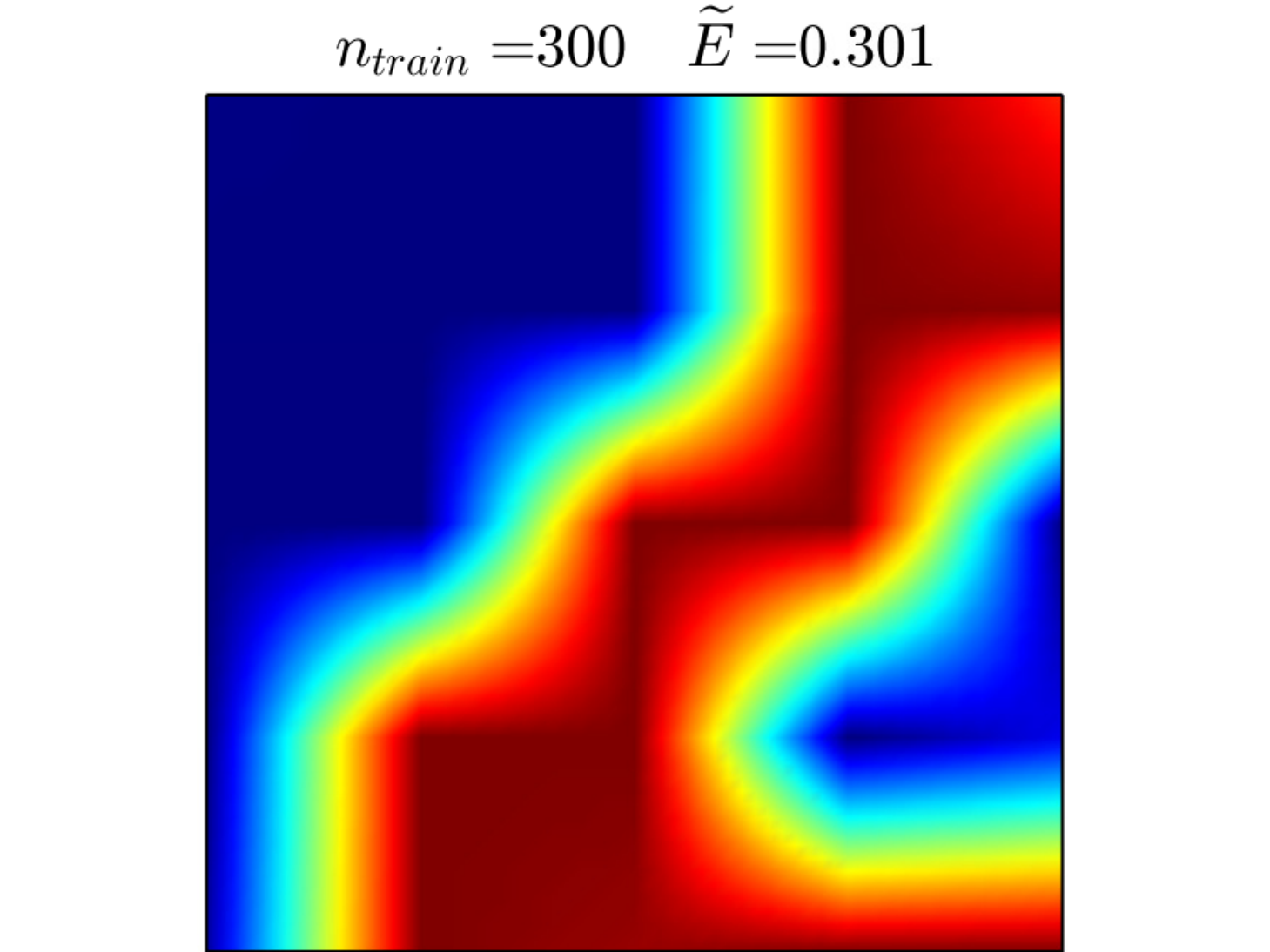}
	\end{minipage}
	\begin{minipage}{0.32\textwidth}\centering
		\xincludegraphics[height=5cm,label={\textbf{d}}]{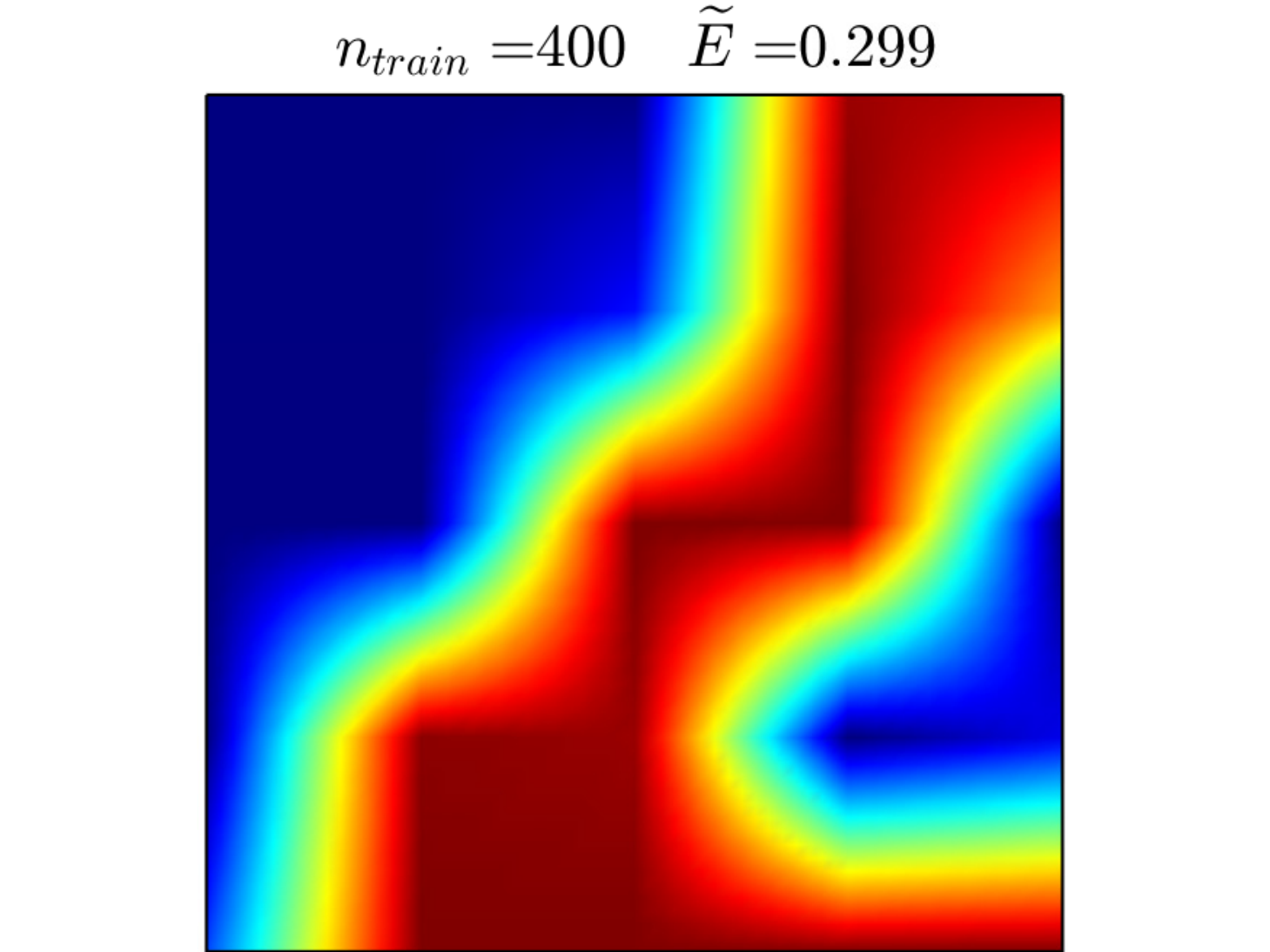}
	\end{minipage}
	\begin{minipage}{0.32\textwidth}\centering
		\xincludegraphics[height=5cm,label={\textbf{e}}]{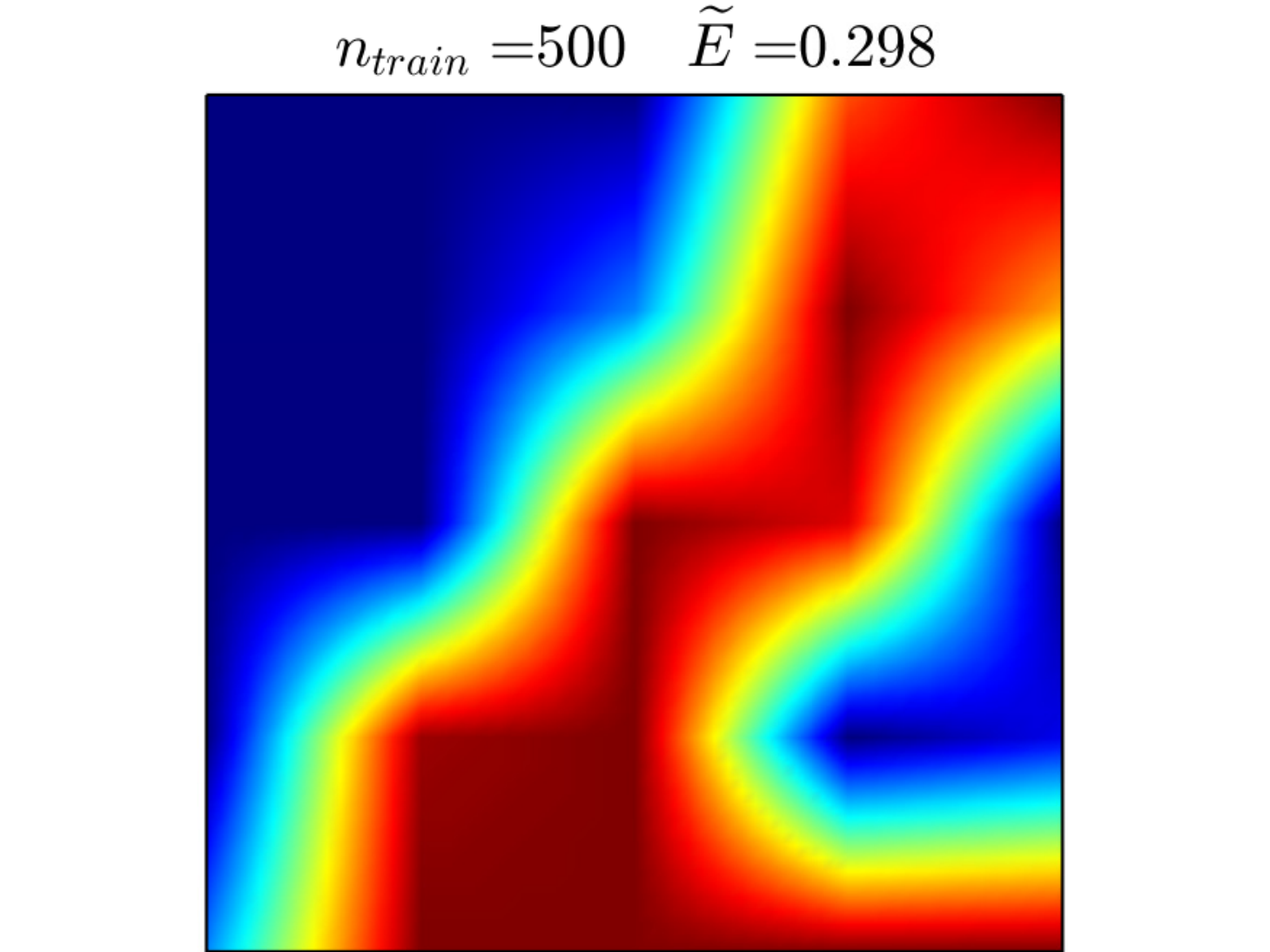}
	\end{minipage}
	\begin{minipage}{0.32\textwidth}\centering
		\xincludegraphics[height=5cm,label={\textbf{f}}]{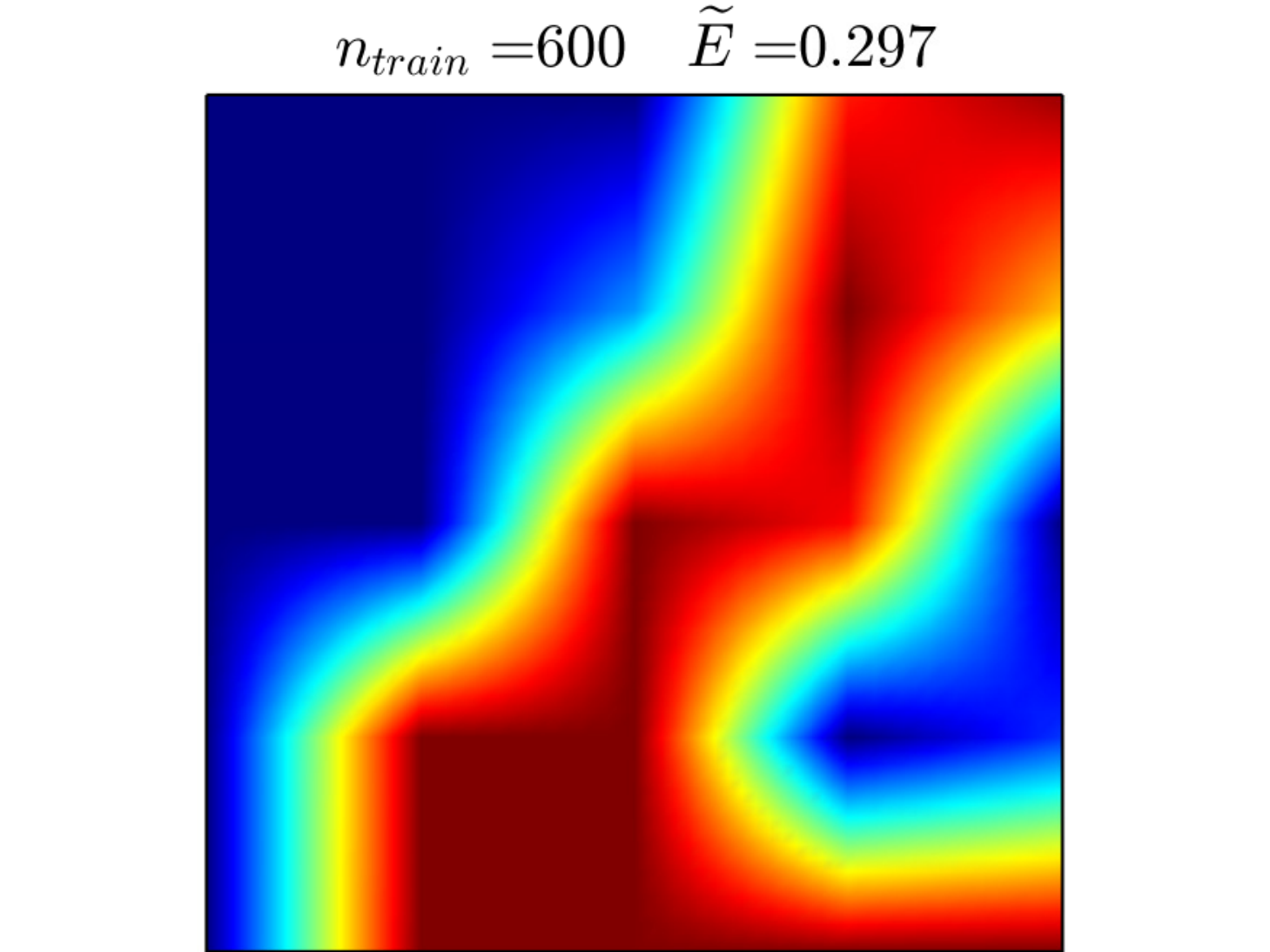}
	\end{minipage}
	\begin{minipage}{0.32\textwidth}\centering
		\xincludegraphics[height=5cm,label={\textbf{g}}]{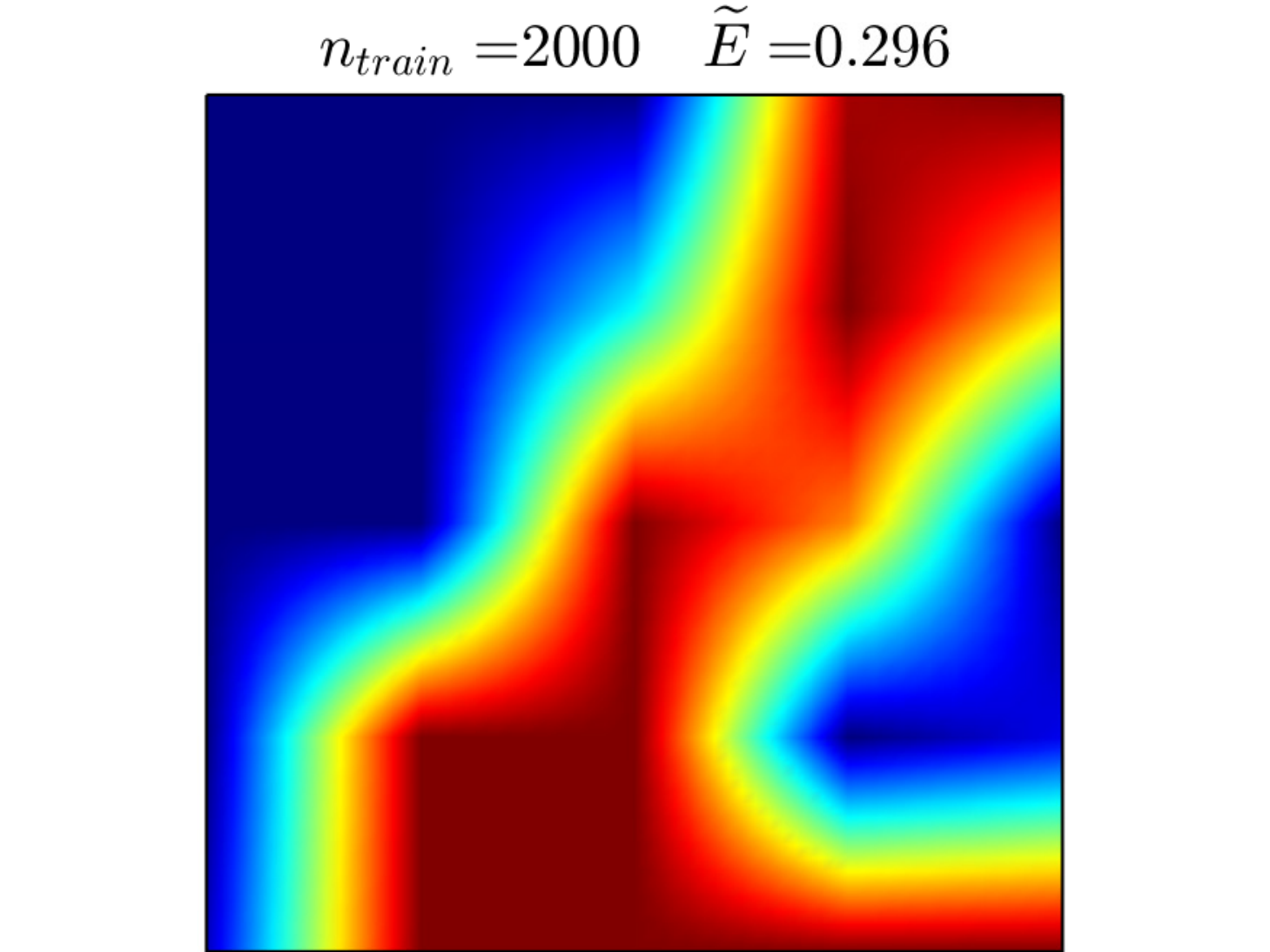}
	\end{minipage}
	\begin{minipage}{0.32\textwidth}\centering
		\xincludegraphics[height=5cm,label={\textbf{h}}]{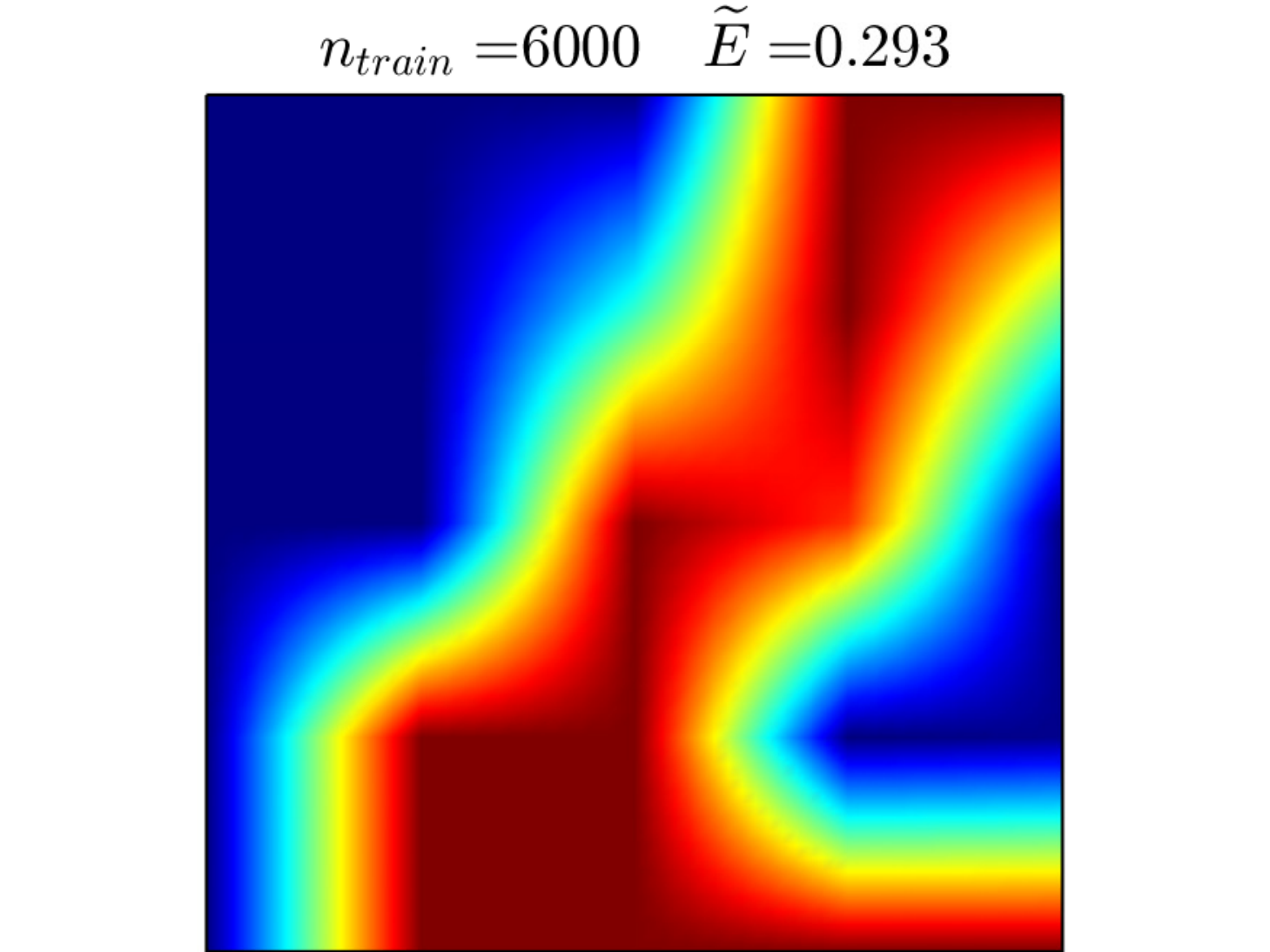}
	\end{minipage}
	\begin{minipage}{0.32\textwidth}\centering
		\xincludegraphics[height=5cm,label={\textbf{i}}]{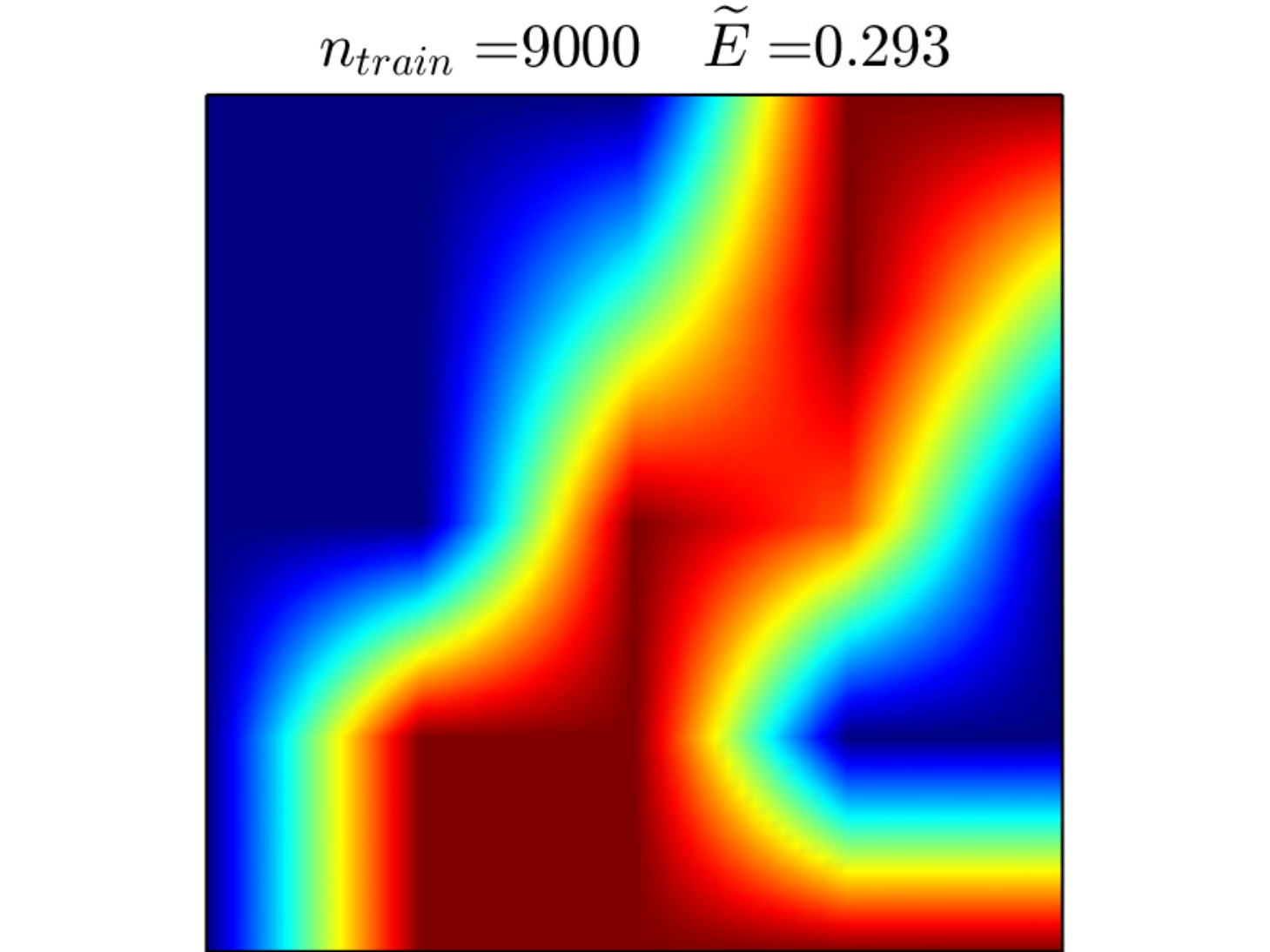}
	\end{minipage}
	\caption{\textbf{Evolution of the solution from SOLO for the compliance minimization problem with 5$\times$5  variables.} Each plot is the best among $n_{train}$ accumulated training data  and the corresponding energy $\widetilde{E}$ is marked. There is no obvious change after hundreds of training samples.}
	\label{figs2}
\end{figure}

\begin{figure}[H]
	\centering
	\begin{minipage}{0.32\textwidth}\centering
		\xincludegraphics[height=5cm,label={\textbf{a}}]{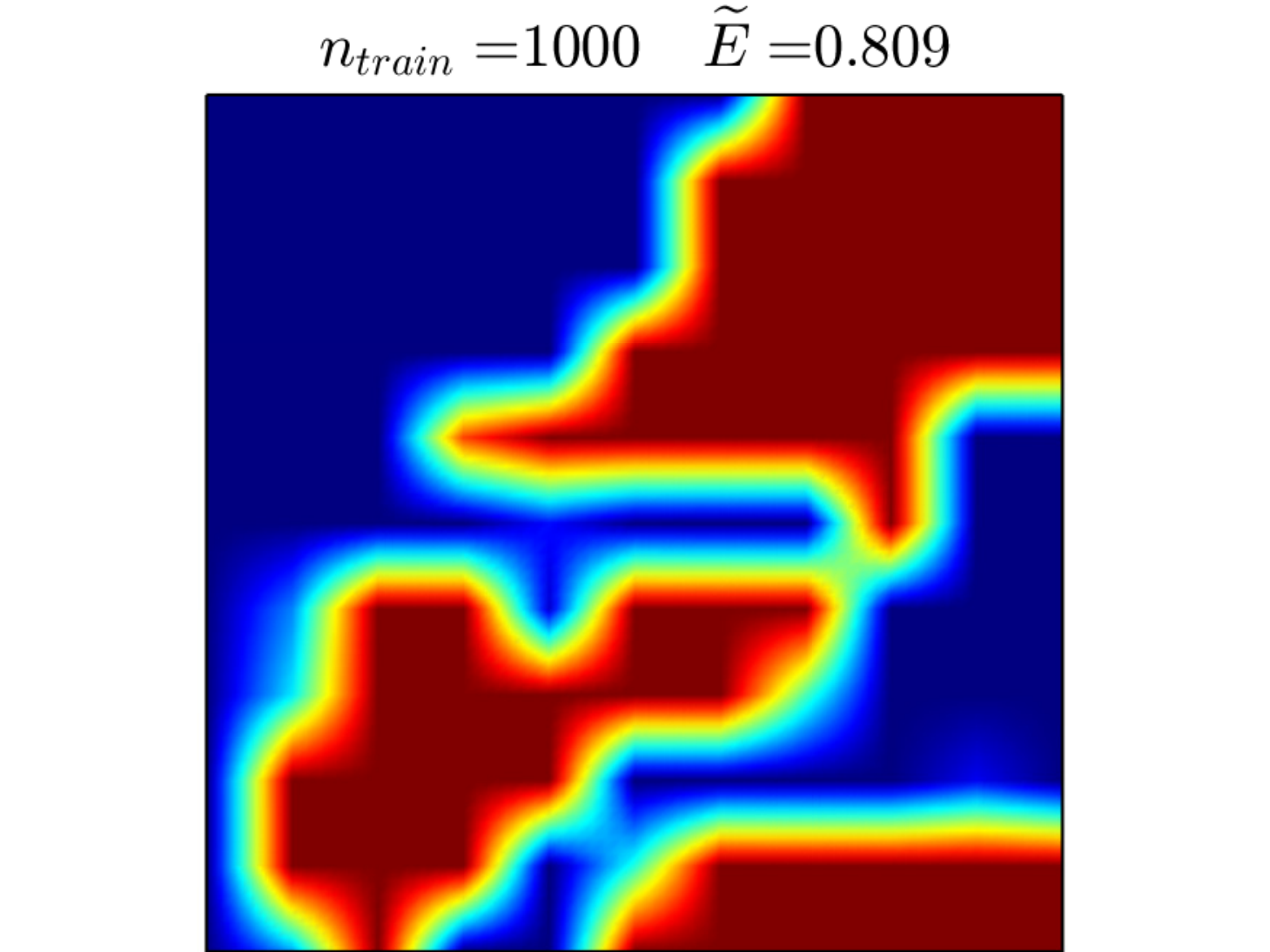}
	\end{minipage}
	\begin{minipage}{0.32\textwidth}\centering
		\xincludegraphics[height=5cm,label={\textbf{b}}]{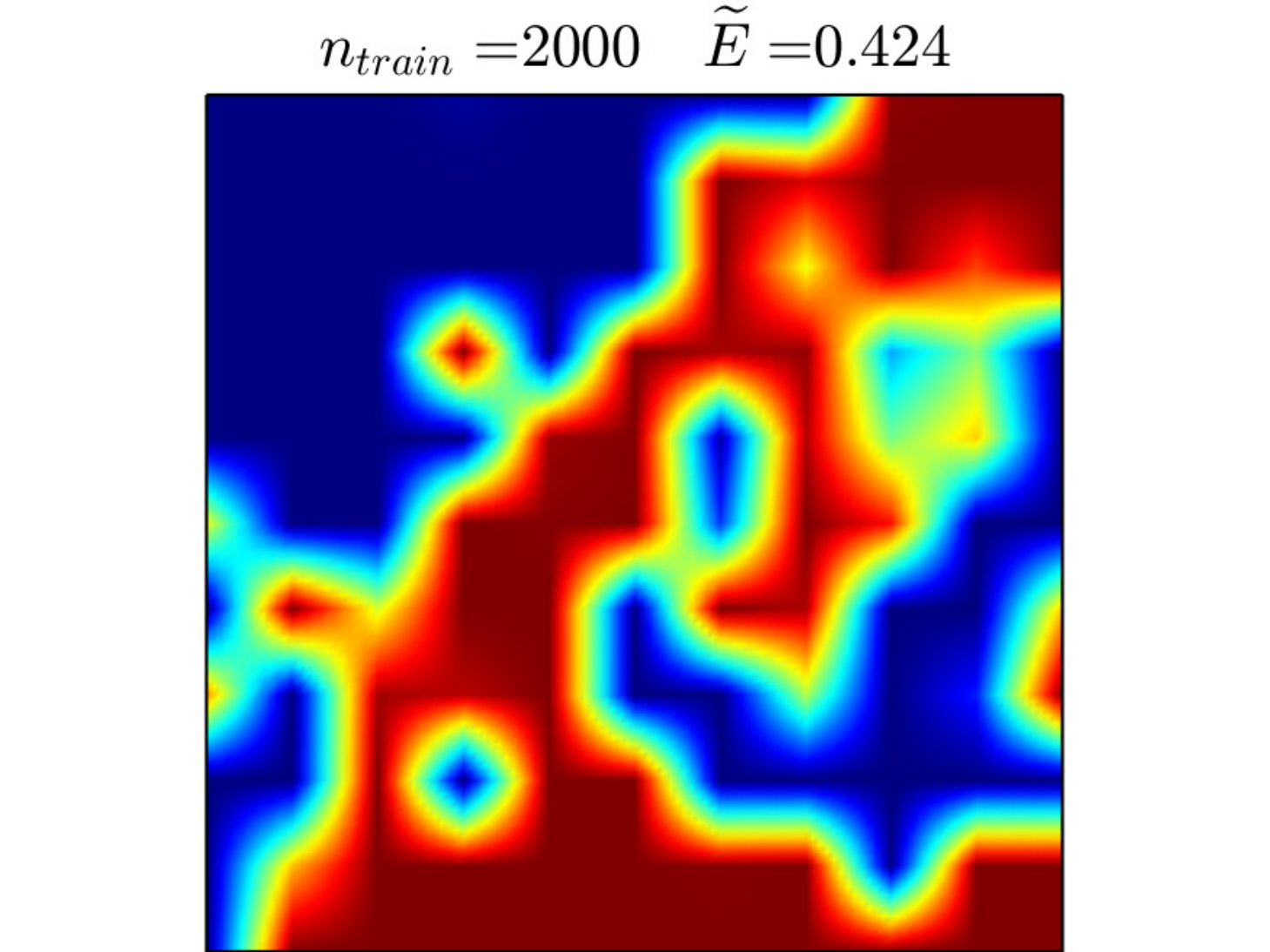}
	\end{minipage}
	\begin{minipage}{0.32\textwidth}\centering
		\xincludegraphics[height=5cm,label={\textbf{c}}]{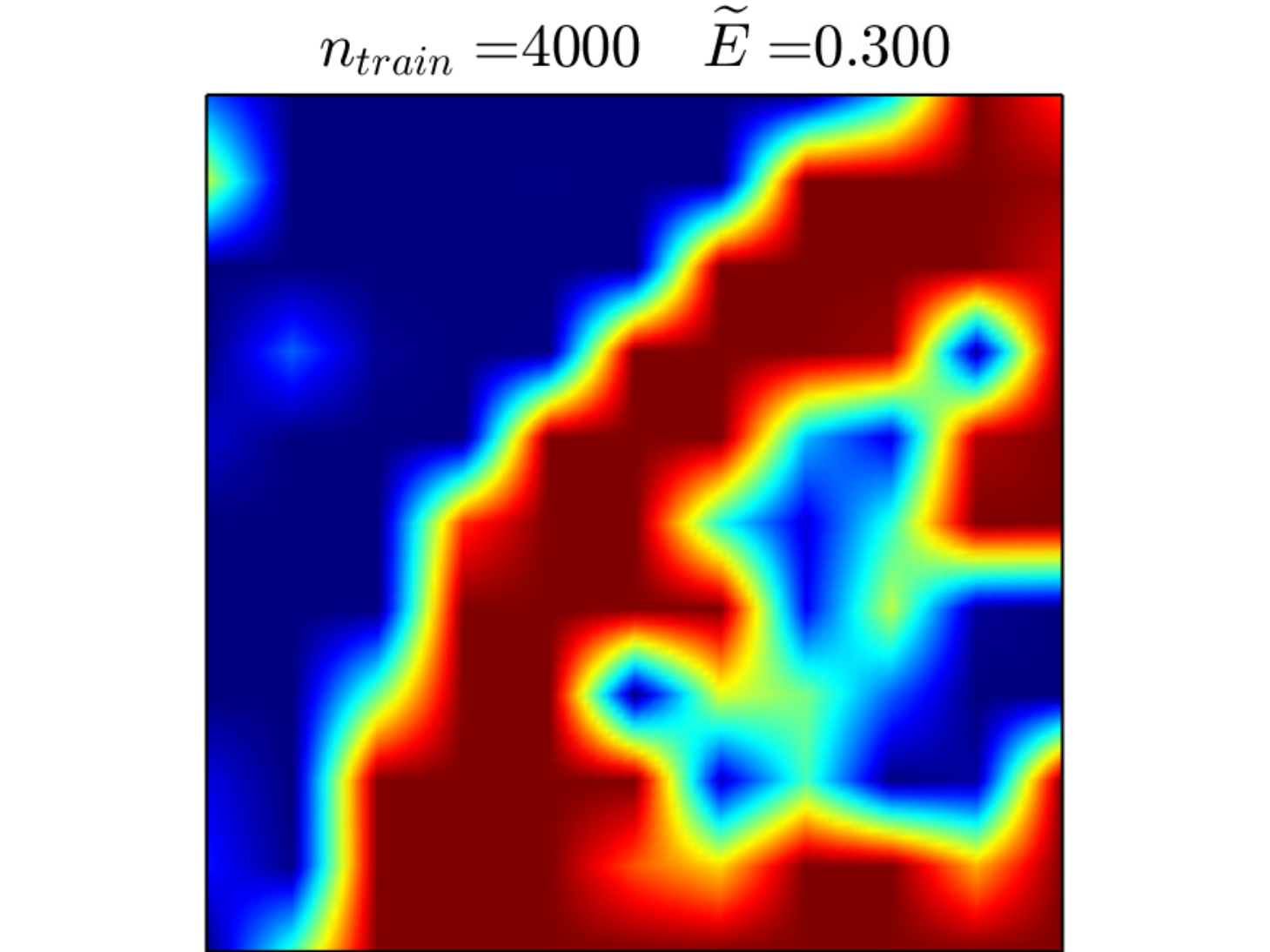}
	\end{minipage}
	\begin{minipage}{0.32\textwidth}\centering
		\xincludegraphics[height=5cm,label={\textbf{d}}]{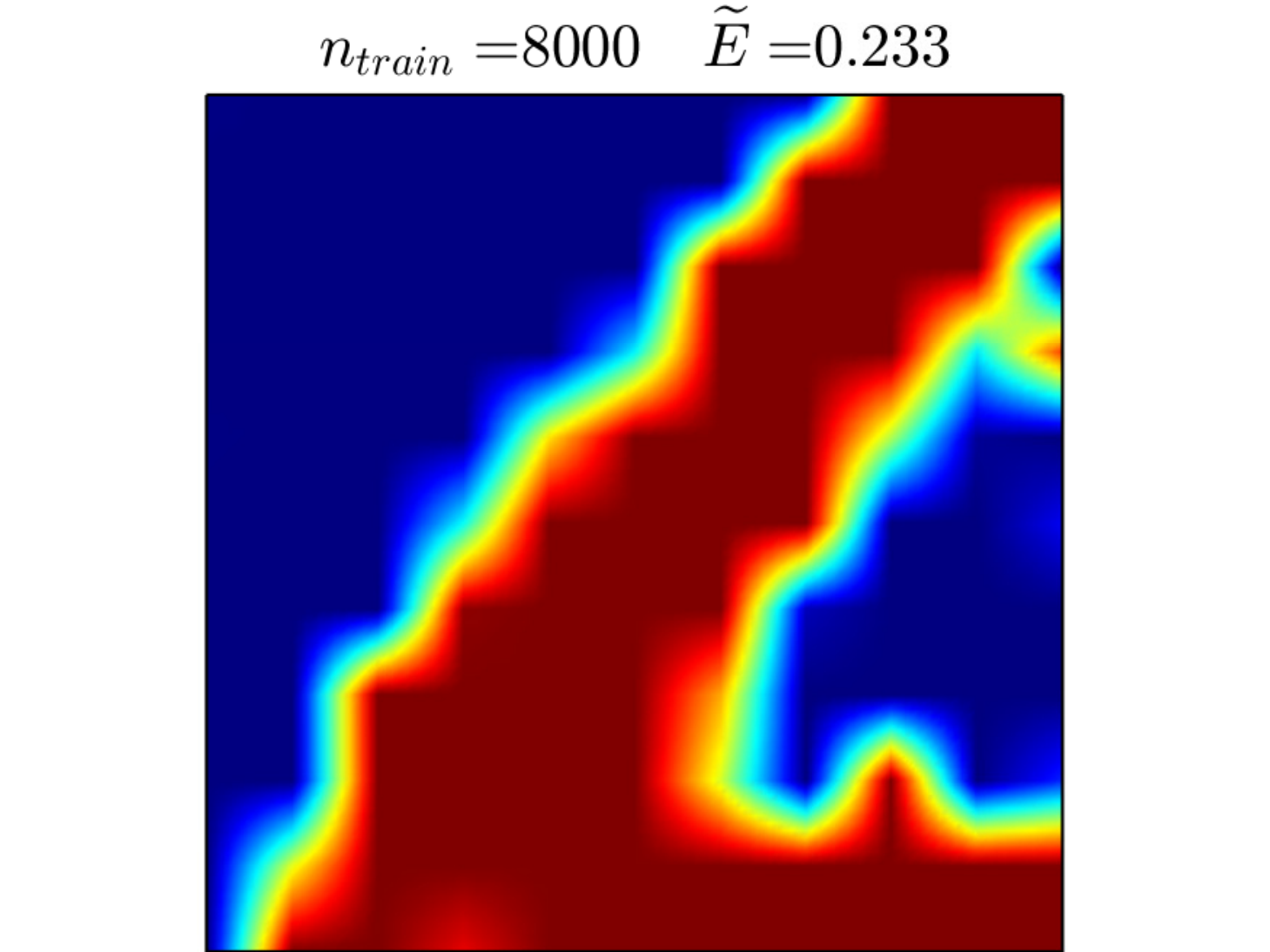}
	\end{minipage}
	\begin{minipage}{0.32\textwidth}\centering
		\xincludegraphics[height=5cm,label={\textbf{e}}]{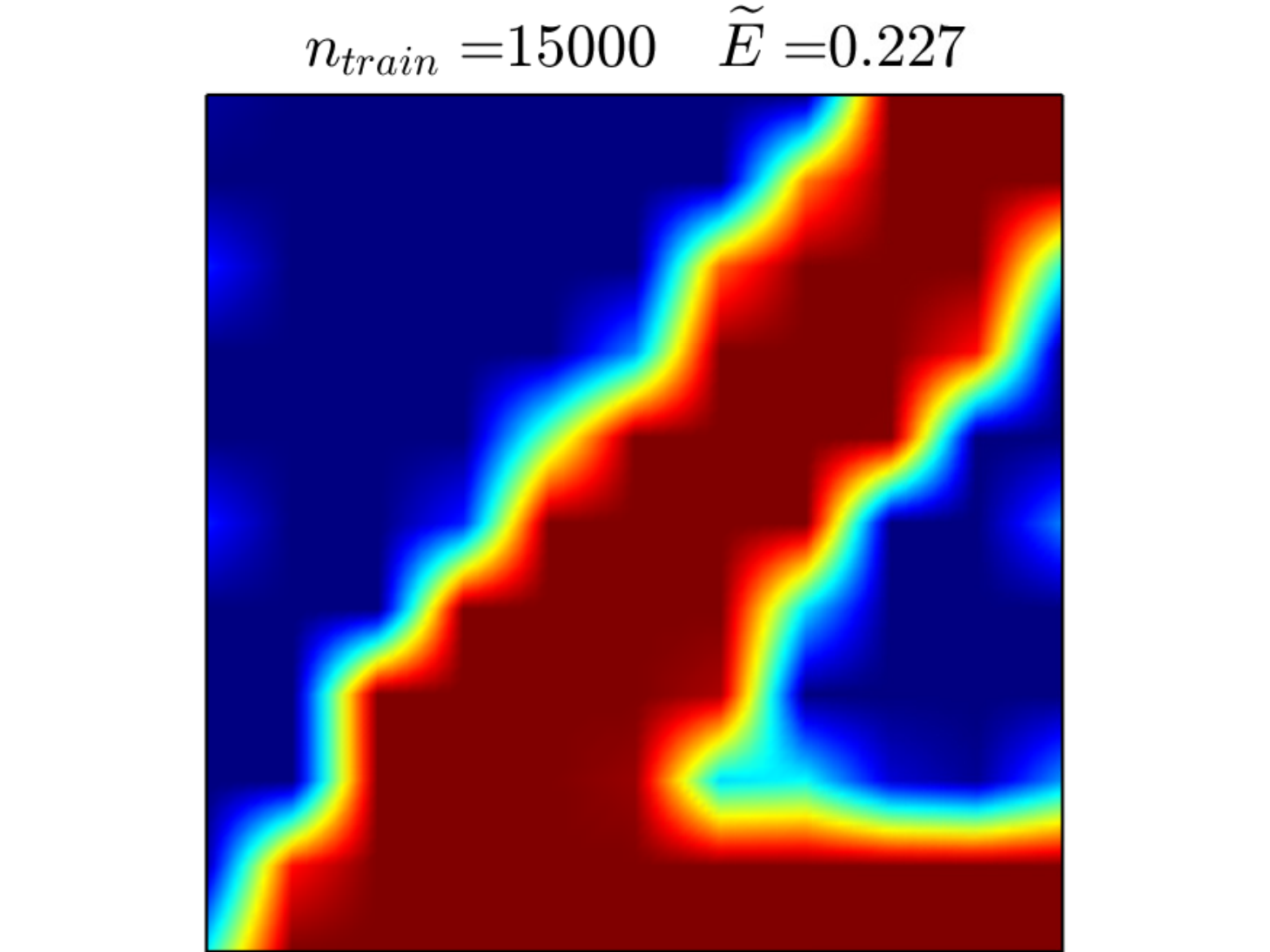}
	\end{minipage}
	\begin{minipage}{0.32\textwidth}\centering
		\xincludegraphics[height=5cm,label={\textbf{f}}]{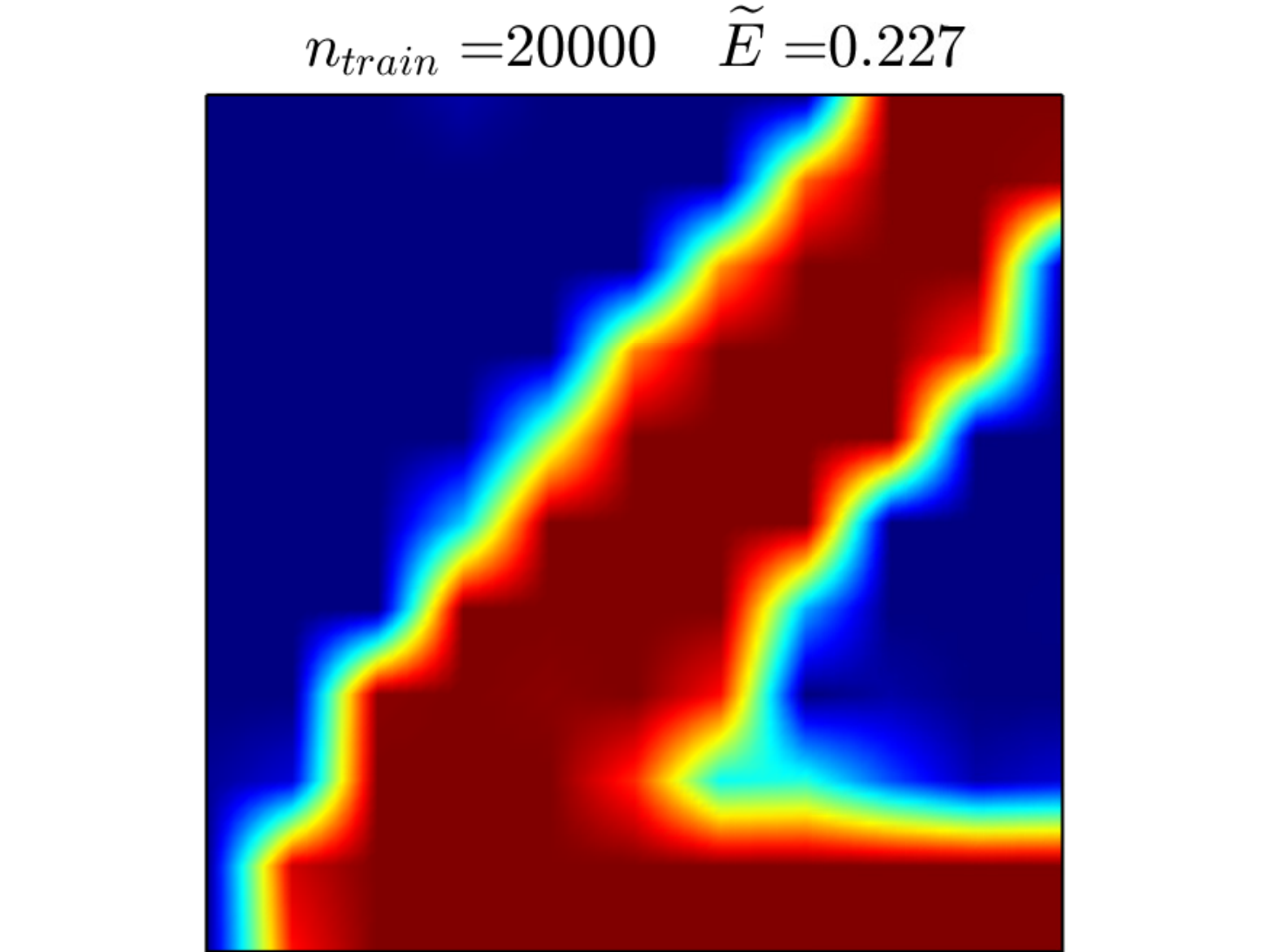}
	\end{minipage}
	\begin{minipage}{0.32\textwidth}\centering
		\xincludegraphics[height=5cm,label={\textbf{g}}]{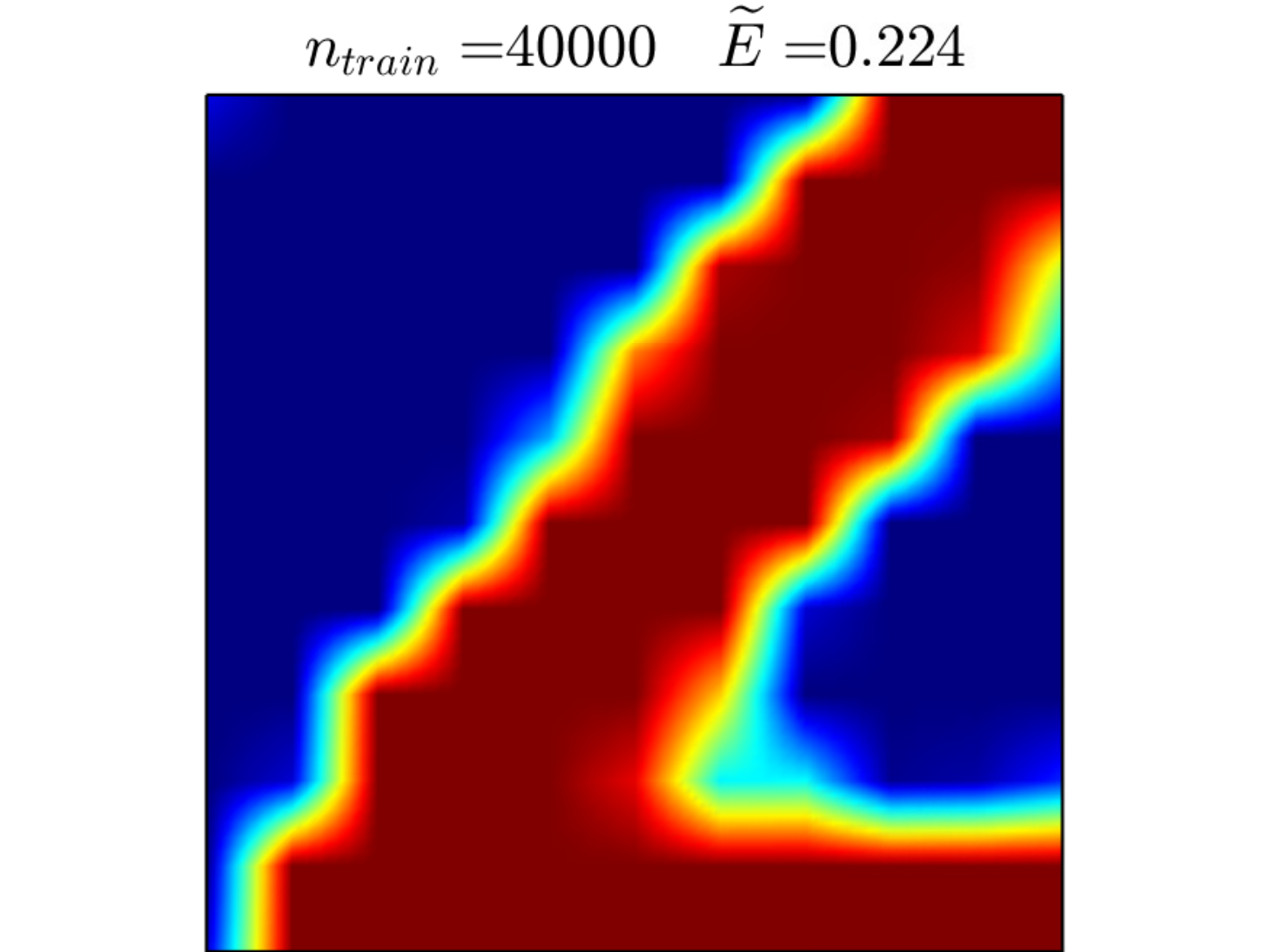}
	\end{minipage}
	\begin{minipage}{0.32\textwidth}\centering
		\xincludegraphics[height=5cm,label={\textbf{h}}]{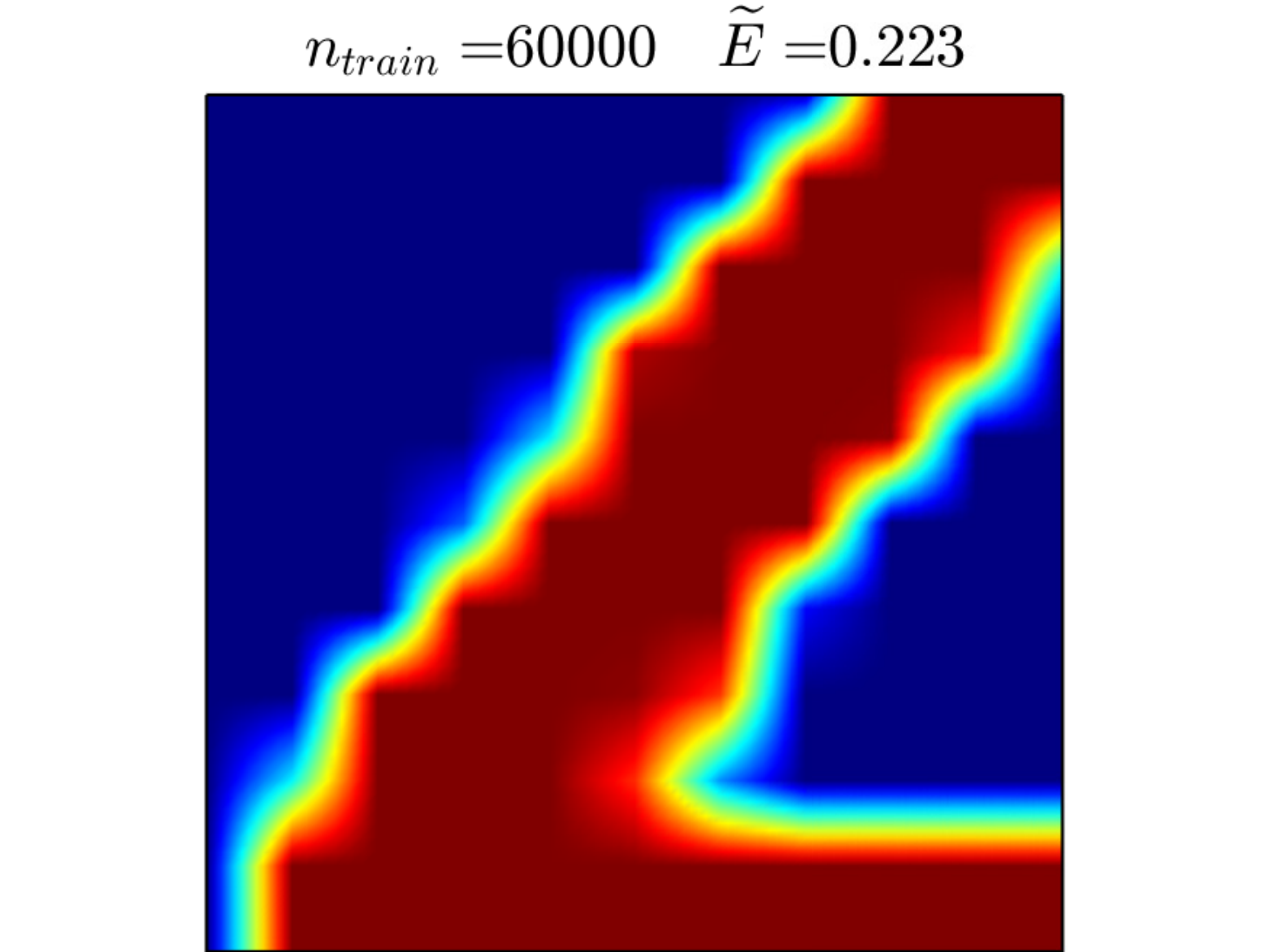}
	\end{minipage}
	\begin{minipage}{0.32\textwidth}\centering
		\xincludegraphics[height=5cm,label={\textbf{i}}]{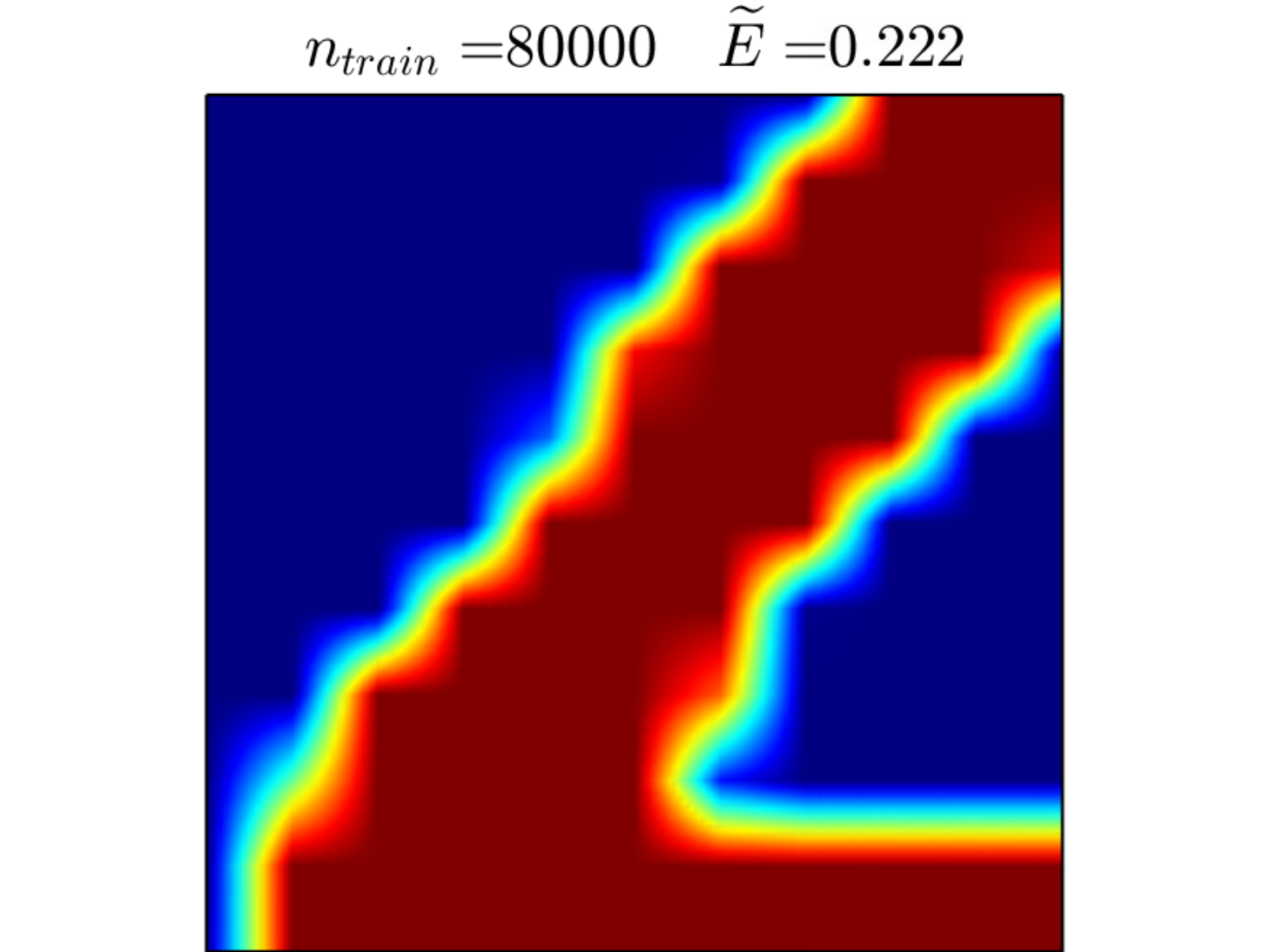}
	\end{minipage}
	\caption{\textbf{Evolution of the solution from SOLO for the compliance minimization problem 11$\times$11 variables.} Each plot is the best among $n_{train}$ accumulated training data  and the corresponding energy $\widetilde{E}$ is marked. There is no obvious change after ten thousand training samples.}
	\label{figs3}
\end{figure}

\begin{figure}[H]
	\centering
	\begin{minipage}{0.33\textwidth}\centering
		\xincludegraphics[width=\textwidth,label={\textbf{a}}]{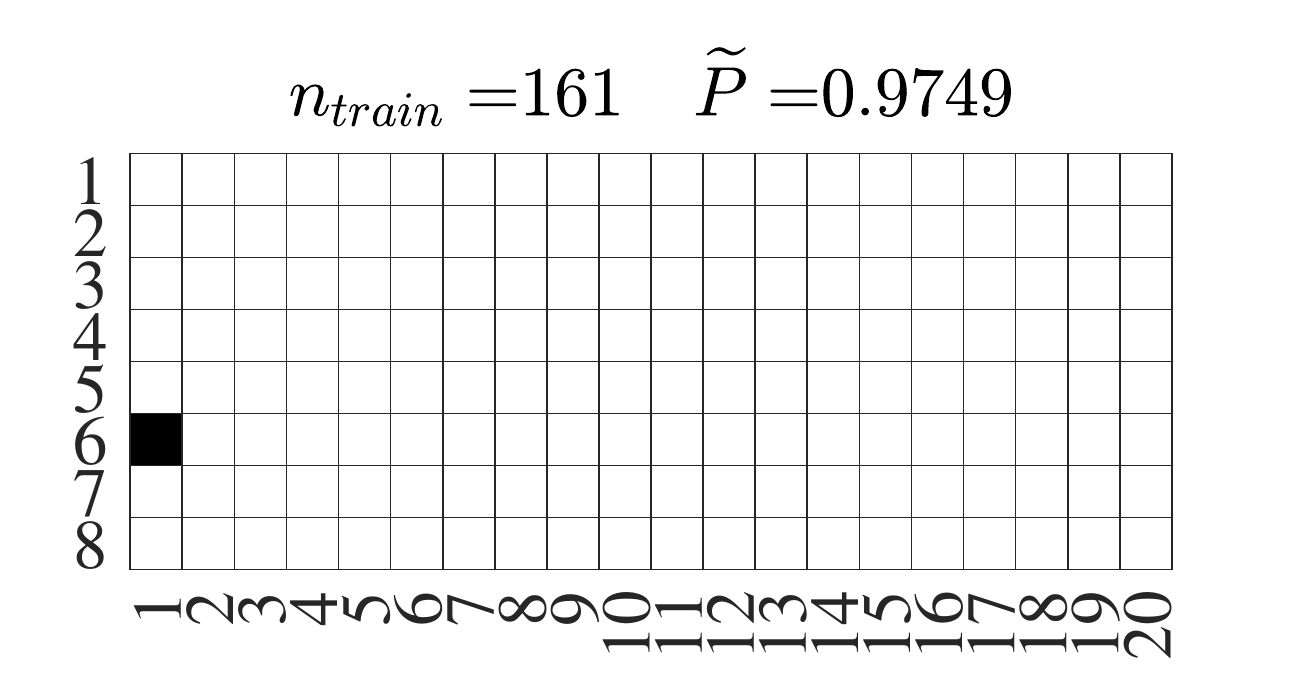}
	\end{minipage}%
	\begin{minipage}{0.33\textwidth}\centering
		\xincludegraphics[width=\textwidth,label={\textbf{b}}]{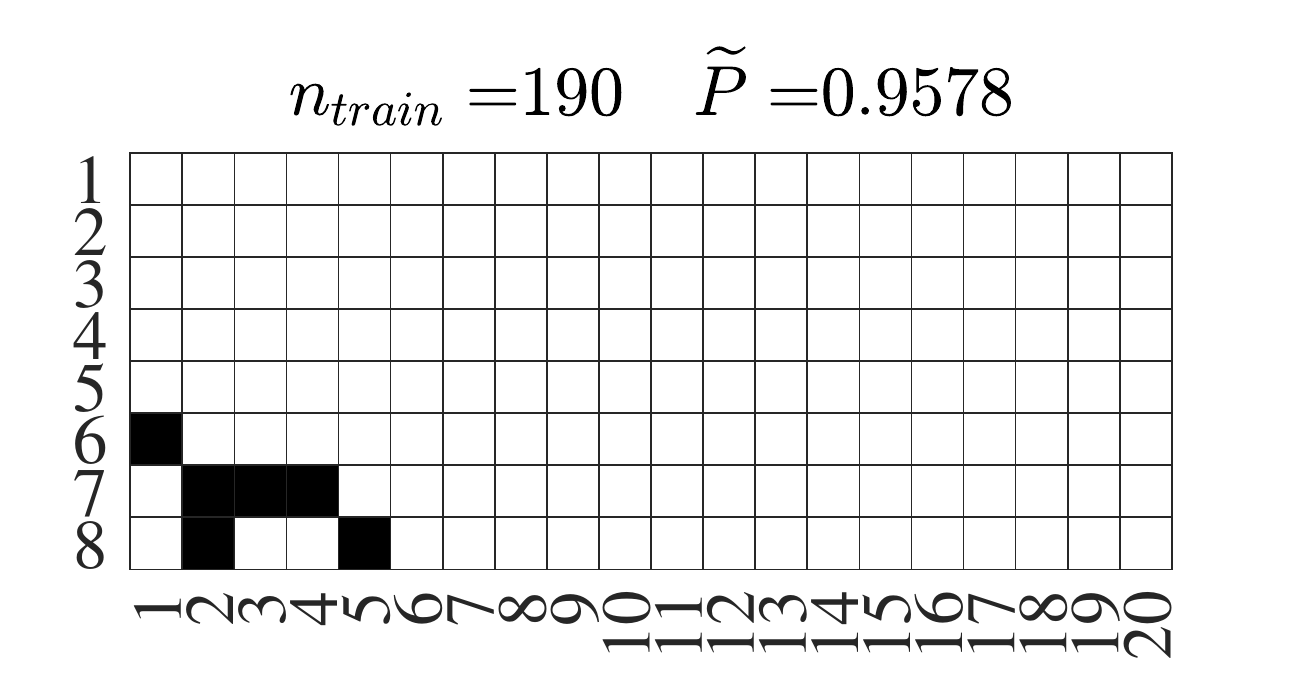}
	\end{minipage}%
	\begin{minipage}{0.33\textwidth}\centering
		\xincludegraphics[width=\textwidth,label={\textbf{c}}]{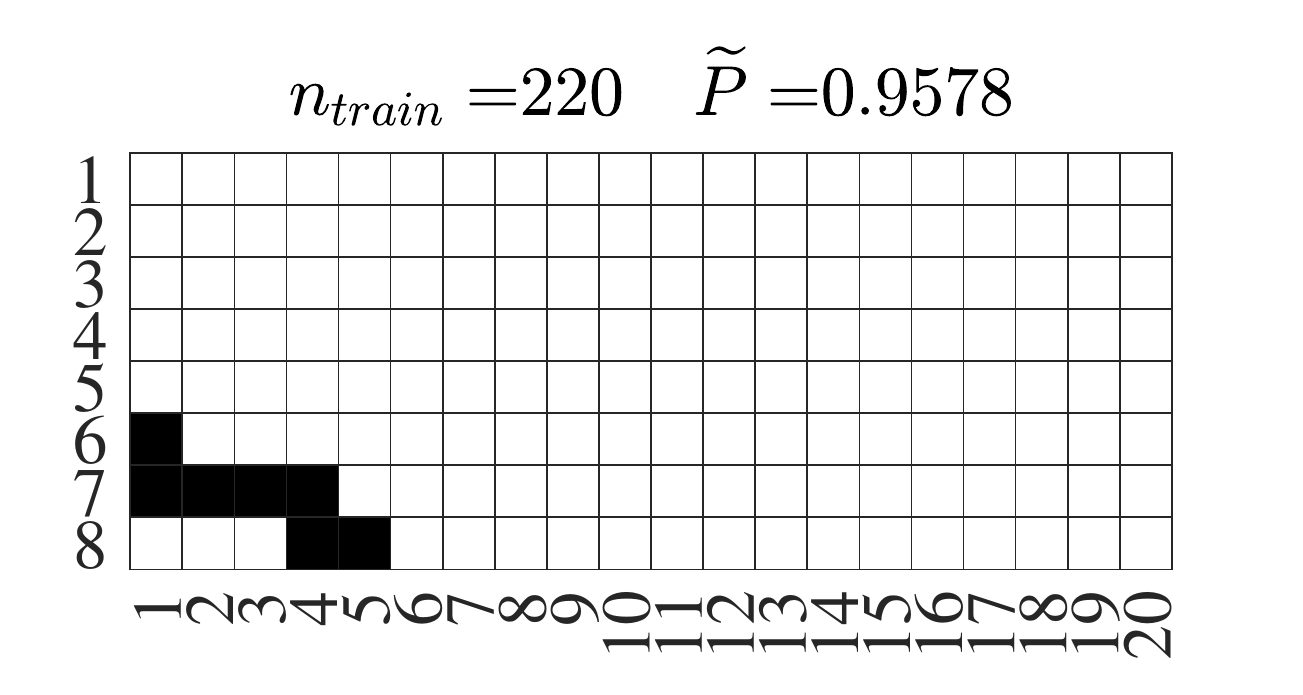}
	\end{minipage}
	\begin{minipage}{0.33\textwidth}\centering
		\xincludegraphics[width=\textwidth,label={\textbf{d}}]{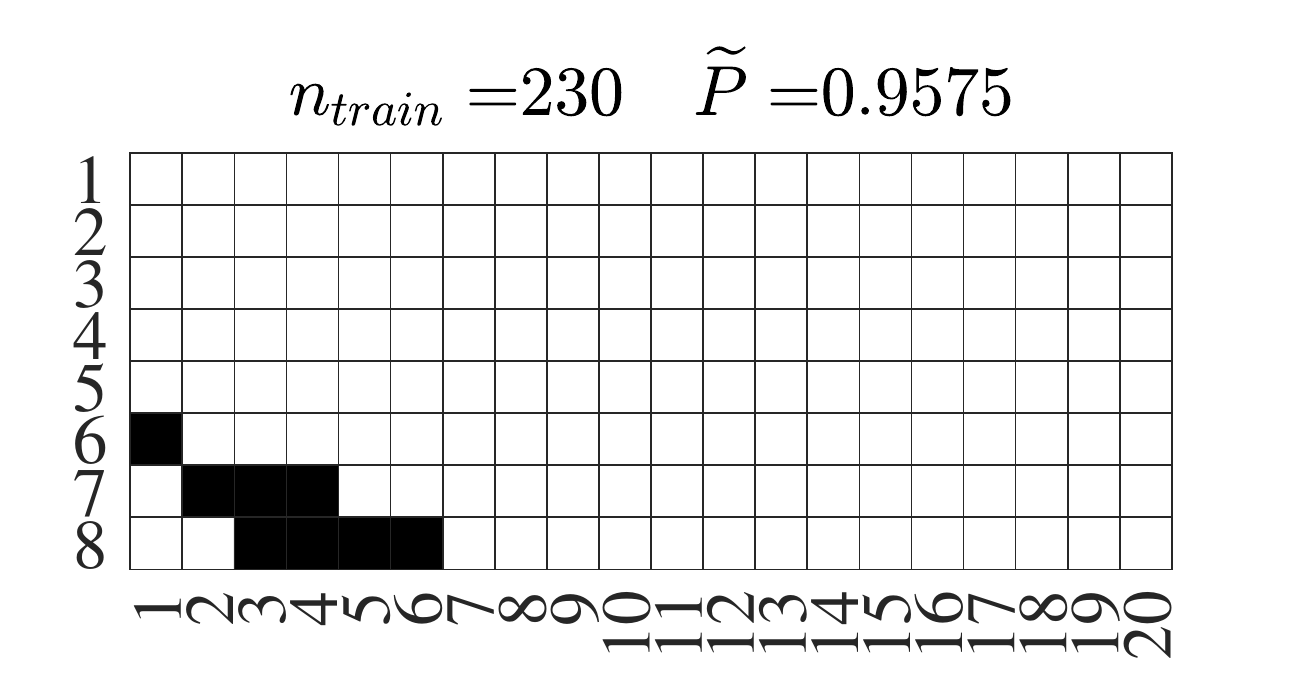}
	\end{minipage}%
	\begin{minipage}{0.33\textwidth}\centering
		\xincludegraphics[width=\textwidth,label={\textbf{e}}]{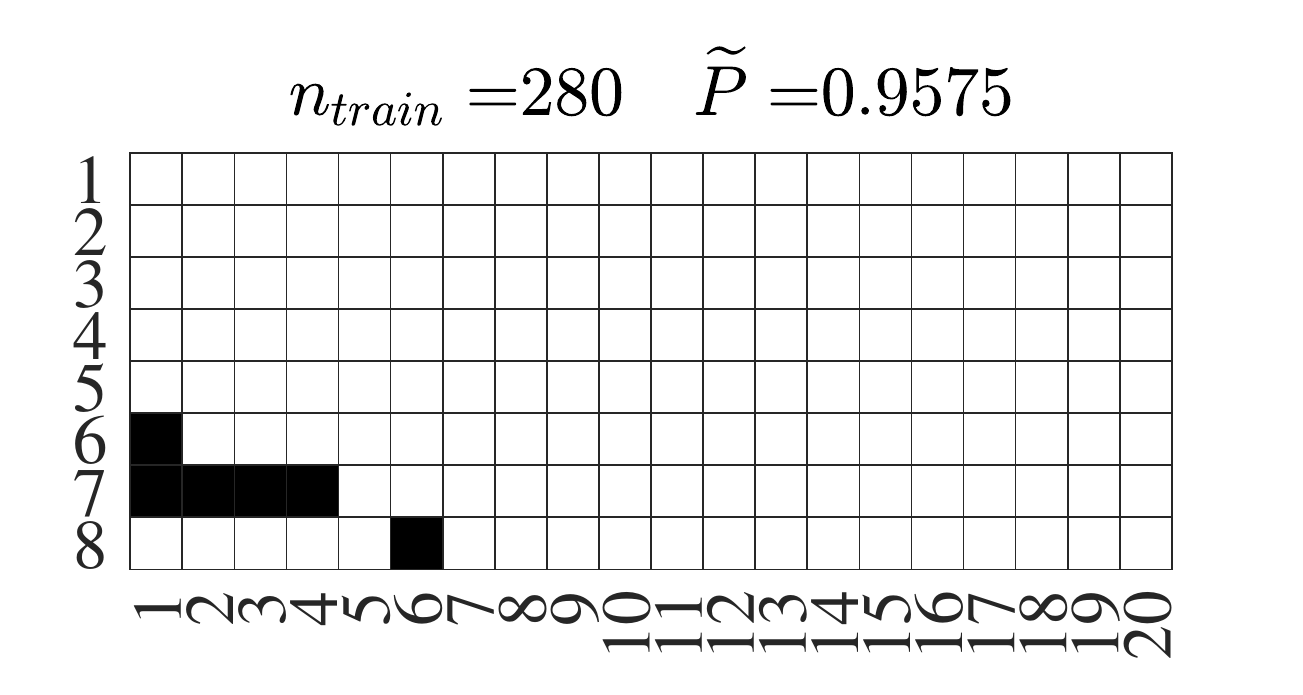}
	\end{minipage}%
	\begin{minipage}{0.33\textwidth}\centering
		\xincludegraphics[width=\textwidth,label={\textbf{f}}]{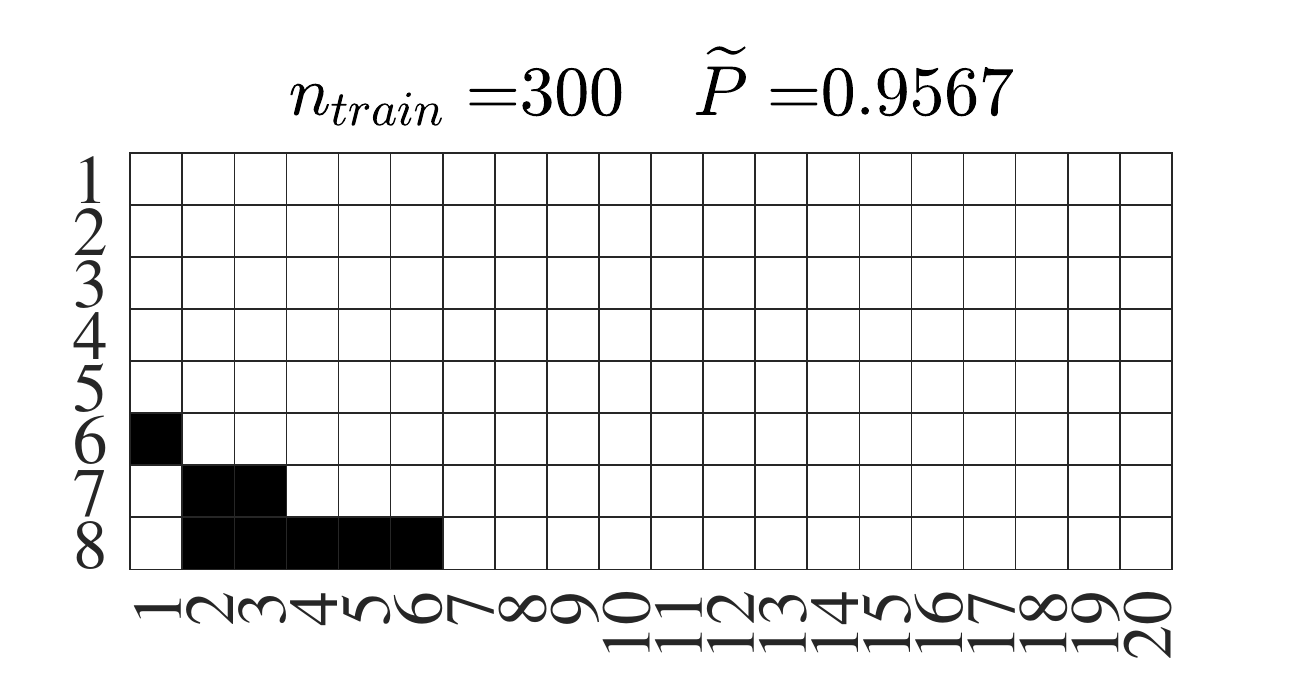}
	\end{minipage}%
	\caption{\textbf{Evolution of the solution from SOLO-G for the fluid-structure optimization problem with 20$\times$8 mesh.} Each plot is the best among $n_{train}$ samples.}
	\label{figs4}
\end{figure}

\begin{figure}[H]
	\centering
	\begin{minipage}{0.5\textwidth}\centering
		\xincludegraphics[width=\textwidth,label={\textbf{a}}]{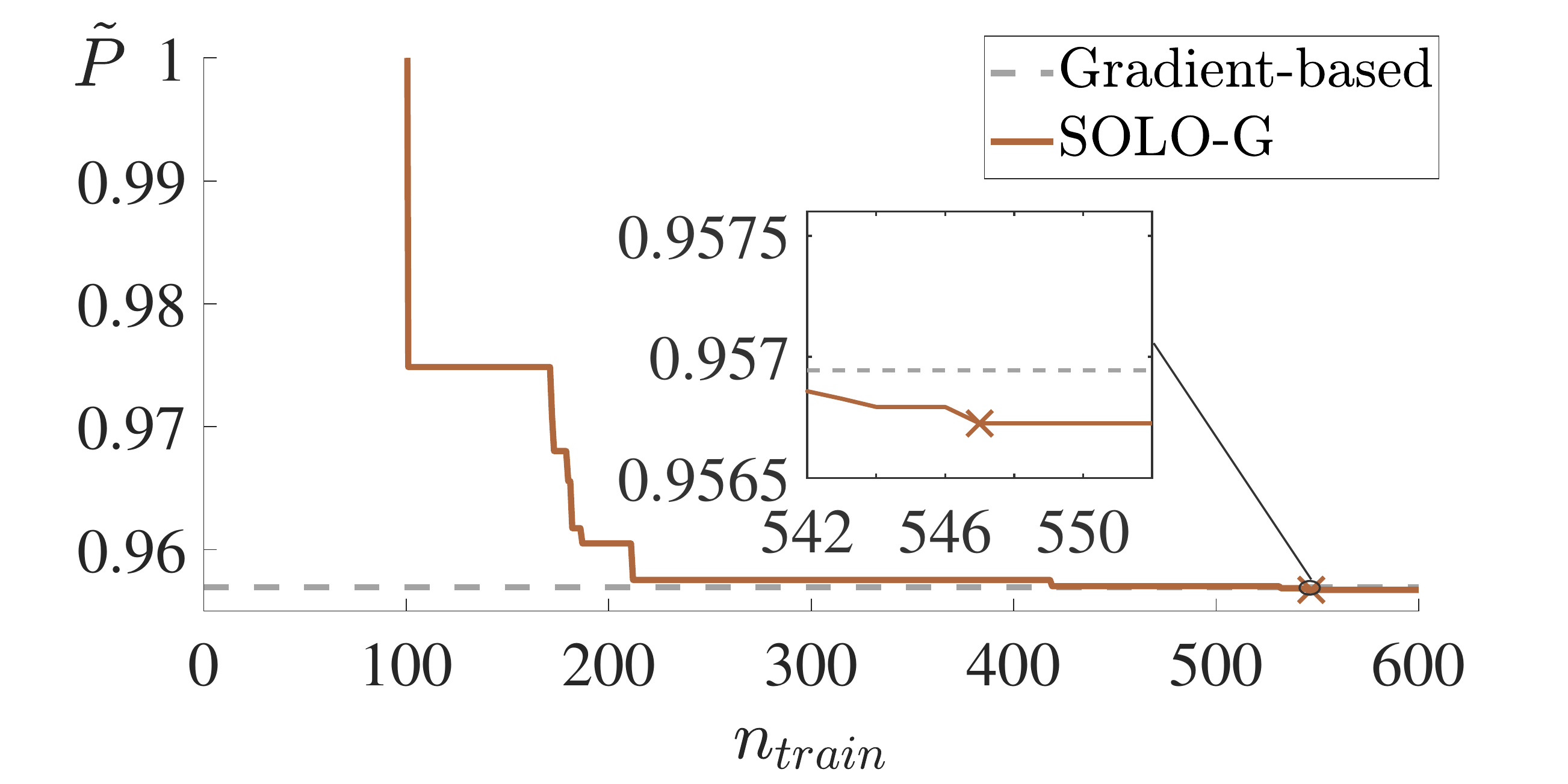}
	\end{minipage}%
	\begin{minipage}{0.5\textwidth}\centering
		\xincludegraphics[width=\textwidth,label={\textbf{b}}]{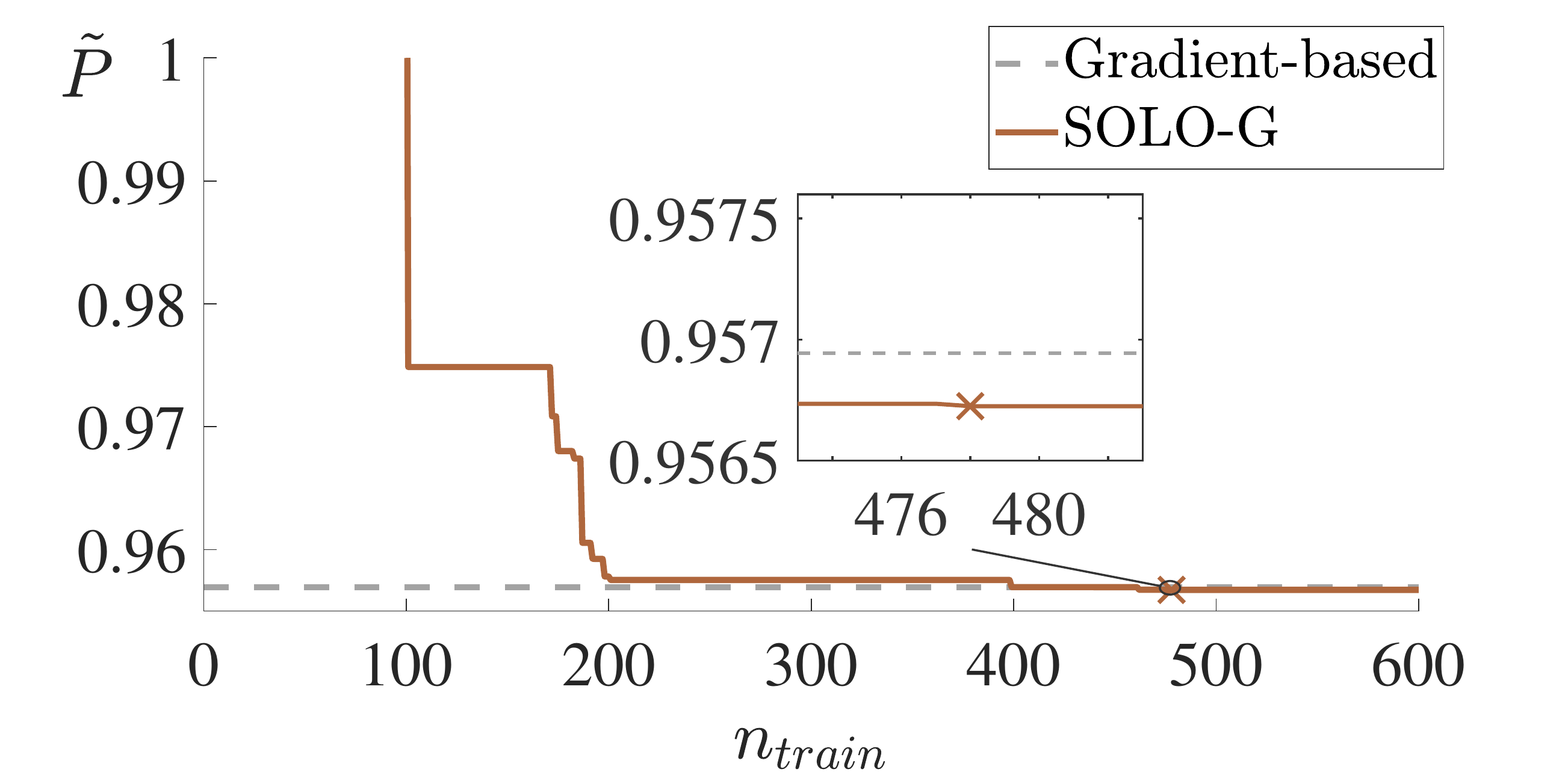}
	\end{minipage}
	\begin{minipage}{0.5\textwidth}\centering
		\xincludegraphics[width=\textwidth,label={\textbf{c}}]{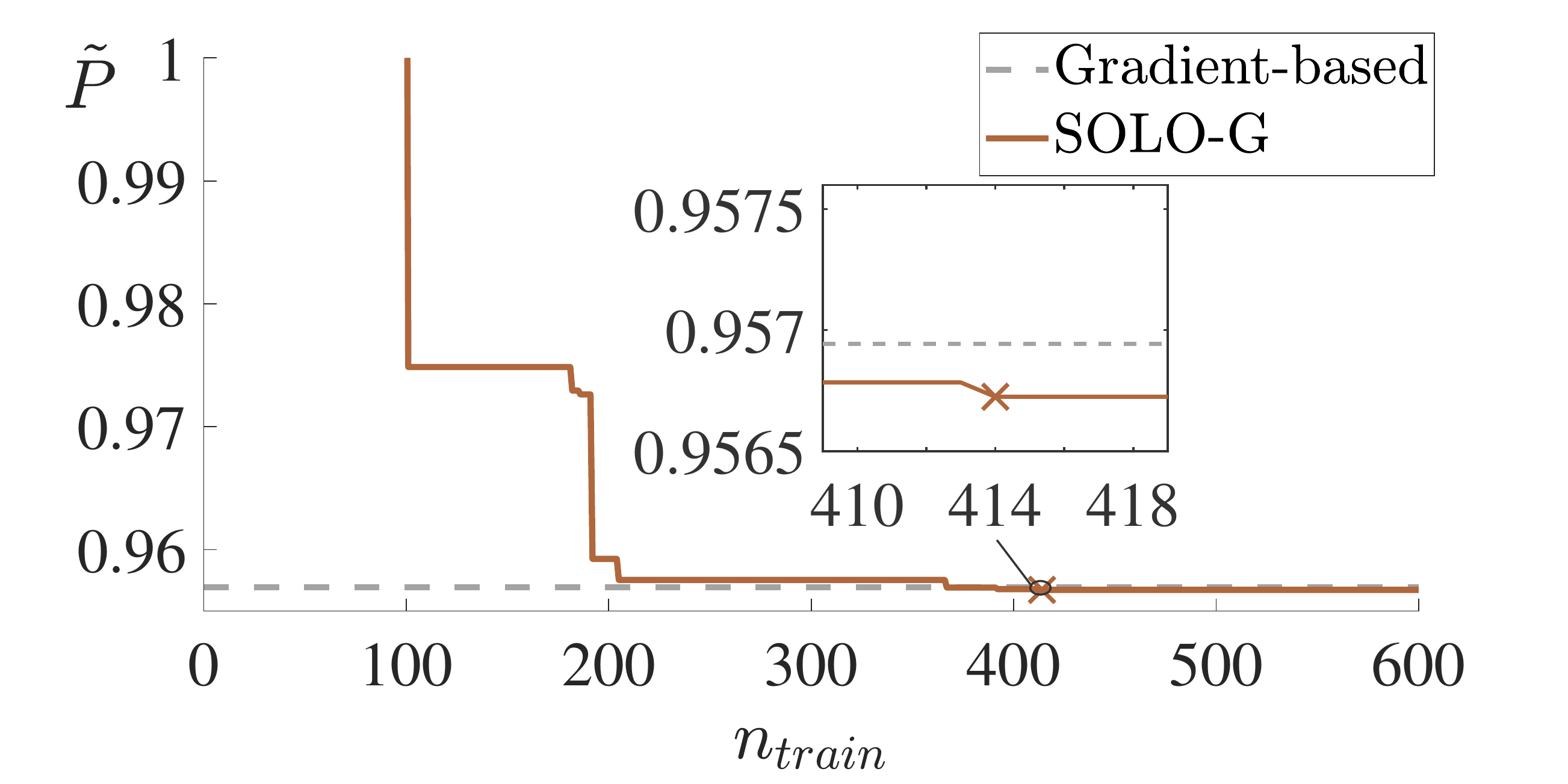}
	\end{minipage}%
	\begin{minipage}{0.5\textwidth}\centering
		\xincludegraphics[width=\textwidth,label={\textbf{d}}]{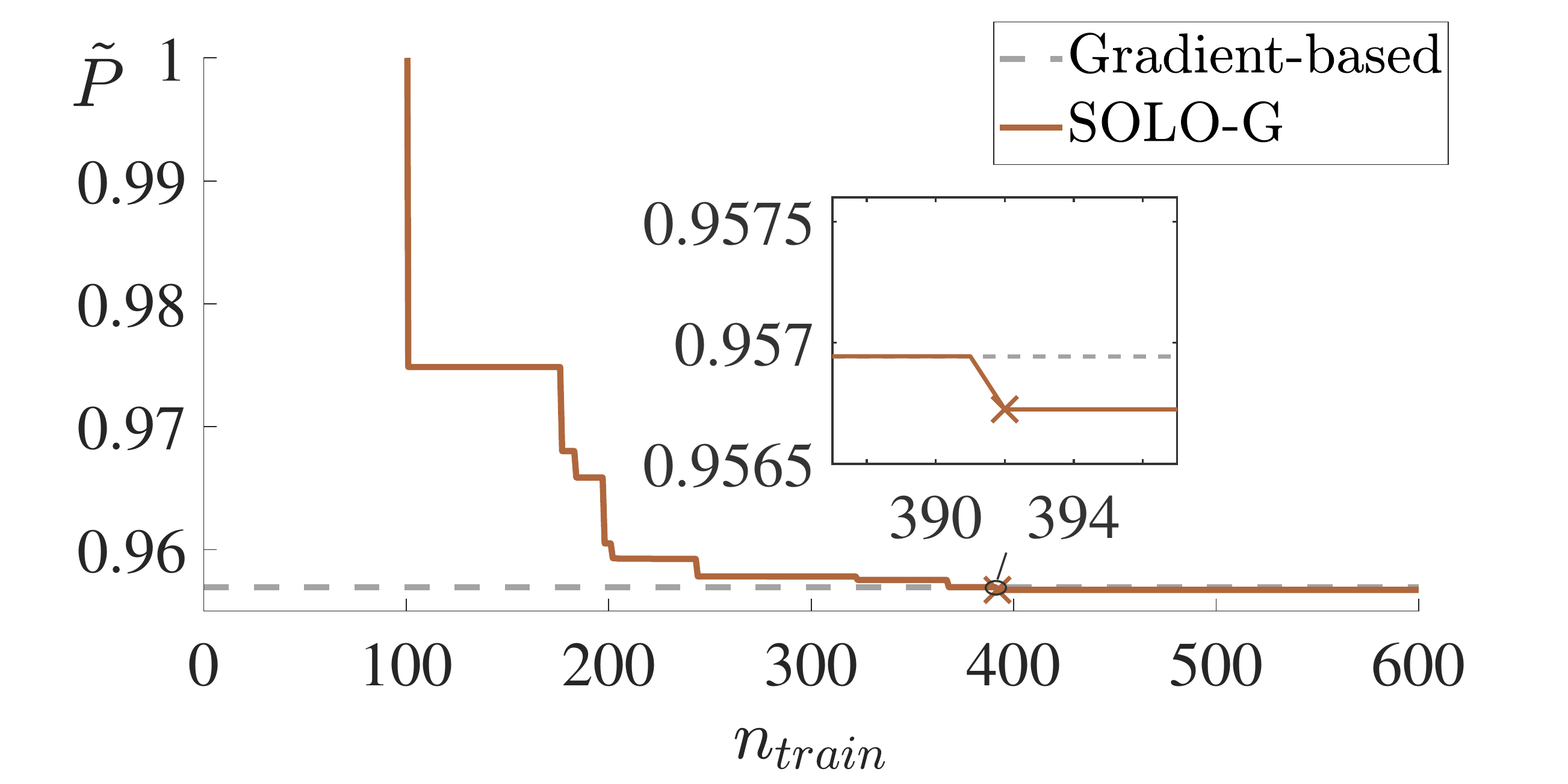}
	\end{minipage}
	\caption{\textbf{Repeating SOLO-G for the fluid-structure optimization problem with 20$\times$8 mesh.} All configurations are the same as Fig. 4b except different random seeds. They obtain the same objective $\widetilde{P}$ despite different convergence rate.}
	\label{figs5}
\end{figure}

\begin{figure}[H]
	\centering
	\begin{minipage}{0.5\textwidth}\centering
		\xincludegraphics[width=\textwidth,label={\textbf{a}}]{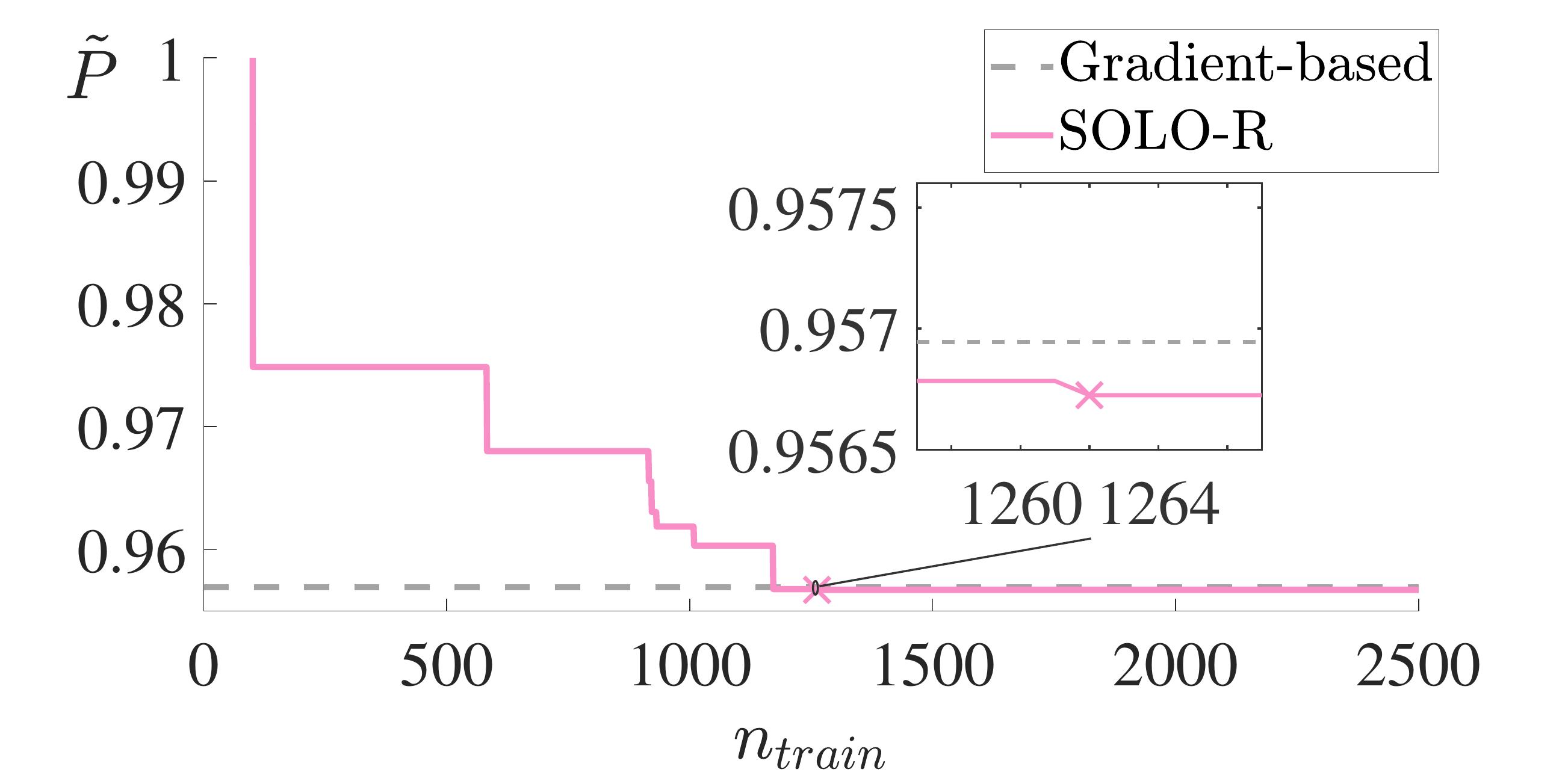}
	\end{minipage}%
	\begin{minipage}{0.5\textwidth}\centering
		\xincludegraphics[width=\textwidth,label={\textbf{b}}]{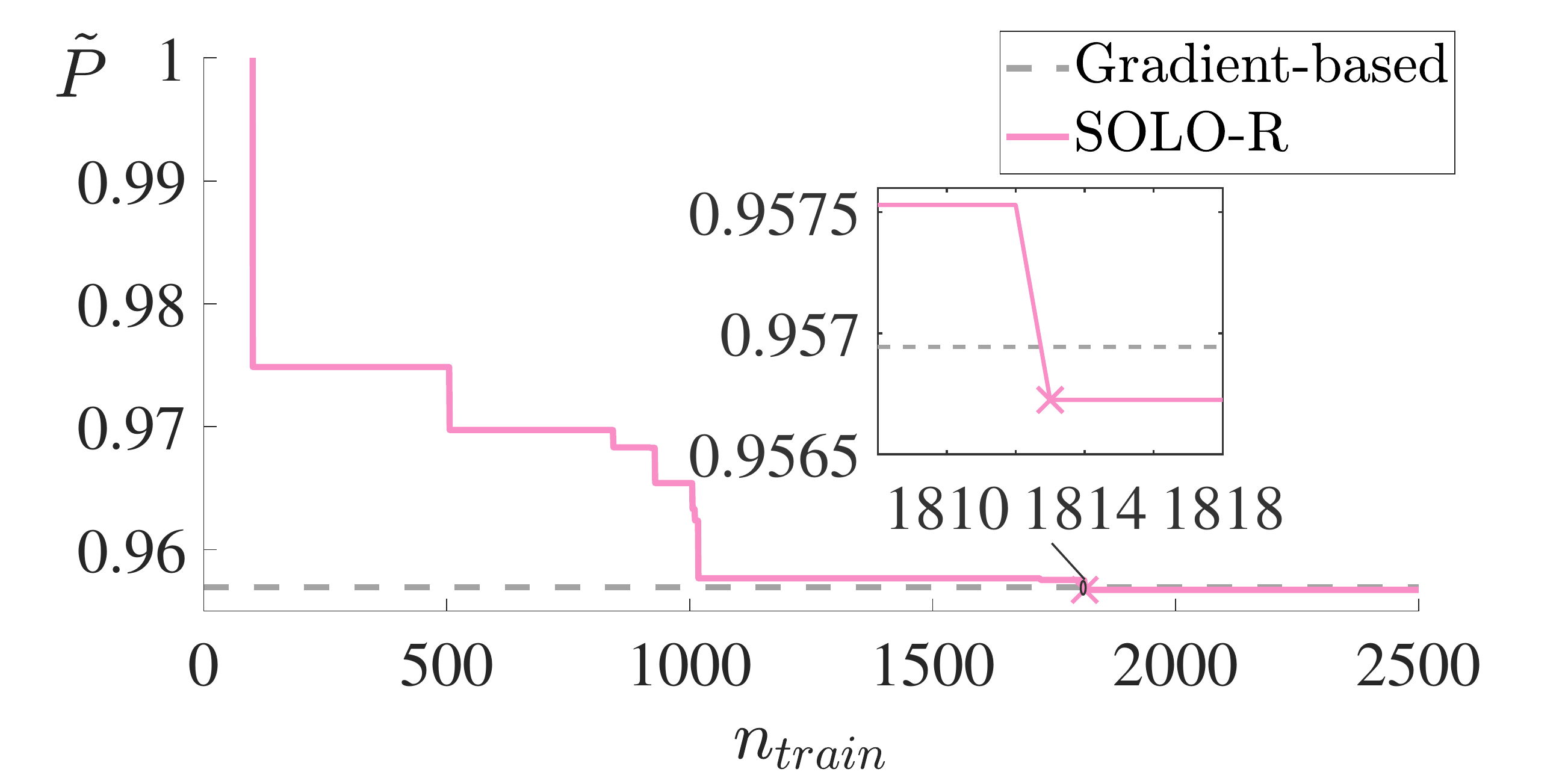}
	\end{minipage}
	\begin{minipage}{0.5\textwidth}\centering
		\xincludegraphics[width=\textwidth,label={\textbf{c}}]{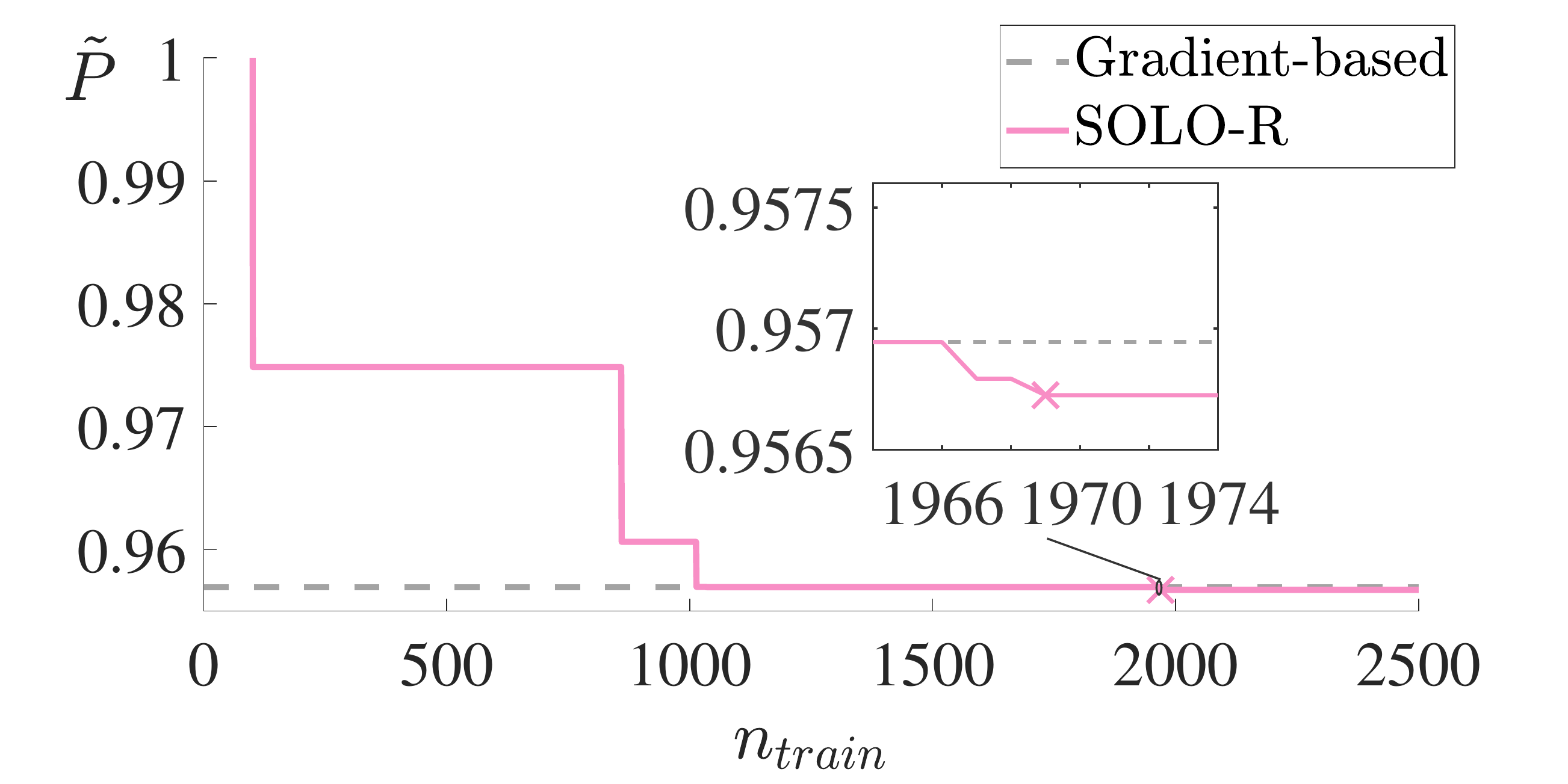}
	\end{minipage}%
	\begin{minipage}{0.5\textwidth}\centering
		\xincludegraphics[width=\textwidth,label={\textbf{d}}]{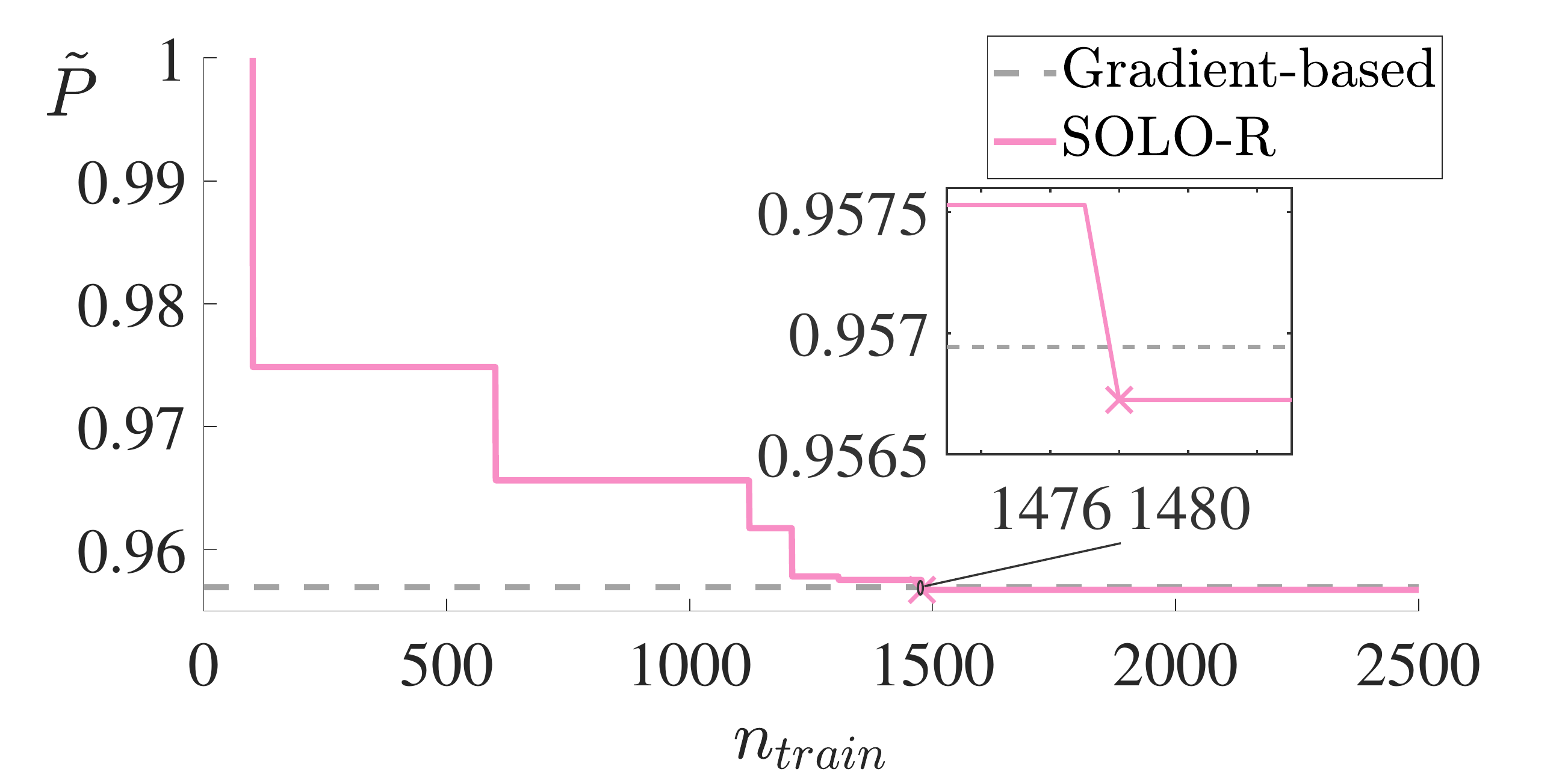}
	\end{minipage}
	\caption{\textbf{Repeating SOLO-R for the fluid-structure optimization problem with 20$\times$8 mesh.} All configurations are the same as Fig. 4b except different random seeds. They obtain the same objective $\widetilde{P}$ despite different convergence rate.}
	\label{figs6}
\end{figure}

\begin{figure}[H]
	\centering
	\begin{minipage}{0.5\textwidth}\centering
		\xincludegraphics[width=\textwidth,label={\textbf{a}}]{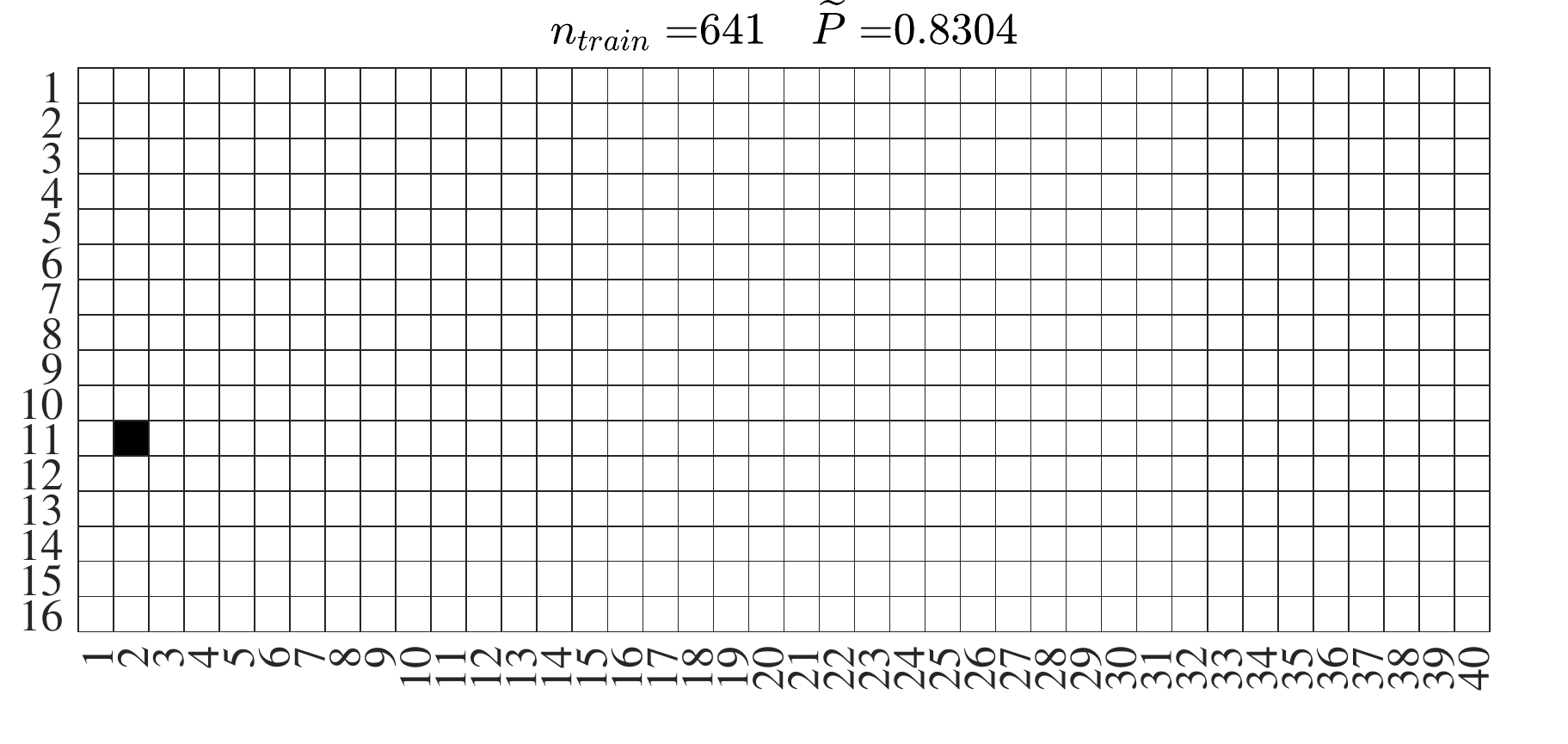}
	\end{minipage}%
	\begin{minipage}{0.5\textwidth}\centering
		\xincludegraphics[width=\textwidth,label={\textbf{b}}]{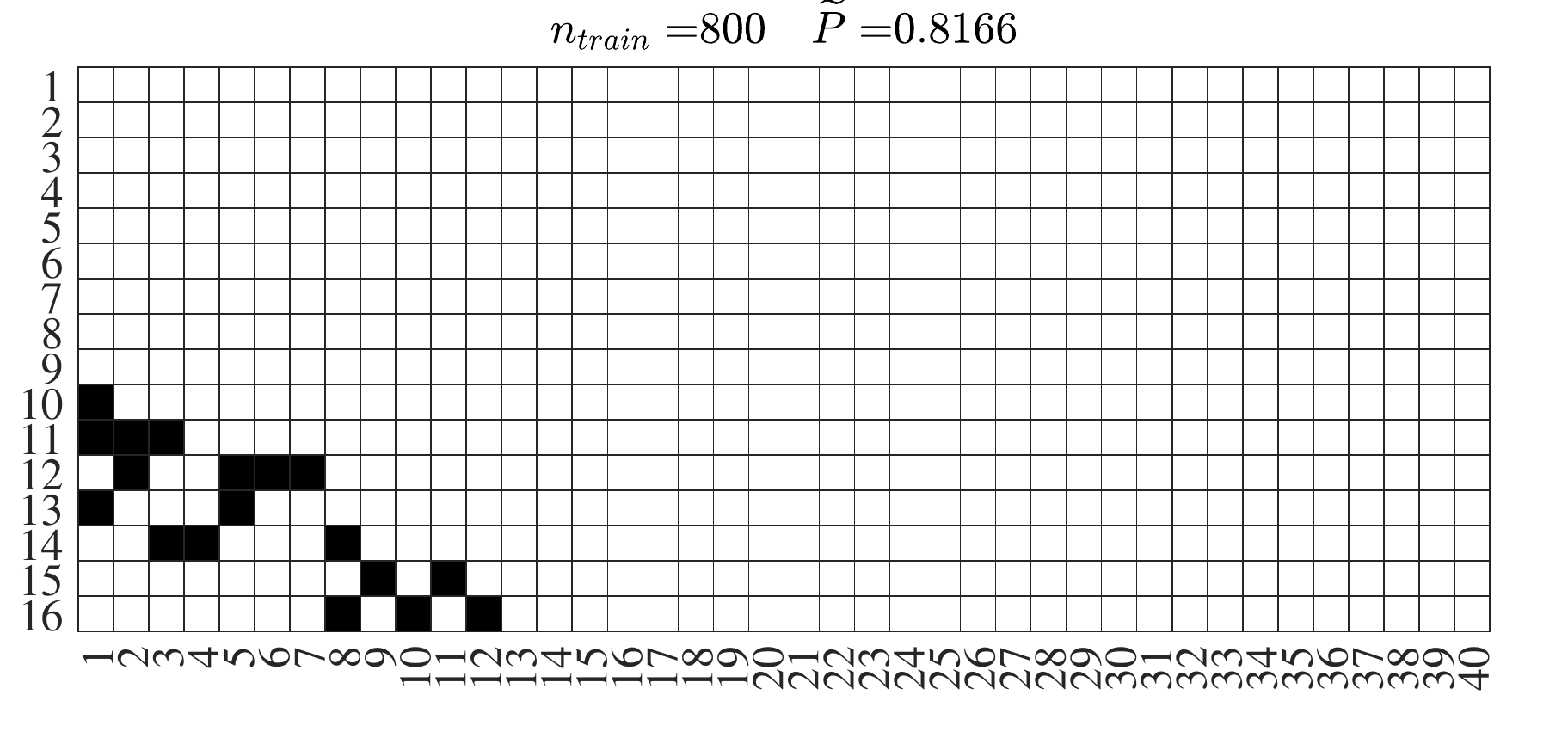}
	\end{minipage}
	\begin{minipage}{0.5\textwidth}\centering
		\xincludegraphics[width=\textwidth,label={\textbf{c}}]{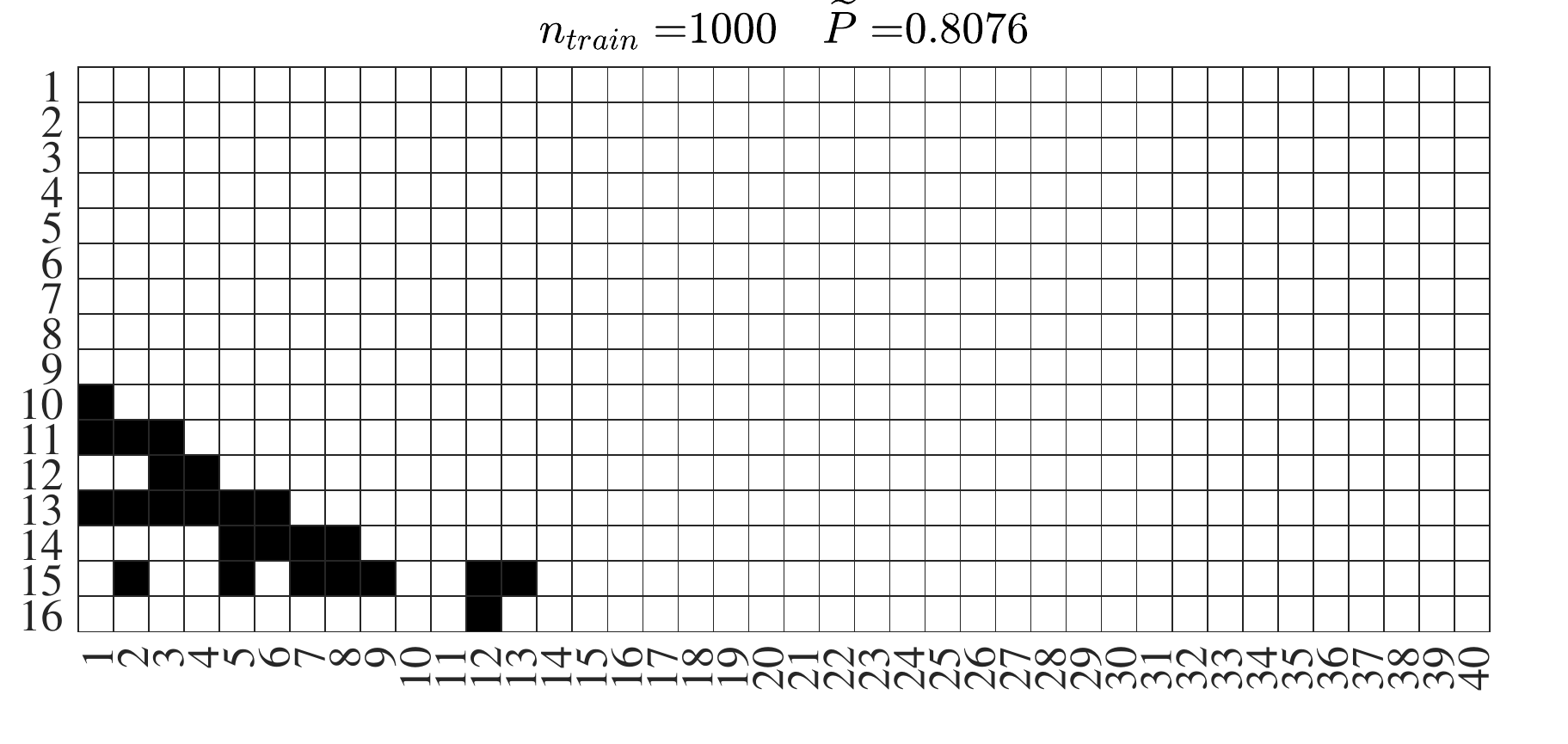}
	\end{minipage}%
	\begin{minipage}{0.5\textwidth}\centering
		\xincludegraphics[width=\textwidth,label={\textbf{d}}]{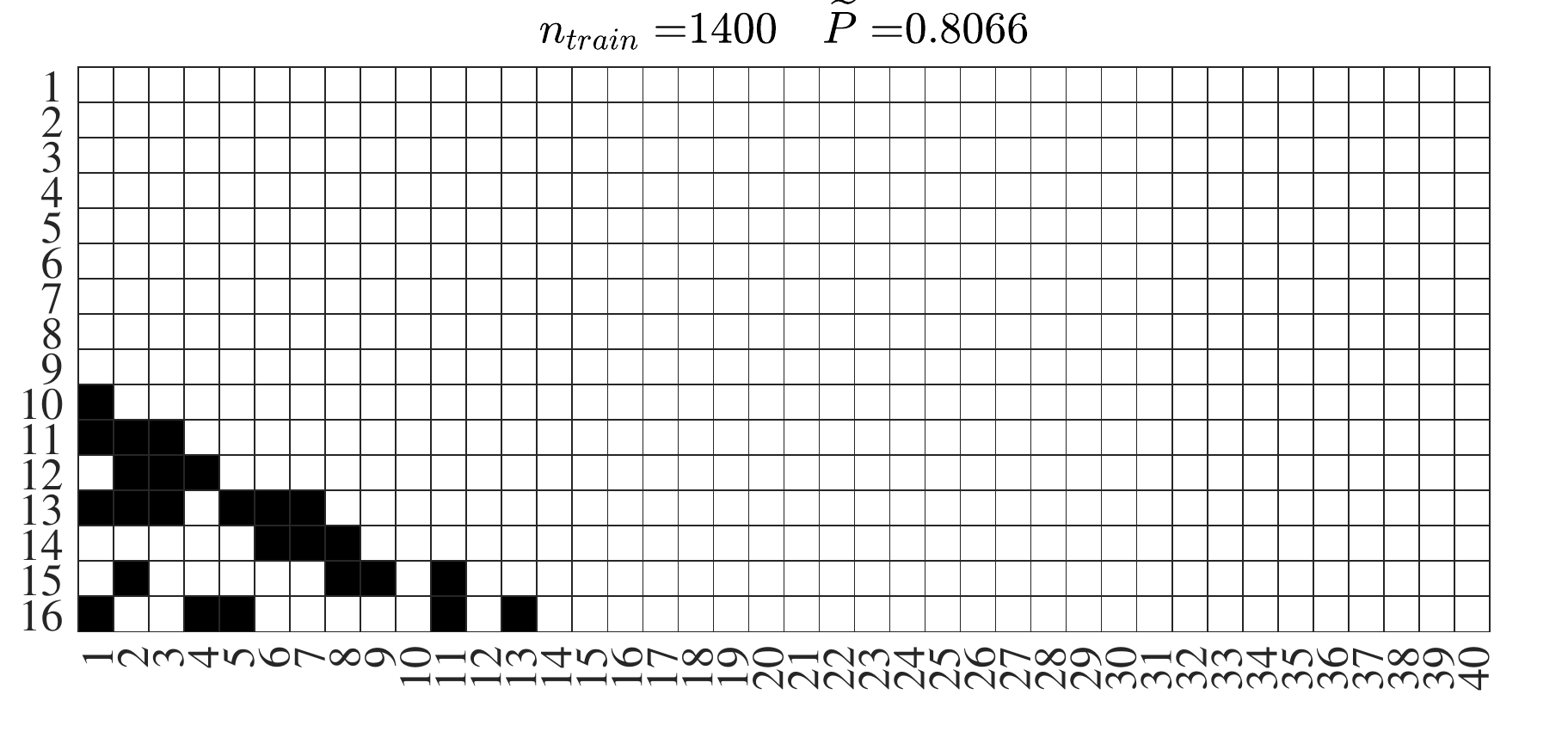}
	\end{minipage}
	\begin{minipage}{0.5\textwidth}\centering
		\xincludegraphics[width=\textwidth,label={\textbf{e}}]{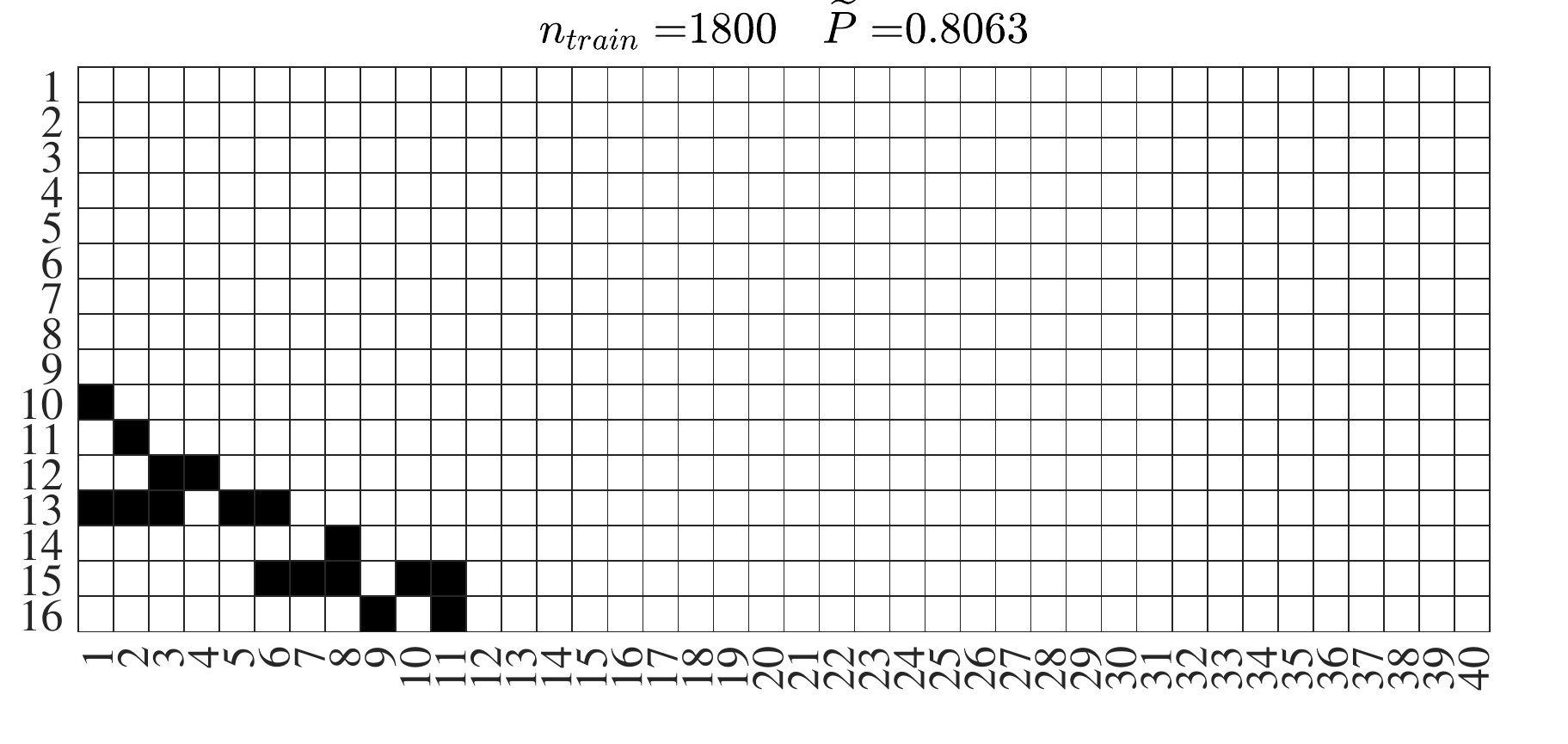}
	\end{minipage}%
	\begin{minipage}{0.5\textwidth}\centering
		\xincludegraphics[width=\textwidth,label={\textbf{f}}]{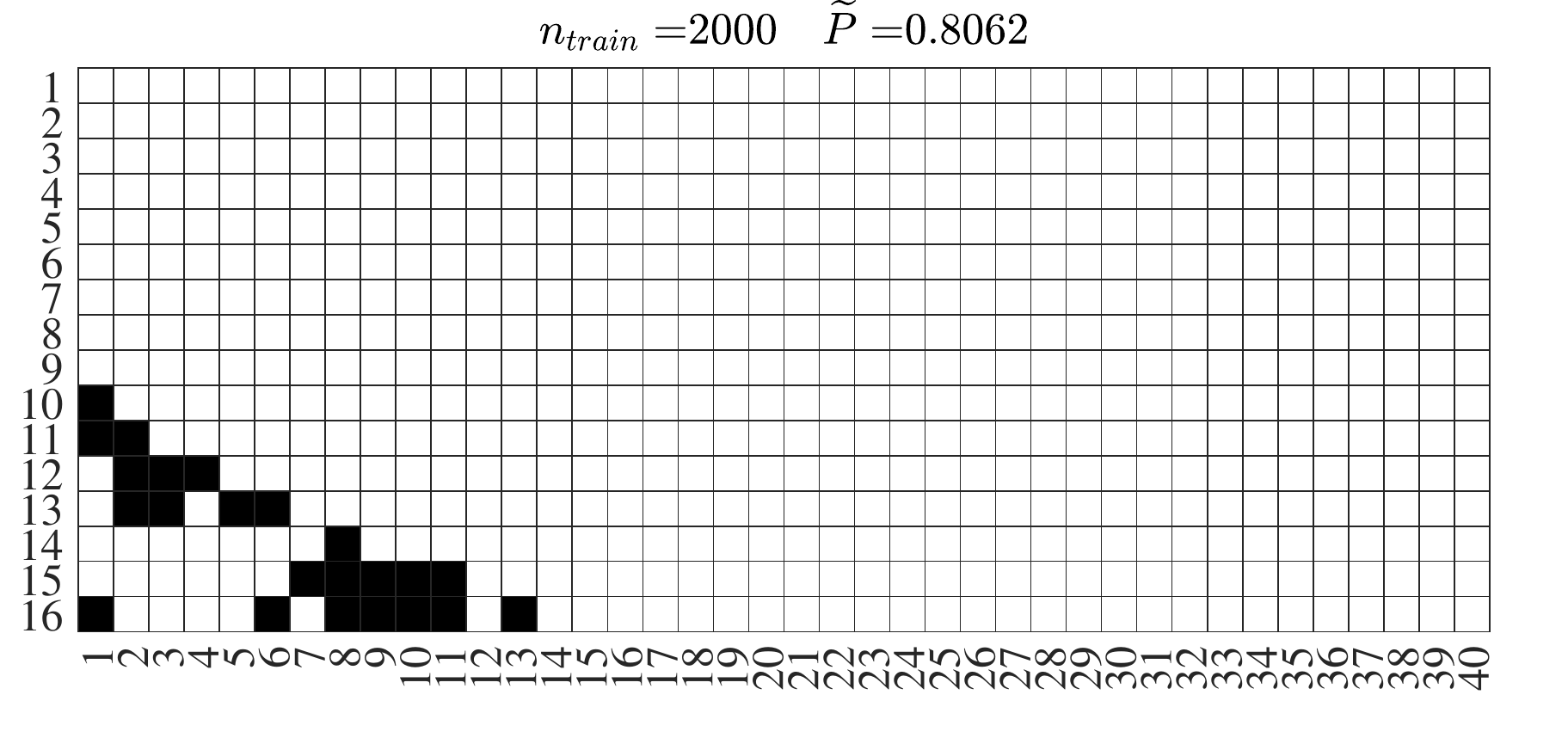}
	\end{minipage}%
	\caption{\textbf{Evolution of the solution from SOLO-G for the fluid-structure optimization problem with 40$\times$16 mesh.} Each plot is the best among $n_{train}$ samples.}
	\label{figs7}
\end{figure}

\begin{figure}[H]
	\centering
	\begin{minipage}{1\textwidth}\centering
		\xincludegraphics[width=0.5\textwidth,label={\textbf{a}}]{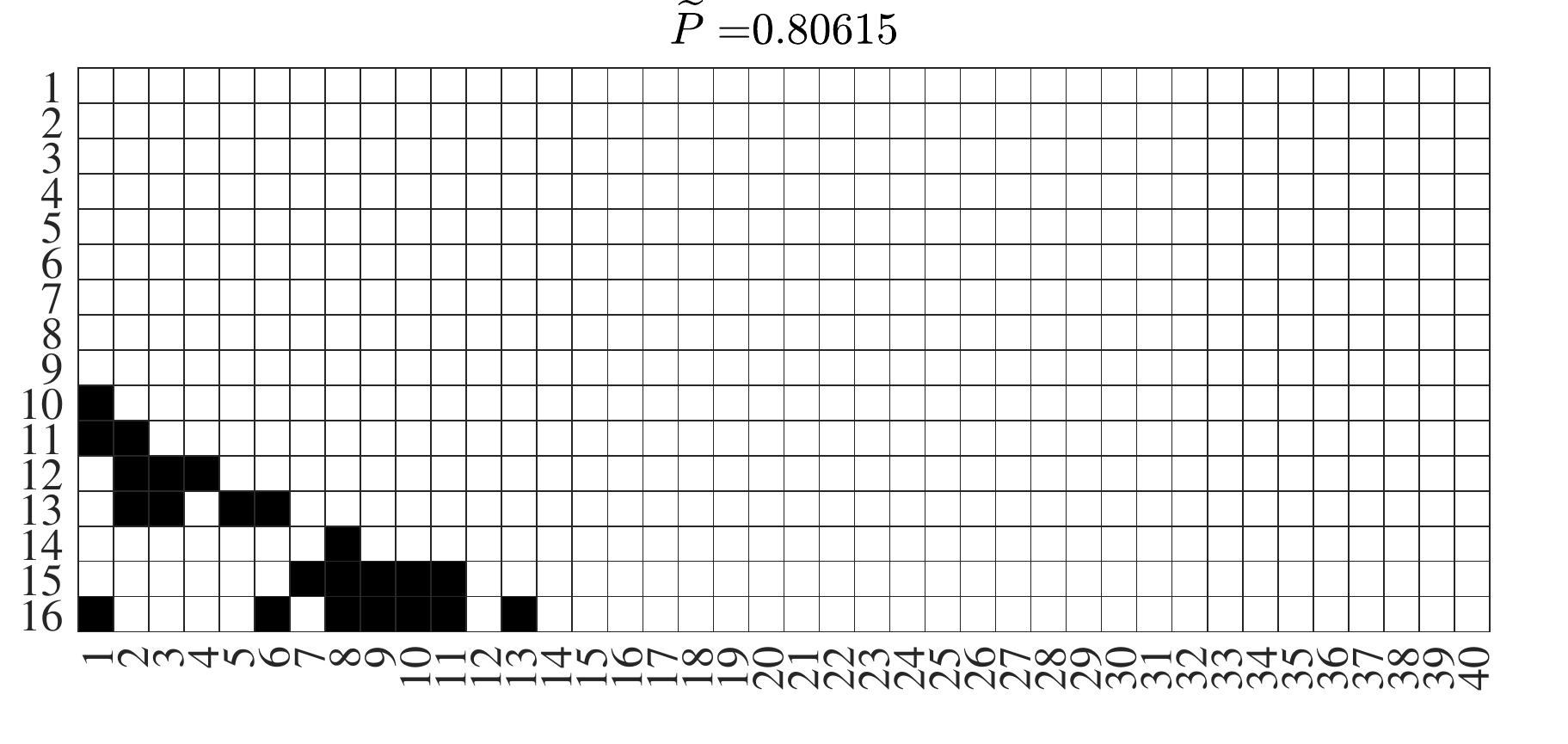}
	\end{minipage}
	\begin{minipage}{0.5\textwidth}\centering
		\xincludegraphics[width=\textwidth,label={\textbf{b}}]{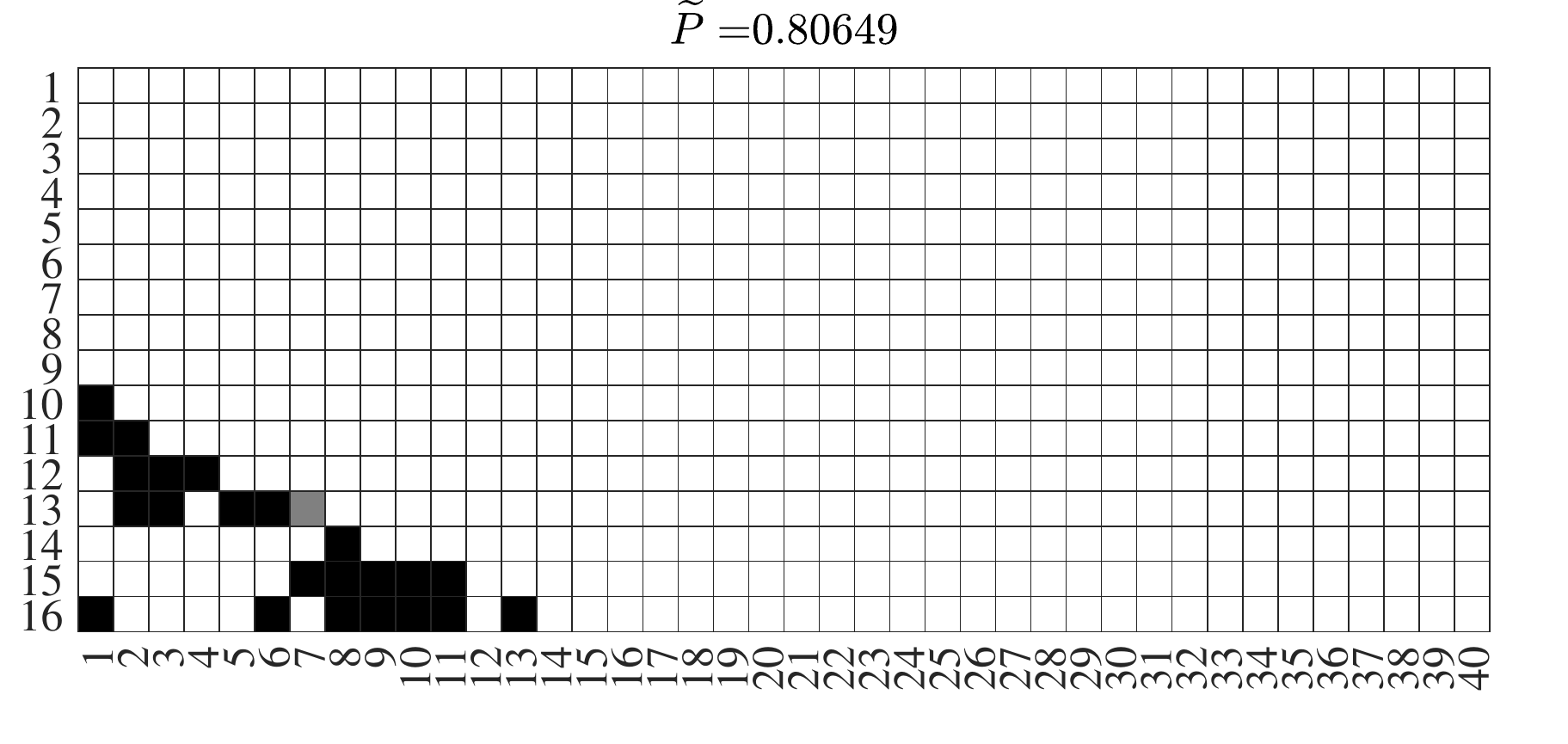}
	\end{minipage}%
	\begin{minipage}{0.5\textwidth}\centering
		\xincludegraphics[width=\textwidth,label={\textbf{c}}]{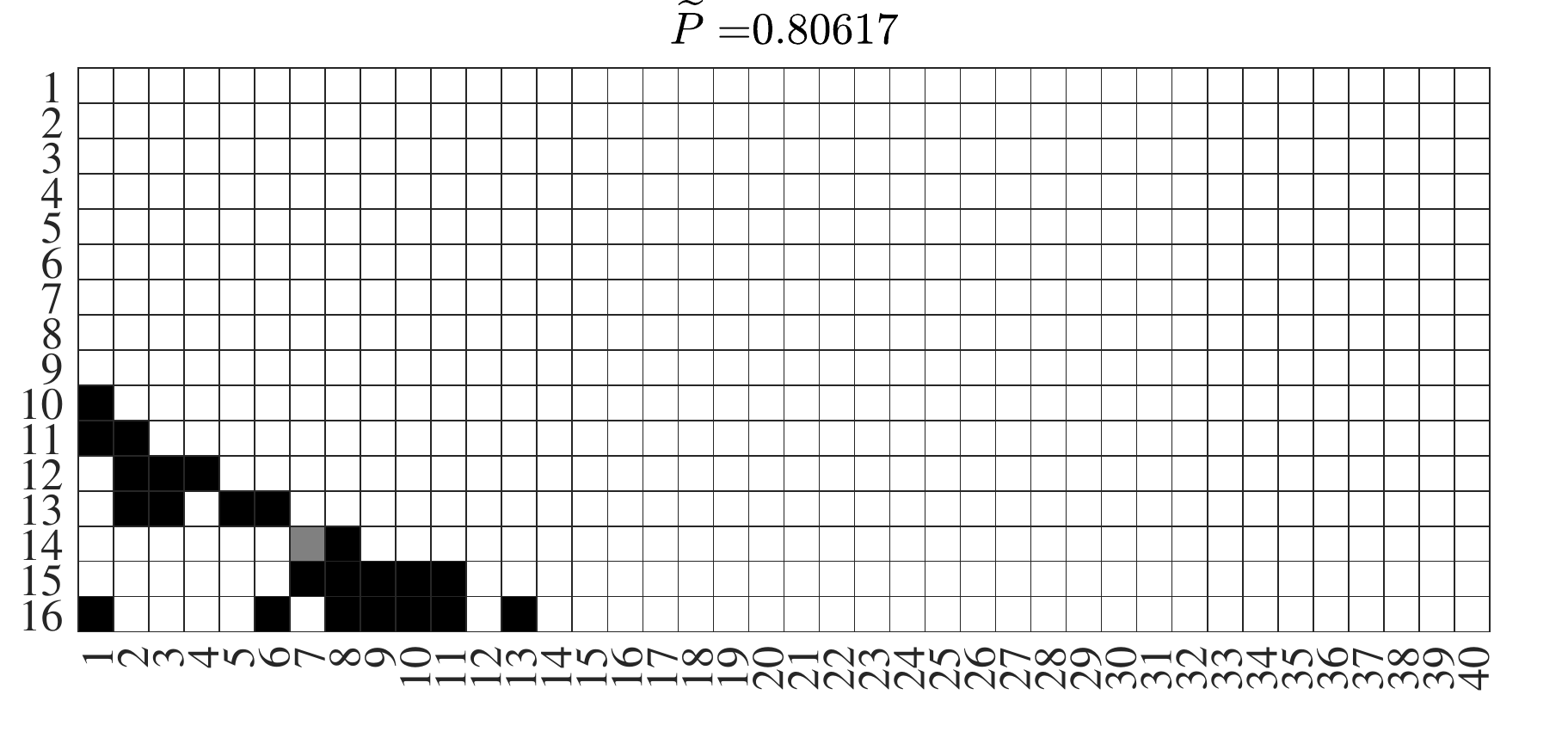}
	\end{minipage}
	\begin{minipage}{0.5\textwidth}\centering
		\xincludegraphics[width=\textwidth,label={\textbf{d}}]{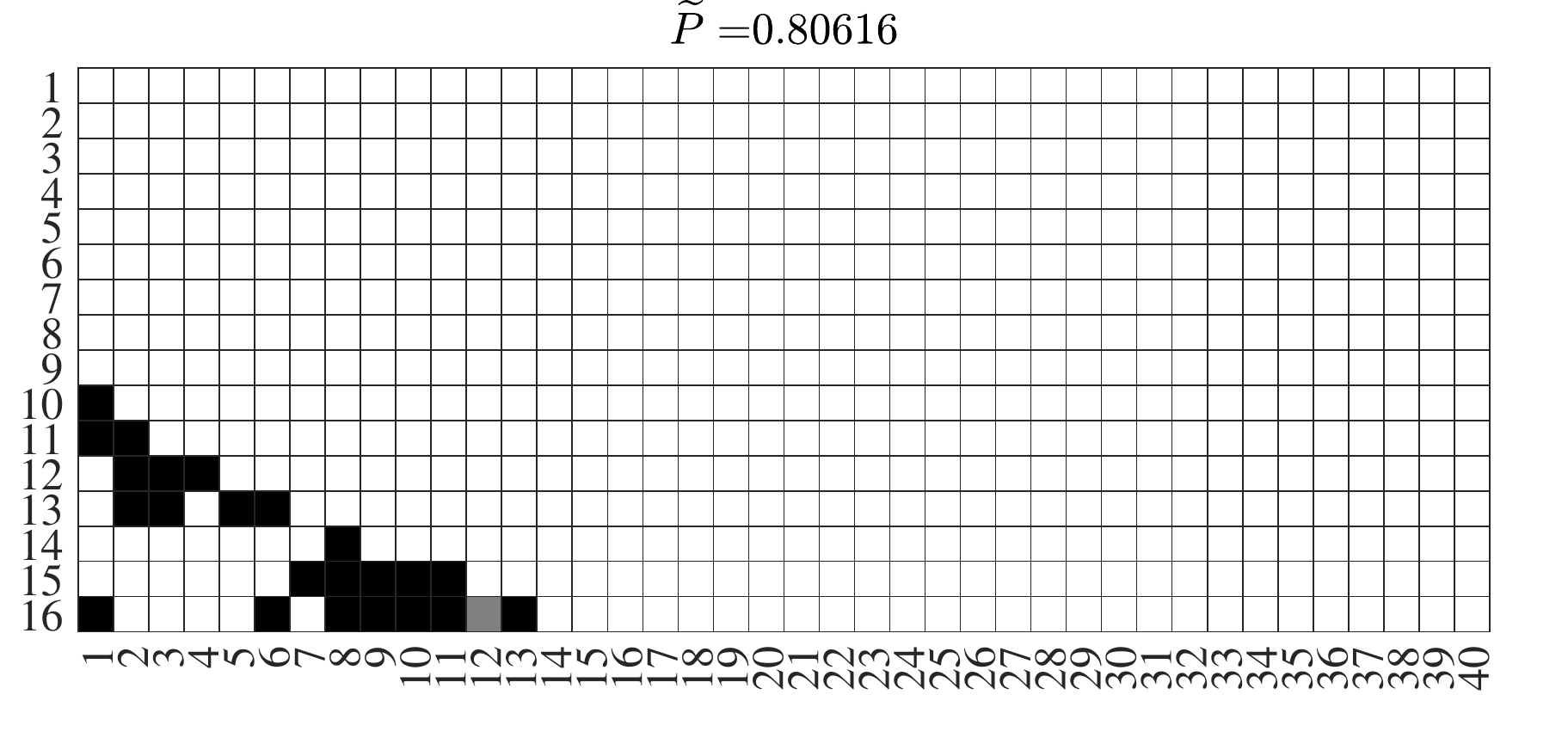}
	\end{minipage}%
	\begin{minipage}{0.5\textwidth}\centering
		\xincludegraphics[width=\textwidth,label={\textbf{e}}]{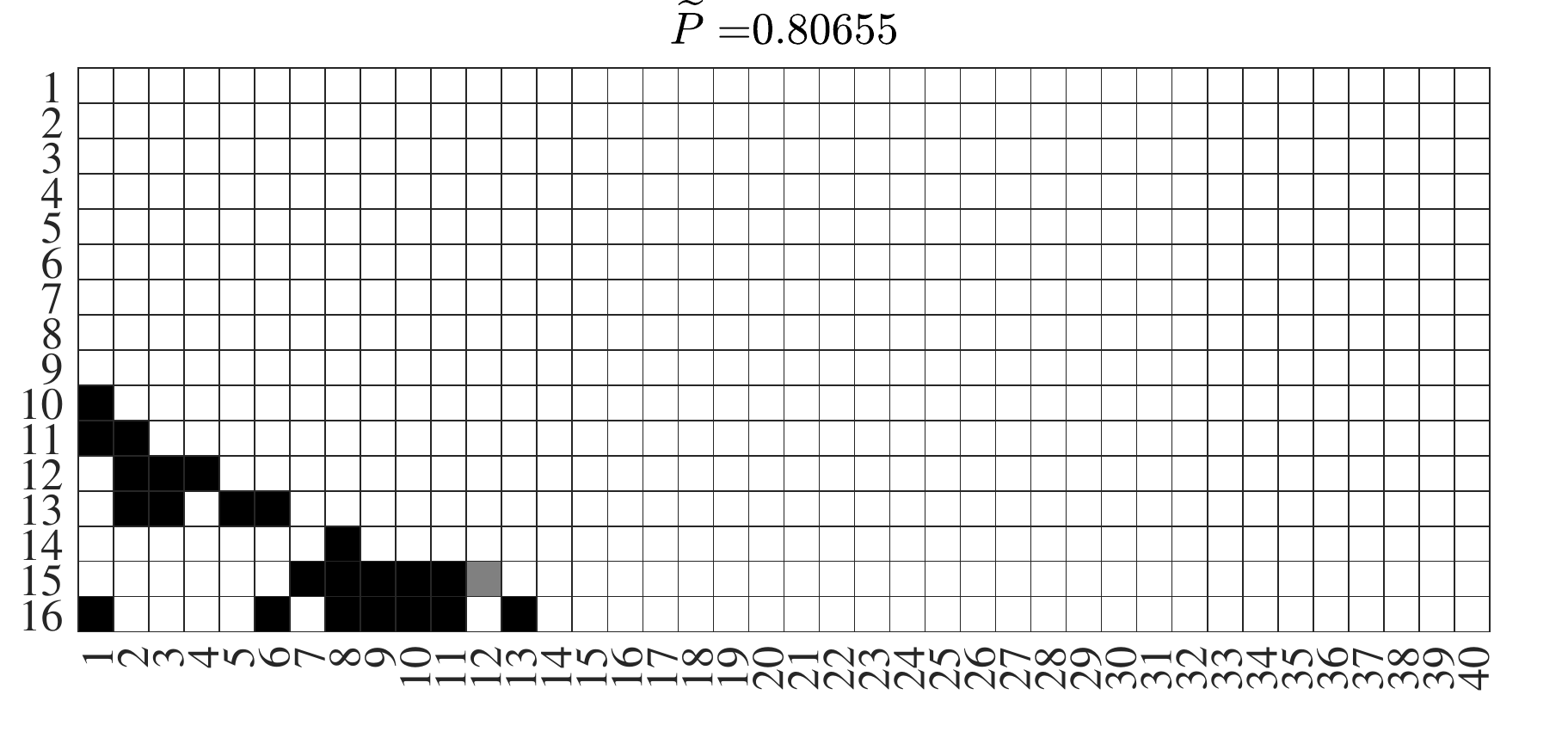}
	\end{minipage}
	\begin{minipage}{0.5\textwidth}\centering
		\xincludegraphics[width=\textwidth,label={\textbf{f}}]{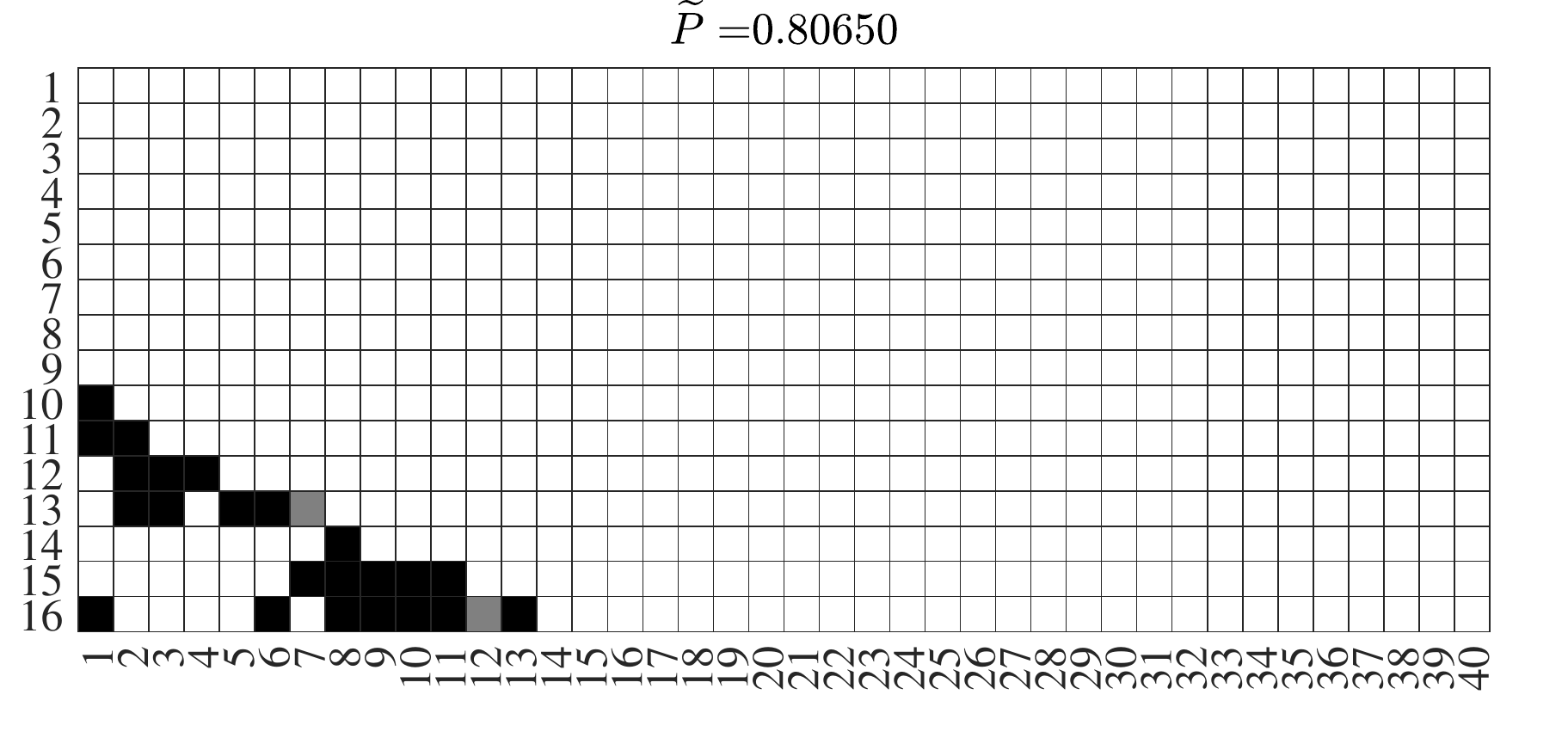}
	\end{minipage}%
	\begin{minipage}{0.5\textwidth}\centering
		\xincludegraphics[width=\textwidth,label={\textbf{g}}]{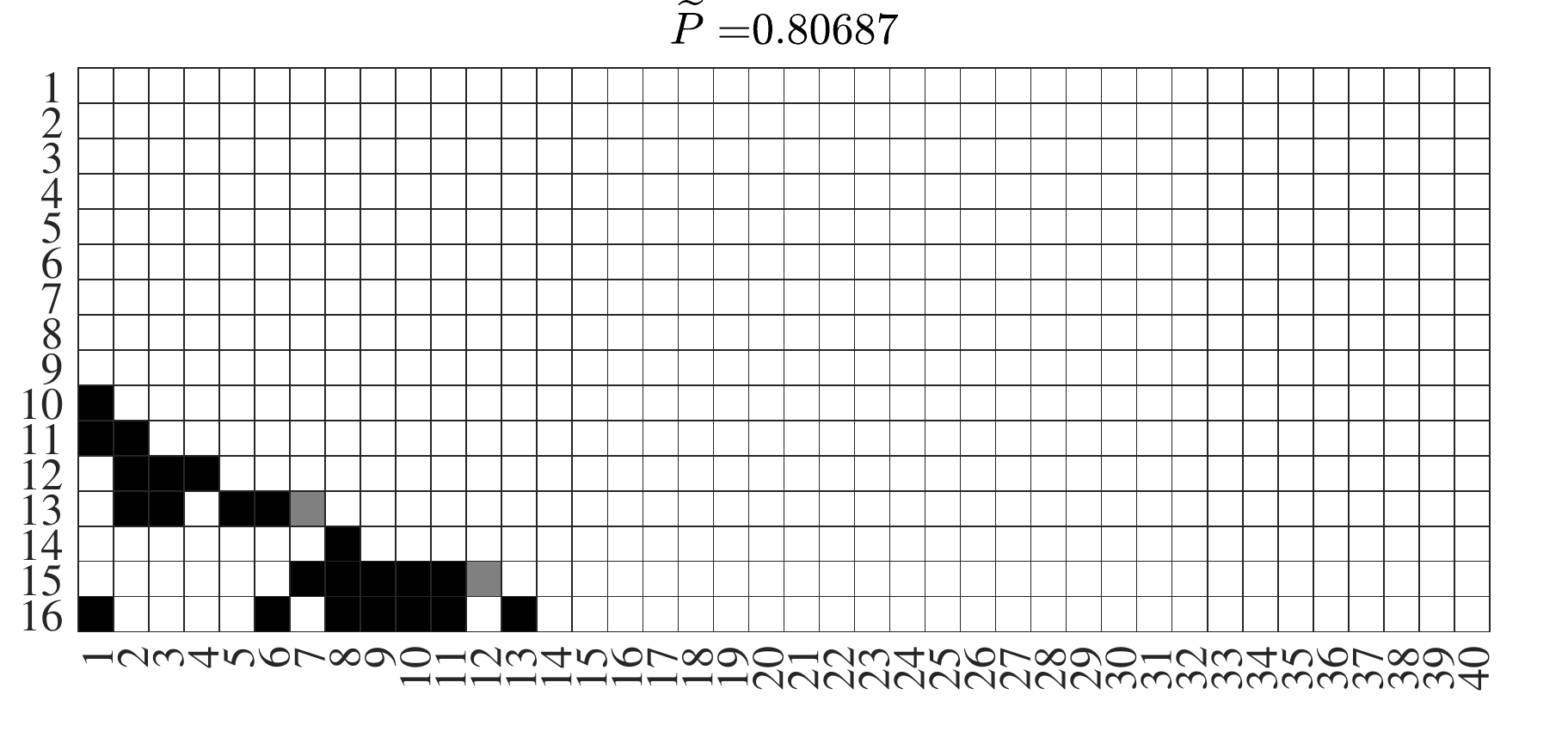}
	\end{minipage}
	\begin{minipage}{0.5\textwidth}\centering
		\xincludegraphics[width=\textwidth,label={\textbf{h}}]{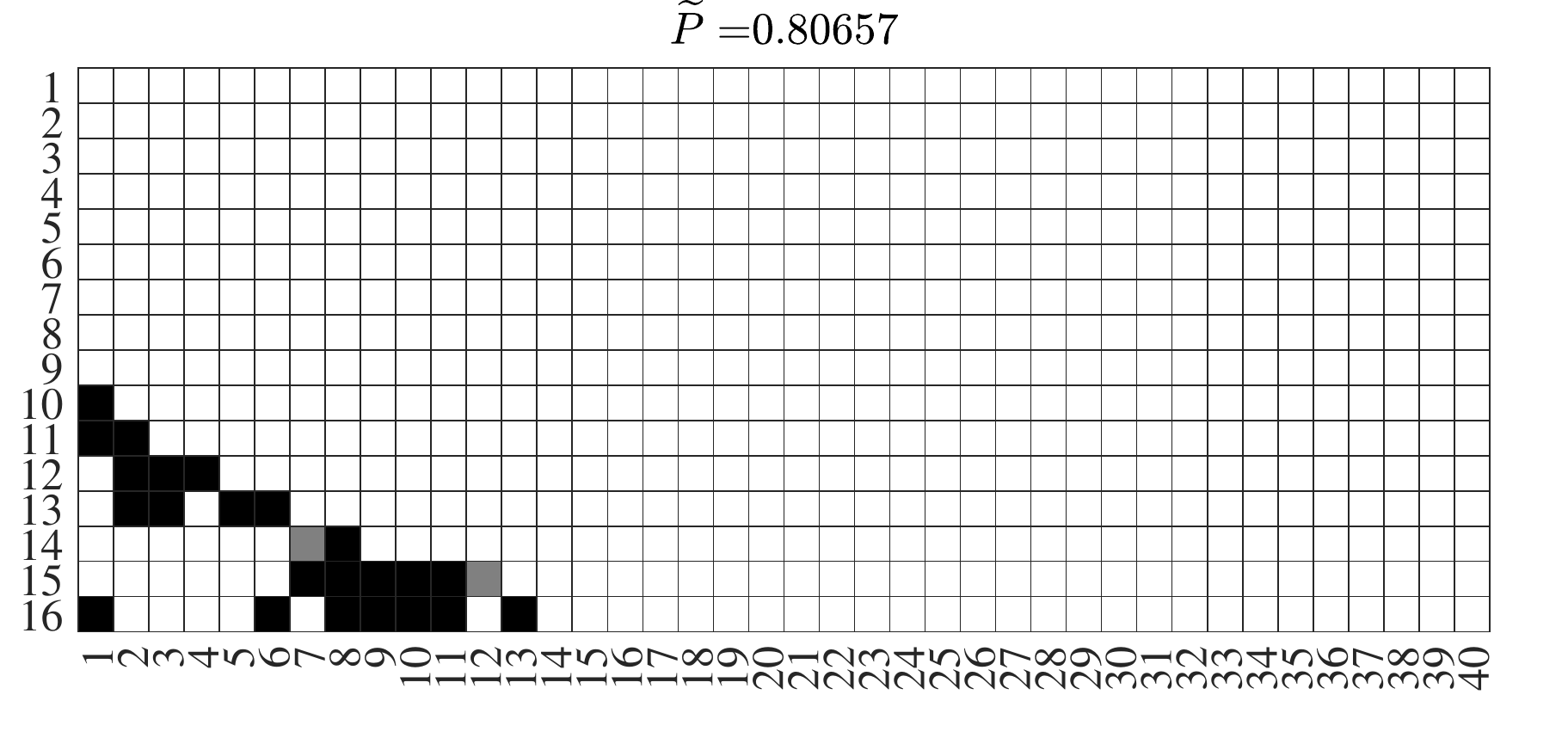}
	\end{minipage}%
	\begin{minipage}{0.5\textwidth}\centering
		\xincludegraphics[width=\textwidth,label={\textbf{i}}]{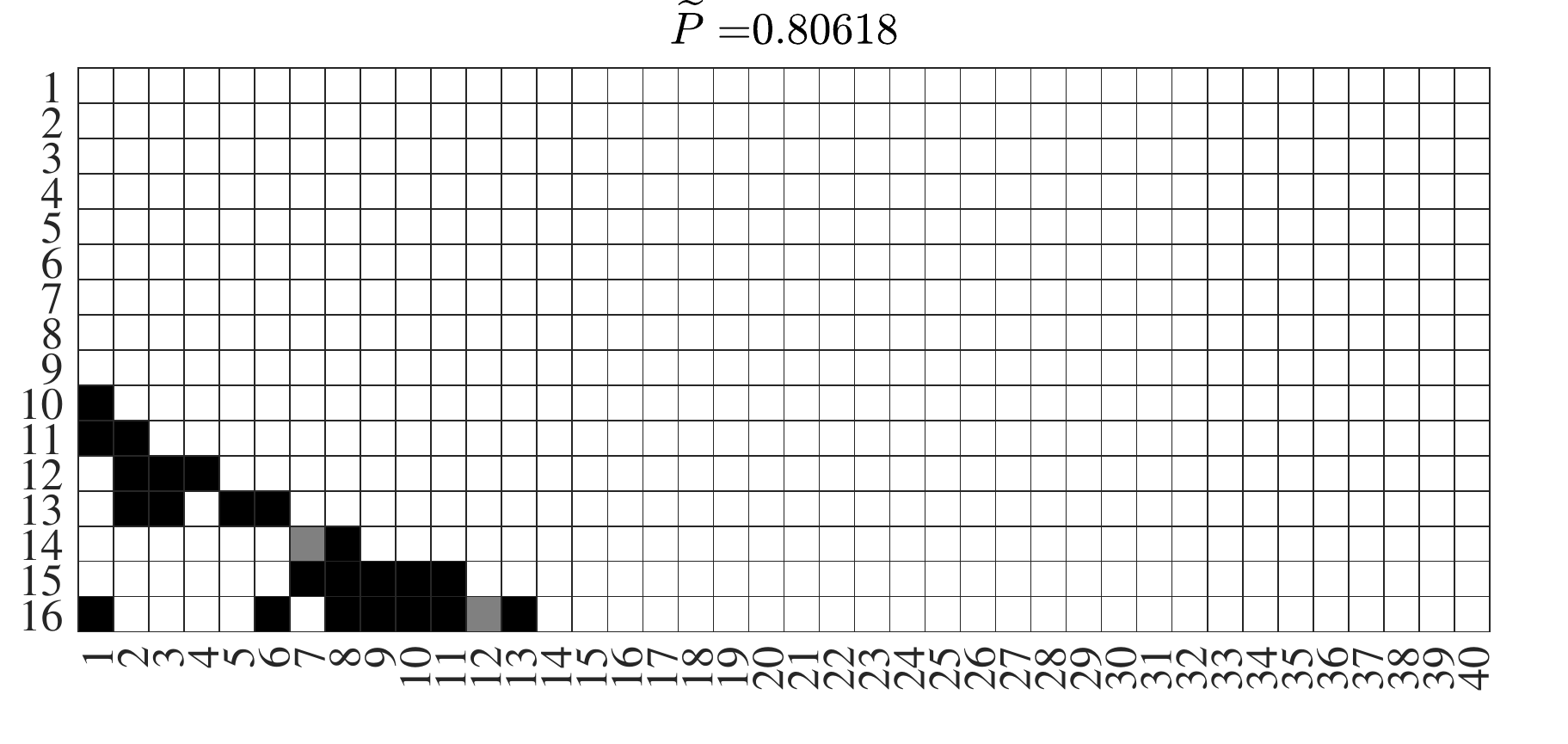}
	\end{minipage}%
	\caption{\textbf{Perturbation of the optimum from SOLO-G for the fluid-structure optimization problem with 40$\times$16 mesh.} Intuitively the ramp should be smooth, yet we observe two gaps in the optimum given by SOLO-G. We try filling the gaps. \textbf{a}, The optimum from SOLO-G. \textbf{b-i}, One or two blocks (gray) are added to fill the gap, with higher $\widetilde{P}$. }
	\label{figs8}
\end{figure}

\begin{figure}[H]
	\centering
	\begin{minipage}{0.5\textwidth}\centering
		\xincludegraphics[width=\textwidth,label={\textbf{a}}]{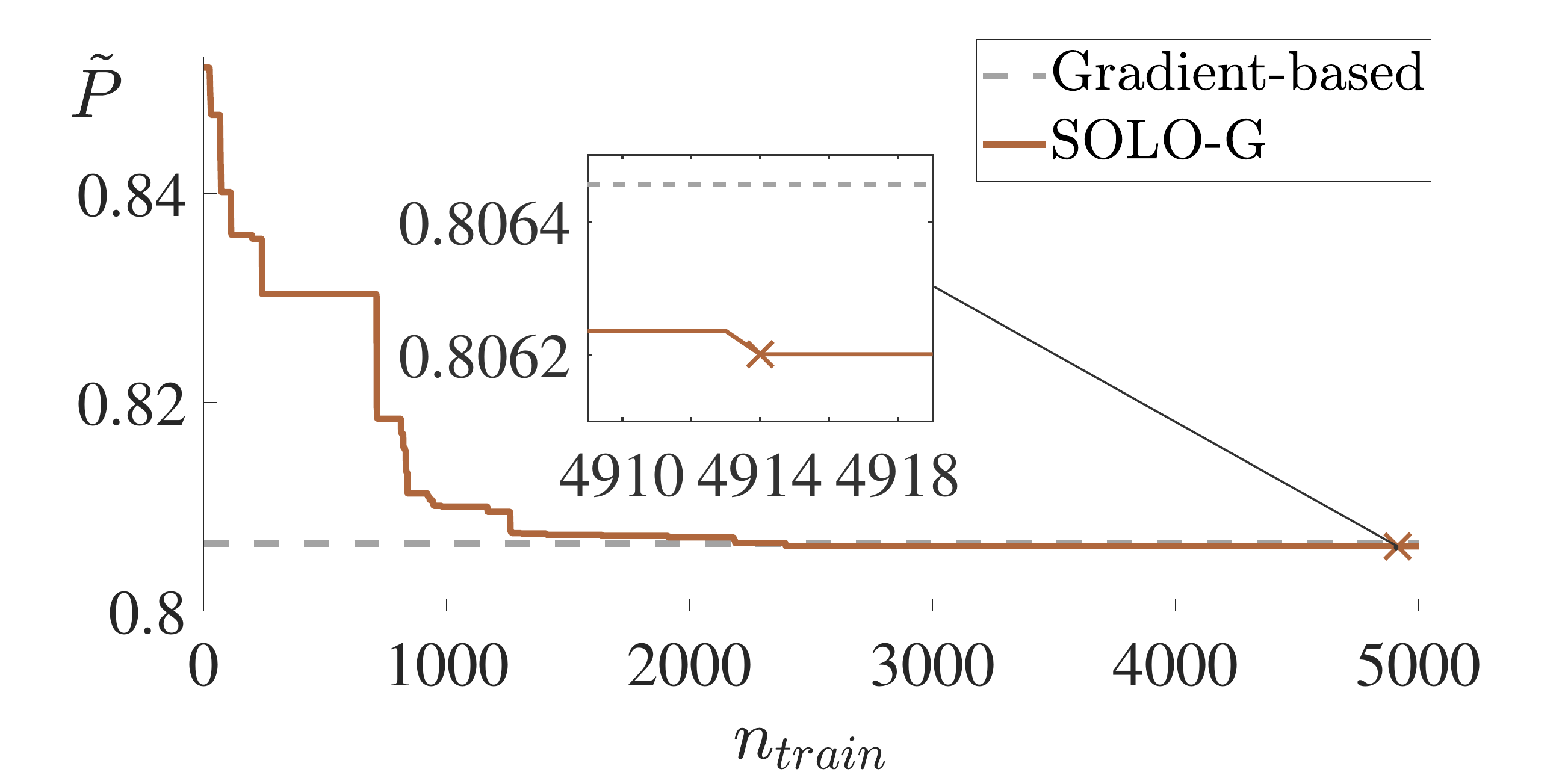}
	\end{minipage}%
	\begin{minipage}{0.5\textwidth}\centering
		\xincludegraphics[width=\textwidth,label={\textbf{b}}]{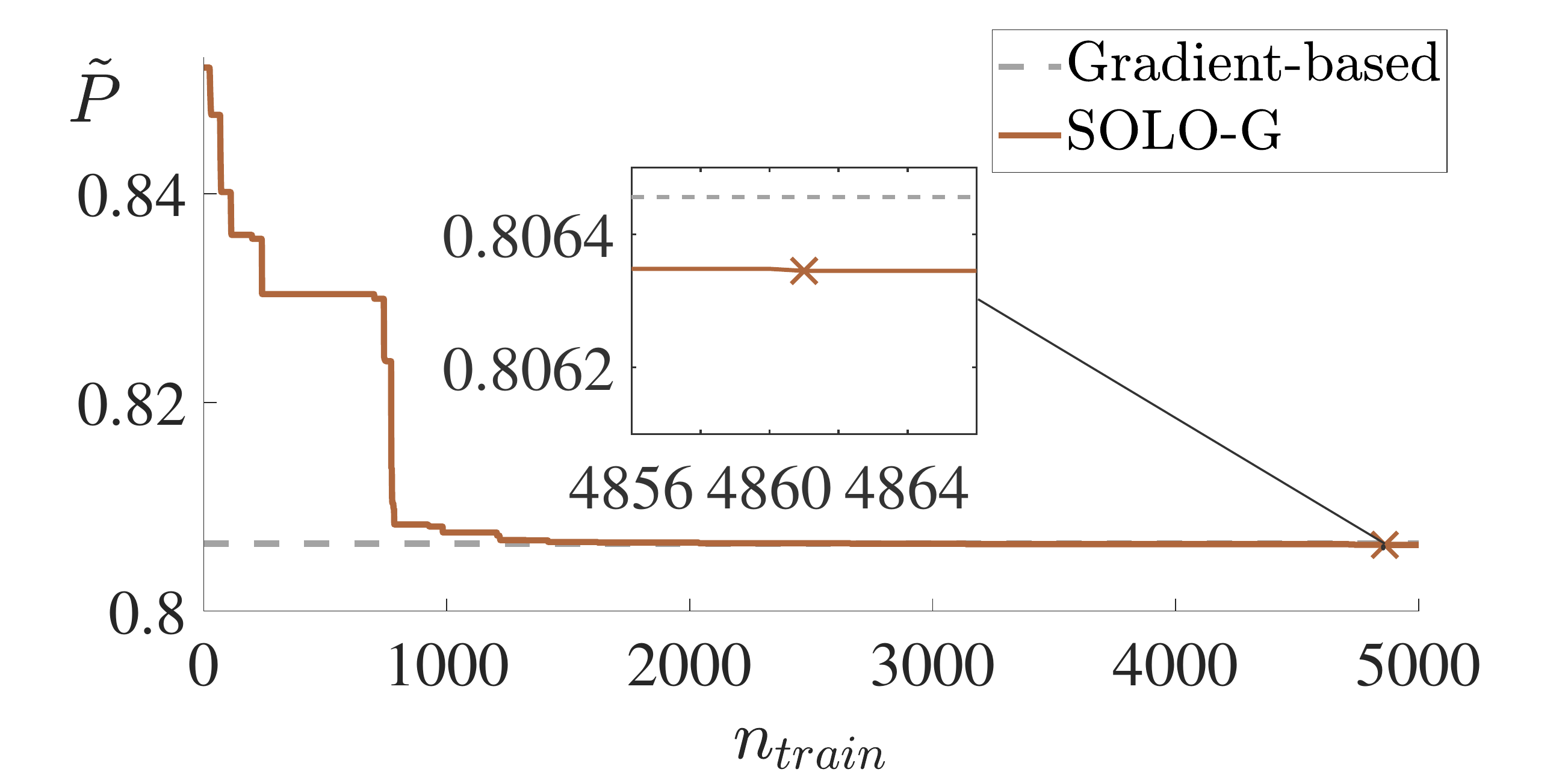}
	\end{minipage}
	\begin{minipage}{0.5\textwidth}\centering
		\xincludegraphics[width=\textwidth,label={\textbf{c}}]{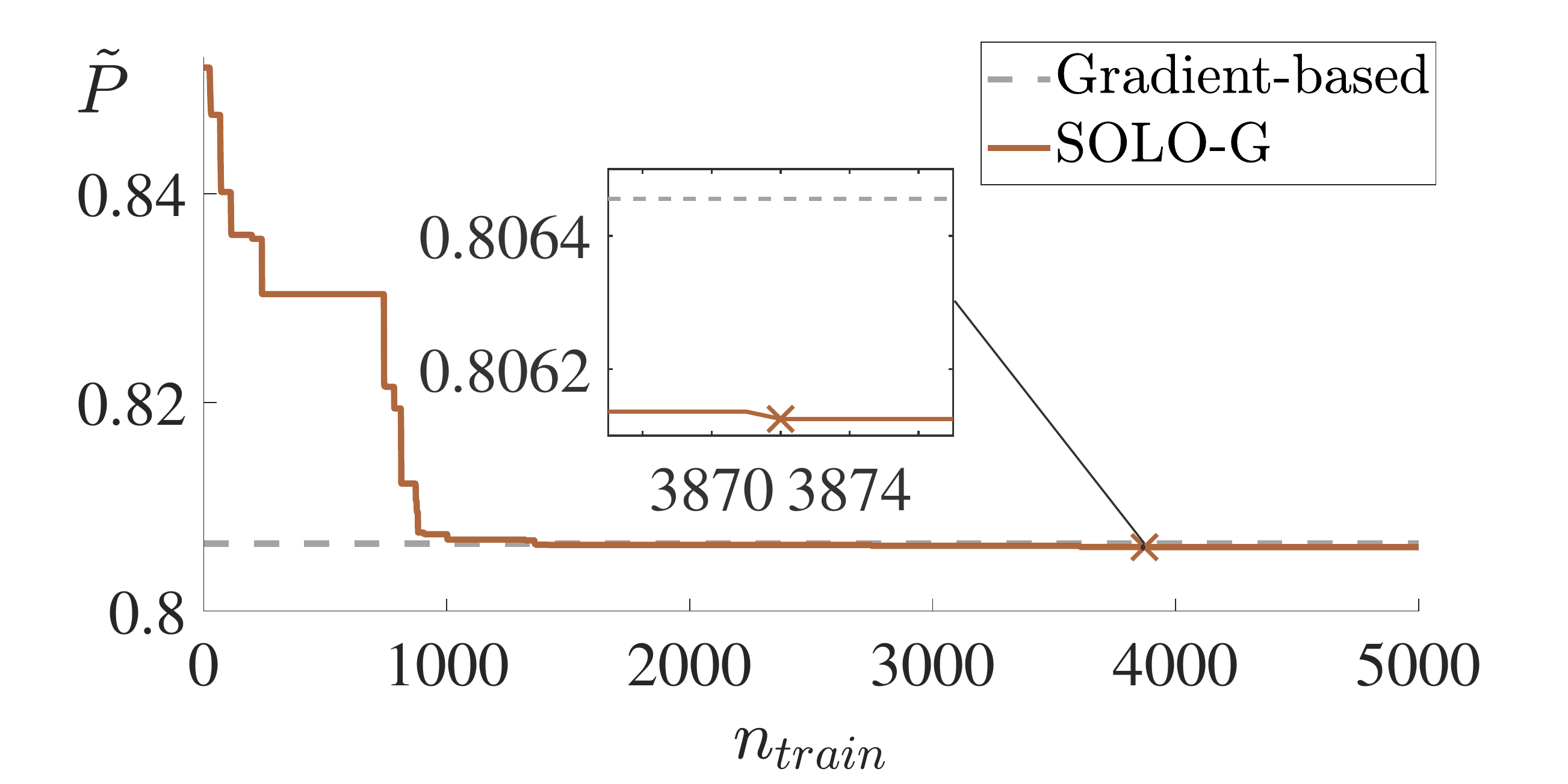}
	\end{minipage}%
	\begin{minipage}{0.5\textwidth}\centering
		\xincludegraphics[width=\textwidth,label={\textbf{d}}]{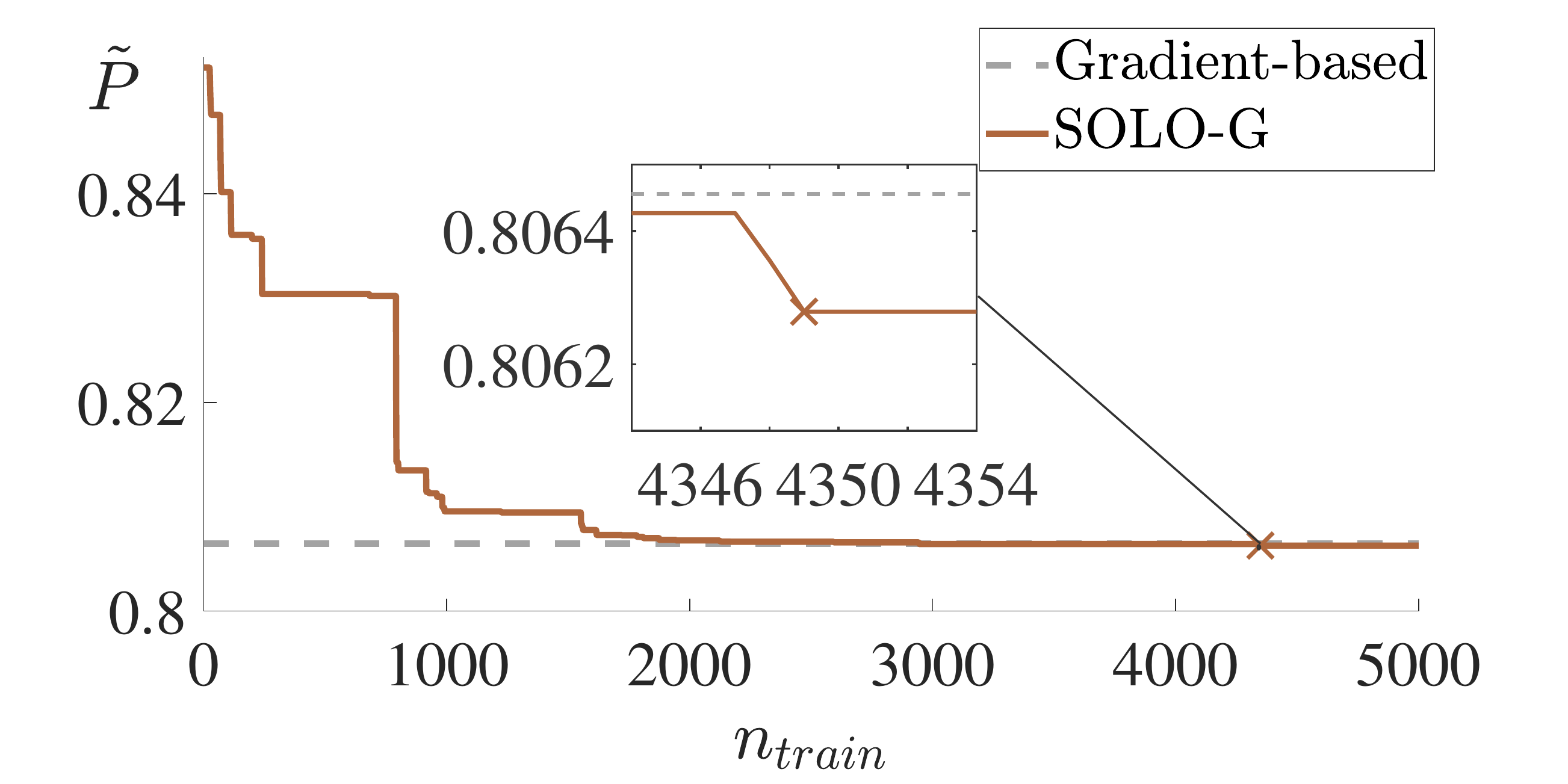}
	\end{minipage}
	\caption{\textbf{Repeating SOLO-G for the fluid-structure optimization problem with 40$\times$16 mesh.} All configurations are the same as Fig. 5b except that different random seeds and higher $n_{trian}$ are used. They all outperform the gradient-based baseline.}
	\label{figs9}
\end{figure}

\begin{figure}[H]
	\centering
	\begin{minipage}{0.33\textwidth}\centering
		\parbox[c][1em]{\textwidth}{\textbf{a} \small{$\widetilde{t}=0.013669$}}
		\includegraphics[width=\textwidth]{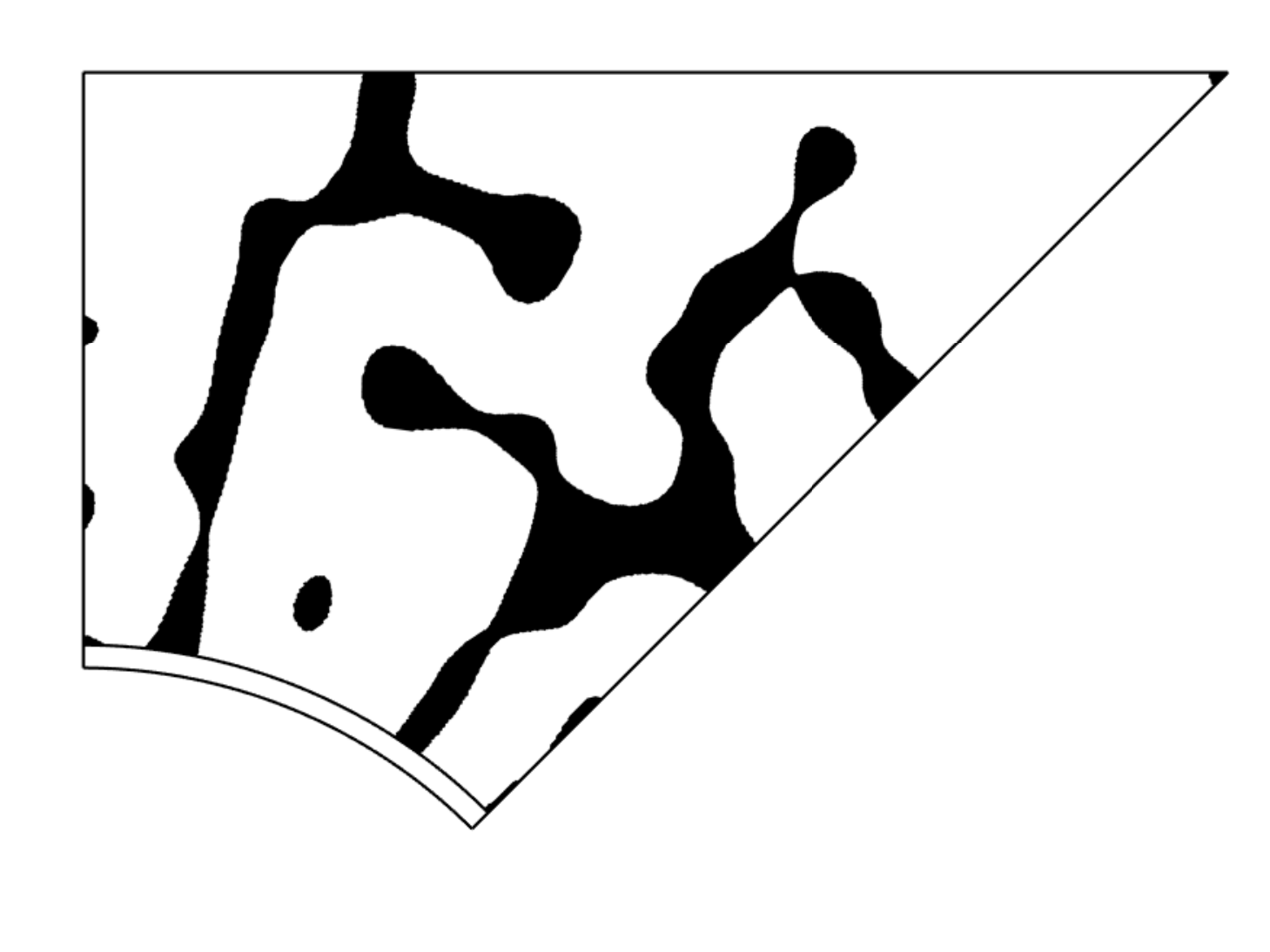}
	\end{minipage}%
	\begin{minipage}{0.33\textwidth}\centering
		\parbox[c][1em]{\textwidth}{\textbf{b} \small{$\widetilde{t}=0.013672$}}
		\includegraphics[width=\textwidth]{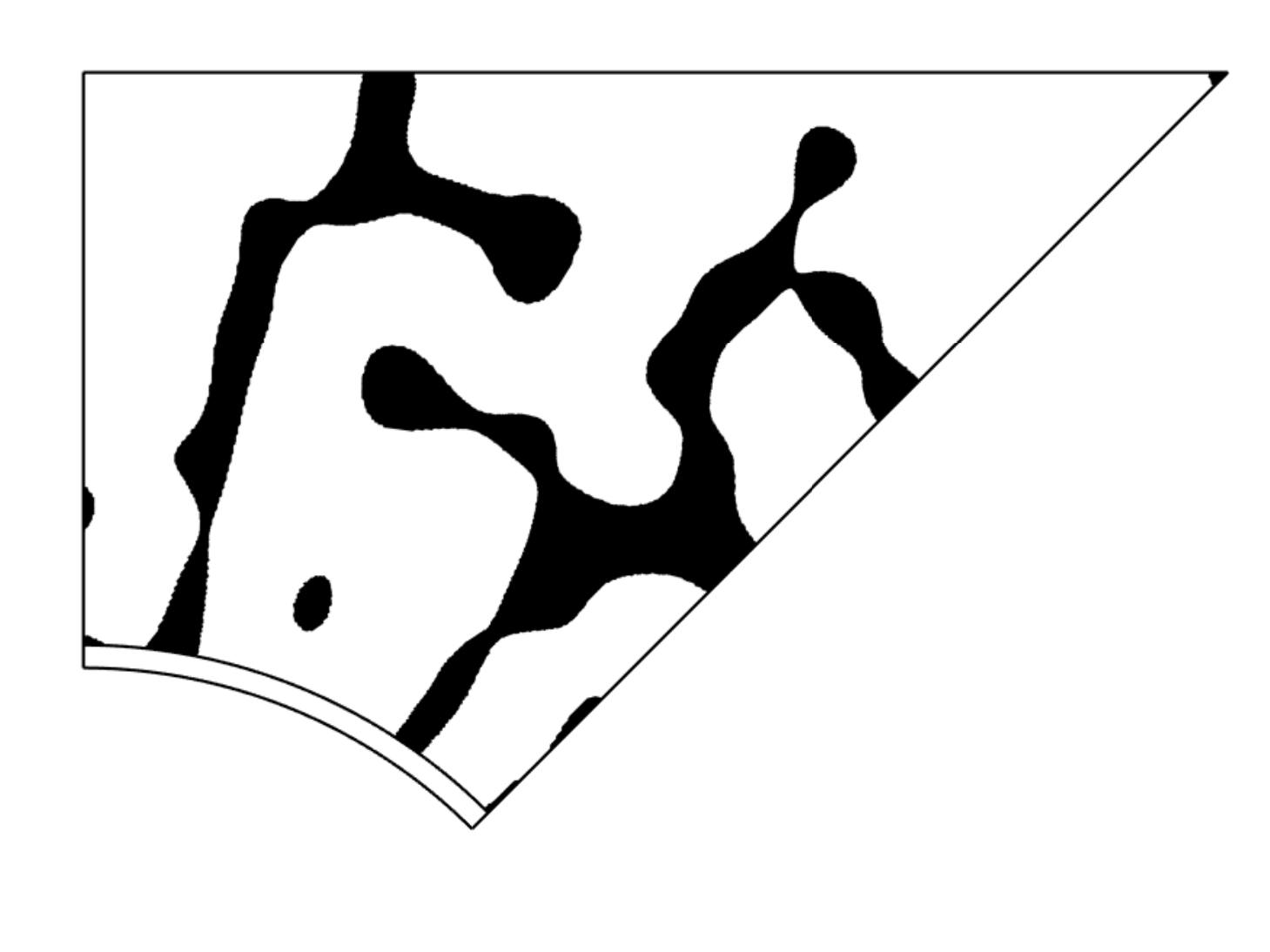}
	\end{minipage}%
	\begin{minipage}{0.33\textwidth}\centering
		\parbox[c][1em]{\textwidth}{\textbf{c} \small{$\widetilde{t}=0.013786$}}
		\includegraphics[width=\textwidth]{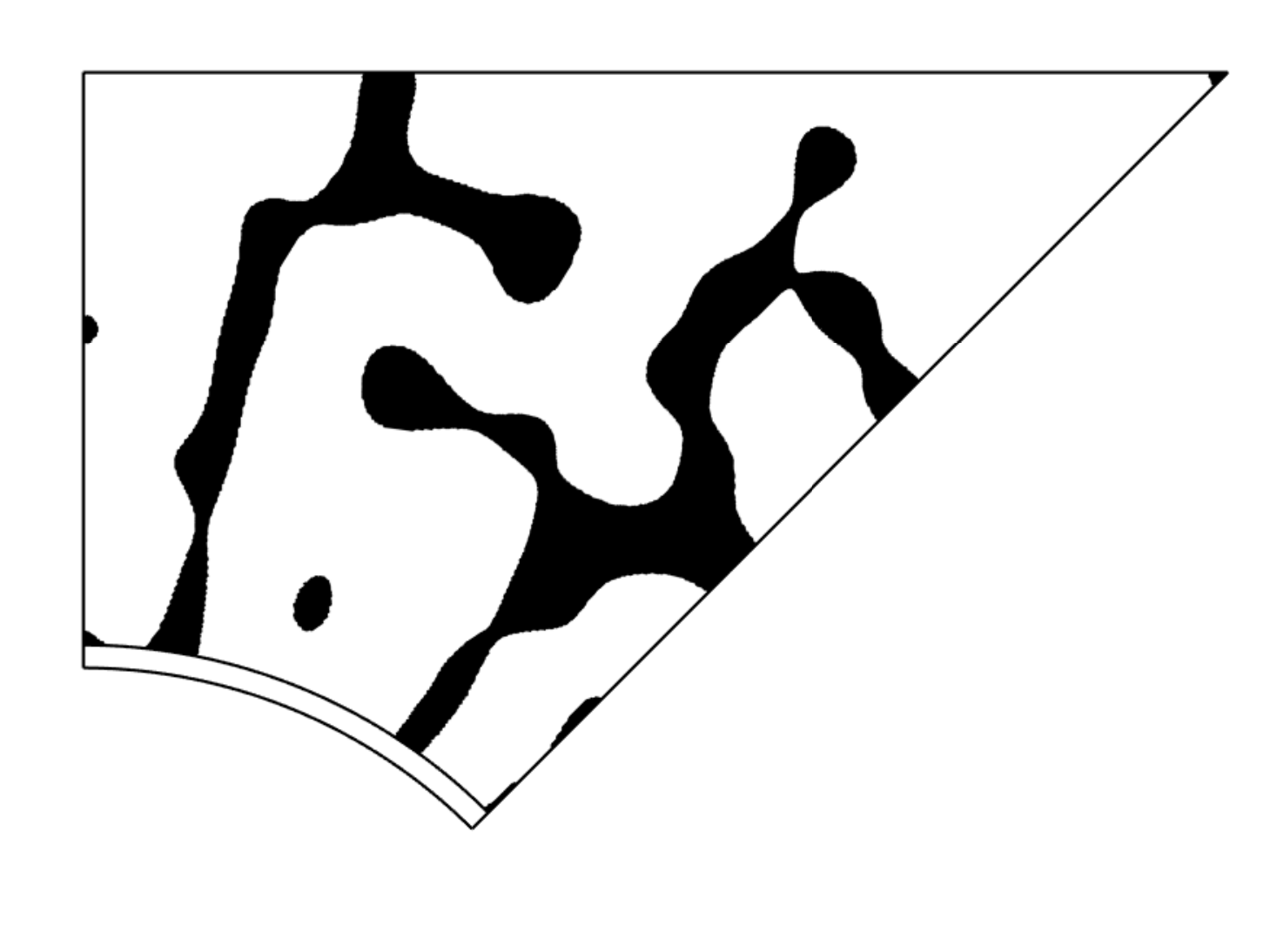}
	\end{minipage}\\
	\begin{minipage}{0.33\textwidth}\centering
		\parbox[c][1em]{\textwidth}{\textbf{d} \small{$\widetilde{t}=0.013894$}}
		\includegraphics[width=\textwidth]{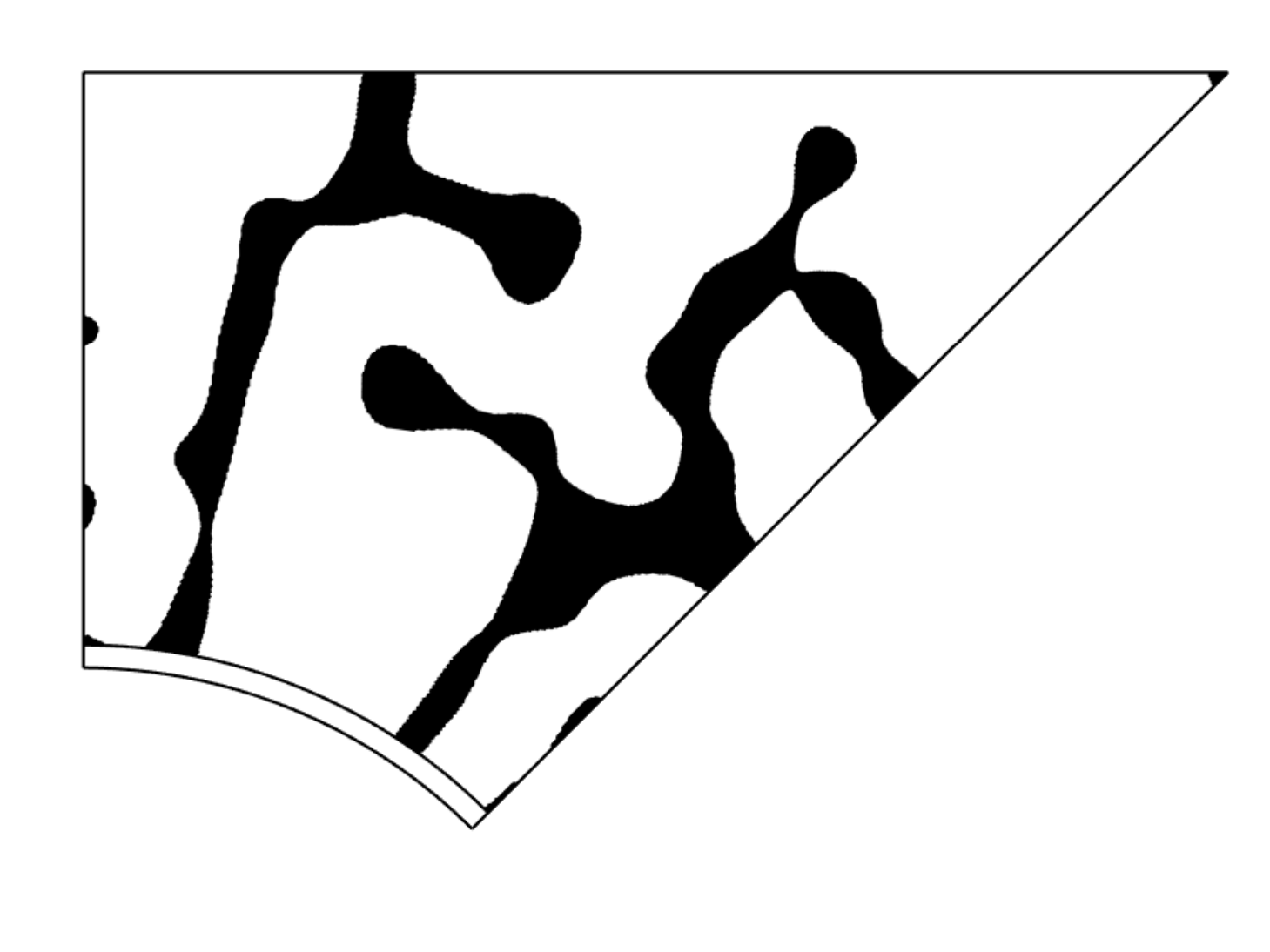}
	\end{minipage}%
	\begin{minipage}{0.33\textwidth}\centering
		\parbox[c][1em]{\textwidth}{\textbf{e} \small{$\widetilde{t}=0.014032$}}
		\includegraphics[width=\textwidth]{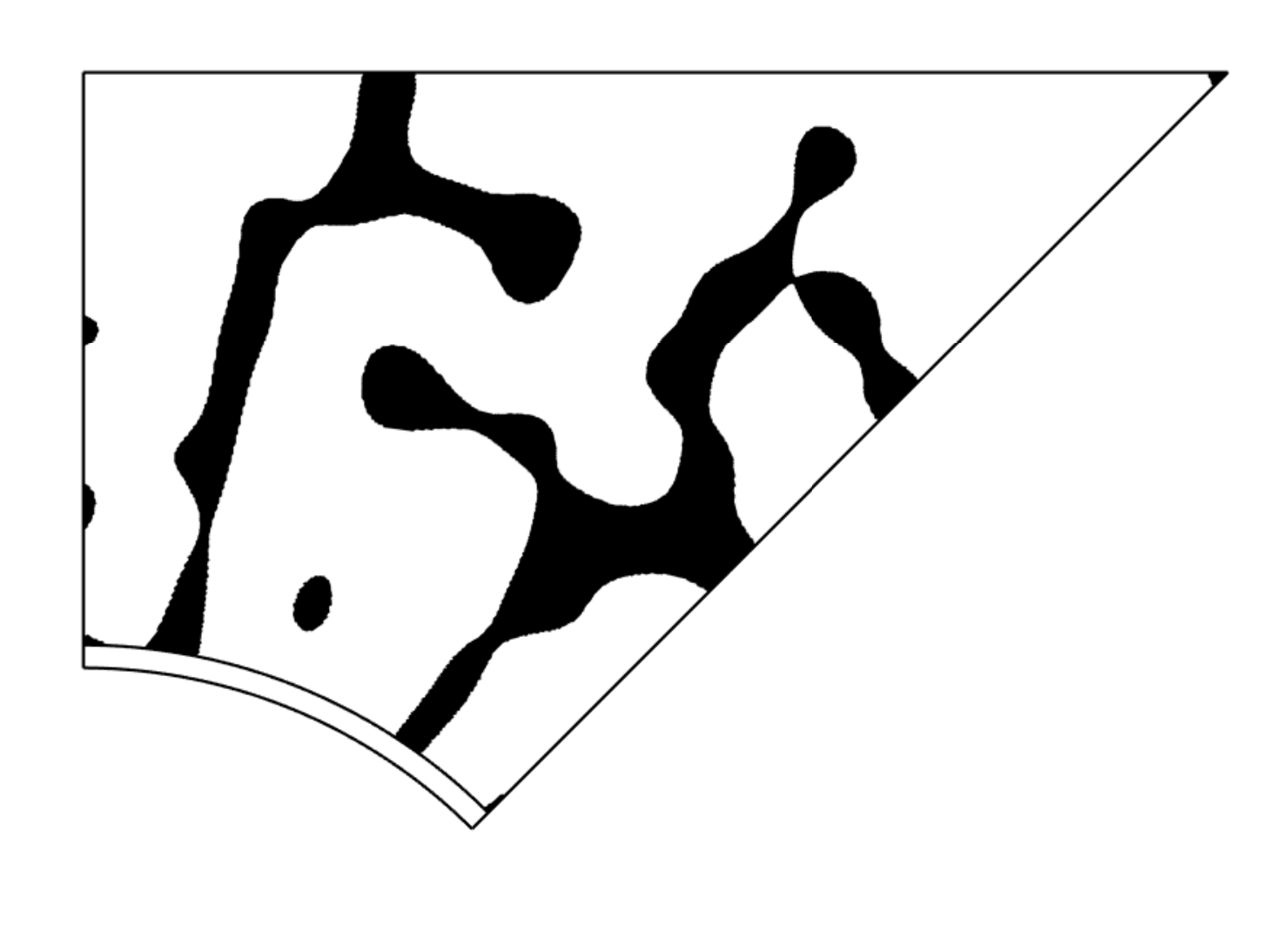}
	\end{minipage}%
	\begin{minipage}{0.33\textwidth}\centering
		\parbox[c][1em]{\textwidth}{\textbf{f} \small{$\widetilde{t}=0.013684$}}
		\includegraphics[width=\textwidth]{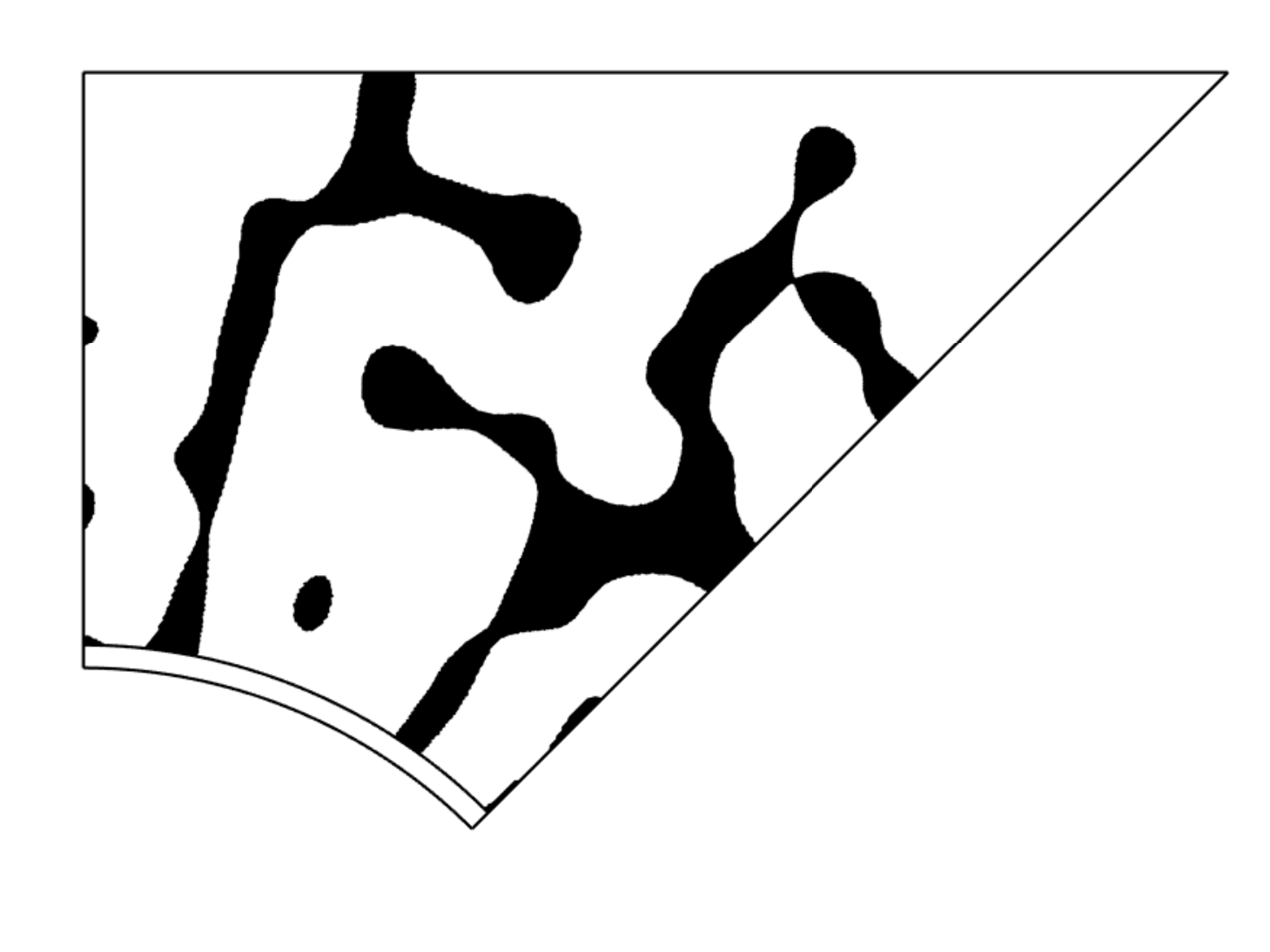}
	\end{minipage}%
	\caption{\textbf{Perturbation of the optimum from SOLO for the heat transfer enhancement problem.} \textbf{a}, The optimum from SOLO. \textbf{b-f}, Copper islands are removed; other copper portions will become thicker to maintain total solid volume. The solution from SOLO gives lowest time $\widetilde{t}$, although some are very close (the difference may even be caused by numerical noise in FEM computation). }
	\label{figs10}
\end{figure}

\begin{itemize}
	\item Supplementary Table \ref{tabs1}: Average wall time within a loop \dotfill \pageref{tabs1}
	\item Supplementary Fig.\ \ref{figs1}: Objective (energy) and prediction error of the compliance minimization problem with 5$\times$5 variables \dotfill \pageref{figs1}
	\item Supplementary Fig.\ \ref{figs2}: Evolution of the solution from SOLO for the compliance minimization problem with 5$\times$5  variables \dotfill \pageref{figs2} 
	\item Supplementary Fig.\ \ref{figs3}: Evolution of the solution from SOLO for the compliance minimization problem 11$\times$11 variables \dotfill \pageref{figs3}
	\item Supplementary Fig.\ \ref{figs4}: Evolution of the solution from SOLO-G for the fluid-structure optimization problem with 20$\times$8 mesh \dotfill \pageref{figs4}
	\item Supplementary Fig.\ \ref{figs5}: Repeating SOLO-G for the fluid-structure optimization problem with 20$\times$8 mesh \dotfill \pageref{figs5}
	\item Supplementary Fig.\ \ref{figs6}: Repeating SOLO-R for the fluid-structure optimization problem with 20$\times$8 mesh mesh \dotfill \pageref{figs6}
	\item Supplementary Fig.\ \ref{figs7}: Evolution of the solution from SOLO-G for the fluid-structure optimization problem with 40$\times$16 mesh \dotfill \pageref{figs7}
	\item Supplementary Fig.\ \ref{figs8}: Perturbation of the optimum from SOLO-G for the fluid-structure optimization problem with 40$\times$16 mesh. \dotfill \pageref{figs8}
	\item Supplementary Fig.\ \ref{figs9}: Repeating SOLO-G for the fluid-structure optimization problem with 40$\times$16 mesh \dotfill \pageref{figs9}
	\item Supplementary Fig.\ref{figs10}: Perturbation of the optimum from SOLO for the heat transfer enhancement problem \dotfill \pageref{figs10}
\end{itemize}

Section \ref{apd: proof}: \nameref{apd: proof} \dotfill \pageref{apd: proof}

\begin{itemize}
	
	\item Section \ref{s2.1}: \nameref{s2.1} \dotfill \pageref{s2.1}
	\item Section \ref{s2.2}: \nameref{s2.2} \dotfill \pageref{s2.2}
	\item Section \ref{s2.3}: \nameref{s2.3} \dotfill \pageref{s2.3}
\end{itemize}


\newpage
\section{Supplementary table and figures}\label{s1}

\begin{table}[htbp]
	\centering
	\caption{\textbf{Average wall time within a loop of SOLO.} There are three major steps in each loop: FEM calculation to obtain corresponding objective function values, DNN training, and optimization which searches for the optimum based on DNN's prediction. We give a very rough estimate on our personal computer (CPU: Intel i7-8086K, GPU: NVidia RTX 2080 Super). \textit{Italic numbers} indicate GPU computing and the others are computed entirely on CPU. Actual running time is sensitive to hardware environment, software packages, parameter setting and so forth. Further, FEM calculation is approximately proportional to the number of additional samples per loop; training time depends on existing training data obtained from previous loops; optimization depends on the number of function evaluations. Similar to other SMBO methods, our surrogate model introduces overhead computation. In the compliance and fluid problems, the overhead is comparable with FEM calculation time, yet it is almost negligible considering the huge benefit of reducing FEM calculations from $10^5\sim10^8$ (see the table) to $10^2\sim10^4$; besides, we chose relatively simple problems and thus each calculation only cost $<0.5$ s for compliance problems and $<6$ s for fluid problems; smaller portion of the overhead is expected for more complicated problems with higher FEM computation time. When the problem becomes more complicated in the heat example, a larger advantage of our method can be observed. In the three truss problems, our focus is on the reduction of FEM calculations rather than computation time since the problems are fast to calculate.}
	\begin{tabular}{lr|rrr}
		\hline
		\multicolumn{1}{c}{\multirow{2}{*}{Problem}} & \multicolumn{1}{c|}{\multirow{1}{*}{Number of}} & \multicolumn{3}{c}{Wall time /s} \\
		&   \multicolumn{1}{c|}{\multirow{1}{*}{additional samples}}     & \multicolumn{1}{l}{FEM} & \multicolumn{1}{l}{Training} & \multicolumn{1}{l}{Optimization (evaluations)} \\
		\hline
		Compliance 5$\times$5 & 100   & 40    & 35    & 70 ($2 \times 10^5$) \\
		Compliance 11$\times$11 & 1000  & 500   & 150   & 1000 ($4 \times 10^6$) \\
		Fluid 20$\times$8 (G) & 10    & 35    & \textit{10}    & \textit{35} ($1 \times 10^8$)  \\
		Fluid 20$\times$8 (R) & 100   & 350   & \textit{20}    & \textit{35} ($1\times 10^8$) \\
		Fluid 40$\times$16 (G) & 10    & 60    & \textit{25}    & \textit{140} ($2 \times 10^8$) \\
		Heat 10$\times$10 & 200    & 40000    & \textit{25}    & \textit{200} ($4 \times 10^8$) \\
		Truss 72 & 10    & \textit{0.02}    & \textit{20}    & \textit{300} ($1 \times 10^9$) \\
		Truss 432 & 50    & \textit{0.05}  & \textit{150}    & \textit{500} ($1 \times 10^9$) \\
		Truss 1008 & 100    & \textit{0.15}    & \textit{1500}    & \textit{1500} ($ 2 \times 10^9$) \\
		\hline
	\end{tabular}%
	\label{tabs1}%
\end{table}%

\newpage

\section{Theory on convergence}\label{apd: proof}
In the main text, we presented a simplified version of convergence (Eq.\ (3)). In this section, we give a detailed description of our theoretical result. We first present the main result (Theorem \ref{thm: main}). Then, we introduce some preliminary definitions and knowledge used in the proof. In the end, we approach the proof.

\subsection{Formulation and theorem}\label{s2.1}
The unknown object function is denoted as $F(\bm{\rho})$, where $\bm{\rho} \in \RR^N$.We denote the domain of $\{\bm{\rho}~|~ 0\le\rho_i\le 1 , 1\le i \le N\}$ as $K$. We suppose the global minimizer $\bm{\rho}^*=\argmin_{\bm{\rho}} F(\bm{\rho})$. 

We consider the total iteration number to be $T$. At iteration $t(1\le t \le T)$, the DNN  is denoted as $f_t(\cdot)$ and we denote the empirical minimizer of this DNN function to be $\hat{\bm{\rho}}^{(t)}$, i.e.
\begin{equation}\label{eq: argmin-em}
	\hat{\bm{\rho}}^{(t)}=\argmin_{\bm{\rho}} f_t(\bm{\rho}).
\end{equation}
Besides, we denote our DNN as a $D$-layer neural network which is formulated as follows:
\begin{align*}
	f_t(\bm{\rho}) = \mathbf{W}_D^\top\sigma(\mathbf{W}_{D-1}\sigma(...\sigma(\mathbf{W}_{1}\bm{\rho}))),
\end{align*}
where $\cW=\{\mathbf{W}_{k}\in\RR^{d_{k-1}\times d_k}~|~k=1,...,D\}, d_0= N$(number of input dimensions), $d_D=1$, and $\sigma(v) = [\max\{v_1,0\},...,\max\{v_d,0\}]^\top$ is the ReLU\cite{nair2010rectified} activation function  for $v\in\RR^{d}$. We further denote $d=\max\{d_i\}$ and the function class of such neural networks as $\cH_f$.

At time step $t$, given the empirical optimal point $\hat{\bm{\rho}}^{(t-1)}$, the additional $m$ training points is generated through the following process:
\begin{align*}
	\bm{\rho}^{(j_t)}= \hat{\bm{\rho}}^{(t-1)} + \bm{\xi}^{(j_t)}, j_t=mt-m+1, mt-m+2, \cdots, mt.
\end{align*} 
Here $\bm{\xi}^{(j)}$ denotes random noise for perturbation. Hence through the iterating process, the sampled points are random variables. At time step $t$, we denote all the realizations of random training data points set as $K_t=\{\bm{\rho}^{(i)}~\big|~ i=1,\cdots, mt\}$.

Now before we proceed, we need to impose some mild assumptions on the problem.
\begin{assumption}\label{asp: bounded}
	We suppose that
	\begin{itemize}
		\item [1)] the spectral norm of the matrices in DNNs are uniformly bounded, i.e., there exists $B_W>0$ s.t. $\norm{\mathbf{W}_k}_2 \le B_W, \forall k =1, \cdots, D$.
		\item [2)] the target function is bounded, i.e., there exists $B_F>0$ s.t. $\norm{F}_{\infty} \le B_F$.
	\end{itemize}
\end{assumption}
1) of Assumption \ref{asp: bounded} is a commonly studied assumption in existing generalization theory literature on deep neural networks \cite{bartlett2017spectrally,neyshabur2017pac,chen2019generalization}. 
2) of Assumption \ref{asp: bounded} assumes $F$ is bounded, which is standard and intuitive since $F$ has a physical meaning.
\begin{assumption}\label{asp: data}
	We assume that for any iteration $t$, $\bm{\xi}^{(j_t)}~(j_t=mt-m+1, \cdots, mt)$ are i.i.d. (independent and identically distributed) perturbation noise. The generated training data $\left\{\bm{\rho}^{(j_{t_1})}\right\}$ are independent of $\left\{\bm{\rho}^{(j_{t_2})}\right\}$ if $t_1 \neq t_2$.
\end{assumption}
The assumption of the i.i.d. properties of noise in Assumption \ref{asp: data} is common in optimization literature \cite{blair1985problem,nemirovski2009robust,lin2019gradient,chen2020computation}. The difference is that in traditional optimization literature noise refers to the difference between the true gradient and the stochastic gradient while the noise here denotes perturbations to generate new samples in each iteration. Note that our Assumption \ref{asp: data} only needs the i.i.d. property of noise, which is weaker than the standard assumptions for stochastic gradient methods which require unbiased property and bounded variance \cite{nemirovski2009robust,lin2019gradient,chen2020computation}. Since our fitting DNN $f_t$s are continuously changing throughout iterations and the empirical minimizers $\hat{\rho}^{(t)}$ are also alternating, it is reasonable for us to assume that the different groups of generated data samples are independent for the ease of theoretical analysis in the sequel. 

We denote the distribution of samples $\{\bm{\rho}^{(j_t)}~\big|~j_t =mt-m+1, mt-m+2, \cdots, mt\}$ as $D_t (1\le t \le T)$, with which we can introduce the following definition.
\begin{definition}
	For a measurable function $f$, we denote 
	\begin{equation}
		\EE_{D_{1:T}} f(\bm{\rho}) = \frac{\sum_{t=1}^T \EE_{\bm{\rho} \sim D_t} f(\bm{\rho})}{T},
	\end{equation}
	where $\EE$ denotes expectation.
\end{definition}
\begin{assumption}\label{asp: infiity-2-relation}
	For any $t$ and $f_t \in \cH_f$,
	\begin{align*}
		\norm{F-f_t}_{\infty}^2 = C(t) \EE_{\bm{\rho} \sim D_{1:t}} (F - f_t)^2,
	\end{align*}
	where $ C(t)$ is a monotonically decreasing function w.r.t.\ iteration number $t$.
\end{assumption}
Assumption \ref{asp: infiity-2-relation} basically describes that the Chebyshev distance of our DNN at time $t$ and $F$ is bounded by a constant number (w.r.t.\ $t$) times the average true loss of $(F-f_t)^2$ till time $t$. This assumption is reasonable in that the the average true loss can be seen as a variant of Euclidean distance between our DNN at time $t$ and $F$.

Eventually we arrive at our main result.
\begin{theorem}\label{thm: main}
	Under Assumptions \ref{asp: bounded}, \ref{asp: data} and \ref{asp: infiity-2-relation}, given iteration number $T$ and any $\delta>0$, for any trained DNN $f_T \in \cH_f$ with empirical MSE training error $\epsilon$ at iteration $T$, we have that with probability at least $1-\delta$ over the joint distribution of $\bm{\rho}^{(1)}, \bm{\rho}^{(2)}, \cdots, \bm{\rho}^{(mT)}$,
	\begin{align*}
		&\left(F(\hat{\bm{\rho}}^{(T)}) - F(\bm{\rho}^*)\right)^2 \\
		&\le 4C(T)\left(\frac{96B^2}{\sqrt{mT}}\sqrt{d^2D \log \left(1+8BB_W^DD\sqrt{mTd}\right)} + 12B^2 \sqrt{\frac{2\log \frac{2}{\delta}}{mT}} +\frac{8}{mT}+ \epsilon\right),
	\end{align*}
\end{theorem}
where $B = \max\{B_F,B_W^D\}$.

\subsection{Preliminaries}\label{s2.2}

Before showing the complete proof, we introduce some definitions and lemmas.

\begin{lemma}[McDiarmid's Inequality\cite{mcdiarmid1998concentration}]\label{lem:McD_ineq}
	Let $X_1, \cdots, X_m \in \cX$ be a set of $m\ge 1$ independent random variables and assume that there exist $c_1, \cdots, c_m >0$ such that $h: \cX^m \rightarrow \RR$ satisfies the following  conditions:
	\begin{align*}
		|h(x_1,\cdots,x_i,\cdots,x_m) - h(x_1,\cdots,x_i',\cdots,x_m)| \le c_i,
	\end{align*}
	for all $i \in [m]$ and any points $x_1,\cdots,x_m,x_i' \in \cX$. Here $x$s are the realizations of $X$s. Let $h(S)$ denote $h(X_1,\cdots,X_m)$, then, for all $s>0$, the following inequality hold:
	\begin{align}
		\PP \left\{h(S) - \EE[h(S)] \ge s\right\} \le \exp\left(\frac{-2s^2}{\sum_{i=1}^m c_i^2}\right), \label{eq: mcdiarmid-1}\\
		\PP \left\{h(S) - \EE[h(S)] \le -s\right\} \le \exp\left(\frac{-2s^2}{\sum_{i=1}^m c_i^2}\right), \label{eq: mcdiarmid-2}
	\end{align}
	where $\PP$ denotes probability and  $\EE$ denotes expectation. 
\end{lemma}
\begin{definition}[Covering Number\cite{mohri2018foundations}]
	Let $(V,\norm{\cdot})$ be a normed space, and $\Theta \subset V$. Vector set $\{V_i\in V|i=1,\cdots,N\}$ is an $\iota$-covering of $\Theta$ if $\Theta \subset \cup_{i=1}^N B(V_i,\iota$) where $B(V_i,\iota)$ denotes the ball with center $V_i$ and radius $\iota$, equivalently, $ \forall \theta \in \Theta, \exists i $ such that $\norm{\theta - V_i} \le \iota$. The covering number is defined as :
	\begin{align*}
		\cN(\Theta,\norm{\cdot},\iota):=\min\left\{n : \exists \iota\text{-covering }  \text{over } \Theta \text{ of size }n\right\}.
	\end{align*}
\end{definition}
\begin{definition}[Rademacher Complexity \& Empirical Rademacher Complexity\cite{bartlett2002rademacher,mohri2018foundations}]\label{def:rad}
	Given a sample $S=\{\mathbf{x}^{(1)},\mathbf{x}^{(2)},\cdots,\mathbf{x}^{(n)}\}$ and a set of real-valued function $\cH$, the \textit{Empirical Rademacher Complexity} is defined as
	\begin{align*}
		\hat{\mathfrak{R}}_n(\cH)=\mathfrak{R}_n(\cH_{|S}):=\frac{1}{n}\EE_{\sigma} \sup_{h \in \cH} \sum_{i=1}^n \sigma_i h(\mathbf{x}^{(i)}),
	\end{align*}
	where sup denotes supremum and the expectation is over the Rademacher random variables $(\sigma_1,\sigma_2,\cdots,\sigma_i,\cdots,\sigma_n)$, which are i.i.d. (independent and identically distributed) with $\PP(\sigma_i=1)=\PP(\sigma_i=-1)=\frac{1}{2}$. The \textit{Rademacher Complexity} is defined as \begin{align*}
		\mathfrak{R}_n(\cH):=\EE_{S}\mathfrak{R}_n(\cH_{|S})=\frac{1}{n}\EE_{S,\sigma} \sup_{h \in \cH} \sum_{i=1}^n \sigma_i h(\mathbf{x}^{(i)}),
	\end{align*}
	which is the expectation of the Empirical Rademacher Complexity over sample $S$.
\end{definition}
\begin{lemma}[Dudley's Entropy Integral Bound\cite{chen2019generalization}]\label{lem: dudley}
	Given a sample $S=\{\mathbf{x}^{(1)},\mathbf{x}^{(2)},\cdots,\mathbf{x}^{(n)}\}$, let $\cH$ be a real-valued function class taking values in  $[0,r]$ for some constant $r$, and assume that zero function ${\bf 0} \in \cH$. Then we have
	\begin{align*}
		\hat{\mathfrak{R}}_n(\cH) \le \inf_{\alpha>0} \left(\frac{4\alpha}{\sqrt{n}} + \frac{12}{n} \int_{\alpha}^{r\sqrt{n}}\sqrt{\log \cN\left(\cH,\iota,\norm{\cdot}_{\infty}\right)}d\iota\right),
	\end{align*}
\end{lemma}

where inf denotes infimum. 
\begin{lemma}(Covering number bound using volume ratio\cite{chen2019generalization})\label{lem: volume-ratio}
	Let $\cW=\{W\in\RR^{a\times b}:\norm{W}_2\le\lambda\}$ be the set of matrices with bounded spectral norm and $\iota$ be given. The covering number $\cN(\cW,\iota,\norm{\cdot}_F)$ is upper bounded by 
	\begin{equation*}
		\cN\left(\cW,\iota,\norm{\cdot}_{\textrm F}\right)\le\left(1+2\cdot\frac{\min\left\{\sqrt{a},\sqrt{b}\right\}\lambda}{\iota}\right)^{ab}.
	\end{equation*}
\end{lemma}

\subsection{Proof}\label{s2.3}
This subsection presents the complete proof of Theorem \ref{thm: main}. We first give a proof sketch.

\paragraph{Proof sketch} We provide a sketch of proof of Theorem~\ref{thm: main} for readers' convenience. First by the property of our algorithm and telescoping, we can get 
\begin{equation}\label{eq: sketch-1}
	\sup_{f_T \in H_f}\left(F\left(\hat{\bm{\rho}}^{(T)}\right) - F\left(\bm{\rho}^*\right)\right)^2 \le 4 \sup_{f_T \in H_f}\norm{F - f_T}_{\infty}^2. 
\end{equation}
\eqref{eq: sketch-1} means that when function $f_T$ can fit the target function $F$ very well, the universal convergence can be guaranteed. By Assumption~\ref{asp: infiity-2-relation}, we can rewrite \eqref{eq: sketch-1} as 
\begin{equation}\label{eq: sketch-2}
	\sup_{f_T \in H_f}\left(F\left(\hat{\bm{\rho}}^{(T)}\right) - F\left(\bm{\rho}^*\right)\right)^2 \le 4C(T) \sup_{f_T \in H_f} \frac{ \sum_{t=1}^T \EE_{\bm{\rho} \sim D_t} (F(\bm{\rho})-f_T(\bm{\rho}))^2}{T}.
\end{equation}
Then we can employ the standard argument of Rademacher Complexity to bound the RHS of~\eqref{eq: sketch-2} and then obtain
\begin{align}\label{eq: phi-bound-by-erm-sketch}
	\sup_{f_T \in H_f}& \frac{ \sum_{t=1}^T \EE_{\bm{\rho} \sim D_t} (F(\bm{\rho})-f_T(\bm{\rho}))^2}{T} \nonumber\\
	& \le 2 \hat{\mathfrak{R}}_{K_T}(\cH_M) +12B^2 \sqrt{\frac{2\log\frac{2}{\delta}}{mT}} + \sup_{f_T \in H_f}\frac{1}{mT}\sum_{i=1}^{mT}\left(F\left(\bm{\rho}^{(i)}\right) - f_T\left(\bm{\rho}^{(i)}\right)\right),
\end{align}
where function class $\cH_M = \{(f_T(\bm{\rho})-F(\bm{\rho}))^2~|~f_T \in \cH_f\}$ and $\sup_{f_T \in H_f}\frac{1}{mT}\sum_{i=1}^{mT}\left(F\left(\bm{\rho}^{(i)}\right) - f_T\left(\bm{\rho}^{(i)}\right)\right)$ can be viewed as the supreme of the training error ($\epsilon$ by our assumption, can be arbitrarily small). Then utilizing Lemma~\ref{lem: dudley}, we have
\begin{align}\label{eq: dudley-sketch}
	\hat{\mathfrak{R}}_{K_T}(\cH_M) \le \frac{4\alpha}{\sqrt{mT}} + \frac{48B^2}{\sqrt{mT}} \sqrt{\log \cN\left(\cH_M,\alpha,\norm{\cdot}_{\infty}\right)},
\end{align}
where $\cN$ denotes the covering number. Then through investigating the Lipschitz property of $f_T$ w.r.t to its parameter set, employing the argument of volume ratio (Lemma~\ref{lem: volume-ratio}) and setting $\alpha$ as $\frac{1}{\sqrt{mT}}$, we can bound the covering number by 
\begin{equation}\label{eq: covering-number-bound-sketch}
	\cN\left(\cH_M,\frac{1}{\sqrt{mT}},\norm{\cdot}_{\infty}\right) \le d^2D \log \left(1+8BDB_W^D\sqrt{mTd}\right).
\end{equation}
Finally combining~\eqref{eq: sketch-2}, \eqref{eq: phi-bound-by-erm-sketch}, \eqref{eq: dudley-sketch} and \eqref{eq: covering-number-bound-sketch}, we get the desired universal convergence result.

Before showing the full proof, we introduce an auxiliary lemma here.
\begin{lemma}\label{lem: indep}
	Under Assumptions \ref{asp: data}, we have
	\begin{enumerate}
		\item [1)] the whole generated data points $\left\{\bm{\rho}^{(i)}~\big|~i=1,2,\cdots,mT\right\}$ are mutually independent. 
		\item [2)] for any $t$, $\left\{\bm{\rho}^{(j_t)}~\big|~j_t=mt-m+1, \cdots, mt \right\}$ are i.i.d.. 
	\end{enumerate}
\end{lemma}
Lemma \ref{lem: indep} is a straightforward result of Assumption~\ref{asp: data}.

Now we can approach the proof of Theorem \ref{thm: main}.
\begin{proof}
	We first bound term $\sup_{f_T \in H_f}(F(\hat{\bm{\rho}}^{(T)}) - F(\bm{\rho}^*))^2$ by telescoping:
	\begin{align}\label{eq: first}
		\sup_{f_T \in H_f}(F(\hat{\bm{\rho}}^{(T)}) &- F(\bm{\rho}^*))^2 \nonumber\\
		&\mathop{\le}^{\rm (i)} \sup_{f_T \in H_f}(F(\hat{\bm{\rho}}^{(T)}) - f_T(\hat{\bm{\rho}}^{(T)}) + f_T(\bm{\rho}^*) -F(\bm{\rho}^*))^2 \nonumber\\
		&\mathop{\le}^{\rm (ii)} \sup_{f_T \in H_f}2\{[F(\hat{\bm{\rho}}^{(T)}) - f_T(\hat{\bm{\rho}}^{(T)})]^2 + [f_T(\bm{\rho}^*) -F(\bm{\rho}^*)]^2 \}\nonumber\\
		&\le 4 \sup_{f_T \in H_f}\norm{F - f_T}_{\infty}^2 \nonumber\\
		&\mathop{=}^{\rm (iii)} 4C(T) \sup_{f_T \in H_f} \frac{ \sum_{t=1}^T \EE_{\bm{\rho} \sim D_t} (F(\bm{\rho})-f_T(\bm{\rho}))^2}{T}.
	\end{align}
	Here (i) comes from Eq.\ \eqref{eq: argmin-em}, (ii) uses the fact that for any real number $x$ and $y$, we have $(x+y)^2 \le 2(x^2 + y^2)$. (iii) arises from Assumption \ref{asp: infiity-2-relation}.
	
	For notational simplicity we further denote 
	\begin{equation}\label{eq: def-Phi}
		\Phi(K_T) =\sup_{f_T \in \cH_f} \left[\EE_{D_{1:T}}(F-f_T)^2 - \hat{\EE}_{K_T} (F - f_T)^2\right],
	\end{equation}
	where $\hat{\EE}_{K_T} (F - f_T)^2 = \frac{1}{mT}\sum_{i=1}^{mT}(F(\bm{\rho}^{(i)}) - f_T(\bm{\rho}^{(i)}))$ corresponds to the empirical MSE loss when training our neural network.
	
	Suppose $K_T'$ and $K_T$ are two samples which are different only in the $k$-th point, namely $K_T=\{\bm{\rho}^{(1)},...,\bm{\rho}^{(k)},...,\bm{\rho}^{(mT)}\}$ and $K_T'=\{\bm{\rho}^{(1)},...,\bm{\rho}^{(k)}{'},...,\bm{\rho}^{(mT)}\}$, we have
	\begin{align*}
		|\Phi (K_T') - \Phi (K_T)| &\le \sup_{f_T \in \cH_f} \left|\hat{\EE}_{K_T} (F - f_T)^2 - \hat{\EE}_{K_T'} (F - f_T)^2\right| \\
		&=\sup_{f_T \in \cH_f} \left|\frac{(F(\bm{\rho}^{(k)}) - f_T(\bm{\rho}^{(k)}))^2}{mT} - \frac{(F(\bm{\rho}^{(k)}{'}) - f_T(\bm{\rho}^{(k)}{'}))^2}{mT}\right| \\
		& \le \frac{8B^2}{mT},
	\end{align*}
	then by Mcdiarmid's Inequality (\eq{eq: mcdiarmid-1} in Lemma \ref{lem:McD_ineq}), we get
	\begin{align}\label{eq:mcdiarmid}
		\PP(\Phi(K_T) - \EE_{K_T}\Phi(K_T) \ge s) &\le \exp \left(\frac{-2s^2}{mT \cdot (\frac{8B^2}{mT})^2}\right) \nonumber\\
		& = \exp \left(\frac{-mTs^2}{32B^4}\right).
	\end{align}
	Given any $\delta>0$, by setting the right hand side of \eqref{eq:mcdiarmid} to be $\frac{\delta}{2}$, we have with probability at least $1-\frac{\delta}{2}$,
	\begin{equation}\label{eq: bound-by-rm}
		\Phi(K_T) \le \EE_{K_T} \Phi(K_T) + 4B^2\sqrt{\frac{2\log\frac{2}{\delta}}{mT}}.
	\end{equation}
	Notice that
	\begin{align}\label{eq: symmetry-1}
		\EE_{K_T} \Phi(K_T) &=\EE_{K_T} \left\{\sup_{f_T \in \cH_f} \left[\EE_{D_{1:T}}(F-f_T)^2 - \hat{\EE}_{K_T} (F - f_T)^2\right]\right\} \nonumber\\
		&=\EE_{K_T} \left\{\sup_{f_T \in \cH_f}  \EE_{K_T'}\left[\hat{\EE}_{K_T'}(F-f_T)^2 - \hat{\EE}_{K_T} (F - f_T)^2\right]\right\}.
	\end{align}
	Here the second equality in Eq.\ \eqref{eq: symmetry-1} is because:
	\begin{align*}
		\EE_{K_T'}\left[\hat{\EE}_{K_T'}(F-f_T)^2\right] =& \frac{1}{mT}\sum_{i=1}^{mT} \EE_{K_T'} \left[F\left(\bm{\rho}^{(i)}\right) - f_T(\bm{\rho}^{(i)})\right]^2 \\
		\mathop{=}^{\rm (i)}& \frac{1}{mT} \left\{\sum_{i=1}^{m} \EE_{\bm{\rho}^{(i)} \sim D_1} \left[F(\bm{\rho}^{(i)}) - f_T(\bm{\rho}^{(i)})\right]^2 \right.\\
		&\left.+ \sum_{i=m+1}^{2m} \EE_{\bm{\rho}^{(i)} \sim D_2} \left[F(\bm{\rho}^{(i)}) - f_T(\bm{\rho}^{(i)})\right]^2 + \cdots \right. \\
		&\left.+\sum_{i=mT-T+1}^{mT} \EE_{\bm{\rho}^{(i)} \sim D_T} \left[F(\bm{\rho}^{(i)}) - f_T(\bm{\rho}^{(i)})\right]^2\right\} \\
		\mathop{=}^{\rm (ii)}& \frac{1}{mT} \left[m\EE_{D_1} (F - f_T)^2 + m\EE_{D_2} (F - f_T)^2 + \cdots +m \EE_{D_T} (F - f_T)^2\right] \\
		=& \EE_{D_{1:T}}(F-f_T)^2 ,
	\end{align*}
	where (i) results from 1) of Lemma \ref{lem: indep} and (ii) comes from 2) of Lemma \ref{lem: indep}.
	
	Further we have
	\begin{align}\label{eq: symmetry-2}
		&\EE_{K_T} \left\{\sup_{f_T \in \cH_f}  \EE_{K_T'}\left[\hat{\EE}_{K_T'}(F-f_T)^2 - \hat{\EE}_{K_T} (F - f_T)^2\right]\right\} \nonumber\\
		\mathop{\le}^{\rm (i)}& \EE_{K_T, K_T'} \sup_{f_T \in \cH_f} \left[\hat{\EE}_{K_T'}(F-f_T)^2 - \hat{\EE}_{K_T} (F - f_T)^2\right] \nonumber\\
		= &\EE_{K_T, K_T'} \sup_{f_T \in \cH_f} \frac{1}{mT} \sum_{i=1}^{mT} \left[(F(\bm{\rho}^{(i)\prime})-f_T(\bm{\rho}^{(i)\prime}))^2 - (F(\bm{\rho}^{(i)})-f_T(\bm{\rho}^{(i)}))^2\right] \nonumber\\
		\mathop{=}^{\rm (ii)} & \EE_{{\bf \sigma},K_T, K_T'} \sup_{f_T \in \cH_f} \frac{1}{mT} \sum_{i=1}^{mT} \sigma_i\left[(F(\bm{\rho}^{(i)\prime})-f_T(\bm{\rho}^{(i)\prime}))^2 - (F(\bm{\rho}^{(i)})-f_T(\bm{\rho}^{(i)}))^2\right] \nonumber\\
		\mathop{\le}^{\rm (iii)} &\EE_{{\bf \sigma},K_T'} \sup_{f_T \in \cH_f} \frac{1}{mT} \sum_{i=1}^{mT} \left[\sigma_i(F(\bm{\rho}^{(i)}{'})-f_T(\bm{\rho}^{(i)}{'}))^2 \right] + \EE_{{\bf \sigma},K_T} \sup_{f_T \in \cH_f} \frac{1}{mT} \sum_{i=1}^{mT} \left[- \sigma_i(F(\bm{\rho}^{(i)})-f_T(\bm{\rho}^{(i)}))^2\right] \nonumber\\
		= &2\EE_{{\bf \sigma},K_T} \sup_{f_T \in \cH_f} \frac{1}{mT} \sum_{i=1}^{mT} \left[\sigma_i(F(\bm{\rho}^{(i)})-f_T(\bm{\rho}^{(i)}))^2\right],
	\end{align}
	where $\sigma_i$ are Rademacher variables (Definition \ref{def:rad}), which are uniformly distributed independent random variables taking values in $\{-1,+1\}$.
	Here (i) and (iii) hold due to the sub-additivity of the supremum function (considering the convexity of supremum function, by Jensen's Inequality, we have for any function $h$, $\sup \int_x h(x)\le\int_x \sup h(x)$ holds). (ii) combines the definition of Rademacher variable $\sigma_i$ and the fact that the expectation is taken over both $K_T$ and $K_{T'}$.
	
	For notational simplity, given any function $f_T\in \cH_f$, we define the non-negative loss function $M(f_T):\bm{\rho} \rightarrow (f_T(\bm{\rho})-F(\bm{\rho}))^2$ and its function class $\cH_{M}=\{M(f_T): f_T \in \cH_f\}$.
	
	Then  combining \eqref{eq: symmetry-1} and \eqref{eq: symmetry-2} we obtain 
	\begin{equation}\label{eq: sym}
		\EE_{K_T} \Phi (K_T) \le 2 \mathfrak{R}_{mT}(\cH_M),
	\end{equation}
	where $\mathfrak{R}_{mT}(\cH_M) = \EE_{\sigma,K_T} \sup_{f_T \in \cH_f} \frac{1}{mT} \sum_{i=1}^{mT} \sigma_i (F(\bm{\rho}^{(i)}) - f_T(\bm{\rho}^{(i)}))^2$ is the Rademacher Complexity (Definition \ref{def:rad}) of $\cH_M$. 
	
	Now, we define  the Empirical Rademacher Complexity of $\cH_M$ as 
	\begin{equation*}
		\hat{\mathfrak{R}}_{K_T}(\cH_M) := \EE_{\sigma} \sup_{f_T \in \cH_f} \frac{1}{mT} \sum_{i=1}^{mT} \sigma_i (F(\bm{\rho}^{(i)}) - f_T(\bm{\rho}^{(i)}))^2.
	\end{equation*}
	Again, suppose $K_T'$ and $K_T$ are two samples which are different only in the $k$-th point, namely $K_T=\{\bm{\rho}^{(1)},...,\bm{\rho}^{(k)},...,\bm{\rho}^{(mT)}\}$ and
	$K_T'=\{\bm{\rho}^{(1)},...,\bm{\rho}^{(k)}{'},...,\bm{\rho}^{(mT)}\}$, we have
	\begin{align*}
		&|\hat{\mathfrak{R}}_{K_T}(\cH_M) - \hat{\mathfrak{R}}_{K_T'}(\cH_M)|\\ 
		& = \left|\EE_{\sigma}\sup_{f_T \in \cH_f} \frac{1}{mT}\sum_{i=1}^{mT} \sigma_i (F(\bm{\rho}^{(i)}) - f_T(\bm{\rho}^{(i)}))^2  - \EE_{\sigma}\sup_{f_T \in \cH_f} \frac{1}{mT}\sum_{i=1}^{mT} \sigma_i (F(\bm{\rho}^{(i)}{'}) - f_T(\bm{\rho}^{(i)}{'}))^2\right| \\
		&\le \sup_{f_T \in \cH_f} \left|\frac{(F(\bm{\rho}^{(k)}) - f_T(\bm{\rho}^{(k)}))^2}{mT} - \frac{(F(\bm{\rho}^{(k)}{'}) - f_T(\bm{\rho}^{(k)}{'}))^2}{mT}\right| \\
		& \le \frac{8B^2}{mT},
	\end{align*}
	then by Mcdiarmid's Inequality (\eq{eq: mcdiarmid-2} in Lemma \ref{lem:McD_ineq}), we get
	\begin{align}\label{eq:mcdiarmid-again}
		\PP(\hat{\mathfrak{R}}_{K_T}(\cH_M) - \mathfrak{R}_{mT}(\cH_M) \le -s) &\le \exp \left(\frac{-2s^2}{mT \cdot (\frac{8B^2}{mT})^2}\right) \nonumber\\
		& =\exp\left(\frac{-mTs^2}{32B^4}\right). 
	\end{align}
	Given any $\delta>0$, by setting the right handside of \eq{eq:mcdiarmid-again} to be $\frac{\delta}{2}$, we have with probability at least $1-\frac{\delta}{2}$,
	\begin{equation}\label{eq: bound-by-erm}
		\mathfrak{R}_{mT}(\cH_M) \le \hat{\mathfrak{R}}_{K_T}(\cH_M) + 4B^2\sqrt{\frac{2\log\frac{2}{\delta}}{mT}}.
	\end{equation}

	Now combining \eqref{eq: bound-by-rm}, \eqref{eq: sym} and \eqref{eq: bound-by-erm}, we get with probability at least $1-\delta$,
	\begin{equation}\label{eq: phi-bound-by-erm}
		\Phi(K_T) \le 2 \hat{\mathfrak{R}}_{K_T}(\cH_M) +12B^2 \sqrt{\frac{2\log\frac{2}{\delta}}{mT}},
	\end{equation}
	here we use the fact that \eqref{eq: bound-by-rm} and \eqref{eq: bound-by-erm} hold with probability $1-\frac{\delta}{2}$ respectively and that $(1-\frac{\delta}{2})^2>1-\delta$.
	
	It is straightforward that $\norm{M(f_T)}_{\infty}\le 4B^2$, then Dudley's Entropy (Lemma \ref{lem: dudley}) gives us 
	\begin{align}\label{eq: dudley}
		\hat{\mathfrak{R}}_{K_T}(\cH_M) &\le \frac{4\alpha}{\sqrt{mT}} + \frac{12}{mT} \int_{\alpha}^{4B^2\sqrt{mT}} \sqrt{\log \cN(\cH_M, \iota, \norm{\cdot}_{\infty})} d\iota \nonumber\\
		&\le \frac{4\alpha}{\sqrt{mT}} + \frac{48B^2}{\sqrt{mT}} \sqrt{\log \cN(\cH_M,\alpha,\norm{\cdot}_{\infty})},
	\end{align}
	where $\cN$ denotes the covering number. We pick $\alpha=\frac{1}{\sqrt{mT}}$, and combine \eqref{eq: first}, \eqref{eq: phi-bound-by-erm} and \eqref{eq: dudley} to get
	\begin{align}\label{eq: last-but-two}
		&\sup_{f_T \in H_f}(F(\hat{\bm{\rho}}^{(T)}) - F(\bm{\rho}^*))^2 \nonumber\\ 
		&\le 4C(T)\left(\frac{96B^2}{\sqrt{mT}}\sqrt{\log \cN\left(\cH_M, \frac{1}{\sqrt{mT}}, \norm{\cdot}_{\infty}\right)} + 12B^2 \sqrt{\frac{2\log \frac{2}{\delta}}{mT}} +\frac{8}{mT}+ \hat{\EE}_{K_T} (F - f_T)^2\right).
	\end{align}
	Next we need to compute the covering number $\cN(\cH_M, \frac{1}{\sqrt{mT}}, \norm{\cdot}_{\infty})$. To derive a tight covering number we investigate the Lispchitz continuity of $f_T$ with respect to the weight matrices $\mathbf{W}_1, \cdots, \mathbf{W}_D$. Consider two neural networks $f_T(\bm{\rho}) = \mathbf{W}_D^\top\sigma(\mathbf{W}_{D-1}\sigma(...\sigma(\mathbf{W}_{1}\bm{\rho})))$ and $f_T'(\bm{\rho})=\mathbf{W}_D'^\top\sigma(\mathbf{W}_{D-1}'\sigma(...\sigma(\mathbf{W}_{1}'\bm{\rho})))$ with different sets of weight matrices, we first notice that 
	\begin{align*}
		\norm{M(f_T)-M(f_T')}_{\infty} &= \sup_{\bm{\rho}}\left|(f_T(\bm{\rho}) - F(\bm{\rho}))^2 - (f_T'(\bm{\rho}) - F(\bm{\rho}))^2\right|\\
		&=\sup_{\bm{\rho}}\left|(f_T(\bm{\rho})+f_T'(\bm{\rho})-2F(\bm{\rho}))(f_T(\bm{\rho}) - f_T'(\bm{\rho}))\right|\\
		&\le 4B \norm{f_T - f_T'}_{\infty}.
	\end{align*}
	Next we  get the bound based on weight matrices. 
	Specifically, given two different sets of matrices $\mathbf{W}_1,\cdots,\mathbf{W}_D$ and $\mathbf{W}'_1,\cdots,\mathbf{W}'_D$, we have
	\begin{align*}
		&\norm{f_T - f_T'}_{\infty}\\
		\le & \norm{ \mathbf{W}_D^\top\sigma(\mathbf{W}_{D-1}\sigma(...\sigma(\mathbf{W}_{1}\bm{\rho})...)) -  (\mathbf{W}'_D)^\top\sigma(\mathbf{W}'_{D-1}\sigma(...\sigma(\mathbf{W}'_{1}\bm{\rho})...))}_2\\
		\le & \norm{ \mathbf{W}_D^\top\sigma(\mathbf{W}_{D-1}\sigma(...\sigma(\mathbf{W}_{1}\bm{\rho})...)) -  (\mathbf{W}'_D)^\top\sigma(\mathbf{W}_{D-1}\sigma(...\sigma(\mathbf{W}_{1}\bm{\rho})...))}_2 \\
		&+ \norm{ (\mathbf{W}'_D)^\top\sigma(\mathbf{W}_{D-1}\sigma(...\sigma(\mathbf{W}_{1}\bm{\rho})...)) -  (\mathbf{W}'_D)^\top\sigma(\mathbf{W}'_{D-1}\sigma(...\sigma(\mathbf{W}'_{1}\bm{\rho})...))}_2 \\
		\le & \norm{\mathbf{W}_D-\mathbf{W}'_D}_2\norm{\sigma(\mathbf{W}_{D-1}\sigma(...\sigma(\mathbf{W}_{1}\bm{\rho})...))}_2 \\
		&+ \norm{\mathbf{W}'_D}_2\norm{\sigma(\mathbf{W}_{D-1}\sigma(...\sigma(\mathbf{W}_{1}\bm{\rho})...)) - \sigma(\mathbf{W}'_{D-1}\sigma(...\sigma(\mathbf{W}'_{1}\bm{\rho})...))}_2.
	\end{align*}
	Note that we have
	\begin{align*}
		\norm{\sigma(\mathbf{W}_{D-1}\sigma(...\sigma(\mathbf{W}_{1}\bm{\rho})...))}_2&\mathop{\le}^{\rm (i)} \norm{\mathbf{W}_{D-1}\sigma(...\sigma(\mathbf{W}_{1}\bm{\rho})...)}_2\\
		&\hspace{-1in}\le \norm{\mathbf{W}_{D-1}}_2\norm{\sigma(...\sigma(\mathbf{W}_{1}\bm{\rho})...)}_2\mathop{\le}^{\rm (ii)}  B_W^{D-1}\norm{\bm{\rho}}_2\mathop{\le}^{\rm (iii)} B_W^{D-1} ,
	\end{align*}
	where (i) comes from the definition of the ReLU activation, (ii) comes from $\norm{\mathbf{W}_k}_2\leq B_W$ and recursion, and (iii) comes from the boundedness of $\bm{\rho}$. Accordingly, we have
	\begin{align}
		\norm{M(f_T) - M(f_T')}_{\infty} &\le
		4B\norm{f_T(\bm{\rho}) - f_T'(\bm{\rho})}_{\infty} \nonumber\\
		&\le 4B( B_W^{D-1}\norm{\mathbf{W}_D-\mathbf{W}'_D}_2 + \norm{\mathbf{W}'_D}_2\norm{\sigma(\mathbf{W}_{D-1}\sigma(...)) - \sigma(\mathbf{W}'_{D-1}\sigma(...))}_2) \nonumber\\
		& \mathop{\le}^{\rm (i)} 4B( B_W^{D-1}\norm{\mathbf{W}_D-\mathbf{W}'_D}_2 + B_W\norm{\mathbf{W}_{D-1}\sigma(...) - \mathbf{W}'_{D-1}\sigma(...)}_2) \nonumber\\
		& \mathop{\le}^{\rm (ii)} 4BB_W^{D-1}\sum\limits_{k=1}^{D}\norm{\mathbf{W}_k - \mathbf{W}_k'}_2, \label{eq: M-W-lip}
	\end{align}
	where (i) comes from the fact that $\forall \mathbf{A}_1, \mathbf{A}_2 \in \RR^{a\times b}, \norm{\sigma( \mathbf{A}_1)-\sigma( \mathbf{A}_2)}_2\leq\norm{ \mathbf{A}_1- \mathbf{A}_2}_2$, and (ii) comes from the recursion. Considering the fact that each $M(f_T)$ corresponds to its parameter set $\mathbf{W}_1, \cdots, \mathbf{W}_D$, we can then derive the covering number of $\cH_M$ by the Cartesian product of the matrix covering of $\mathbf{W}_1,...,\mathbf{W}_D$:
	\begin{align}
		\cN(\cH_M, \|\cdot\|_{\infty}, \iota) &\mathop{\le}^{\rm (i)} \prod_{k=1}^D \cN\left(\mathbf{W}_k,\frac{\iota }{4BB_W^{D-1}D},\|\cdot\|_2\right) \nonumber\\
		&\mathop{\le}^{\rm (ii)} \prod_{k=1}^D \cN\left(\mathbf{W}_k,\frac{\iota }{4BB_W^{D-1}D},\|\cdot\|_{\textrm F}\right)\nonumber\\
		&\mathop{\le}^{\rm (iii)}\left( 1+ \frac{8BB_W^DD \sqrt{d}}{\iota}\right)^{d^2D}.\label{eq: covering-nn}
	\end{align}
	Here (i) utilizes the fact that if  $\forall k=1,2,\cdots,D$, matrix set $$\left\{\mathbf{V}_{k,j_k}\in\RR^{d_{k-1}\times d_k}~\bigg|~j_k=1,2,\cdots,\cN\left(\mathbf{W}_k,\frac{\iota }{4BB_W^{D-1}D},\|\cdot\|_2\right)\right\}$$ is a $\frac{\iota }{4BB_W^{D-1}D}-$covering of set $\{\mathbf{W}_k~|~\norm{\mathbf{W}_k}_2\le B_W\}$, then by \eqref{eq: M-W-lip} we have function set 
	$$\left\{ \mathbf{V}_{D,j_D}^\top\sigma(\mathbf{V}_{D-1,j_{D-1}}\sigma(...\sigma(\mathbf{V}_{1,j_1}\bm{\rho})...))~\bigg|~1\le j_k\le \cN\left(\mathbf{W}_k,\frac{\iota }{4BB_W^{D-1}D},\|\cdot\|_2\right), \forall 1\le k\le D\right\}$$ 
	is an $\iota-$covering of $\cH_M$. (ii) comes from the fact that for any matrix $W$ we have $\norm{W}_2\le\norm{W}_{\textrm F}$, and (iii) employs Lemma \ref{lem: volume-ratio}. Plugging  \eqref{eq: covering-nn} into \eqref{eq: last-but-two}, we get 
	\begin{align}\label{eq: last-but-one}
		&\sup_{f_T \in H_f}(F(\hat{\bm{\rho}}^{(T)}) - F(\bm{\rho}^*))^2 \nonumber\\ \le& 4C(T)\left(\frac{96B^2}{\sqrt{mT}}\sqrt{d^2D \log \left(1+8BDB_W^D\sqrt{mTd}\right)} + 12B^2 \sqrt{\frac{2\log \frac{2}{\delta}}{mT}} +\frac{8}{mT}+ \hat{\EE}_{K_T} (F - f_T)^2\right) .
	\end{align}
	Since we consider the empirical MSE training loss to be less than $\epsilon$, i.e.,
	\begin{equation}\label{eq: mse}
		\hat{\EE}_{K_T} (F - f_T)^2 \le \epsilon,
	\end{equation}
	so by plugging \eqref{eq: mse} into \eqref{eq: last-but-one}, we get the desired result.
\end{proof}

\end{document}